\documentclass[aps,rmp,a4paper,twocolumn,showpacs]{revtex4}

\usepackage{hyperref}
\usepackage{dcolumn}   % needed for some tables
\usepackage{amsmath}
\usepackage{graphicx}
\usepackage{bm}
\usepackage{multirow}
\usepackage[caption=false]{subfig}

\def\simge{\mathrel{%
   \rlap{\raise 0.511ex \hbox{$>$}}{\lower 0.511ex \hbox{$\sim$}}}}
\def\simle{\mathrel{
   \rlap{\raise 0.511ex \hbox{$<$}}{\lower 0.511ex \hbox{$\sim$}}}}
\def\pythia{{\sc pythia}}
\def\alpgen{{\sc alpgen}}
\def\hOOO{\hphantom{000}}
\def\metvec{\mathbin{\vec{E\mkern - 11mu/}_T}}
\def\mptvec{\mathbin{\vec{p\mkern - 8mu/}_T}}
\def\mstat{\mathrm{(stat)}}
\def\mstatJ{\mathrm{(stat+JES)}}
\def\msys{\mathrm{(syst)}}
\def\theo{\mathrm{(theo)}}
\def\dilep{\ell\ell}
\def\ljets{\ell+\mathrm{jets}}
\def\mjets{\MET+\mathrm{jets}}

\newcommand {\fb} {~\rm{fb}^{-1}}

\newcommand{\pt}{p_T}
\newcommand{\Et}{E_T}

\newcommand {\GeVc} {{\rm GeV/c}}

\newcommand {\GeV} {{\rm GeV/c}^2}

\newcommand{\MET}{\mbox{$E\kern-0.60em\raise0.10ex\hbox{/}_{T}$}}
\newcommand{\TET}{\mbox{$T\kern-0.60em\raise0.10ex\hbox{/}_{T}$}}
\newcommand{\HET}{\mbox{$H\kern-0.60em\raise0.10ex\hbox{/}_{T}$}}

\newcommand{\W}{W}
\newcommand{\Z}{Z}
\newcommand{\WW}{\W\W}
\newcommand{\WZ}{\W\Z}
\newcommand{\ZZ}{\Z\Z}
\newcommand{\WZdecay}{WZ \rightarrow \ell^{\prime}\nu_{\ell^{\prime}} \ell\ell}
\newcommand{\HWW}{H\rightarrow WW^{(\ast)}}
\newcommand{\HZZ}{H\rightarrow ZZ^{(\ast)}}

\newcommand{\ppbar}{p \overline{p}}
\newcommand{\eebar}{e^{+} e^{-}}
\newcommand{\ttbar}{t \overline{t}}
\newcommand{\qqbar}{q \overline{q}}
\newcommand{\bbbar}{bb}
\newcommand{\ccbar}{c \overline{c}}

\newcommand{\Dzero} {D0}

\newcommand{\intL}{\int {\cal L}~dt}
\newcommand{\METrel}{\mbox{$E\kern-0.60em\raise0.10ex\hbox{/}_{T,rel}$}}
\newcommand{\MetDeltaPhi}{\Delta\phi_{E\kern-0.50em\raise0.08ex\hbox{\scriptsize /}_{T},(\ell{\rm,jet)}}}

\begin{document}

\bibpunct{[}{]}{;}{n}{,}{,}

%%%%%%%%%%%%%%%%%%%%%%%%%%%%%%%%%%%%%%%%%%%%%%%%%%%%%%%%%%%%%%%%%%%%%%%%%%%%
%
% Front matter
%
%%%%%%%%%%%%%%%%%%%%%%%%%%%%%%%%%%%%%%%%%%%%%%%%%%%%%%%%%%%%%%%%%%%%%%%%%%%%

\title{Tests of the Standard Electroweak Model at the Energy Frontier}
\author{John D. Hobbs
  \footnote{Electronic address: John.Hobbs@stonybrook.edu}}
\affiliation{Department of Physics and Astronomy, State University of
New York, Stony Brook, New York 11794, USA}

\author{Mark S. Neubauer
\footnote{Electronic address: msn@illinois.edu} }
\author{Scott Willenbrock
\footnote{Electronic address: willen@illinois.edu}}
\affiliation{Department of Physics, University of Illinois at
Urbana-Champaign, Urbana, Illinois, 61801, USA}

%%%%%%%%%%%%%%%%%%%%%%%%%%%%%%%%%%%%%%%%%%%%%%%%%%%%%%%%%%%%%%%%%%%%%%%%%%%%
%
% Abstract
%
%%%%%%%%%%%%%%%%%%%%%%%%%%%%%%%%%%%%%%%%%%%%%%%%%%%%%%%%%%%%%%%%%%%%%%%%%%%%

\begin{abstract}
In this review, we summarize tests of standard electroweak (EW) theory  at
the highest available energies as a precursor to the Large Hadron
Collider (LHC) era. Our primary focus is on the published results (as
of March 2010) from proton-antiproton collisions at $\sqrt{s}=1.96$
TeV at the Fermilab Tevatron collected using the CDF and D0
detectors. This review is very timely since the LHC scientific program
is nearly underway with the first high-energy ($\sqrt{s}=7$ TeV)
collisions about to begin. After presenting an overview of the EW
sector of the standard model, we provide a summary of current
experimental tests of EW theory. These include gauge boson properties
and self-couplings, tests of EW physics from top quark sector, and
searches for the Higgs boson.
\end{abstract}

%%%%%%%%%%%%%%%%%%%%%%%%%%%%%%%%%%%%%%%%%%%%%%%%%%%%%%%%%%%%%%%%%%%%%%%%%%%%
%
% PACS
%
%%%%%%%%%%%%%%%%%%%%%%%%%%%%%%%%%%%%%%%%%%%%%%%%%%%%%%%%%%%%%%%%%%%%%%%%%%%%

%12.15.Ji   Applications of electroweak models to specific processes 
%13.85.Rm   Limits on production of particles
%14.70.Fm   W bosons 
%14.70.Hp   Z bosons 
%14.80.Bn   Standard-model Higgs bosons 

\maketitle
\tableofcontents

%%%%%%%%%%%%%%%%%%%%%%%%%%%%%%%%%%%%%%%%%%%%%%%%%%%%%%%%%%%%%%%%%%%%%%%%%%%%
%
% Introduction
%
%%%%%%%%%%%%%%%%%%%%%%%%%%%%%%%%%%%%%%%%%%%%%%%%%%%%%%%%%%%%%%%%%%%%%%%%%%%%

\section{INTRODUCTION}
\label{sec:intro}

The goal of particle physics is to explain the nature of the Universe
at its most fundamental level, including the basic constituents of
matter and their interactions. The standard model of particle
physics (SM) is a quantum field theory based on the $SU(3)_{\rm C} \otimes
SU(2)_{\rm L} \otimes U(1)_{Y}$ gauge symmetry group which describes
the strong, weak, and electromagnetic interactions among fundamental
particles. This theory has been the focus of intense scrutiny by
experimental physicists,  most notably at high energy particle
colliders, over the last three decades. It has been demonstrated to
accurately describe fundamental particles and their interactions up to 
$O(100)$ GeV, with the existence of non-zero neutrino masses and
mixing being the only known exception. Despite the success of the SM,
there are many reasons to believe that the SM is an effective theory which is
only valid up to $\sim$1 TeV.  Some of the shortcomings of the SM will
be described in Section~\ref{sec:overview}. 

Particle physics is embarking on a unique, and possibly defining,
period in its history with the start of particle collisions at the
Large Hadron Collider (LHC) at CERN. At the LHC, bunches of protons
will be collided with a planned 14 TeV of center-of-mass energy,
creating conditions that existed only a tiny fraction of a second
after the big bang. This is seven times the center of mass energy of
collisions at the Fermilab Tevatron. For the first time, physicists
will be able to directly probe the TeV energy scale in the laboratory,
where new physics beyond the SM, with the potential to revolutionize our
understanding of the Universe, could be apparent. We can only
speculate about what form this will take. Is is a tantalizing prospect
that on the horizon is a revolution in our understanding of the
Universe that includes a more complete theory of particles and their
interactions which may explain dark matter and dark energy.

The purpose of this review is to present tests of the electroweak (EW)
sector of the SM ($SU(2)_{\rm L} \otimes U(1)_{Y}$) at the highest
available energies as a precursor to the LHC era. Our focus is on
published results from collider data collected using the CDF
\cite{Aaltonen:2007ps} and D0 \cite{Dzero} detectors during Run II at
the Fermilab Tevatron as it relates to our understanding of
electroweak interactions and spontaneous symmetry breaking. After an
overview of electroweak theory (Section~\ref{sec:overview}), we
present current results on gauge boson properties and self-couplings
(Section~\ref{sec:gaugeBosons}), tests of electroweak physics from top
quark physics (Section~\ref{sec:top}), and searches for the Higgs
boson (Section~\ref{sec:higgs}).

%%%%%%%%%%%%%%%%%%%%%%%%%%%%%%%%%%%%%%%%%%%%%%%%%%%%%%%%%%%%%%%%%%%%%%%%%%%%
%
% Overview
%
%%%%%%%%%%%%%%%%%%%%%%%%%%%%%%%%%%%%%%%%%%%%%%%%%%%%%%%%%%%%%%%%%%%%%%%%%%%%

\section{OVERVIEW}
\label{sec:overview}

The SM is an extremely successful theory of the strong, weak, and
electromagnetic interactions.  It is based on three generations of
quarks and leptons, interacting via an $SU(3)_{\rm C} \otimes
SU(2)_{\rm L} \otimes U(1)_{Y}$ gauge symmetry. The $SU(2)_{\rm L}
\otimes U(1)_{Y}$ symmetry is spontaneously broken to
electromagnetism, $U(1)_{EM}$, by the vacuum-expectation value of the
Higgs field. Given this field content and gauge symmetry, the most
general theory that follows from writing down every term of dimension
four or less is the SM.

In the SM, as usually understood, neutrinos are exactly massless, and
do not mix.  Since neutrino mixing has been definitively observed, we
must go beyond the SM in order to describe this phenomenon.  There are
two ways to do this. One way is to extend the field content of the model
by adding additional fermion and/or Higgs fields (e.g. a right-handed
neutrino or a Higgs triplet). The other way is to extend the SM by
adding operators of dimensionality greater than four. There is only
one operator of dimension five allowed by the gauge symmetries
\cite{Weinberg:1979sa},
\begin{equation}
{\cal L}_5 =\frac{c^{ij}}{\Lambda}(L^{iT}\epsilon\phi) C (\phi^T\epsilon L^j) +
h.c.
\end{equation}
where $L^i$ is the lepton doublet field of the $i^{th}$ generation and
$\phi$ is the Higgs doublet field (the $2\times 2$ matrix $\epsilon$
and the $4\times 4$ charge-conjugation matrix $C$ are present to
ensure invariance under $SU(2)_{\rm L}$ and Lorentz transformations, 
respectively). When the Higgs doublet acquires a vacuum-expectation
value, $\langle\phi\rangle = (0,v/\sqrt 2)$ ($v=246$ GeV),  this term
gives rise to a (Majorana) mass for neutrinos,
\begin{equation}
{\cal L}_5 =-\frac{1}{2} M_\nu^{ij}\nu^{iT} C  \nu^j +
h.c.
\end{equation}
where $M_\nu^{ij}=c^{ij}v^2/\Lambda$ is the neutrino mass matrix.  Due to
the tiny inferred masses of neutrinos, the scale $\Lambda$ lies around
$10^{15}$ GeV, assuming $c^{ij}$ is not much less than order unity.

There are other indications that the SM is not a complete description
of nature, most of them related to gravitation and cosmology. Even
with massive neutrinos included, the SM particles only constitute
4.6\% of the present universe, with the remainder in mysterious dark
matter (23\%) and dark energy (72\%).  Neither dark matter and nor
dark energy are accommodated in the SM.  There is no adequate
mechanism for baryogenesis (the observed excess of baryons over
antibaryons) or inflation, which is the simplest explanation of the
observed temperature fluctuations in the cosmic microwave background.
The SM also provides no explanation of the strong CP problem: the lack
of observed CP violation in the strong interaction, which is allowed
by the SM.

If physics beyond the SM lies at an energy scale less than 1 TeV, then
we should be able to observe it directly at high-energy colliders. If
it lies at a scale greater than 1 TeV, then we can parametrize its
effects via higher-dimension operators, suppressed by inverse powers
of the scale of new physics, $\Lambda$, exactly as in the case of
neutrino masses described above.  Other than the dimension-five
operator responsible for neutrino masses, the lowest-dimension
operators are of dimension six, and are therefore suppressed by two
inverse powers of $\Lambda$.  If $\Lambda$ is of order $10^{15}$ GeV,
as suggested by neutrino masses, then these operators are so
suppressed that they are unobservable, with the possible exception of
baryon-number violating operators that mediate nucleon decay.  However,
there could be more than one scale of new physics, and if this scale
is not much greater than 1 TeV, its effects could be observable via
dimension six operators. Operators of dimension greater than six are
suppressed by even more inverse powers of $\Lambda$ and can be
neglected.

There are many dimension-six operators allowed by the SM gauge
symmetry \cite{Buchmuller:1985jz}.  There are three ways to detect the
presence of these operators.  The first is to observe phenomena that
are absolutely forbidden (or extremely suppressed) in the SM, such as
nucleon decay. The second is to make measurements with such great
precision that the small effects of the dimension-six operators
manifest themselves. The third is to do experiments at such high
energy, $E$, that the effects of these operators, of order
$(E/\Lambda)^2$, become large. If $E>\Lambda$ then one must abandon
this formalism, because operators of arbitrarily high dimensionality
become significant; however, the new physics should then be directly
observable. If no effects beyond the SM are observed, then one can
place bounds on the coefficients of the dimension-six operators,
$c/\Lambda^2$, where $c$ is a dimensionless number. These bounds apply
only to the product $c/\Lambda^2$, not to $c$ and $\Lambda^2$
separately; in fact, there could even be two different scales of new
physics involved ($\Lambda_1\Lambda_2$ in place of $\Lambda^2$).

This approach to physics beyond the SM, dubbed an
effective-field-theory approach \cite{Weinberg:1979sa}, has the
advantage of being model independent. Whatever new physics lies at the
scale $\Lambda$, it will induce dimension-six operators, whose only
dependence on the new physics lies in their coefficients,
$c/\Lambda^2$.  Another advantage of this approach is that it is
universal; it can be applied both to tree-level and loop-level
processes, and any ultraviolet divergences that appear in loop
processes can be absorbed into the coefficients of the operators.
Thus one need not make any {\it ad hoc} assumptions about how the
ultraviolet divergences are cut off. This effective-field-theory
approach thus provides an excellent framework to parametrize physics
beyond the SM \cite{DeRujula:1991se,Hagiwara:1993ck}. 

Hadron colliders contribute to the study of the electroweak
interactions in three distinct ways.  Firstly, because they operate at
the energy frontier, hadron colliders are uniquely suited to searching
for the effects of dimension six operators that are suppressed by a
factor of $(E/\Lambda)^2$.  Secondly, they are able to contribute to
the precision measurement of a variety of electroweak processes, most 
notably to the measurement of the $W$ boson mass and the top quark
mass.  Thirdly, they are able to search for new particles associated
with the electroweak interactions, in particular the Higgs boson.
These three virtues of hadron colliders will manifest themselves
throughout this review. 

\subsection{Electroweak Interactions}
\label{sec:overview_gsw}

The electroweak theory is a spontaneously broken gauge theory based on
the gauge group $SU(2)_{\rm L} \otimes U(1)_{Y}$.  There are three
parameters that describe the theory: the gauge couplings $g$ and $g'$,
and the order parameter of spontaneous symmetry breaking, $v$. In the
SM, this order parameter is the vacuum expectation value of a
fundamental Higgs field.  These parameters are not measured directly,
but rather inferred from precision electroweak measurements.  The
three measurements that are used to fix these parameters are the Fermi
constant $G_F$ determined from the muon lifetime formula; the fine
structure constant $\alpha$, determined from a variety of low-energy
experiments; and the $Z$ boson mass $M_Z$.  With these three inputs,
the predictions of all other electroweak processes can be calculated,
at least at tree level.

The level of precision of electroweak measurements is such that a
tree-level analysis is insufficient, and one must go to at least one
loop. At this level, one finds that predictions depend also on the top
quark mass and the Higgs boson mass, since these particles appear in
loops. In fact, a range for the top quark mass was correctly predicted
by precision electroweak data before the top quark was discovered, and
the measured mass falls into this range.  We are now following the
same tack with the Higgs boson.  Remarkably, the precision electroweak
data imply that the Higgs boson mass is not far above the experimental
lower bound of $m_H>114~\GeV$, which means that it may be accessible at
the Tevatron as well as the LHC.

The electroweak interaction has many other parameters as well.  Along
with the top quark mass, there are the masses of all the other quarks
and leptons, as well as the elements of the Cabbibo-Kobayashi-Maskawa
(CKM) quark-mixing matrix and the Maki-Nakagawa-Sakata (MNS)
lepton-mixing matrix.  Most of these mixing parameters are not
measured at the energy frontier, with one exception: the CKM element
$V_{tb}$ that describes the coupling of a $W$ boson to a top and
bottom quark. The only direct measurement of this parameter comes from
electroweak production of the top quark at hadron colliders via a
process known as single-top production, discussed in
Section~\ref{sec:top_singleTop}.

\subsection{Electroweak Symmetry Breaking}
\label{sec:overview_higgs}

The strong and electroweak forces are gauge theories, based on the
groups $SU(3)_C$ and $U(1)_{EM}$, respectively. The associated gauge
bosons, the gluon and the photon, are massless as a consequence of the 
gauge symmetry.  We know that the interactions of electroweak bosons
with fermions as well as with themselves are also governed by a gauge
theory, with gauge group $SU(2)_{\rm L} \otimes U(1)_{Y}$. Why, then,
are the electroweak bosons, $W^\pm$ and $Z$, not massless, as would be
expected of gauge bosons? In the SM, the answer is that the
electroweak symmetry is spontaneously broken, and that the electroweak
gauge bosons acquire mass through the Higgs mechanism. This is the
most plausible explanation of why the interactions appear to be those
of a gauge theory, despite the fact that the gauge bosons are not
massless. However, this argument leaves completely open the question
of how (and why) the electroweak symmetry is broken.

The simplest model of electroweak symmetry breaking (EWSB), which is
also the original proposal, is based on a fundamental scalar field
that is an electroweak doublet carrying hypercharge $Y=1/2$.  The
potential for this scalar field is chosen such that its minimum is at
nonzero field strength.  This breaks the electroweak symmetry to
$U(1)_{EM}$, as desired.  This simple model, which can be criticized
on several grounds, has withstood the test of time.  It predicts that
there is a scalar particle, dubbed the Higgs boson, of unknown mass
but with definite couplings to other particles.  The discovery of this
Higgs particle is one of the driving ambitions of particle physicists,
and was a primary motivation for the LHC.

As mentioned in the previous section, this simple model is consistent
with precision electroweak data with a Higgs particle close to the
present experimental lower bound of $m_H>114~\GeV$. This consistency
does not rule out more exotic possibilities, however, such as two (or
more) Higgs doublets, Higgs singlets and triplets, composite Higgs
bosons, and other alternative models of electroweak symmetry
breaking.

%%%%%%%%%%%%%%%%%%%%%%%%%%%%%%%%%%%%%%%%%%%%%%%%%%%%%%%%%%%%%%%%%%%%%%%%%%%%
%
% Electroweak Gauge Bosons
%
%%%%%%%%%%%%%%%%%%%%%%%%%%%%%%%%%%%%%%%%%%%%%%%%%%%%%%%%%%%%%%%%%%%%%%%%%%%%

\section{ELECTROWEAK GAUGE BOSONS}
\label{sec:gaugeBosons}

In the SM, the $\W$ and $\Z$ bosons mediate the weak force and acquire
mass through the Higgs mechanism, as described in
Section~\ref{sec:overview}. The $\W$ boson was discovered in 1983 in
$\ppbar$ collisions at the CERN SPS by the UA1 and UA2 experiments
\cite{Arnison:1983rp,Banner:1983jy}, with discovery of the $\Z$ boson
soon to follow~\cite{z-disco-ua1,z-disco-ua2}. The discovery of these
gauge bosons at CERN represents a dramatic validation of
Glashow-Salam-Weinberg (GSW) theory which predicted the existence of
neutral currents mediated by a new gauge boson, the $\Z$ boson, and
predicted the $\W$ bosons to describe nuclear $\beta$-decay
and. Together with the massless photon, these comprise the gauge
bosons of the electroweak interaction. High precision studies of the
$Z$ boson properties made by the LEP collaborations and the
SLD~collaboration~\cite{Z-Pole} using $e^+e^-$ collisions have
provided stringent tests of electroweak theory.

The $\W$ and $\Z$ bosons are copiously produced in $\ppbar$ collisions
at the Fermilab Tevatron due to their large production cross sections
at $\sqrt{s}=1.96$ TeV and the high integrated luminosity data sets
available from the CDF and D0 experiments during Run II. Detailed
measurement of the $\W$ and $\Z$ properties at the Tevatron is not
only important to further test GSW theory and the EWSB mechanism in
the SM but also to search for new physics beyond the SM using the
highest energy collisions currently available. We summarize the
current Tevatron measurements of $\W$ and $\Z$ properties in
Section~\ref{sec:gaugeBosonsIntro}
through~\ref{sec:gaugeBosons_Z_Afb}.

The production of heavy vector boson pairs ($\WW$, $\WZ$, and $\ZZ$)
is far less common than inclusive $\W$ and $\Z$ production. While a
$\W$ boson is produced in every 3 million $p\bar{p}$ collisions and a
$\Z$ boson in every 10 million, the production of a $WW$ pair is a
once in 6 billion event, $\WZ$ a once in 20 billion event, and $\ZZ$ a
once in 60 billion event! Diboson production is sensitive to the
triple gauge couplings (TGCs) between the bosons themselves via an
intermediate virtual boson. The boson TGCs are an important
consequence of the non-Abelian nature of the SM electroweak gauge
symmetry group. At the highest accessible energies available at the
Fermilab Tevatron, diboson production provides a sensitive probe of
new physics, including anomalous trilinear gauge couplings, new
resonances such as the Higgs boson, and large extra dimensions
\cite{Kober:2007bc}. Recent results on diboson production from the
Tevatron are discussed in Section~\ref{sec:gaugeBosons_dibosons}.

\subsection{Heavy Gauge Boson Production}
\label{sec:gaugeBosonsIntro}
\begin{figure}
\unitlength=0.33\linewidth
\includegraphics[width=0.20\textwidth]{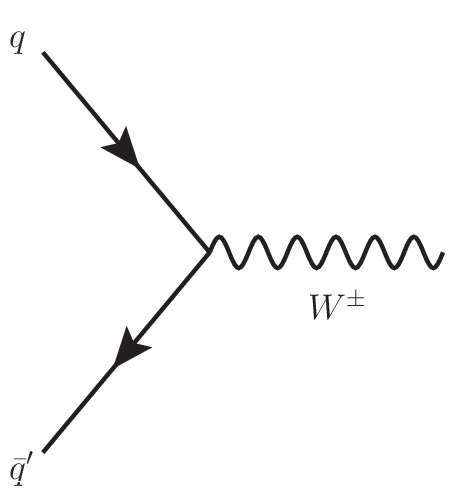}
\put(-0.65,0.0){(a)} 
\unitlength=0.33\linewidth
\includegraphics[width=0.20\textwidth]{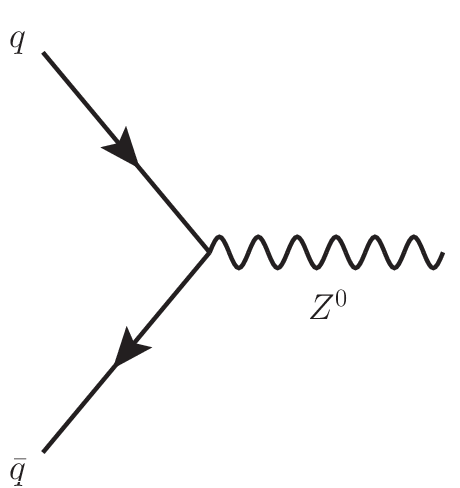}
\put(-0.65,0.0){(b)} 
\caption{Leading order (a) $\W$ and (b) $\Z$ boson production via
  quark-antiquark annihilation. The production is dominated by $q = u,
  d$ valence quarks. The $\gamma^*$ contribution interferes with the
  $\Z$ diagram but is not shown.}
\label{fig:gaugeBosons_feyn_w_z}
\end{figure}

In high energy proton-antiproton collisions at the Fermilab Tevatron,
heavy vector bosons ($V \equiv W, Z$) are produced at tree-level via
quark-antiquark annihilation ($q\bar{q} \rightarrow V$) as shown in
Fig.~\ref{fig:gaugeBosons_feyn_w_z}. At high transverse
momentum\footnote{Throughout this paper, ``transverse'' is taken to be
  in a plane perpendicular to the beam directions, and unless
  otherwise noted, quantities with a ``T'' subscript, e.g $E_T$, are
  values projected onto this plane. Also, charge conjugation is
  assumed throughout.}
, the leading-order QCD subprocesses are $q\bar{q} \rightarrow Vg$ and
$qg \rightarrow Vq$. The production properties of heavy gauge bosons
provide tests of perturbative QCD and, under certain circumstances,
information about quark and gluon momentum distributions within the
proton and antiproton.

The heavy vector boson production cross section in a $p\overline{p}$
collision is given by
\begin{equation*}
  \sigma(V) = \int \sigma_{0,V} f(x_q) \overline{f}(x_{\overline{q}}) dx_q
  dx_{\overline{q}}
\end{equation*}
in which $\sigma_{0,V}$ is the cross section for production of a
vector boson by a quark, antiquark pair with Feynman $x$ values of
$x_q$ and $x_{\overline{q}}$ respectively, and $f(x_q)$ and
$\overline{f}(x_{\overline{q}})$ are the parton distribution functions
for the proton and antiproton.  As defined here, contributions from
pure $\gamma*$ and $Z/\gamma*$ interference terms are not included but
are accounted for in comparisons with theory in the measurements we
describe.

The remainder of this section includes a summary of current $\W$ and
$\Z$ cross section measurements that the Tevatron and their
comparisons with theory. As previously mentioned, these measurements of
$\W$ and $\Z$ boson production cross sections primarily a test QCD
rather than electroweak theory. We include them here since detailed
study of heavy gauge boson states are central to tests of electroweak
theory and, therefore, it is important consider how well their
production in $\ppbar$ collisions is understood.

The cross section times branching fraction of $W$ and $Z$ bosons is
measured in the fully-leptonic decay channels $W\rightarrow \ell\nu$
and $Z\rightarrow \ell\ell$, where $\ell \equiv e,\mu,\tau$. While the
final states involving $\tau$-leptons are important for many reasons, we
restrict ourselves to the $e$ and $\mu$ final states for the cross 
section discussion since these give the highest precision
measurements. The fully-leptonic decay channels are chosen over
hadronic channels for these measurements, since the latter suffer from
large backgrounds to due to the hadronic decay of jets produced by QCD
processes.

\subsubsection{$W$ and $Z$ Cross Sections}
\label{sec:gaugeBosons_xsect}

Using a next-to-next to leading order (NNLO) prediction~\cite{b-Z-xsec-theo} 
calculated with the MRST2004 NNLO parton distribution
function~\cite{b-Z-mrst} and the SM branching fractions for the $W$
and $Z$ bosons into fully-leptonic final states, the cross-section
times branching fractions are calculated to be
$$
\sigma(\ppbar\rightarrow W)\times B(W\to\ell\nu) = 2687\pm54
  \ \mathrm{pb}.
$$
and
$$
\sigma(\ppbar\rightarrow Z)\times B(Z\to\ell\ell) = 251.9^{+5.0}_{-11.8}
  \ \mathrm{pb}.
$$
The uncertainties are a combination of the MRST uncertainties and the
difference between the central value above and that computed using the
CTEQ6.1M parton distribution functions~\cite{b-Z-cteq}.

The $W$ and $Z$ cross sections times branching fractions to
fully-leptonic final states have been measured by CDF
\cite{Abulencia:2005ix} using $\intL = 72$ pb$^{-1}$. The results for
electron and muon channels combined are
\begin{eqnarray*}
  \lefteqn{\sigma(\ppbar\rightarrow W)\times B(W\to\ell\nu) = }  \\ 
  & & {} 2749
  \pm 10 \rm{(stat.)}
  \pm 53 \rm{(syst.)}
  \pm 165 \rm{(lumi.)}
  \ \mathrm{pb}
\end{eqnarray*}
and
\begin{eqnarray*}
  \lefteqn{\sigma(\ppbar\rightarrow Z)\times B(Z\to\ell\ell) = } \\
  & & {} 254.9 
  \pm 3.3 \rm{(stat.)} 
  \pm 4.6 \rm{(syst.)} 
  \pm 15.2 \rm{(lumi.)}
  \ \mathrm{pb}.~~~~~~`
\end{eqnarray*}

The measurements of the $Z$ cross section include additional
contributions from $\gamma*$ and $Z/\gamma*$ interference which give
events that are experimentally indistinguishable from the $Z$
process. The size of these contributions depends on the $Z/\gamma*$
mass range considered. For the range $60\ \mathrm{GeV/c}^2 \le
M(Z/\gamma*) < 130\ \mathrm{GeV/c}^2$ these contributions increase the
cross section by a factor $1.019 \pm 0.001$ relative to the $Z$-only 
cross section, and for the mass range $66\ \mathrm{GeV/c}^2 \le
M(Z/\gamma*) < 116\ \mathrm{GeV/c^2}$, the cross section is increased by a
factor $1.004 \pm 0.001$.

A precision measurement of the ratio $R$ of $W$ and $Z$ cross section times
branching fraction given by
$$
R = \frac{\sigma(\ppbar\rightarrow W)\times
  B(W\to\ell\nu)}{\sigma(\ppbar\rightarrow Z)\times B(Z\to\ell\ell)}
$$
can be used to test the SM. For example, new high mass resonances
decaying to either $W$ or $Z$ bosons could lead to a deviation of the
measured value of $R$ from the SM expectation.

Important systematic uncertainties such as the integrated luminosity
uncertainty cancel in the measurement of $R$. The ratio $R$ has been
measured by CDF \cite{Abulencia:2005ix} using $\intL = 72$ pb$^{-1}$
to be 
$$
R = 10.84 \pm 0.15 {\rm (stat.)} \pm 0.14 {\rm (syst.)}
$$
This measurement has a precision of 1.9\% and is consistent with SM
expectation at NNLO of 10.69 $\pm$ 0.08 \cite{b-Z-xsec-theo}.

A summary of the results are shown in Tab.~\ref{t-zxsec}. 
\begin{table*}
  \begin{ruledtabular}
  \begin{tabular}{cccccc}
                & Integrated    & Data  & Predicted    & & \\ 
    Channel     & Luminosity    & Yield & Background   &
    $A\times\epsilon$   & Measured $\sigma\times$ Br (pb) \\ \hline \\

    $W\to e\nu$    & 72 pb$^{-1}$  & 37584  &  $1762\pm300$   &
    $0.1795^{+0.0034}_{-0.0038}$ & $2.771 \pm 0.014 ^{+0.062}_{-0.056}
    \pm 0.166$~nb \\ 

    $W\to \mu\nu$    & 72 pb$^{-1}$  & 31722  &  $3469\pm151$   &
    $0.1442^{+0.0031}_{-0.0034}$ & $2.722 \pm 0.015 ^{+0.066}_{-0.061}
    \pm 0.163$~nb \\

    $e\nu$ and $\mu\nu$    & ---  & ---  &  ---   &
    --- & $2.749 \pm 0.010 \pm 0.053 \pm 0.165$~nb \\ \\
 
    $Z\to ee$    & 72 pb$^{-1}$  & 4242  &  $62\pm18$   &
    $0.2269^{+0.0047}_{-0.0048}$ & $255.8 \pm 3.9 ^{+5.5}_{-5.4}
    \pm 15.3$~pb \\
    
    $Z\to\mu\mu$    & 72 pb$^{-1}$  & 1785  &  $13\pm13$   &
    $0.0992^{+0.0028}_{-0.0031}$ & $248.0 \pm 5.9 ^{+8.0}_{-7.2} \pm
    14.8$~pb \\ 

    $ee$ and $\mu\mu$&     ---       &  ---  &    ---       &     ---
    & $254.9 \pm 3.3 \pm 4.6 \pm 15.2$~pb  \\ \hline
  \end{tabular}
  \end{ruledtabular}
  \caption{The measured integrated luminosity, yield, background, signal
     acceptance times efficiency and resulting $\sigma\times\mathrm{Br}$ for
     each of the four $Z$ cross section times branching ratio analyses
     by the CDF Collaboration in \cite{Abulencia:2005ix}. The
     combination of the dielectron and dimuon channels using the BLUE
     method is also shown. The $Z/\gamma*$ results have been corrected
     to $Z$ only as described in the text.\label{t-zxsec}}
\end{table*}
The individual results are in good agreement with each other and with
the prediction from theory.

Even with the moderate sample sizes of these measurements, the $W$ and
$Z$ cross section results are limited by the systematic uncertainty
from the luminosity measurements. Because of this, further improvement
in these measurements is not anticipated. CDF and D0 Collaborations
continue to use $Z$-boson production to measure experimental
efficiencies acceptances and cross-checks on temporal or instantaneous
luminosity dependence of detector response. The updated cross section
results could be reinterpreted as measurements of the integrated
luminosity.

\subsection{$\W$ Boson Mass}
\label{sec:gaugeBosons_Wmass}

\begin{figure}
  \includegraphics[width=0.6\linewidth]{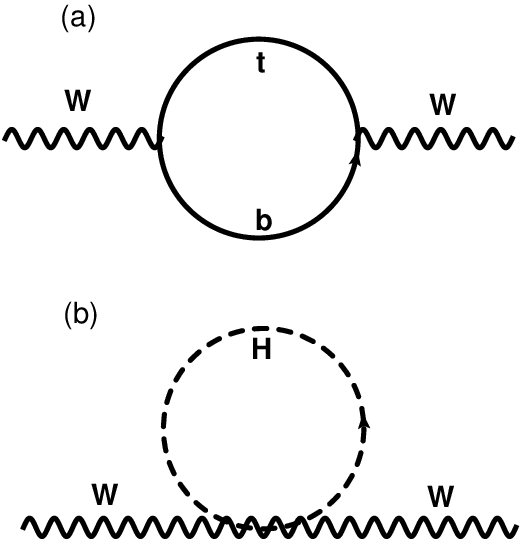}
  \caption{Feynman diagrams showing corrections to the $W$ boson mass
    from (a) the top quark and (b) the SM Higgs boson. \label{f-mwcorr}}
\end{figure}

\begin{figure}
  \includegraphics[width=\linewidth]{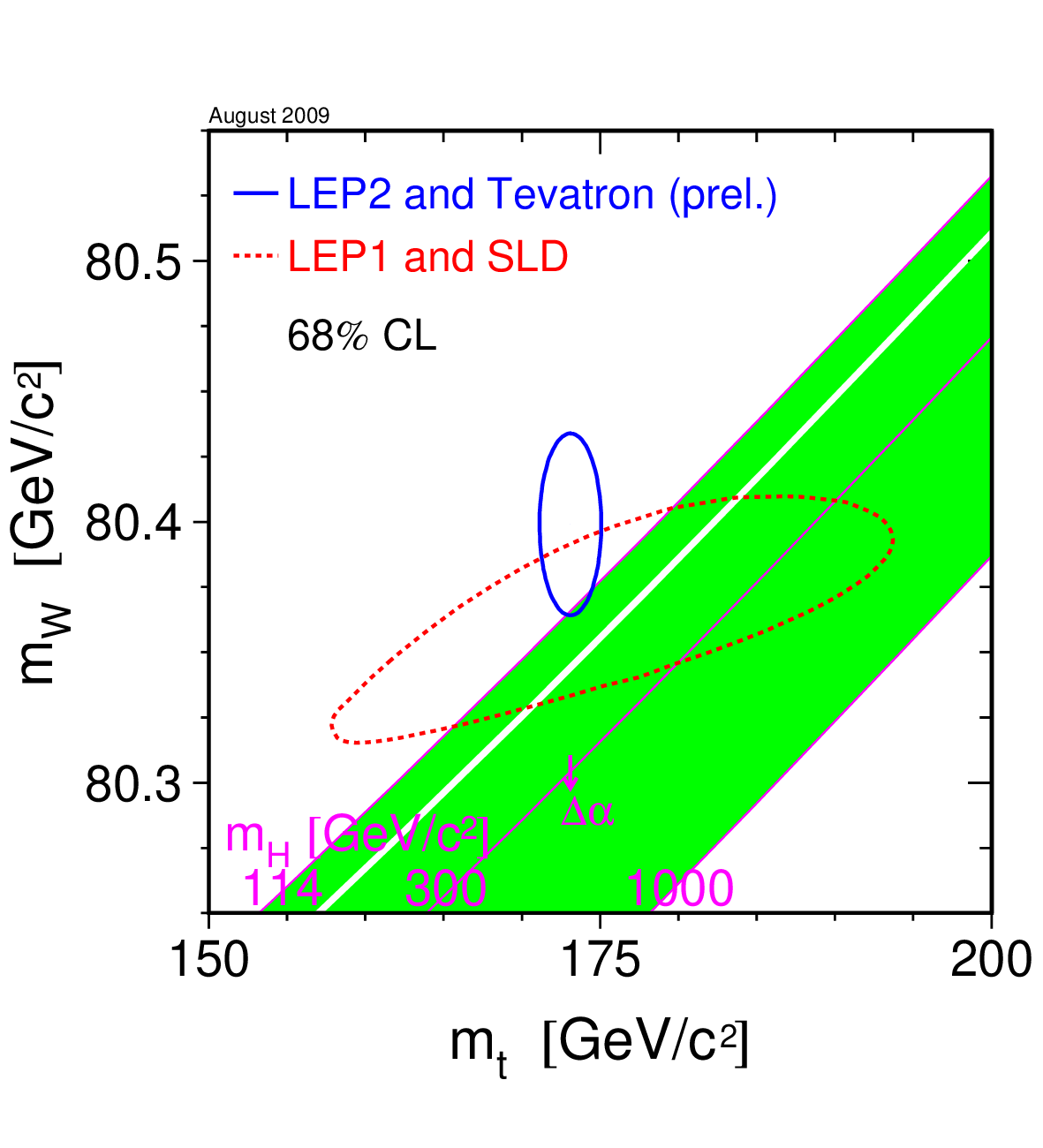}
  \caption{The measured top quark and $W$ boson masses and a band of 
    allowed Higgs boson masses.\cite{Alcaraz:2009jr}  This includes
    the recent results summarized in the text.\label{f-mwmt}}
\end{figure}

At tree level, the $W$ boson mass is fully-determined by the
electromagnetic fine structure constant, the weak Fermi coupling and
the cosine of the weak mixing angle.  When higher-order EW
corrections, like those shown in Fig.~\ref{f-mwcorr}, are included,
the expression is modified to~\cite{Sirlin:1980nh}
\begin{equation}
 M_W^2 = \frac{\hbar^3\pi\alpha_{EM}}{\sqrt{2}G_F}
   \frac{1}{(1-M_W^2/M_Z^2)(1-\Delta r)}
\end{equation}
in which $M_W$ (and later $\Gamma_W$) correspond to the parameters of a 
Breit-Wigner distribution with an $s$-dependent width.
The term $\Delta r$ includes the effects of radiative corrections and
depends on $M_t^2$ and $\log M_H$, where $M_t$ and $M_H$ are the top
quark and Higgs boson masses, respectively. Measurement of the $W$
boson mass can be used to constrain the allowed Higgs boson mass. The
precisions of the $W$ boson and top quark mass measurements are
currently the limiting factors in the indirect constraint on the Higgs
boson mass and are shown in Fig.~\ref{f-mwmt}. Earlier measurements of
the $W$ boson mass have been made by the LEP
experiments~\cite{Schael:2006mz,:2008xh,Achard:2005qy,Abbiendi:2005eq}
and CDF~\cite{b:CDF-mWI} and D0~\cite{b:D0-mWI} in Run I of the
Tevatron. 

Signal-to-background and resolution considerations dictate use of the
$W\to e\nu$ and $W\to\mu\nu$ decays modes for the $W$ boson mass
measurement at the Tevatron. The momentum component of the neutrino
along the beam direction cannot be inferred in $\ppbar$ collision
events, so the $W$ invariant mass cannot be reconstructed from its
decay products and other variables are used to determine the
mass. Three variables are used: (1) the lepton ($e$ or $\mu$)
transverse momentum $p_T^\ell$, (2) the (inferred) neutrino transverse
momentum $p_T^\nu$ and (3) the transverse mass, $m_T \equiv \sqrt{p_T^\ell
  p_T^\nu[1-\cos(\delta\phi)]}$ in which $\delta\phi$ is the
lepton-neutrino opening angle in the plane transverse to the
beam. Although these variables are highly correlated, their systematic
uncertainties are dominated by different sources, so the results are
combined taking their statistical and systematic correlations into account.

The event decay kinematics in the transverse plane are fully
characterized by two quantities: (1) the lepton transverse momentum
$\vec{p}_T^\ell$ and (2) the transverse momentum of the hadronic
recoil $\vec{u}_T$ required to balance the transverse momentum of the
the $W$. The hadronic recoil is defined as the vector sum of all
energy measured in the calorimeter excluding that deposited by the
lepton. From these two measurements, the neutrino transverse momentum
is inferred: $-\vec{p}_T^\nu = \vec{p}_T^\ell + \vec{u_T}$.  In
practice, the recoil is computed using the event transverse missing
energy ($\MET$) measured in the calorimeter after removing the
contribution to calorimeter energy associated with the lepton. The $W$
boson mass is determined by generating predicted distributions
(templates) of the three measurement variables for a range of input
$W$ boson mass hypotheses. These are generated using dedicated fast
Monte Carlo simulation programs, and the mass is determined by
performing a binned maximum likelihood fit of these templates to the
distributions observed in data. The Run II measurements use a
blinding technique in which an unknown offset is added to the value
returned from the fits until the analyses are finished.  At that
point, the offset is removed to reveal the true value. The results
from the different fit variables ($p_T^\ell,\ p_T^\nu,\ m_T$) and
final states ($e$ or $\mu$) are combined using the BLUE
algorithm~\cite{b-BLUE,b-BLUE2}.

Both D0 and CDF have reported mass measurements using Run II data. The
CDF result~\cite{Aaltonen:2007ps} uses $\intL = 0.2$~fb$^{-1}$ and
results are reported for $W\to e\nu$ and $W\to\mu\nu$ decays.  The D0
result~\cite{Abazov:2009cp} uses $\intL = 1.0$~fb$^{-1}$ and results
are reported for the $W\to e\nu$ decay mode only. Candidate events are
selected by requiring a single high-$p_T$ isolated charged lepton and
large $\MET$.  Tab.~\ref{t-mwYield} shows the kinematic selection
requirements, event yields and background fraction for the $W$ event
selections.
\begin{table}
  \begin{ruledtabular}
  \begin{tabular}{lccc}
              & \multicolumn{2}{c}{CDF}                        &  D0 \\ 
              & \hOOO$W\to e\nu$\hOOO & \hOOO$W\to\mu\nu$\hOOO & \hOOO$W\to e\nu$\hOOO \\ \hline
   $\intL$    &   0.2~fb$^{-1}$       &  0.2~fb$^{-1}$         & 1.0~fb$^{-1}$ \\ 
  $E_T^\ell$, cal  &   $>30$ GeV      &       ---              &  $>25$ GeV    \\
  $p_T^\ell$, trk  &   $>18\ \GeVc$   &  $>30\ \GeVc$          &  $>10\ \GeVc$ \\
 $|\eta_\ell|$&                       &                        &  $<1.05$      \\
     $\MET$   &   $>30\ \GeVc$        &  $>30\ \GeVc$          &  $>25\ \GeVc$ \\
      $u_T$   &   $<15\ \GeVc$        &  $<15\ \GeVc$          &  $<15\ \GeVc$ \\
      Yield   &    63964              &   51128                &  499830       \\
   Background &    7.5\%              &   1.1\%                &   4.0\%       \\
%    {\bf ??? Z ????} & & & \\
   \end{tabular}
   \end{ruledtabular}
   \caption{The integrated luminosity, kinematic selection, event
     yield and backgrounds for the CDF and D0 $W$ boson mass analyses.
     The background is given as a percentage of the total
     yield.\label{t-mwYield}}
\end{table}

The backgrounds include $Z\to\ell\ell$ events in which one lepton
escapes identification, $WW$ diboson events, and misidentification
backgrounds in which the lepton is either a jet misidentified as an
electron or a muon from semileptonic decay of hadrons in which the
rest of the associated hadronic jet is not reconstructed.  An
additional source of events are the sequential decays $W\to\tau\nu\to\
e\nu\nu\nu$ and $W\to\tau\nu\to\mu\nu\nu\nu$. The CDF analysis treats
these as signal while the D0 analysis incorporates these into the
background template distributions. 

The {\em in situ} calibration of charged particle momenta (CDF) and
calorimetric measurement of electron energy (CDF, D0) is of crucial
importance to this result. The CDF analysis uses a calibration of the
tracker momentum scale ($p$) determined from dimuon and dielectron
decays of $J/\psi$, $\Upsilon$ and $Z$ particles. This calibration is
then transferred to the calorimeter energy measurement ($E$) using the
$E/p$ ratio. A final improvement is made for the $W\to e\nu$ mode by
incorporating an additional calorimeter calibration based on $Z\to ee$
decays. The D0 analysis uses calorimeter energy measurements, and the
calibration is based on the mass reconstructed in $Z\to ee$ events and
a detailed simulation of the calorimeter response. For both
experiments, this calibration is the dominant source of systematic
uncertainty. Other sources of systematic uncertainty arise from
trigger efficiency, lepton identification efficiency, correlation
(in)efficiency such as occurs when the hadronic recoil is near the
charged lepton, backgrounds, electroweak and strong contributions to
the production and decay model and the parton distribution functions.

\begin{figure}
  \includegraphics[width=0.97\linewidth]{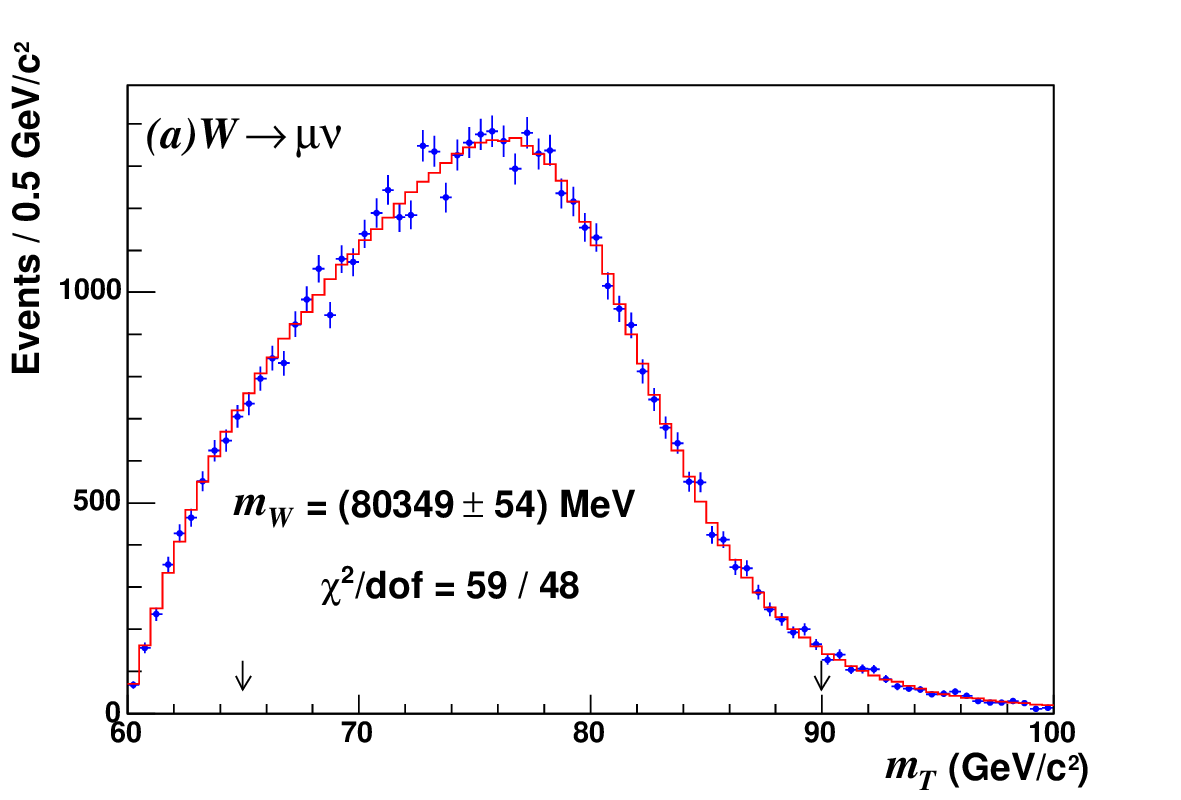}
  \includegraphics[width=0.97\linewidth]{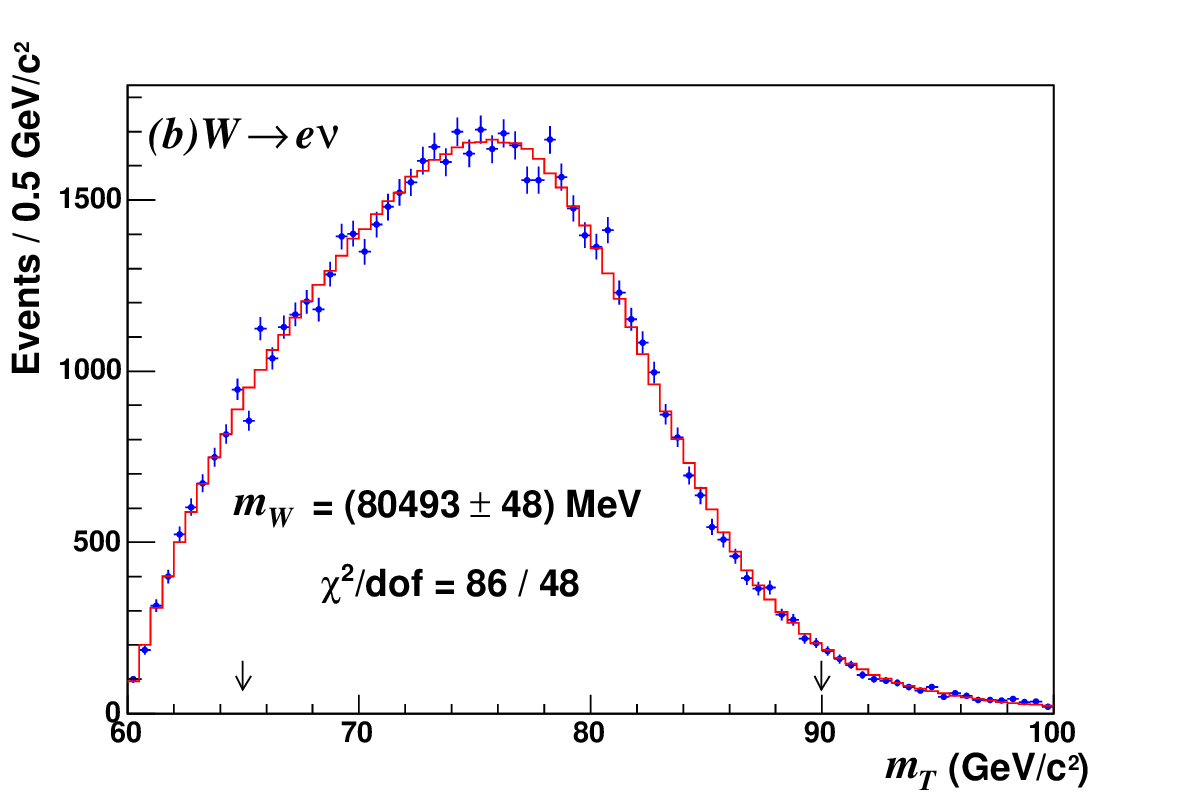}
  \includegraphics[width=0.93\linewidth]{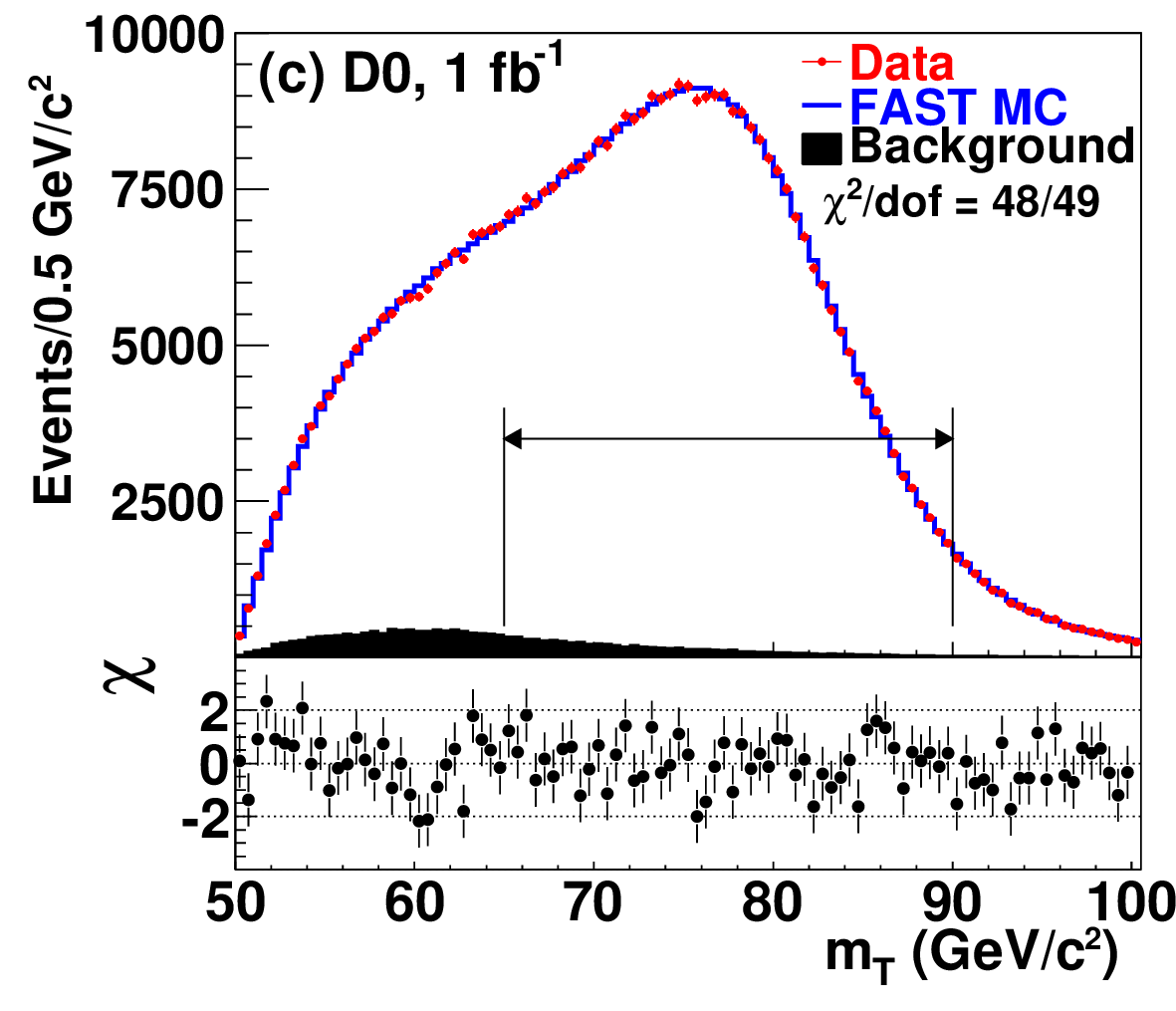}
  \caption{The $m_T$ distributions for (a) the CDF $W\to\mu\nu$ channel,
    (b) the CDF $W\to e\nu$ channel and (c) the D0 $W\to e\nu$ channel.
    \label{f-mwres}}
\end{figure}

\begin{figure}
  \includegraphics[width=\linewidth]{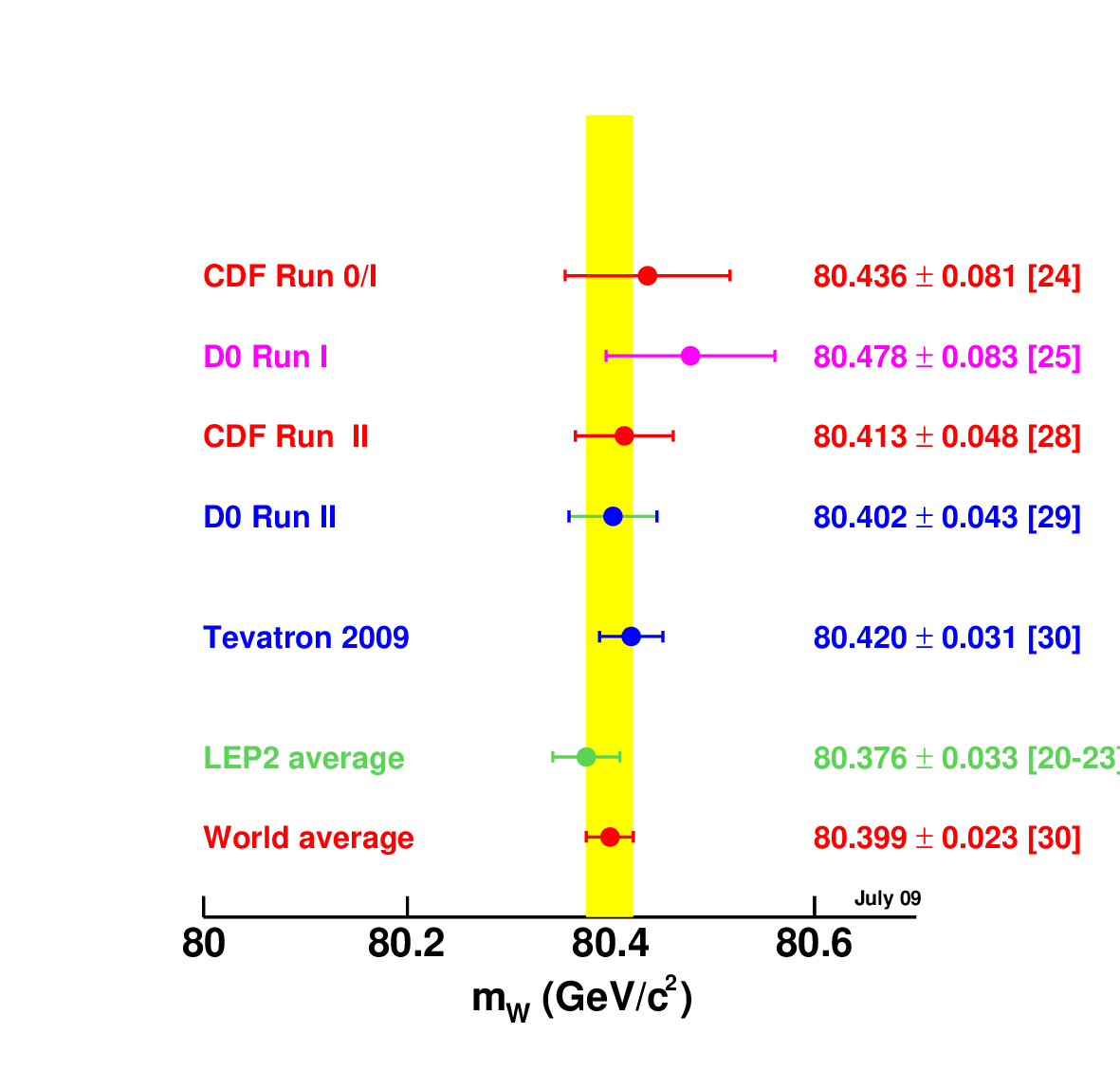}
  \caption{Ideogram of previous measurements, the Run II CDF and D0 
    measurements and the new world average. The Ref. is shown to the
    right of each measurement as [XX]. \label{f-mwwa}}
\end{figure}

The $m_T$ distributions for each channel are shown in
Fig.~\ref{f-mwres}, and the results from each channel and the
combinations of the channels for each experiment are shown in
Tab.~\ref{t-mw-answ}.  The systematic uncertainties are dominated by
the lepton energy calibration.  This contributes 17~MeV uncertainty to
the CDF $p_T^\ell$ and $p_T^\nu$ channels, 30~MeV to the CDF $m_T$
channel and 34~MeV to each D0 channel.
\begin{table}
\def\hWeC{\hphantom{CDF $W\to e\nu$,}}
\def\hWeD{\hphantom{D0 $W\to e\nu$,}}
\def\hWm{\hphantom{CDF $W\to\mu\nu$,}}
  \begin{ruledtabular}
   \begin{tabular}{lllc}
               & \hOOO\hOOO & \hOOO $M_W$~(MeV)\hOOO & $\chi^2/$dof \\ \hline
   CDF $W\to e\nu$& $m_T$   & $80493 \pm 48 \pm 39$  & 86/48 \\
    \hWeC         & $p_T^e$ & $80451 \pm 58 \pm 45$  & 63/62 \\
    \hWeC         & $\MET$  & $80473 \pm 57 \pm 54$  & 63/62 \\ \hline
    Combined     &          & $80477 \pm 62$         &       \\ \hline
    CDF $W\to\mu\nu$ & $m_T$& $80349 \pm 54 \pm 27$  & 59/48 \\
    \hWm         & $p_T^\mu$& $80321 \pm 66 \pm 40$  & 72/62 \\
    \hWm         &  $\MET$  & $80396 \pm 66 \pm 46$  & 44/63 \\ \hline
    Combined      &         & $80352 \pm 60$         &       \\ \hline
    CDF $W\to e\nu + \mu\nu$ & & $80413 \pm 34 \pm 34$  &  \\ \hline\hline
    D0 $W\to e\nu$ &   $m_T$& $80401 \pm 23 \pm 37$  & 48/49 \\
    \hWeD          & $p_T^e$& $80400 \pm 27 \pm 40$  & 39/31 \\
    \hWeD          &$\MET$  & $80402 \pm 23 \pm 43$  & 32/31 \\ \hline
    D0 Combined    &       & $80401 \pm 21 \pm 38$  &        \\
   \end{tabular}
   \caption{The individual CDF and D0 $W$ boson mass results and their
     combinations. When two uncertainties are given, the first is the
     statistical uncertainty and the second is the systematic
     uncertainty. \label{t-mw-answ}}
  \end{ruledtabular}
\end{table}
These results are shown along with previous measurements in
Fig.~\ref{f-mwwa}. The world average
combination~\cite{:2009nu,Alcaraz:2009jr} has been updated with these
measurements using the BLUE~\cite{b-BLUE,b-BLUE2} algorithm including
correlations. The result is 
$$
  M_W = 80.420 \pm 0.031\ \mathrm{GeV/c}^2.%,\ World\ Average}.
$$
Because the systematic uncertainties are dominated by the statistical
precision of the calibrations determined from control data samples,
the systematic uncertainty in future measurements is expected to
improve as the integrated luminosity increases.  The ultimate limiting
systematic is expected to be that introduced by the parton
distribution functions.  In the current results, this contributes an
uncertainty of 11~MeV to all channels with a 100\% correlation among
the channels.

\subsection{$\W$ Width}
\label{sec:gaugeBosons_W_width}

Like the $W$ boson mass, the width $\Gamma_W$ is also predicted by the
SM. It is given by
\begin{equation}
  \Gamma_W = [3 + 2N_C(1+\frac{\alpha_S}{\pi})]\frac{G_FM_W^3}{6\sqrt{2}\pi}(1+\delta)
\end{equation}
in which $N_C=3$ is the number of colors, $1+\alpha_S/\pi$ is the QCD
correction factor to first order, and $\delta =
2.1$\%~\cite{Rosner:1993rj} is an EW correction factor. Direct
measurements of $\Gamma_W$ were made by CDF~\cite{Affolder:2000mt} and 
D0~\cite{Abazov:2002xj} using Tevatron Run I data and
combined~\cite{Abazov:2002xj}.  Measurements were also made by the
experiments at LEP~\cite{Schael:2006mz,:2008xh,Achard:2005qy,Abbiendi:2005eq}.

Because the $W$ boson mass is distributed according to a Breit-Wigner,
there is a tail at large mass values.  The $W$ boson width result is
obtained using the $m_T$ distribution in a region where the shape and
event yield are dominated by events from the high mass region of the
Breit-Wigner with limited impact from detector resolution effects.  As
for the $W$ boson mass measurement, a binned likelihood comparison of
the observed spectrum to templates generated for different $W$ boson
widths is used to extract the numerical value. 

\begin{figure}
  \includegraphics[width=0.48\textwidth]{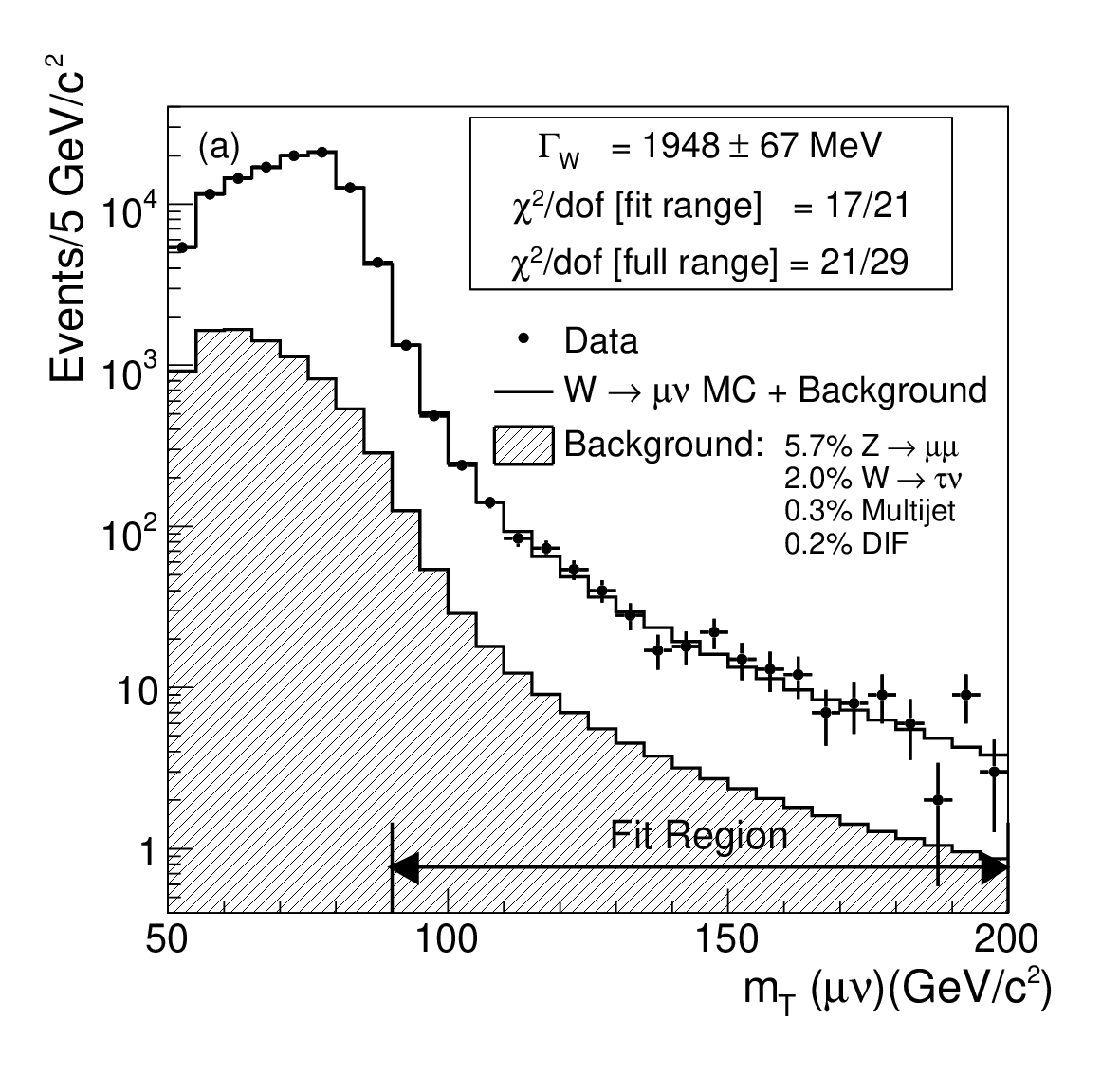}
  \includegraphics[width=0.48\textwidth]{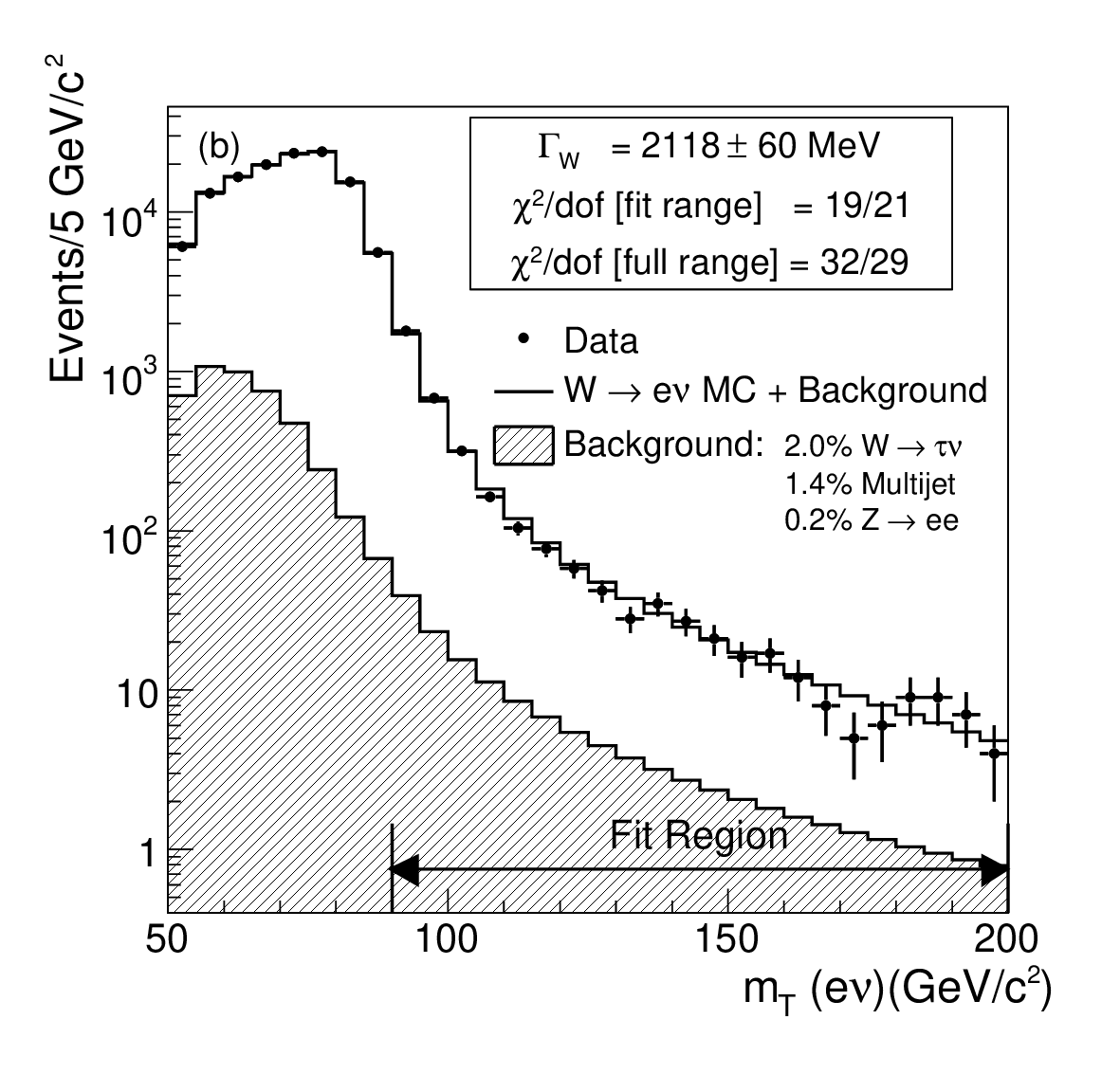}
  \includegraphics[width=0.44\textwidth]{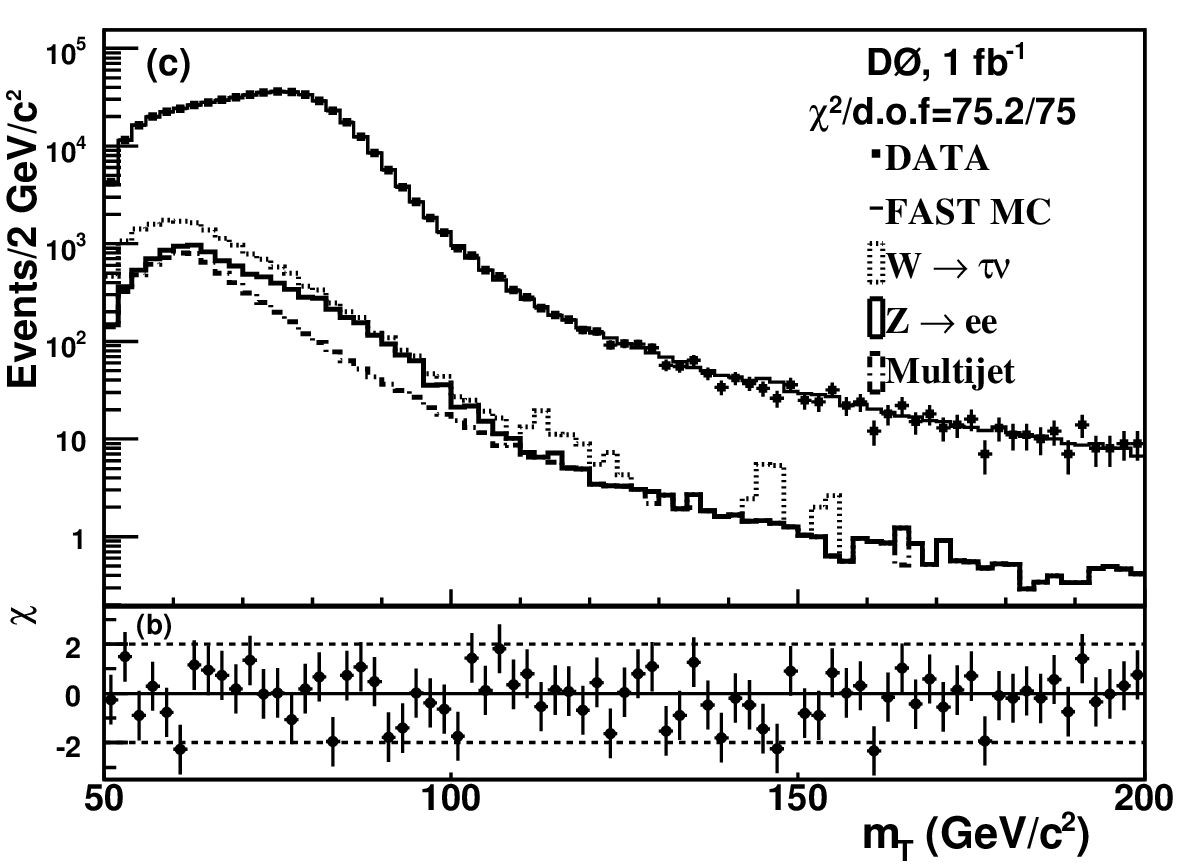}
  \caption{The $m_T$ distributions used for the $\Gamma_W$
    measurement.  (a) The CDF $W\to\mu\nu$ channel. (b) The CDF $W\to e\nu$
     channel.  (c) The D0 $W\to e\nu$ channel.\label{f-width}}
\end{figure}

\begin{figure}
  \includegraphics[width=0.5\textwidth]{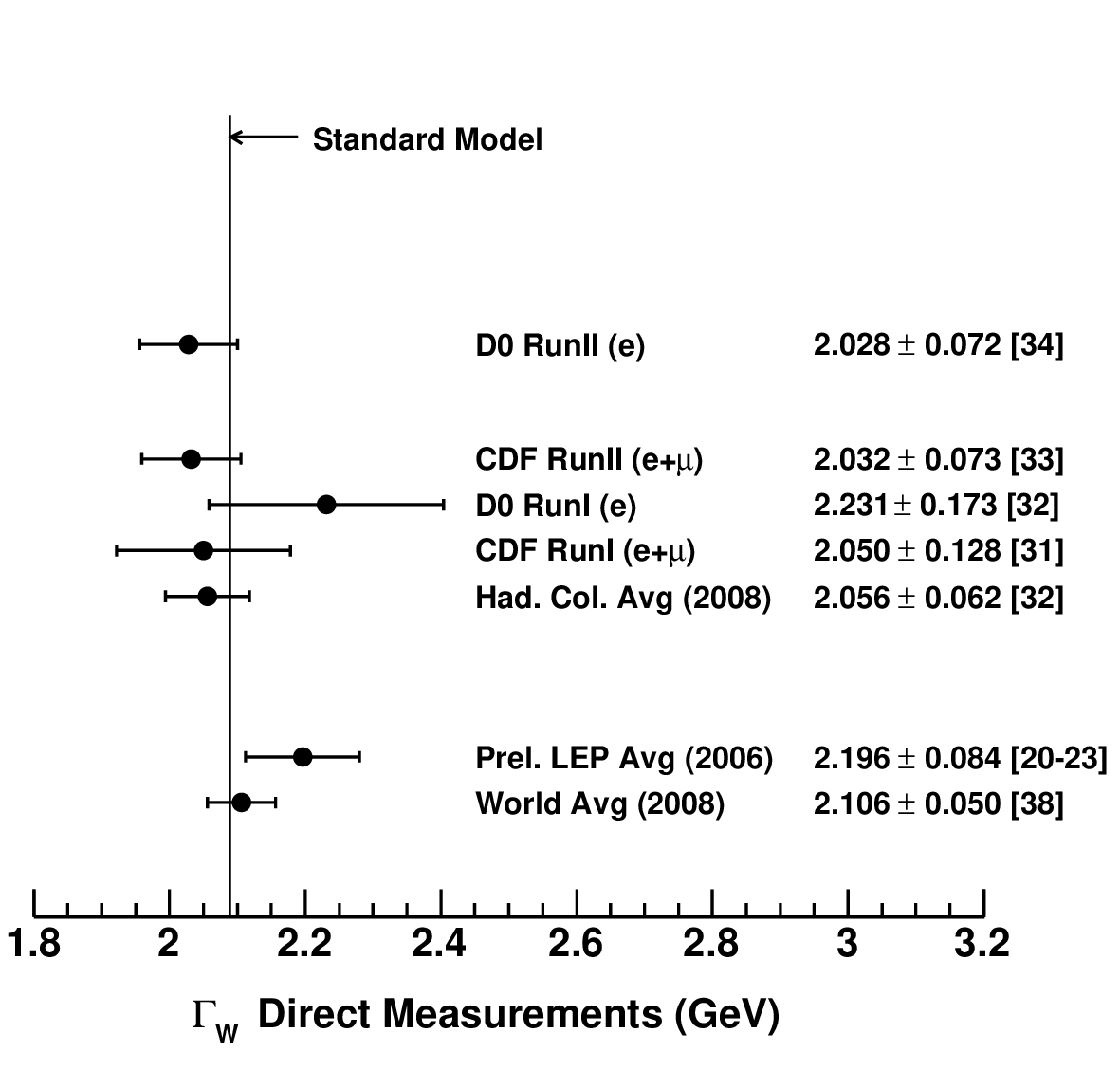}
  \caption{Summary of the $\Gamma_W$ measurements.  The Hadron Collider 
    Average and World Average results do not include the 
    D0~\cite{Collaboration:2009vsa} result.\label{f-gwansw}}
\end{figure}

The $\W$ width was measured by CDF \cite{Aaltonen:2007mg} and D0
\cite{Collaboration:2009vsa} using Run II data.  The CDF result uses a
data set of $\intL = 0.35$~fb$^{-1}$ and the electron and muon final
states.  The D0 result uses $\intL = 1$~fb$^{-1}$ and the electron
final state.  The D0 data set is the same one used for the $W$ boson
mass measurement. Similar dedicated simulations and processing were
used for the $W$ width measurement as were used for the $W$ boson mass
measurement.  D0 used a different hadronic recoil
procedure~\cite{b:d0HadNIM}.  The new procedure gives a $W$ boson mass
result consistent with the standard method.  Fig.~\ref{f-width}
shows the high mass region of the $m_T$ distribution for the three
channels analyzed: the CDF $W\to\mu\nu$ channel, the CDF $W\to e\nu$
channel and the D0 $W\to e\nu$ channel.
The $\Gamma_W$ results are given in Tab.~\ref{t-gwansw} and shown in
Fig.~\ref{f-gwansw}.  The systematic uncertainties are dominated by
hadronic recoil scale and resolution uncertainties.  These contribute
54(49)~MeV for the CDF $W\to e\nu\ (W\to\mu\nu)$ channel and 41~MeV to
the D0 result.  Other important sources include the lepton scale
uncertainty and background uncertainty which are one half to two
thirds the size of the hadronic recoil uncertainty.
\begin{table*}
  \begin{ruledtabular}
   \begin{tabular}{lcccc}
      Channel       &  Yield  & Fit Range($\GeV$) & $\Gamma_W$ (MeV)   & $\chi^2/$ndof \\ \hline
  CDF $W\to\mu\nu$  &  2619   & $90<m_T<200$      & $1948\pm67\pm71$   &    17/21      \\
  CDF $W\to e\nu$   &  3436   & $90<m_T<200$      & $2118\pm60\pm79$   &    19/21     \\ \hline
  CDF Combined      &         &                   & $2032\pm45\pm57$   &              \\ \hline\hline
  D0  $W\to e\nu$   &  5272   & $100<m_T<200$     & $2028\pm39\pm61$   &   75.2/75    \\
   \end{tabular}
   \caption{The $\Gamma_W$ measurements. For the result, the first uncertainty is the
      statistical uncertainty and the second is the systematic uncertainty.
      \label{t-gwansw}}
  \end{ruledtabular}
\end{table*}

\subsection{Forward-backward Asymmetry, $A_{FB}$}
\label{sec:gaugeBosons_Z_Afb}
Production of $Z$ bosons at the Tevatron is dominated by the process
$\qqbar\to Z/\gamma^*$ in which $q = u, d$ are proton valence
quarks. The SM couplings of the $Z$ and $\gamma$ to quarks depend on
the quark charge $Q$, the isospin $I_3$, and the sine of the weak
mixing angle $\sin^2\theta_W$. The differential cross section as a
function of the direction of the fermion resulting from the
$Z/\gamma^*$ decay is given by
\begin{equation}
 \frac{d\sigma}{d\cos\theta} = a(1+\cos^2\theta) + b\cos\theta
\end{equation}
in which $\theta$ is the angle of the fermion from the $Z/\gamma^*$
decay measured relative to the incoming quark direction in the
$Z/\gamma^*$ rest frame.  The relative $Z$ and $\gamma^{*}$ contributions
to the cross section vary as a function of the $Z/\gamma^*$ mass, and
differences in the $Z$ and $\gamma$ couplings to quarks result in
different angular distributions for the decay products for up-type
($I_3 = +1/2$) and down-type ($I_3= -1/2$) quarks. Together, these two
effects produce mass and flavor dependence in the coefficients $a$ and
$b$ which can be calculated assuming the SM.

The forward-backward $Z/\gamma^*$ production charge asymmetry is
defined as 
\begin{equation*}
   A \equiv \frac{\sigma_+ - \sigma_-}{\sigma_+ + \sigma_-}
\end{equation*}
in which $\sigma_+$ and $\sigma_-$ are the integrated cross sections
for the cases $\cos\theta>0$ and $\cos\theta<0$ respectively. The
asymmetry extracted experimentally is given by
\begin{equation}
  A_{FB} = \frac{N_+ - N_-}{N_+ + N_-}
\end{equation}
in which $N_+$ and $N_-$ are the acceptance, efficiency and background
corrected fermion yields in the forward $(\cos\theta>0$) and backward
$(\cos\theta<0)$ directions respectively.  Measuring the asymmetry
rather than differential cross sections allows cancellation of many
systematic uncertainties, particularly those affecting the overall
normalization.

Measurements of the asymmetry as a function of dielectron mass have
been made by both CDF~\cite{b-zasymm-cdf} and D0~\cite{b-zasymm-d0}
using the dielectron final state. The CDF result uses a sample with
$\intL=72$~pb$^{-1}$, and the D0 measurement uses
$\intL=1.1$~fb$^{-1}$. The selection criteria are similar to those for
the $Z$ cross section measurements although a larger dielectron mass
range was selected for the asymmetry measurements. Two experimental
issues of particular importance to these measurements are (1)
controlling asymmetries in either detector acceptance or selection
efficiency as a function of dielectron mass and (2) limiting the
impact of electron charge misidentification.

\begin{figure}
  \includegraphics[width=\linewidth]{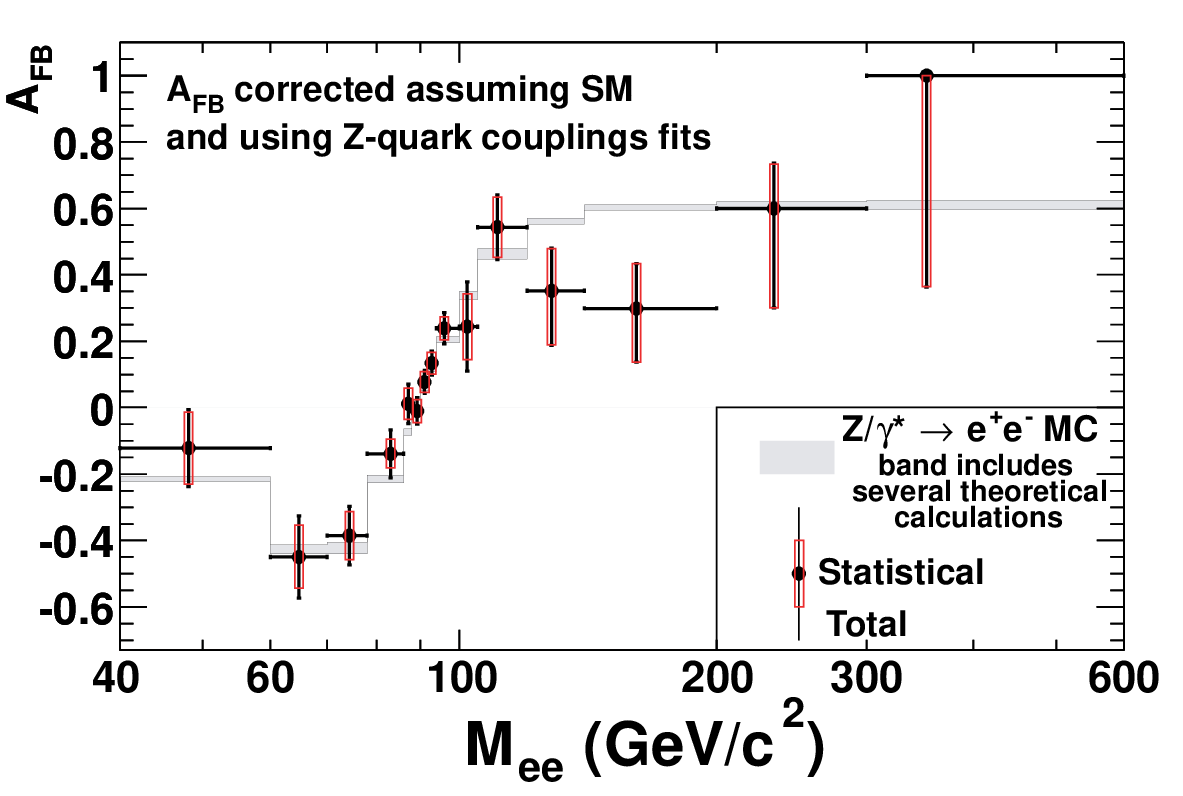}
  \caption{Dielectron forward-backward asymmetry as a function of dielectron
    mass(CDF).\label{f-zasymm-cdf}}
\end{figure}

\begin{figure}
  \includegraphics[width=\linewidth]{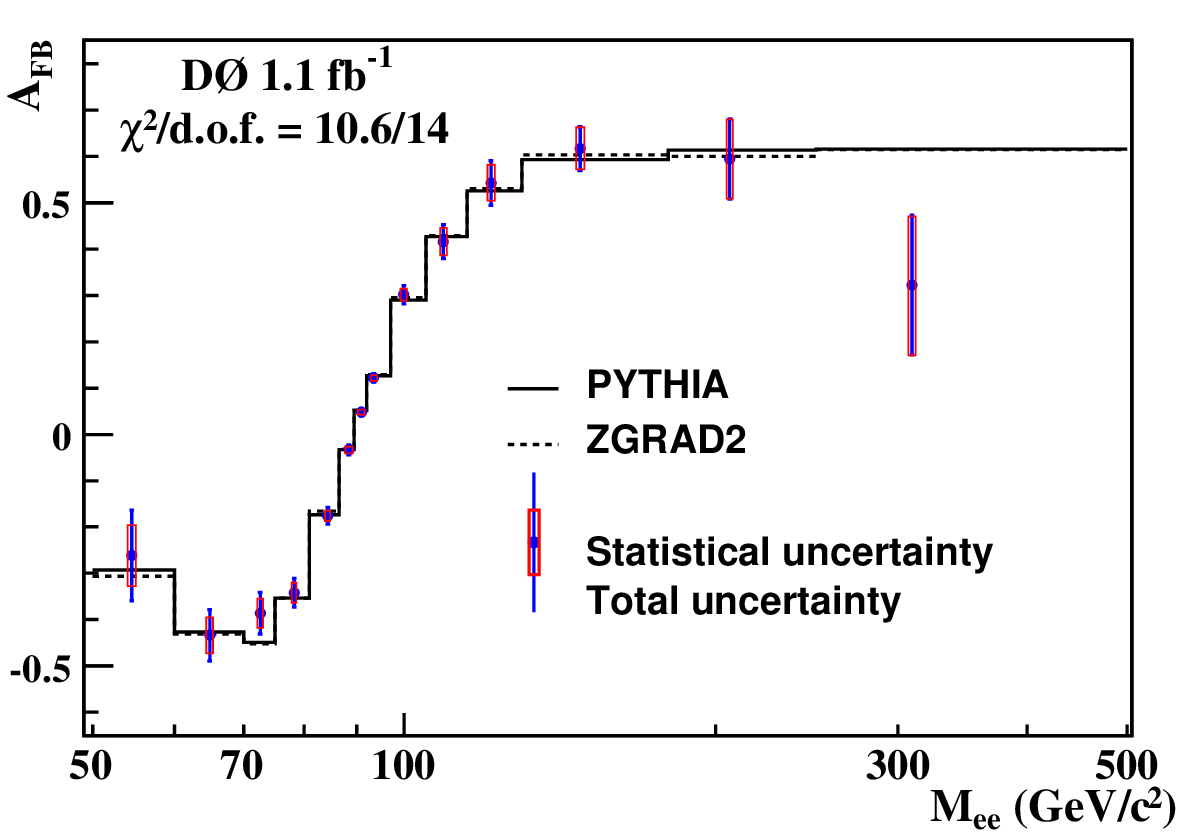}
  \caption{Dielectron forward-backward asymmetry as a function of dielectron
    mass (D0).\label{f-zasymm-d0}}
\end{figure}

In Tab.~\ref{t-zasymm-d0}, the dielectron mass range and the predicted
and measured values of $A_{FB}$ for each mass bin from the D0
measurements are shown, and Figs.~\ref{f-zasymm-cdf}
and~\ref{f-zasymm-d0} show the measured asymmetries and the SM
predictions as a function of mass for the CDF and D0 results
respectively.

\begin{table}
  \begin{ruledtabular}
  \begin{tabular}{c|ccc}
    Dielectron Mass & \multicolumn{3}{c}{$A_{FB}$} \\
  Range (GeV/c)$^2$& \ \ Pythia \ \ & \ \ ZGrad \ \ & Measured \\ \hline
     $50 - 60$     &  $-0.293$  & $-0.307$  & $-0.262 \pm 0.066 \pm 0.072$ \\
     $60 - 70$     &  $-0.426$  & $-0.431$  & $-0.434 \pm 0.039 \pm 0.040$ \\
     $70 - 75$     &  $-0.449$  & $-0.452$  & $-0.386 \pm 0.032 \pm 0.031$ \\
     $75 - 81$     &  $-0.354$  & $-0.354$  & $-0.342 \pm 0.022 \pm 0.022$ \\
     $81 - 86.5$   &  $-0.174$  & $-0.166$  & $-0.176 \pm 0.012 \pm 0.014$ \\
     $86.5 - 89.5$ &  $-0.033$  & $-0.031$  & $-0.034 \pm 0.007 \pm 0.008$ \\
     $89.5 - 92$   &  $~~0.051$ & $~~0.052$ & $~~0.048 \pm 0.006 \pm 0.005$ \\
     $92 - 97$     &  $~~0.127$ & $~~0.129$ & $~~0.122 \pm 0.006 \pm 0.007$ \\
     $97 - 105$    &  $~~0.289$ & $~~0.296$ & $~~0.301 \pm 0.013 \pm 0.015$ \\
     $105 - 115$   &  $~~0.427$ & $~~0.429$ & $~~0.416 \pm 0.030 \pm 0.022$ \\
     $115 - 130$   &  $~~0.526$ & $~~0.530$ & $~~0.543 \pm 0.039 \pm 0.028$ \\
     $130 - 180$   &  $~~0.593$ & $~~0.603$ & $~~0.617 \pm 0.046 \pm 0.013$ \\
     $180 - 250$   &  $~~0.613$ & $~~0.600$ & $~~0.594 \pm 0.085 \pm 0.016$ \\
     $250 - 500$   &  $~~0.616$ & $~~0.615$ & $~~0.320 \pm 0.150 \pm 0.018$ \\
  \end{tabular}
  \caption{The expected and measured asymmetries as a function of dielectron
    mass (D0). For the measured values, the first uncertainty is statistical
    and the second is systematic.\label{t-zasymm-d0}}
\end{ruledtabular}
\end{table}

Using these measurements and the SM prediction for the coefficients
$a$ and $b$, $\sin^2\theta_W^{eff}$ can be determined.  Here
$\theta_W^{eff}$ is the weak mixing angle including higher order
corrections. The current world average is
\begin{equation*}
\sin^2\theta_W^{eff} = 0.23149 \pm 0.00013
\end{equation*}
using the $\overline{MS}$ scheme~\cite{b-pdg08}. Among the measurements used 
for the world average are
two, the charge asymmetry for $b$-quark production~\cite{Z-Pole} from
LEP and SLD and the measurement from NuTeV~\cite{b-nutev}, which
differ from the world average by more than two standard deviations.

\begin{figure}
  \includegraphics[width=0.95\linewidth]{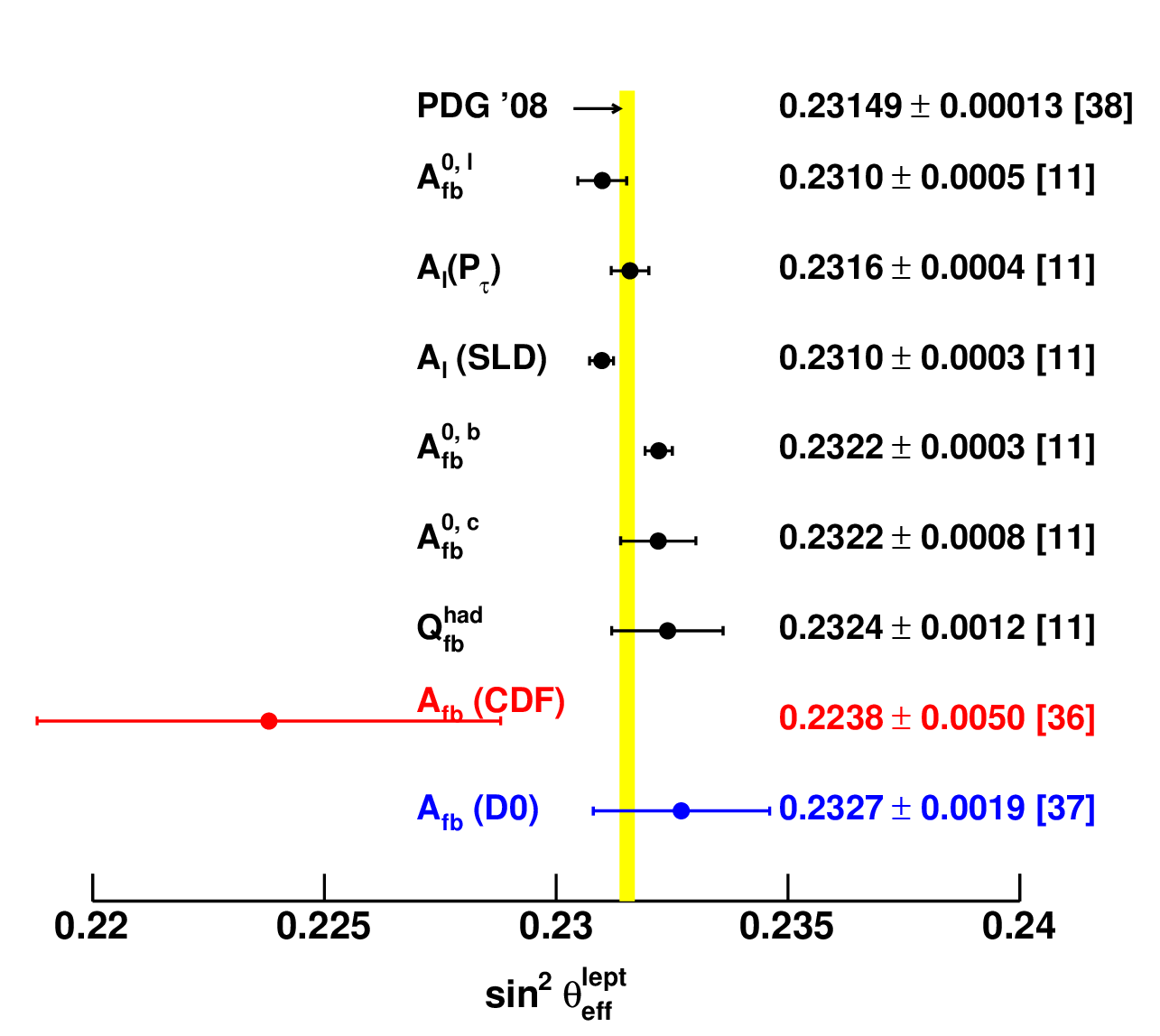}
  \caption{Comparison of the Tevatron asymmetry results with those from other
    experiments.\label{f-zasymm}}
\end{figure}

The values for $\sin^2\theta_W^{eff}$ extracted using fits to the CDF
and D0 $A_{FB}$ distributions are
\begin{equation*}
  \sin^2\theta_W^{eff} = 0.2238\pm 0.0040\ \mathrm{(stat)} \pm 0.0030\ \mathrm{(syst)}
\end{equation*}
for CDF and
\begin{equation*}
  \sin^2\theta_W^{eff} = 0.2327\pm 0.0018\ \mathrm{(stat)} \pm 0.0006\ \mathrm{(syst)}
\end{equation*}
for D0. Fig.~\ref{f-zasymm} shows these results compared to other
measurements.  The results from D0 are comparable in precision to
other measurements for light quarks. The current Tevatron results are
limited by sample statistics, but by the end of the Tevatron running,
CDF and D0 are expected to have the most precise measurements of
$\sin^2\theta_W^{eff}$ for light quarks.

\begin{figure}
  \includegraphics[width=\linewidth]{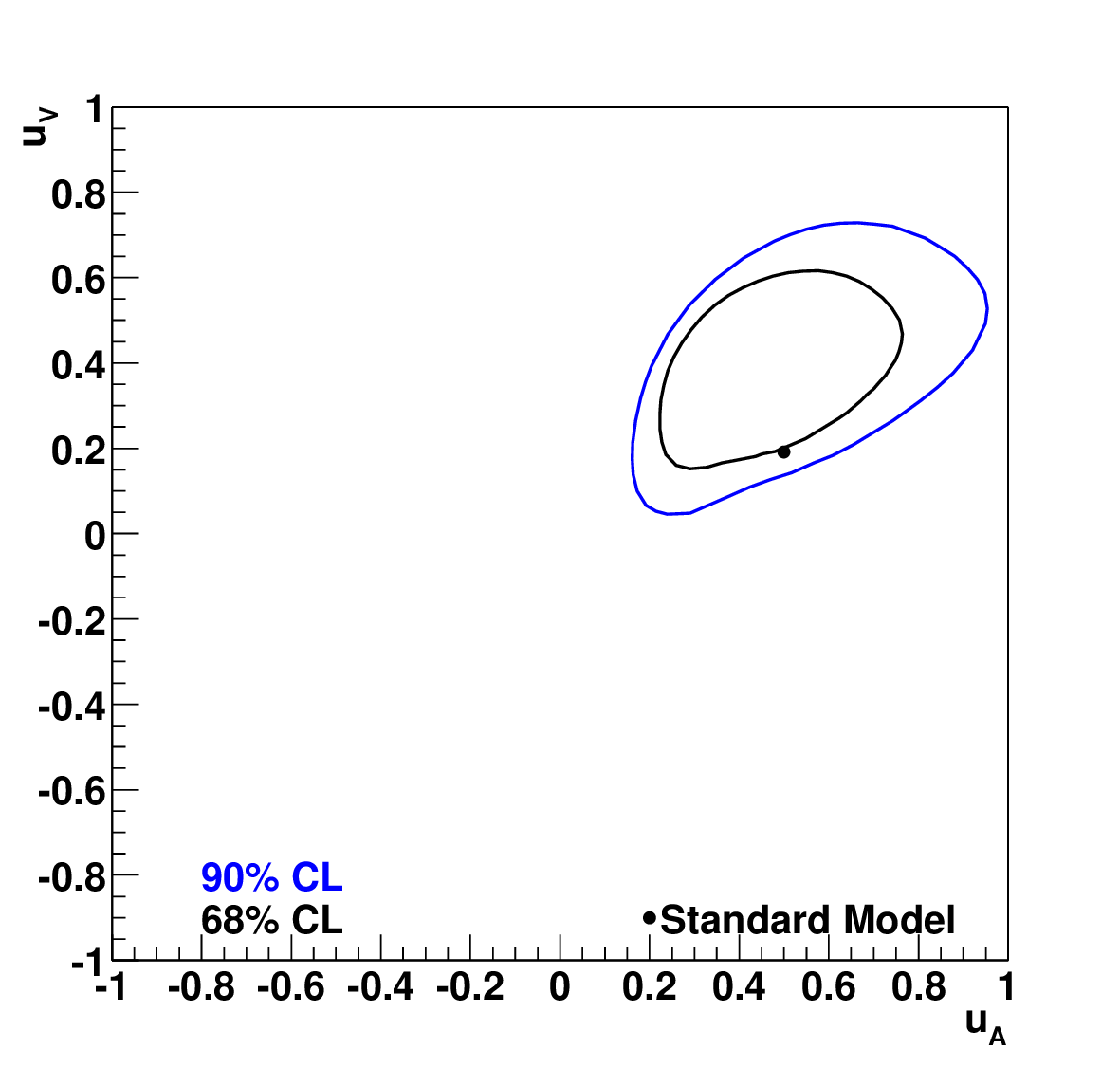}
  \includegraphics[width=\linewidth]{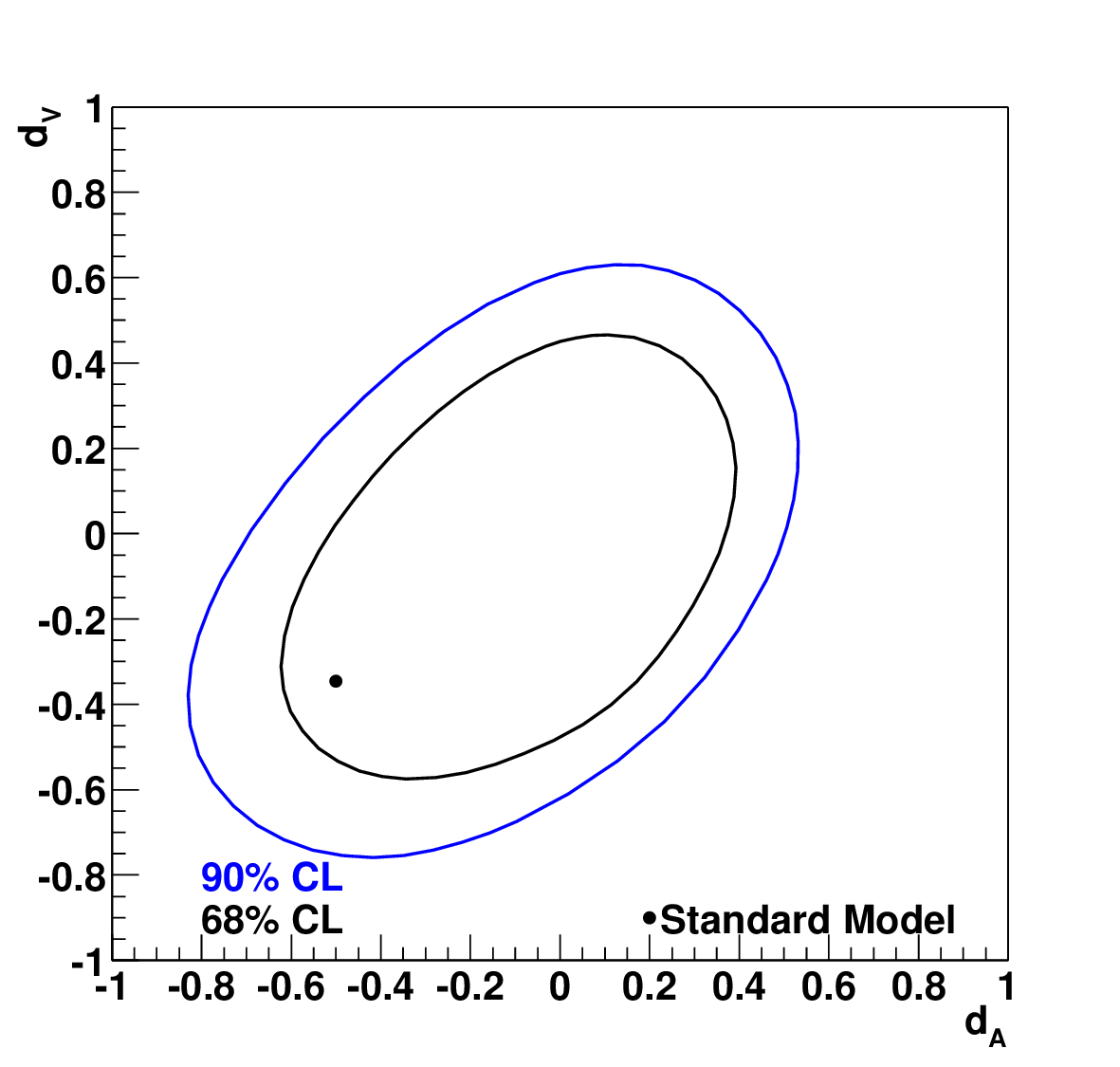}
  \caption{The $Zuu$ and $Zdd$ coupling constants.~\cite{b-zasymm-cdf}
     \label{f-zasymm-ud}}
\end{figure}

CDF also removed the assumption of SM quark couplings and determined
values from a four parameter fit of $A_{FB}$ measurements to the SM
prediction as function of the vector and axial vector couplings for
$u$ and $d$ quarks.  The fit has a $\chi^2/\mathrm{dof} = 10.4/11$,
and the resulting coupling values are shown (with the SM values) in
Fig.~\ref{f-zasymm-ud}.  No evidence of deviation from the SM is
observed.

\subsection{Dibosons}
\label{sec:gaugeBosons_dibosons}

\subsubsection{Trilinear Gauge Couplings (TGCs)}
\label{sec:gaugeBosons_dibosons_aTGC}

\begin{figure}
\includegraphics[width=0.3\textwidth]{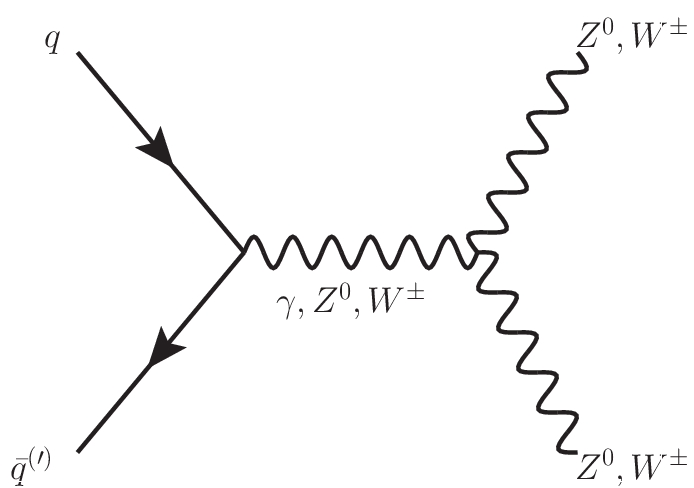}
\caption{Leading order diagram for diboson production via
  quark-antiquark annihilation involving the trilinear gauge coupling.}
\label{fig:gaugeBosons_dibosons_feyn_diboson_generic}
\end{figure}

The non-Abelian nature of the gauge theory describing the electroweak
interactions leads to a striking feature of the theory. In quantum
electrodynamics, the photons carry no electric charge and thus lack
photon-to-photon couplings and do not self-interact. In contrast, the
weak vector bosons carry weak charge and do interact amongst
themselves through trilinear and quartic gauge boson
vertices. Fig.~\ref{fig:gaugeBosons_dibosons_feyn_diboson_generic}
shows the tree-level diagram for diboson production involving
trilinear gauge couplings.

The SM Lagrangian that describes the $WWV (V=Z,\gamma)$ interaction is
given by 
\begin{eqnarray*}
{\cal L}^{\rm SM}_{WWV} &=& i g_{WWV} [(W^+_{\mu\nu}W^{-\mu}-W^{+\mu}W^-_{\mu\nu})V^\nu \\
& & {} + W^+_\mu W^-_\nu V^{\mu\nu}]
\end{eqnarray*}
where $W^{\mu}$ denotes the $W$ field, $W_{\mu\nu}=\partial_\mu W_\nu
- \partial_\nu W_\mu$, $V_{\mu\nu}=\partial_\mu V_\nu
- \partial_\nu V_\mu$, the overall couplings are $g_{WW\gamma}=-e$ and
$g_{WWZ}=-e\cot\theta_W$, and $\theta_W$ is the weak mixing angle
\cite{Hagiwara:1986vm}. At tree level in the SM, the trilinear boson
couplings involving only neutral gauge bosons ($\gamma$ and $Z$)
vanish because neither the photon nor the $Z$ boson carry electric
charge or weak hypercharge.

A common approach used to parameterize the low energy effects from
high-scale new physics is the effective Lagrangian approach that
involves additional terms not present in the SM Lagrangian
\cite{Hagiwara:1986vm}. This approach is convenient because it allows 
for diboson production properties measured in experiments to be
interpreted as model-independent constraints on anomalous coupling
parameters which can be compared with the predictions of new physics
models.

A general form for the $WWV$ Lorentz-invariant interaction Lagrangian
with anomalous coupling parameters $g_1^V$, $\kappa_V$, and
$\lambda_V$ is given by \cite{Hagiwara:1986vm,PhysRevD.41.2113}
\begin{eqnarray*}
{\cal L}^{\rm eff}_{WWV} &=& i g_{WWV} [ g_1^V (W^+_{\mu\nu}W^{-\mu}-W^{+\mu}W^-_{\mu\nu})V^\nu  \\ 
& & {} +~\kappa_V W^+_\mu W^-_\nu V^{\mu\nu} \\
& & {} +~\frac{\lambda_V}{M_W^2}W^{+\nu}_\mu W^{-\rho}_\nu V^\mu_\rho]
\end{eqnarray*}
Note that ${\cal L}^{\rm eff}$ reduces to ${\cal L}^{\rm SM}$ for the
values $\lambda_\gamma = \lambda_Z = 0$ and $g_1^\gamma = g_1^Z =
\kappa_\gamma = \kappa_Z = 1$. Deviations from the SM values of the
coupling parameters are denoted by $\Delta g_1^V$, $\Delta
\kappa_V$, and $\Delta \lambda_V$. We have assumed that $C$ and $P$
are conserved in the interaction Lagrangian. There is no reason to
believe that this assumption is valid unless the physics that leads to
anomalous couplings respects these symmetries. It is straightforward
to include additional terms that violate $C$ and $P$, but we refrain
from doing so in order to keep the discussion simple.

Electromagnetic gauge invariance requires $\Delta g_1^\gamma = 0$. The
$W$ boson magnetic moment $\mu_W$ and the electric quadrupole moment
$Q_W$ are related to the coupling parameters by
$$
\mu_W = \frac{e}{2m_W}(1 + \kappa_\gamma + \lambda_\gamma) 
$$
and
$$
Q_W = -\frac{e}{m^2_W}(\kappa_\gamma - \lambda_\gamma)
$$

The anomalous couplings (aside from $g_1^\gamma$) are usually assumed
to have some dependence on an energy scale (form factors) which
suppresses them at large scales to avoid violation of tree-level
unitarity in the diboson production amplitude
\cite{PhysRevD.37.1775,Baur1988383}. The parameterization generally
used for the energy dependence of a given coupling parameter $\alpha$
is
$$
\alpha(\hat{s}) = \frac{\alpha_0}{(1+\hat{s}/\Lambda^2)^2}
$$
where $\sqrt{\hat{s}}$ is the partonic center-of-mass collision
energy, $\alpha_0$ is the value of the coupling parameter in the limit
$\hat{s}\rightarrow 0$, and $\Lambda$ is the cutoff scale.

The $\sqrt{\hat{s}}$ distribution used in the measurements described
in this Section is obtained through Monte Carlo simulation of the
collision physics. With the substantially increased diboson
statistics that will be available at the LHC, anomalous TGC searches
can be reported as a function of $\sqrt{\hat{s}}$ in diboson decay
channels resuting in fewer than two neutrinos, where the
$\sqrt{\hat{s}}$ can be estimated on an event-by-event basis. This
approach would lead to improved sensitivity and less dependence on
ad-hoc form factors as compared to the standard approach.

When reporting coupling limits from hadron collider data, the value of
$\Lambda$ is taken to be close to the hadron collision energy; even
large variations (e.g. 50\%) of $\Lambda$ around this scale have
minimal impact on the results. Physically, the scale $\Lambda$ can be
considered the scale at which the new physics responsible for the
anomalous coupling is directly accessible (e.g. through pair
production of new particles). This approach is different from the
effective field theory approach discussed in
Section~\ref{sec:overview}, where the coefficients of higher-dimension
operators are constants. While in the same spirit as effective field
theory, the effective Lagrangian approach to anomalous couplings is
different in practice. In particular, the form factors invoked in the
effective Lagrangian approach are unnecessary in an effective field
theory approach.

In the presence of new physics, neutral TGCs (those involving
only $\gamma$ and $Z$ bosons) can contribute to $Z\gamma$ and $ZZ$
production. As previously described, neutral TGCs are anomalous by
their very nature since these couplings are absent in the SM. For each
of the diboson final states $Z\gamma$ and $ZZ$, one can follow an
analogous procedure to the anomalous charged TGCs (those involing a
$W$ boson) by writing down the most general effective Lagrangian that
respects Lorentz invariance and electromagnetic gauge invariance
\cite{Baur:1992cd,Baur:2000ae}. Using prescriptions detailed in
\cite{Baur:1992cd} and \cite{Baur:2000ae}, the effective Lagrangians
introduce anomalous coupling parameters 
$h_{i0}^V (V=\gamma,Z~{\rm and}~i=3,4)$ and $f_{j0}^V (V=\gamma,Z~{\rm
  and}~j=4,5)$ respectively, which can be constrained through an
analysis of $Z\gamma$ and $ZZ$ production in high-energy collider 
data. It is important to note that, under the assumption of on-shell
$Z$ bosons, the $Z\gamma Z$ couplings contributing to $Z\gamma$
production and $ZZ\gamma$ couplings contributing to $ZZ$ production
are completely independent \cite{Baur:2000ae}.

In general, the effect on observables from turning on anomalous TGCs
are correlated. When we refer to ``1D limits,'' we refer to the limits
derived on one parameter when the others are set to their SM values.

There are a few important differences regarding the study of diboson
physics in particle collisions at LEP, Tevatron, and the LHC that are
worth pointing out at this stage:

\begin{itemize}
\item At LEP, $e^+ e^-$ collisions occur at a well-defined energy that
  is set by the accelerator. Therefore, the center-of-mass energy is
  known with good precision and there are no form factors in anomalous
  coupling analyses.
\item In $e^+ e^-$ collisions, the initial state has zero electric
  charge. Therefore, exclusive states with net charge, such as $WZ$
  and $W\gamma$, cannot be produced at LEP. The $WW$ and $ZZ$
  states can and have been produced and studied at LEP. A measurement
  of the $WW$ cross section over a scan in beam energy dramatically
  illustrates the existence of the $WWZ$ coupling in electroweak
  theory \cite{QuiggReview}.
\item At hadron colliders, $\sqrt{s}$ is fixed for long periods of
  time (defining different periods of the accelerator operation that
  change very infrequently) but $\sqrt{\hat{s}}$ varies collision by
  collision. Any anomalous couplings are likely to be $\sqrt{\hat{s}}$
  dependent. The form factor ansatz used to cut off the anomalous
  coupling parameters at large $\sqrt{\hat{s}}$ to preserve S-matrix
  unitary of the amplitude reflects this dependence. For this reason,
  it is reasonable to expect that, once a sufficient amount of
  integrated luminosity has been acquired, the higher-energy reach
  afforded by high-energy hadron collisions will lead to better
  sensitivity to anomalous couplings as compared to the limits from 
  LEP, despite larger backgrounds in a typical hadron collision
  event. In other words, hadron collisions at the Tevatron and LHC
  sample events with a larger average $\sqrt{\hat{s}}$ as compared to
  LEP collision energies and it is exactly those high $\sqrt{\hat{s}}$
  events that are most sensitive to effects of anomalous couplings
  from new physics at higher energy scale.
\item Because the Tevatron is a $p\bar{p}$ collider, the production
  cross sections for $W^+Z$ and $W^-Z$ are equal. The same is true for
  $W^+\gamma$ and $W^-\gamma$ production. When produced in a $pp$
  collider such as the LHC, positive and negative net charge diboson
  states have different production cross sections
  (e.g. $\sigma(pp\rightarrow W^+Z) > \sigma(pp\rightarrow W^-Z)$).
\end{itemize}

\subsubsection{$\W\gamma$}
\label{sec:gaugeBosons_dibosons_Wgamma}

The $W\gamma$ final state observed at hadron colliders provides a
direct test of the $WW\gamma$ TGC. Anomalous $WW\gamma$ couplings lead
to an enhancement in the production cross section and an excess of
large $\Et$ photons. Both CDF and D0 have published measurements of
the $W\gamma$ cross section using leptonic decays of the $W$ bosons
and $\intL=0.2 \fb$ \cite{Acosta:2004it,Abazov:2005ni}. The signature
of the $W\gamma$ signal is an isolated high $\Et$ lepton, an isolated
high $\Et$ photon, and large $\MET$ due to the neutrino from the $W$
decay. The dominant background is from $W$+jets where a jet mimics an
isolated photon. A lepton-photon separation requirement in $\eta-\phi$
space of $\Delta R = \sqrt{(\Delta\eta)^2 + (\Delta\phi)^2} > 0.7$ is
made by both CDF and D0 to suppress events with final-state radiation
of the photon from the outgoing lepton and to avoid collinear
singularities in theoretical calculations. A kinematic requirement on
photon $\Et$ of $\Et > 7 (8)$~GeV is made by CDF (D0 ) in the analysis.

CDF measures
\begin{eqnarray*}
\lefteqn{\sigma(p\bar{p}\rightarrow W\gamma + X) \times
BR(W\rightarrow l\nu)} \\
& & {} = 18.1~\pm~1.6 \rm{(stat.)}~\pm~2.4
\rm{(syst.)}~\pm~1.2 \rm{(lum.)}~\rm{pb}
\end{eqnarray*}
\cite{Acosta:2004it} in agreement
with the next-to-leading order (NLO) theoretical expectation ($\Et >
7$ GeV) \cite{Baur:1992cd} of $19.3~\pm~1.4$ pb. D0
measures 
\begin{eqnarray*}
\lefteqn{\sigma(p\bar{p}\rightarrow W\gamma + X) \times
BR(W\rightarrow l\nu)} \\
& & {} = 14.8~\pm~1.6 \rm{(stat.)}~\pm~1.0 \rm{(syst.)}
\pm 1.0 \rm{(lum.)}~\rm{pb}
\end{eqnarray*}
\cite{Abazov:2005ni} also in agreement with
the NLO expectation ($\Et >8$ GeV) \cite{Baur:1992cd} of
$16.0~\pm~0.4$ pb. Tab.~\ref{tbl:wg_xsect} summarizes the $W\gamma$
cross section measurement results.

\begin{table}
\begin{ruledtabular}
\begin{tabular}{lr@{$\,\pm\,$}lr@{$\,\pm\,$}lr@{$\,\pm\,$}lr@{$\,\pm\,$}l}
& \multicolumn{4}{c}{D0 Analysis} &
\multicolumn{4}{c}{CDF Analysis} \\
& \multicolumn{2}{c}{$e\nu\gamma$} & 
  \multicolumn{2}{c}{$\mu\nu\gamma$}
& \multicolumn{2}{c}{$e\nu\gamma$} & 
  \multicolumn{2}{c}{$\mu\nu\gamma$} \\ \hline \\
$\int{\cal L} \ dt$ (fb$^{-1}$) & \multicolumn{2}{c}{0.16} & 
             \multicolumn{2}{c}{0.13} &
             \multicolumn{2}{c}{0.20 (0.17)} & 
             \multicolumn{2}{c}{0.19 (0.18)} \\
\hline
$W +$ jet(s) & 59 & 5 & 62 & 5 
& 60 & 18 & 28 & 8 \\
$\ell e X$ & 1.7 & 0.5 & 0.7 & 0.2 
& \multicolumn{2}{l}{~~~~~~$-$} & \multicolumn{2}{l}{~~~~~~$-$} \\
$W\gamma \rightarrow \tau\nu\gamma$ & 0.42 & 0.02 & 1.9 & 0.2 
& 1.5 & 0.2 & 2.3 & 0.2 \\ 
$Z\gamma \rightarrow \ell\ell\gamma$ &
\multicolumn{2}{l}{~~~~~~$-$} & 6.9 & 0.7 
& 6.3 & 0.3 & 17.4 & 1.0 \\
\hline
Total Bkg. & 61 & 5 & 71 & 5 
& 67 & 18 & 47 & 8 \\ 
$N_{\rm observed}$ & \multicolumn{2}{c}{112} & \multicolumn{2}{c}{161} 
& \multicolumn{2}{c}{195} & \multicolumn{2}{c}{128} \\
$\sigma\times$ BR (pb) & 13.9 & 3.4 & 15.2 & 2.5
& 19.4 & 3.6 & 16.3 & 2.9 \\
Theory $\sigma\times$ BR
& \multicolumn{4}{c}{$16.0\pm 0.4$} 
& \multicolumn{4}{c}{$19.3\pm 1.4$}
\end{tabular}
\end{ruledtabular}
\caption{\label{tbl:wg_xsect}Summary of the
  $W\gamma\rightarrow\ell\nu\gamma$ CDF \cite{Acosta:2004it} and D0
  \cite{Abazov:2005ni} cross section analyses compared with the theory
  \cite{Baur:1992cd}. For the CDF analysis, the luminosities with and
  without parentheses correspond to central and forward leptons,
  respectively. The theory cross section is the NLO calculation (in
  pb) from \cite{Baur:1992cd}.}
\end{table}

\begin{figure}
  \includegraphics[width=0.96\linewidth]{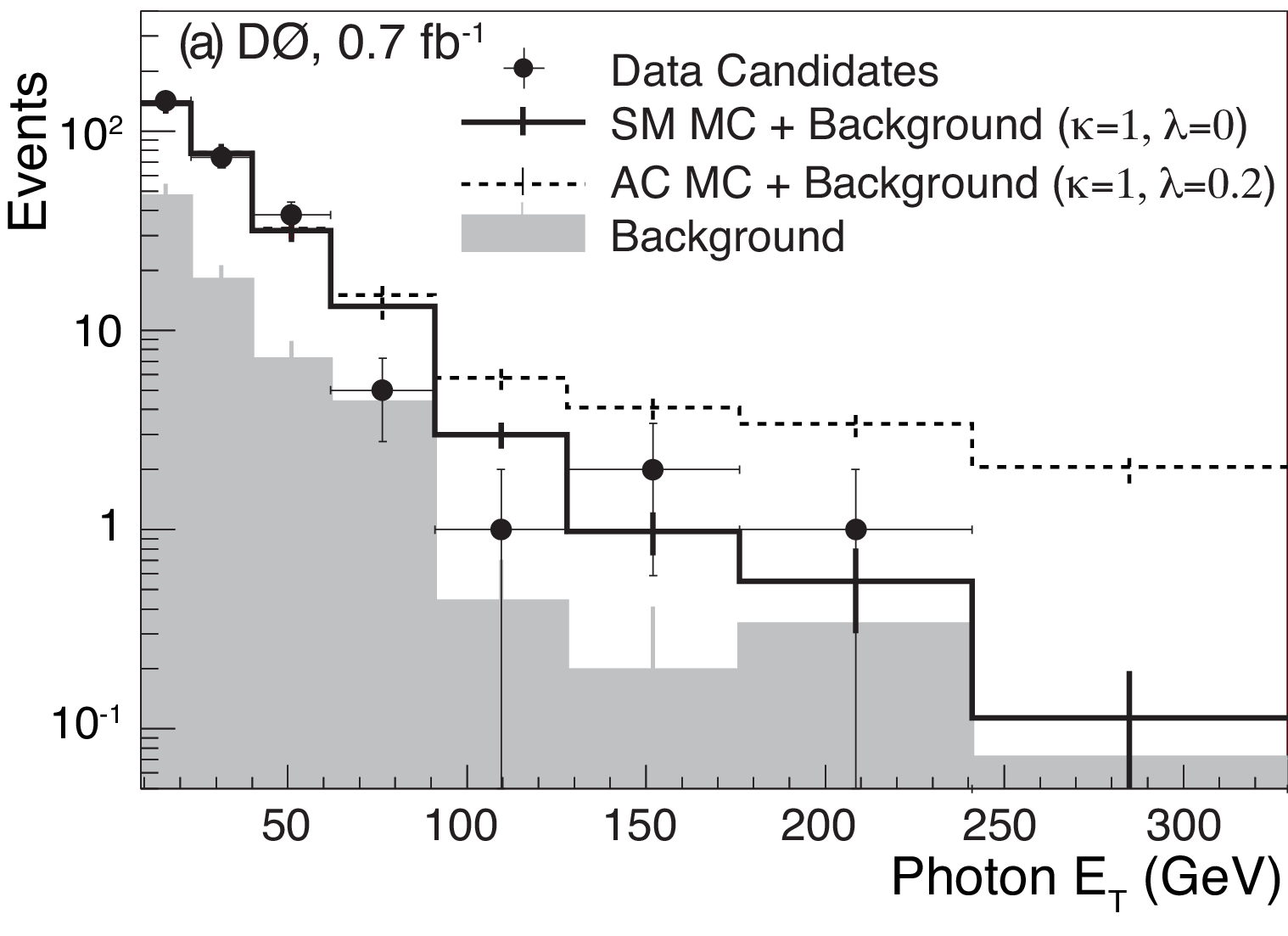}\\
  \includegraphics[width=0.96\linewidth]{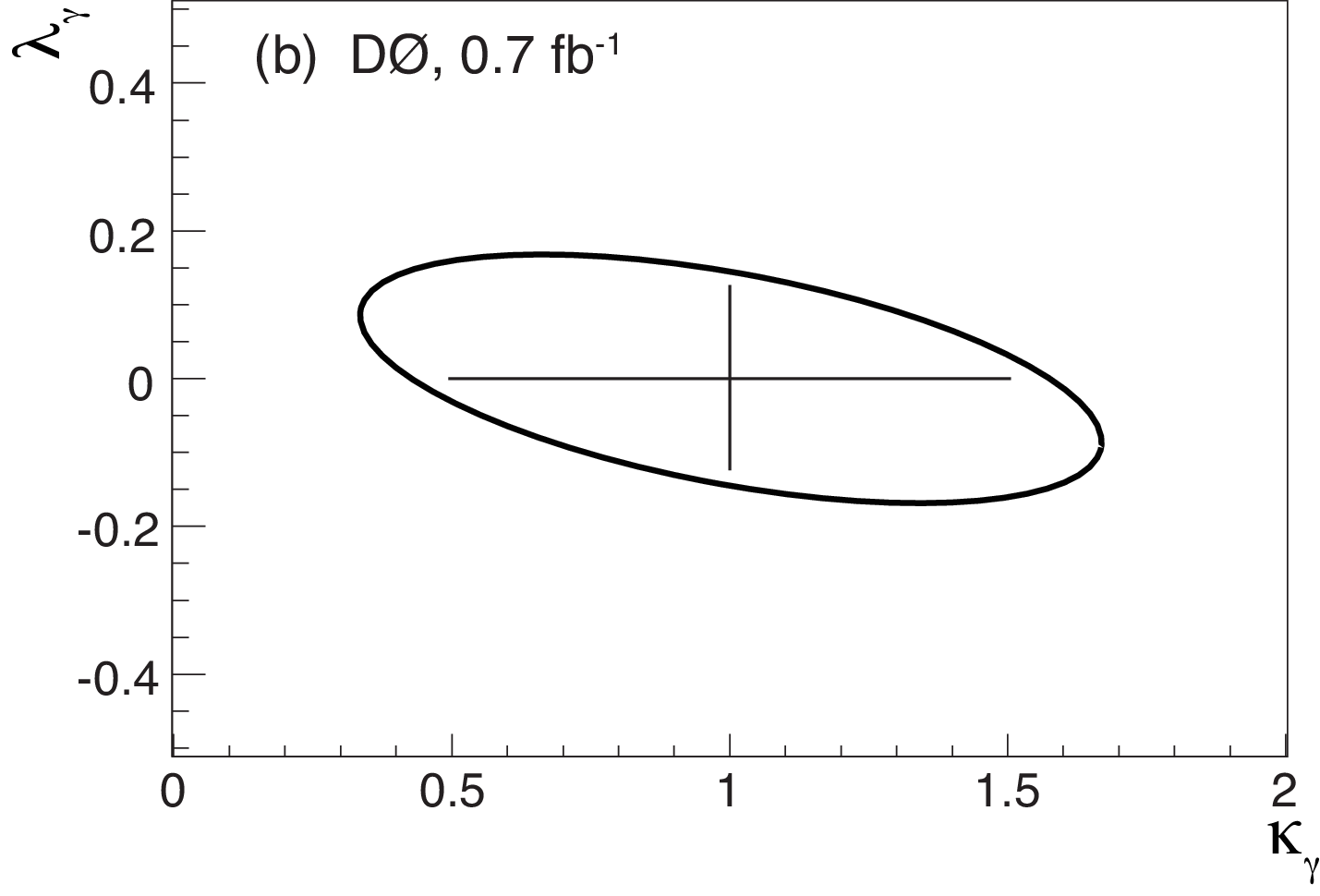}
%\begin{tabular}{cc}
%  \multirow{1}{*}[1.0in]{(a)} & \includegraphics[width=0.96\linewidth]{eps/gaugeBosons/0803.0030-photonEt_combo_finalopt.eps}\\
%  \multirow{1}{*}[1.0in]{(b)} & \includegraphics[width=0.96\linewidth]{eps/gaugeBosons/0803.0030-2d_limits.eps}
%\end{tabular}    
  \caption{The (a) photon $E_T$ distribution for the D0 $W\gamma$
    analysis \cite{Abazov:2008vja}. The photon $E_T$ distribution of
    events are used to constrain anomalous coupling parameters shown
    in (b). \label{fig:wg-dzero}}
\end{figure}

Both of the CDF and D0 $W\gamma$ cross section measurement are
consistent with the SM expectations at NLO. In a more recent analysis
\cite{Abazov:2008vja}, D0 uses four times more integrated luminosity
as compared to \cite{Abazov:2005ni} and adds photons reconstructed in
their endcap calorimeters ($1.5 < |\eta_{det}| < 2.5$) to search for
anomalous $WW\gamma$ couplings based on the observed photon $\Et$
spectrum \cite{Abazov:2008vja} for photons with $E_T > 9$
GeV. Additionally, the three-body transverse mass of the 
photon, lepton and $\MET$ must exceed 120 $\GeVc$ (110 $\GeVc$) for
the electron (muon) channel in order to suppress final state
radiation. The photon $E_T$ spectrum and anomalous TGC limits are
shown in Fig.~\ref{fig:wg-dzero}. A LO simulation \cite{Baur:1989gk}
of the $W\gamma$ signal is used with NLO corrections
\cite{Baur:1993ir} applied to the photon $E_T$ spectrum. The
one-dimensional limits at 95\% confidence level (CL) are $-0.51<\Delta
\kappa_\gamma<0.51$ and $-0.12<\lambda_\gamma<0.13$ for
$\Lambda = 2.0$ TeV.

The SM $W\gamma$ production involves interference between the
amplitudes for a photon radiated off of an incoming quark (QED initial
state radiation) and the photon produced from the $WW\gamma$
vertex. This interference leads to a zero amplitude for the SM in the
photon angular distribution
\cite{PhysRevD.20.1164,PhysRevLett.43.746,PhysRevD.23.2682,PhysRevLett.49.966}.
In $W\gamma$
production, the radiation amplitude zero (RAZ) manifests itself as a
dip at $\sim  -1/3$ in the charged-signed rapidity difference $Q_\ell
\times \Delta y$ between the observed photon and the charged lepton
from decay of the $W$ boson \cite{Baur:1994sa}. Experimentally, the
pseudorapidity difference $\Delta \eta$ is used in place of the
rapidity difference $\Delta y$, since it involves only the production
angle $\theta$ with respect to the beam line ($\eta =
-\ln(\tan(\theta/2))$) and is a very good approximation to $\Delta y$
in the limit of massless particles. Using the same data they used to
limits on $WW\gamma$ anomalous TGCs, D0 made a first detailed study of
the $Q_\ell \times \Delta y$ to search for the RAZ effect
\cite{Abazov:2008vja}.

\begin{figure}
\includegraphics[width=0.96\linewidth]{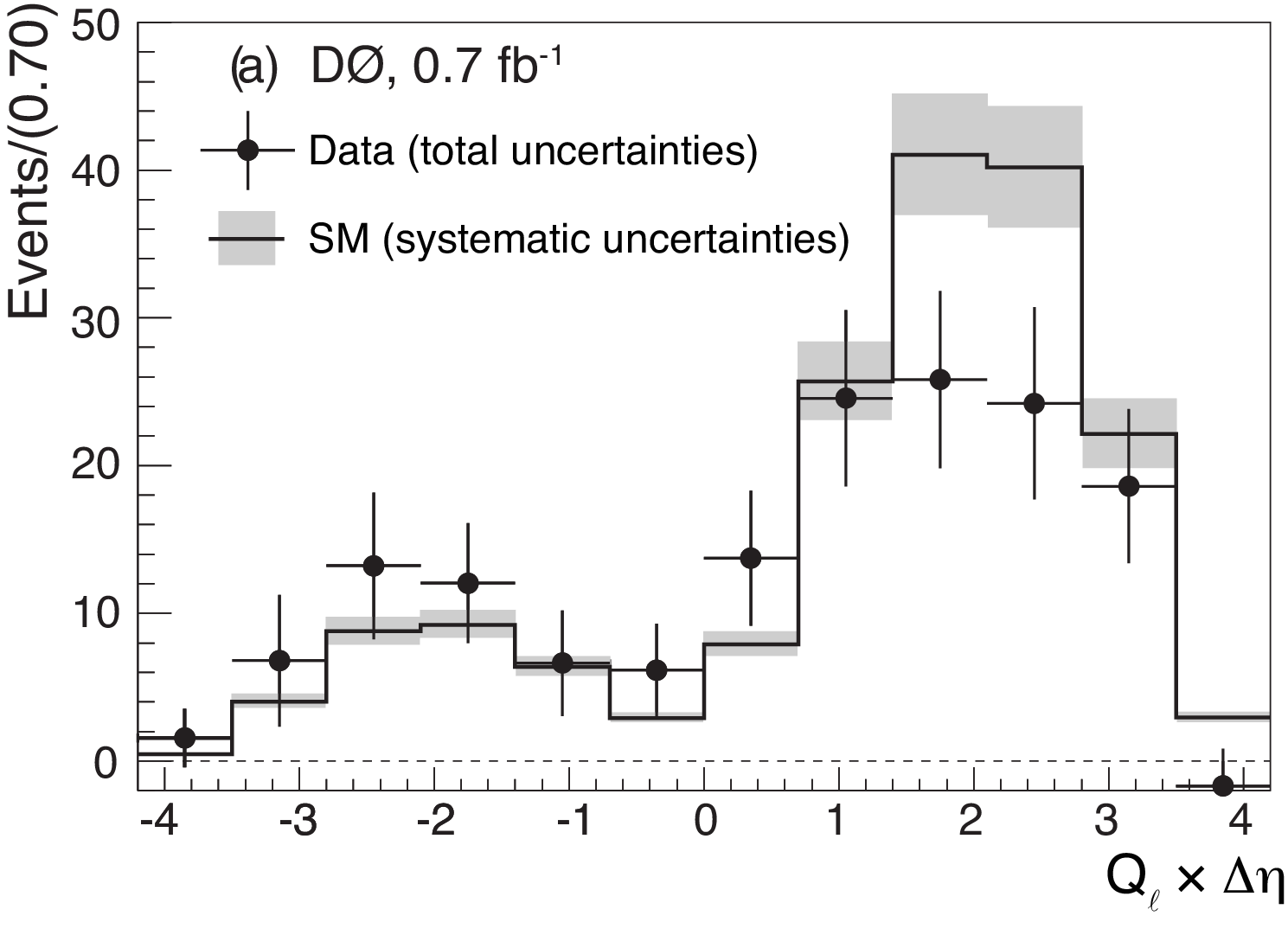}\\
\includegraphics[width=0.96\linewidth]{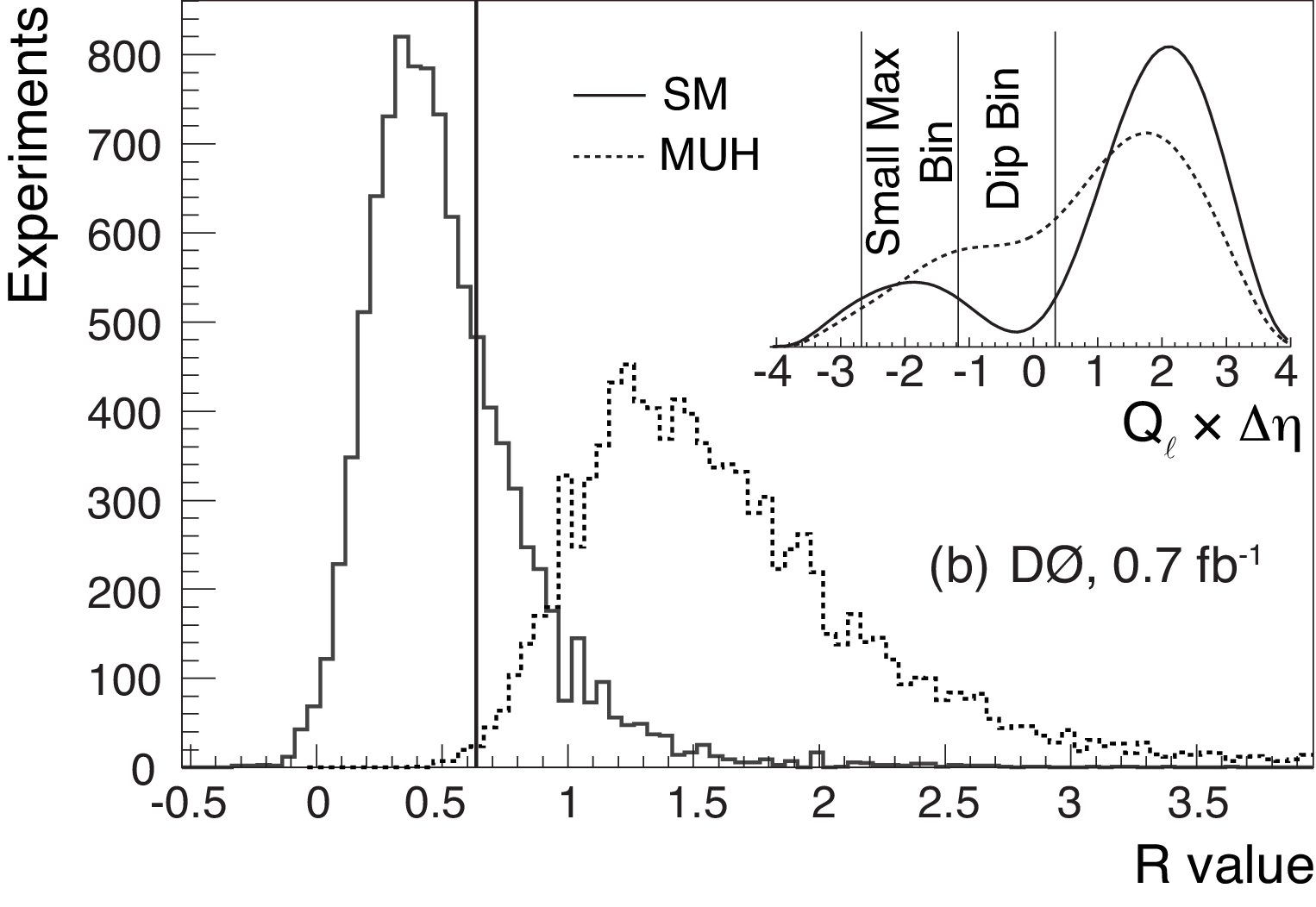}
%\begin{tabular}{cc}
%  \multirow{1}{*}[1.0in]{(a)} & \includegraphics[width=0.96\linewidth]{eps/gaugeBosons/0803.0030-c_rapDiffBkgSub_combo_97.eps}\\
%  \multirow{1}{*}[1.0in]{(b)} & \includegraphics[width=0.96\linewidth]{eps/gaugeBosons/0803.0030-overlayEnsemble_Rtest.eps}
%\end{tabular}    
  \caption{(a) The background-subtracted charge-signed rapidity difference
    of events is shown in along with the SM expectation from
    \cite{Abazov:2008vja}. Consideration of the full covariance
    matrix gives a 15\% $\chi^2$ probability 
    for compatibility between the data and the SM expectation. (b)
    Distributions of the R-test statistic for the SM (solid line) and
    ``no-dip'' (dashed line) hypothesis pseudo-experiments from
    \cite{Abazov:2008vja}. The data result is 0.64 shown by the
    vertical line where only 45 out of 10$^4$ no-dip pseudo-experiment
    had a value smaller as small or smaller than 0.64. The R-test
    statistic is defined as the ratio of events in the ``Dip Bin'' to
    the ``Small Max Bin'', as shown in the inset where the solid line
    is the SM expectation and the dashed line is for the no-dip
    hypothesis. \label{fig:wg-dzero-raz}}
\end{figure}

Fig.~\ref{fig:wg-dzero-raz}(a) shows the $Q_\ell \times \Delta y$
distribution of data compared with the SM expectation, which has a
15\% $\chi^2$ probability for compatibility between the data and the
SM expectation, demonstrating a reasonable level of agreement. To
specifically investigate the dip region around $Q_\ell \times \Delta y
= -1/3$ indicative of the RAZ effect, a simple test statistic $R$ is
constructed which is the ratio of the number events observed in a bin
including the dip region to the number of events observed in a bin
with more negative $Q_\ell \times \Delta y$ where a maximum is
expected, based on Monte Carlo, from the SM including the total D0
acceptance in this analysis. A value of $R = 0.64$ is observed in the
data and it is determined that 28\% of SM pseudo-experiments give a
value as large or larger than 0.64.

A particular anomalous coupling hypothesis corresponding to
$\kappa_\gamma = 0$, $\lambda_\gamma = -1$ is chosen to test against as a
null hypothesis since it is a specific model which leads to a $Q_\ell
\times \Delta y$ distribution without a dip as shown in the inset of
Fig.~\ref{fig:wg-dzero-raz}(b). Only the shape is tested such that
the null ``no-dip'' distribution is normalized to the yield expected
from the SM. The $R$ distribution for the SM and no-dip hypotheses are
shown in Fig.~\ref{fig:wg-dzero-raz}(b) along with the value (0.64)
observed in the data. From this analysis, the no-dip hypothesis is
excluded at 2.6$\sigma$ Gaussian significance.

\subsubsection{$\Z\gamma$}
\label{sec:gaugeBosons_dibosons_Zgamma}

As mentioned in Section~\ref{sec:gaugeBosons_dibosons_aTGC}, photons
do not directly couple to $Z$ bosons at tree-level in the
SM. Therefore, observation of such a coupling would constitute
evidence for new phenomena. The $Z\gamma$ final state at hadron
colliders involves a combination of $ZZ\gamma$ and $Z\gamma\gamma$
couplings. Both CDF and D0 have made measurements of the $Z\gamma$
cross section in leptonic decay channels of the $Z$ boson. The
signature of the $Z\gamma$ signal is two isolated high $\Et$ charged
leptons having the same flavor and opposite charge with invariant mass
consistent with decay of a $Z$ boson, and an isolated high $\Et$
photon. The dominant background is from $Z$+jets where a jet mimics an 
isolated photon. As in the $W\gamma$ analyses described in
Section~\ref{sec:gaugeBosons_dibosons_Wgamma}, a lepton-photon 
separation requirement is imposed. A requirement of $\Delta R > 0.7$
is made by both CDF and D0. Kinematic requirements on the photon $\Et$
of $\Et > 7 (8) \GeV$ and the dilepton invariant mass of $M_{\ell\ell}
> 40 (30)\ \GeV$ are made by CDF (D0 ) in the analysis. 

\begin{figure}
  \includegraphics[width=0.96\linewidth]{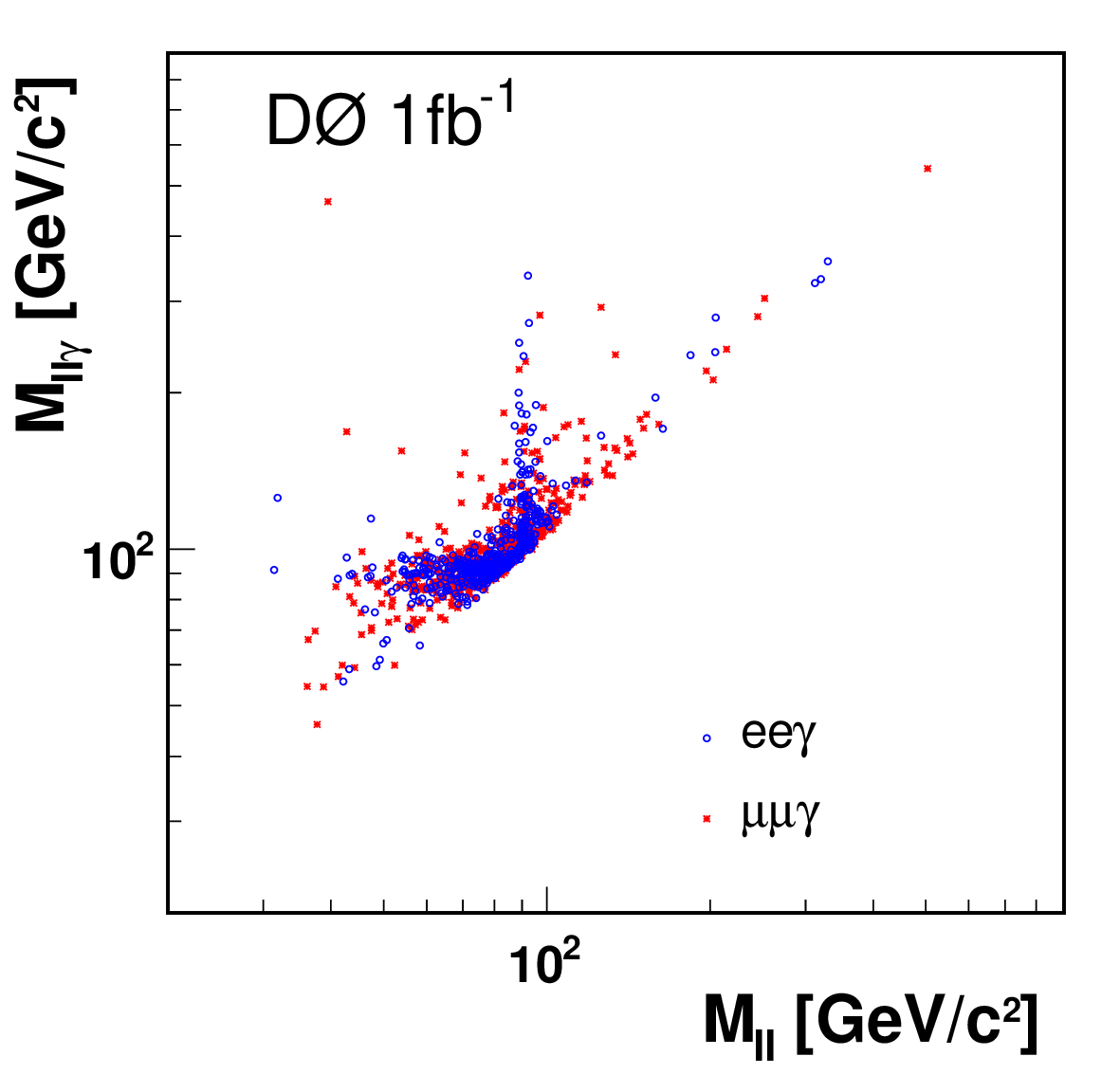}
  \caption{The three-body mass $M_{\ell\ell\gamma}$ versus the
    dilepton mass $M_{\ell\ell}$ for $Z\gamma$ candidate events in the
    D0 analysis \cite{Abazov:2007wy} \label{fig:zg-dzero-2d}}
\end{figure}

Using $\intL=0.2\fb$, CDF measures
\begin{eqnarray*}
\lefteqn{\sigma(p\bar{p}\rightarrow Z\gamma + X) \times BR(Z\rightarrow ll)} \\
& & {} = 4.6~\pm~0.5 \rm{(stat.)}~\pm~0.2 \rm{(syst.)}~\pm~0.3
\rm{(lum.)}~\rm{pb}
\end{eqnarray*}
\cite{Acosta:2004it} in agreement with the NLO theoretical expectation
using Ref. \cite{Baur:1992cd} of $4.5~\pm~0.3$ pb using the CDF
acceptance. Using $\intL=1.1\fb$, D0 measures 
\begin{eqnarray*}
\lefteqn{\sigma(p\bar{p}\rightarrow Z\gamma + X) \times BR(Z\rightarrow ll)} \\
& & {} = 5.0~\pm~0.3 \rm{(stat.+syst.)}~\pm~0.3 \rm{(lum.)}~\rm{pb}
\end{eqnarray*}
\cite{Abazov:2007wy} also in agreement with the NLO expectation
using the generator described in \cite{PhysRevD.57.2823} of
$4.7~\pm~0.2$ pb and the D0 acceptance. Tab.~\ref{tbl:zg_xsect}
summarizes the $Z\gamma$ cross section measurement
results. Fig.~\ref{fig:zg-dzero-2d} shows the three-body mass
$M_{\ell\ell\gamma}$ versus the dilepton mass $M_{\ell\ell}$ for
$Z\gamma$ candidate events in the D0 analysis \cite{Abazov:2007wy}.

\begin{table}
\begin{ruledtabular}
\begin{tabular}{lr@{$\,\pm\,$}lr@{$\,\pm\,$}lr@{$\,\pm\,$}lr@{$\,\pm\,$}l}
& \multicolumn{4}{c}{D0 Analysis} &
\multicolumn{4}{c}{CDF Analysis} \\
& \multicolumn{2}{c}{$ee\gamma$} & 
  \multicolumn{2}{c}{$\mu\mu\gamma$}
& \multicolumn{2}{c}{$ee\gamma$} & 
  \multicolumn{2}{c}{$\mu\mu\gamma$} \\ \hline \\
$\int{\cal L} \ dt$ (fb$^{-1}$) & \multicolumn{2}{c}{1.1} & 
             \multicolumn{2}{c}{1.0} &
             \multicolumn{2}{c}{0.20 (0.17)} & 
             \multicolumn{2}{c}{0.19 (0.18)} \\
$Z +$ jets background & 55 & 8 & 61 & 9 
& 2.8 & 0.9 & 2.1 & 0.6 \\
$N_{\rm observed}$ & \multicolumn{2}{c}{453} & \multicolumn{2}{c}{515} 
& \multicolumn{2}{c}{36} & \multicolumn{2}{c}{35} \\
$\sigma\times$ BR (pb) 
& \multicolumn{4}{c}{$5.0~\pm~0.4$}
& 4.8 & 0.9 & 4.4 & 0.8 \\
Theory $\sigma\times$ BR
& \multicolumn{4}{c}{$4.7~\pm~0.2$} 
& \multicolumn{4}{c}{$4.5~\pm~0.3$}
\end{tabular}
\end{ruledtabular}
\caption{\label{tbl:zg_xsect}Summary of the
  $Z\gamma\rightarrow\ell\ell\gamma$ CDF \cite{Acosta:2004it} and D0
  \cite{Abazov:2007wy} cross section analyses compared with the theory
  \cite{PhysRevD.57.2823}. For the CDF analysis, the luminosities with and
  without parentheses correspond to central and forward leptons,
  respectively. The theory cross section is the NLO calculation (in
  pb) from \cite{PhysRevD.57.2823}.}
\end{table}

Using $\intL=3.6\fb$, the D0 collaboration has observed (5.1$\sigma$)
the $Z\gamma\rightarrow \nu\nu\gamma$ process for the first time at a
hadron collider \cite{Abazov:2009cj} and used these events to search
for $Z\gamma\gamma$ and $ZZ\gamma$ couplings that are absent at
tree-level in the SM. The experimental signature for
$Z\gamma\rightarrow \nu\nu\gamma$ is a high energy photon and large
$\MET$. In the analysis, events are required to have a single photon
candidate with $\Et > 90$ GeV and $\MET > 70$ $\GeVc$. The primary
backgrounds are from $W\rightarrow e\nu$ and events unrelated to the
collision in which muons from beam halo or cosmic rays produce
energetic photons through bremsstrahlung. The $W\rightarrow e\nu$ is 
suppressed by removing events with high-$p_T$ tracks and the
non-collision events are removing using available $z_0$ production
information from the EM calorimeter and pre-shower detectors. A
summary of the background estimates and observed events is shown in
Tab.~\ref{tab:zg-dzero-bkg}. The measured cross section is
\begin{eqnarray*}
\lefteqn{\sigma(p\bar{p}\rightarrow Z\gamma + X) \times
  BR(Z\rightarrow \nu\nu)} \\
& & {} = 32~\pm~9 \rm{(stat.+syst.)}~\pm~2 \rm{(lum.)}~\rm{fb} 
\end{eqnarray*}
for the photon $\Et > 90$ GeV, consistent with the NLO cross section
of $39~\pm~4$ fb \cite{PhysRevD.57.2823}.

\begin{table}
\begin{ruledtabular}
\begin{tabular}{lr}
              & Number of events \\ \hline
$W \to e\nu$            & $9.67 \pm 0.30(\rm stat.) \pm 0.48(\rm syst.)$\\
non-collision		& $5.33 \pm 0.39(\rm stat.) \pm 1.91(\rm syst.)$\\
$W/Z$ + jet             & $1.37 \pm 0.26(\rm stat.) \pm 0.91(\rm syst.)$\\
$W\gamma$		& $0.90 \pm 0.07(\rm stat.) \pm 0.12(\rm syst.)$\\ \hline
Total background	& $17.3 \pm 0.6(\rm stat.) \pm 2.3(\rm syst.)$\\
$N_{\nu\bar\nu\gamma}^{\rm SM}$ & $33.7 \pm 3.4$\\\hline
$N_{\rm obs}$		& 51 \\
\end{tabular}
\end{ruledtabular}
\caption{Summary of background estimates, and the number of observed
  and SM predicted events for the D0 analysis of the
  $Z\gamma\rightarrow \nu\nu\gamma$ channel
  \cite{Abazov:2009cj}.\label{tab:zg-dzero-bkg}}
\end{table}

\begin{figure}
\includegraphics[width=0.96\linewidth]{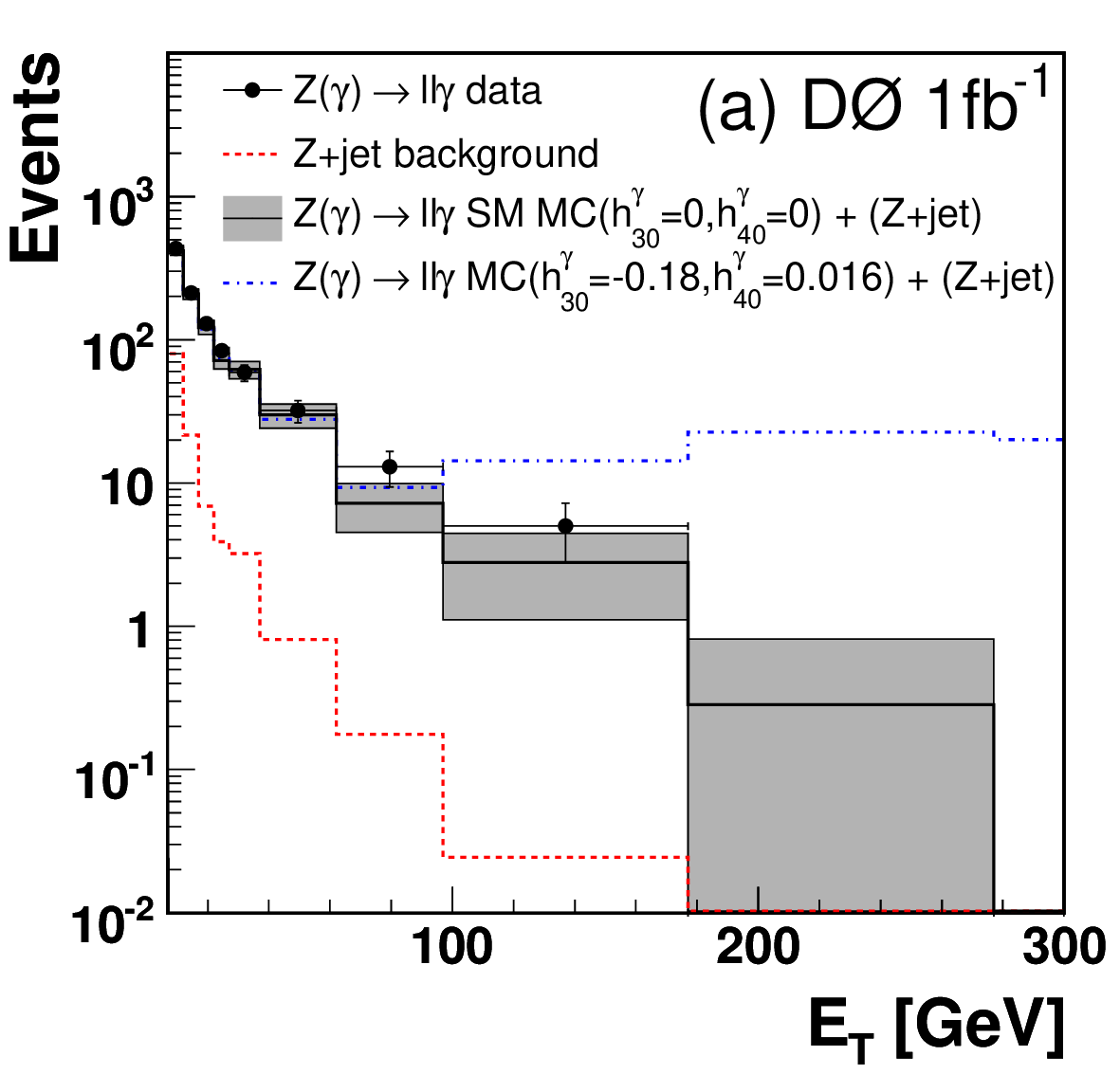}\\
\includegraphics[width=0.96\linewidth]{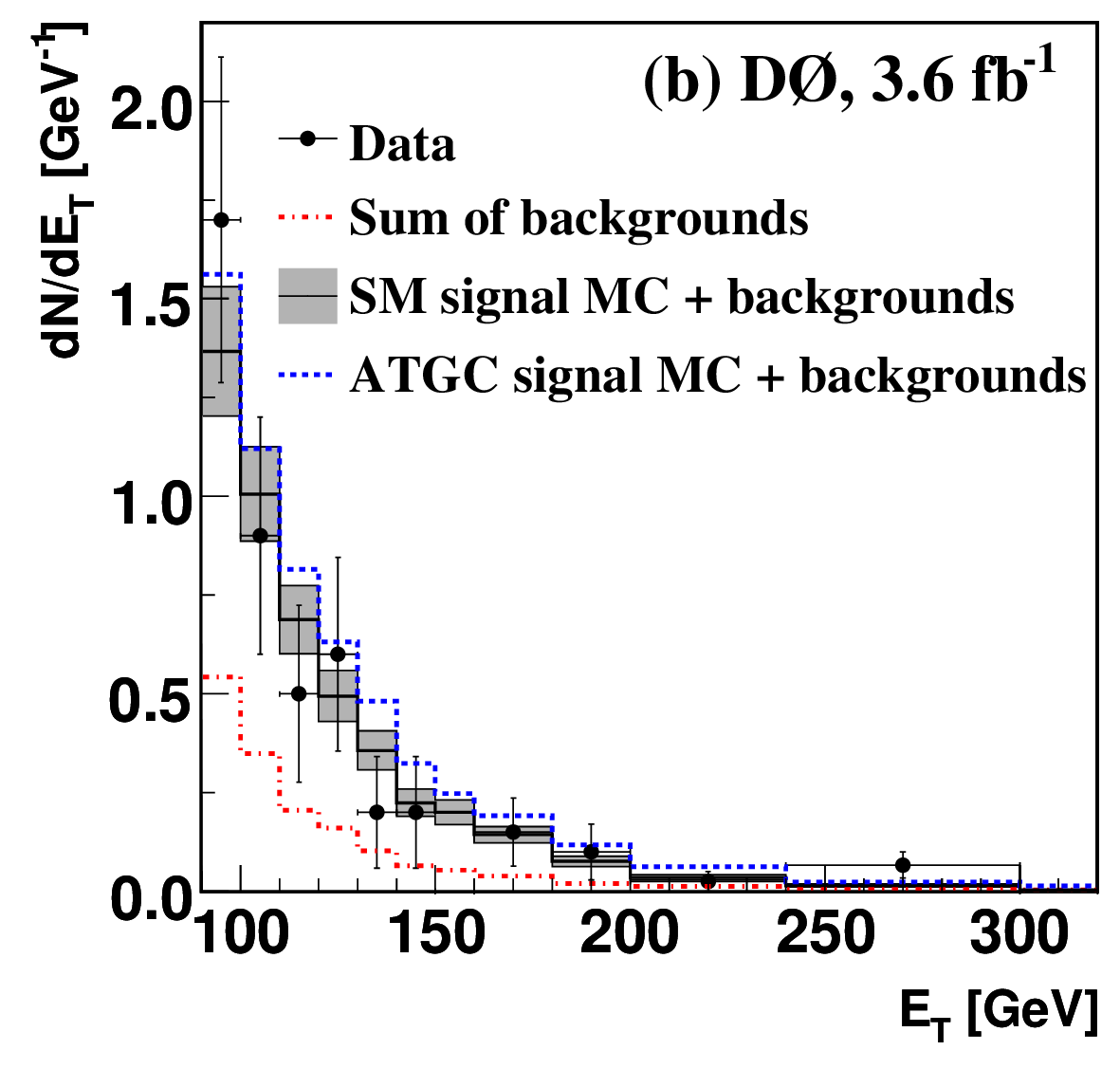}
%\begin{tabular}{cc}
%  \multirow{1}{*}[1.0in]{(a)} & \includegraphics[width=0.96\linewidth]{eps/gaugeBosons/0705.1550-fig04.eps}\\
%  \multirow{1}{*}[1.0in]{(b)} & \includegraphics[width=0.96\linewidth]{eps/gaugeBosons/0902.2157-fig01.eps}
%\end{tabular}    
  \caption{The observed photon $\Et$ spectrum along with the SM
    expectation and possible anomalous TGC scenarios for (a)
    $Z\gamma\rightarrow \ell\ell\gamma$ \cite{Abazov:2007wy} and (b)
    $Z\gamma\rightarrow \nu\nu\gamma$ \cite{Abazov:2009cj}
    channels.\label{fig:zg-dzero-photonEt}}
\end{figure}

\begin{figure}
  \includegraphics[width=0.48\linewidth]{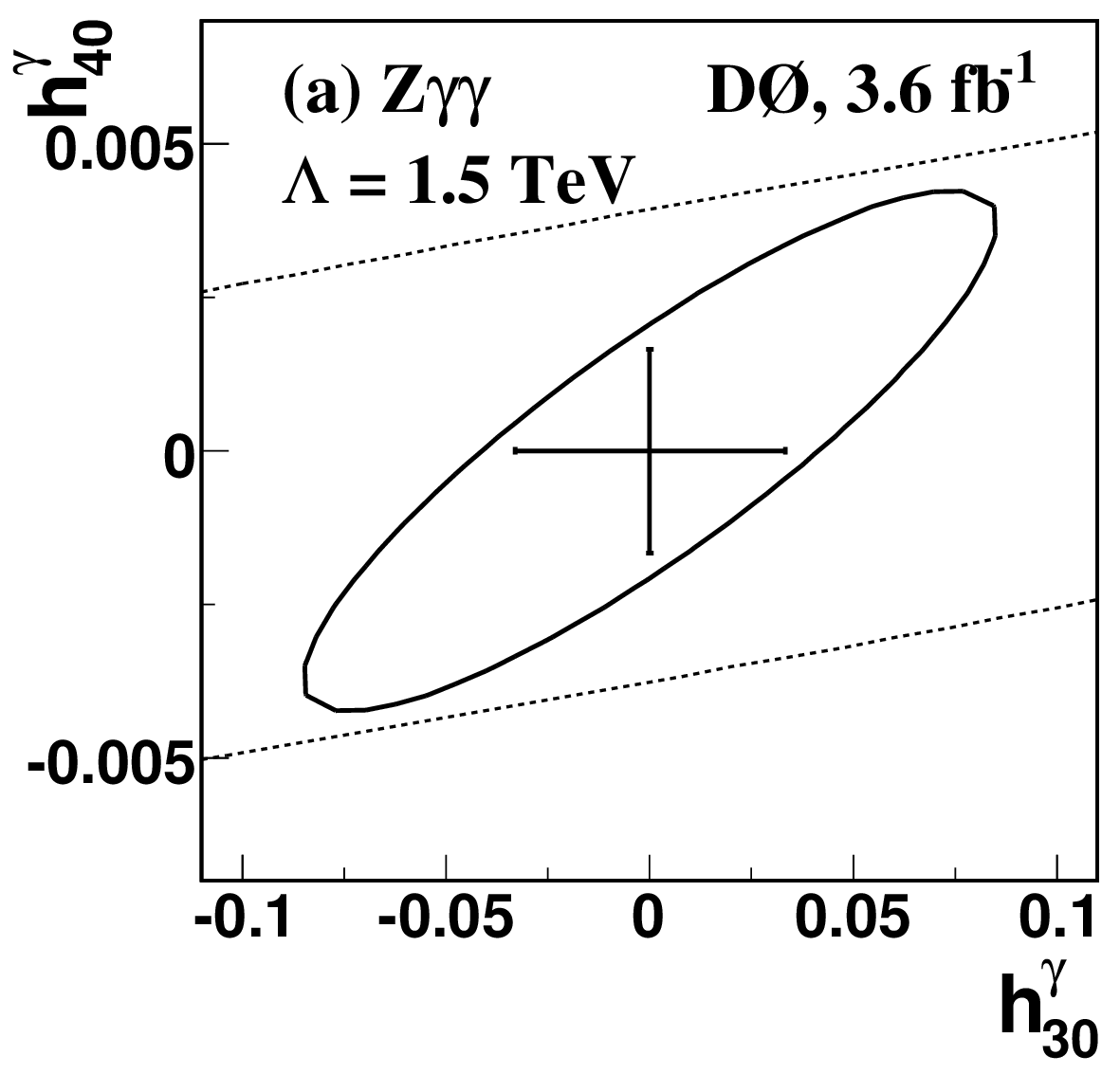}
  \includegraphics[width=0.48\linewidth]{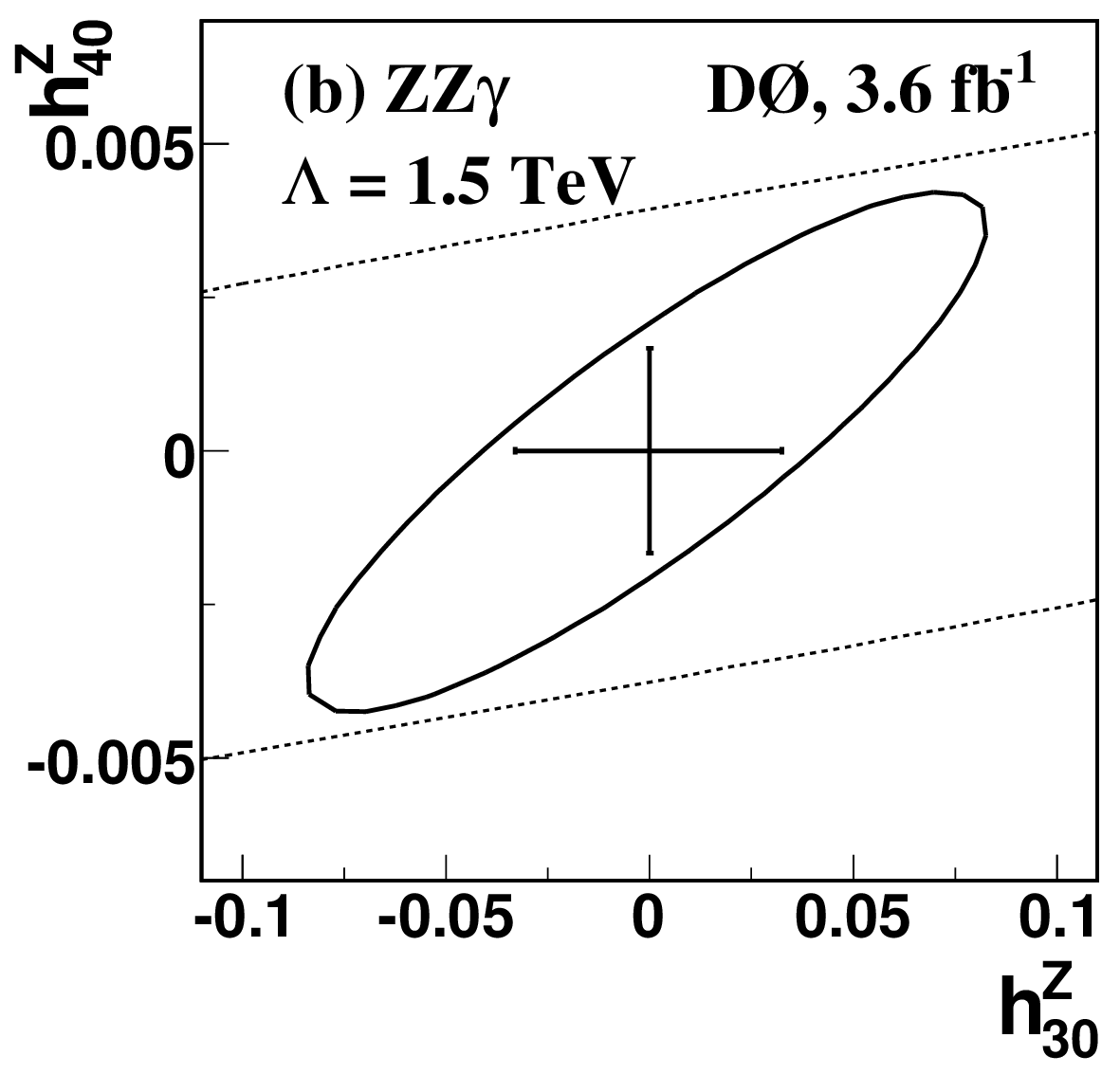}
  \caption{Two-dimensional bounds (ellipses) at 95\% CL on
    CP-conserving (a) $Z\gamma\gamma$ and (b) $ZZ\gamma$
    couplings. The crosses represent the one-dimensional bounds at the
    95\% CL setting all other couplings to zero. The dashed lines
    indicate the unitarity limits for $\Lambda =
    1.5$~TeV \cite{Abazov:2009cj}. \label{fig:zg-dzero-atgc}} 
\end{figure}

Both of the CDF and D0 $Z\gamma$ cross section measurements are
consistent with the SM expectations at NLO. D0 searches for anomalous
$Z\gamma\gamma$ and $ZZ\gamma$ couplings using the observed photon
$\Et$ spectrum in the $Z\gamma\rightarrow \ell\ell\gamma$
\cite{Abazov:2007wy} and $Z\gamma\rightarrow \nu\nu\gamma$
\cite{Abazov:2009cj}. A LO simulation
\cite{Baur:1992cd,PhysRevD.57.2823} of the $Z\gamma$ signal is used
with NLO corrections \cite{PhysRevD.57.2823} applied to the photon
$E_T$ spectrum. The observed photon $\Et$ spectrum along with the SM
expectation and possible anomalous TGC scenarios for two channels is
shown in Fig.~\ref{fig:zg-dzero-photonEt}. The $Z\gamma\rightarrow
\ell\ell\gamma$ and $Z\gamma\rightarrow \nu\nu\gamma$ combined limits
on the CP-conserving $Z\gamma\gamma$ and $ZZ\gamma$ couplings are
shown in Fig.~\ref{fig:zg-dzero-atgc}. The 1D combined limits at
95\% CL are $|h_{30}^\gamma | < 0.033$,  $|h_{40}^\gamma | < 0.0017$,
$|h_{30}^Z | < 0.033$, and $|h_{40}^Z | < 0.0017$ for
$\Lambda = 1.5$ TeV.

\subsubsection{$\WW$}
\label{sec:gaugeBosons_dibosons_WW}

Production of $W$ boson pairs in hadron (and lepton) collisions
involves both the $WW\gamma$ and $WWZ$ couplings. First evidence for
$W$ boson pair production was reported by CDF using Tevatron Run I
data~\cite{PhysRevLett.78.4536}. This process was later measured with
greater significance by D0 and CDF using $\intL=224-252 $~pb$^{-1}$
and $\intL=184 $~pb$^{-1}$ respectively, from Run II of the
Tevatron~\cite{PhysRevLett.94.151801,PhysRevLett.94.211801}. At LEP,
$WW$ production has been extensively studied and stringent limits on 
anomalous TGC were determined. At the Tevatron, much higher $WW$
invariant masses are probed compared to LEP because of the higher
accessible energies. Also, the $WW$ final state is a promising
discovery channel for the Higgs boson at both the Tevatron and the
LHC. In hadron collisions, the production of $W$ boson pairs is most
easily observed in the fully leptonic decay mode
$WW\rightarrow\ell\nu\ell\nu$. The experimental signature of the $WW$ 
signal in leptonic decay is two isolated high $\Et$ charged leptons
with opposite charge and large $\MET$ from the neutrinos. 

Both CDF and D0 have measured the $WW$ production cross section in
fully-leptonic decay and use their data to search for anomalous
$WW\gamma$ and $WWZ$ couplings. In both analyses, the dominant
backgrounds are from $t\bar{t}$, Drell-Yan, other diboson decays, and
$W$+jets where the jet fakes an isolated lepton. In the CDF analysis
\cite{Aaltonen:2009us}, $t\bar{t}$ is suppressed by requiring no
reconstructed jets with $\Et > 15$ GeV and $|\eta| < 2.5$. In the D0
analysis \cite{Abazov:2009ys}, the $\pt$ of the $WW$ system, estimated
from the observed charged lepton momenta and the $\MET$, is required
to be small ($<$20 $\GeVc$ ($ee$), 25 $\GeVc$ ($e\mu$), or 16
$\GeVc$($\mu\mu$)) in order to suppress $t\bar{t}$ decays.

The strategy for measuring the cross section for $p\bar{p}\rightarrow
WW + X$ differ between the CDF and D0 analyses. In the D0
analysis, the $WW$ signal yield is determined from counting the number
of events in excess of the expected SM backgrounds using $\intL=1.1 \fb$,
as shown in Tab.~\ref{tbl:dzero-ww}. D0 measures 
\begin{eqnarray*}
\lefteqn{\sigma(p\bar{p}\rightarrow WW + X)} \\ 
& & {} = 11.5~\pm~2.1\rm{(stat.+syst.)}~\pm~0.7 \rm{(lum.)}~\rm{pb} 
\end{eqnarray*}
\cite{Abazov:2009ys}, in agreement with the NLO expectation of
$12.0~\pm~0.7$ pb \cite{Campbell:1999ah}.

\begin{table}
\begin{ruledtabular}
\begin{tabular}{lr@{$\,\pm\,$}lr@{$\,\pm\,$}lr@{$\,\pm\,$}l}
Process & \multicolumn{2}{c}{$ee$} & \multicolumn{2}{c}{$e\mu$} & \multicolumn{2}{c}{$\mu\mu$} \\
\hline
$Z/\gamma^*\to ee/\mu\mu$	& 0.27 & 0.20 	& 2.52 & 0.56 	& 0.76 & 0.36 \\
$Z/\gamma^*\to\tau\tau$		& 0.26 & 0.05 	& 3.67 & 0.46 	& \multicolumn{2}{c}{---} \\
$t\bar{t}$			& 1.10 & 0.10	& 3.79 & 0.17	& 0.22 & 0.04 \\
$WZ$				& 1.42 & 0.14 	& 1.29 & 0.14 	& 0.97 & 0.11 \\
$ZZ$				& 1.70 & 0.04	& 0.09 & 0.01 	& 0.84 & 0.03 \\
$W\gamma$			& 0.23 & 0.16 	& 5.21 & 2.97 	& \multicolumn{2}{c}{---} \\
$W+\text{jet}$			& 6.09 & 1.72 	& 7.50 & 1.83	& 0.12 & 0.24 \\
Multijet			& 0.01 & 0.01	& 0.14 & 0.13 	& \multicolumn{2}{c}{---} \\
\hline
$WW\to\ell\ell^\prime$		&10.98 & 0.59 	&39.25 & 0.81  	& 7.18 & 0.34 \\
$WW\to\ell\tau/\tau\tau\to\ell\ell^\prime$& 1.40 & 0.20  & 5.18 & 0.29 	& 0.71 & 0.10 \\
\hline
Total expected			&23.46 & 1.90 	&68.64 & 3.88 	&10.79 & 0.58 \\ 
Data		& \multicolumn{2}{c}{22} & \multicolumn{2}{c}{64} & \multicolumn{2}{c}{14} \\
\end{tabular}
\caption{\label{tbl:dzero-ww}
Numbers of signal and background events expected and number of events
observed after the final event selection in each channel for the D0
$WW$ cross section measurement \cite{Abazov:2009ys}.
}
\end{ruledtabular}
\end{table}

In the CDF analysis, the $WW$ signal yield is extracted from a fit to
the distribution of a matrix element likelihood ratio ($LR_{WW}$)
discriminant for events using $\intL=3.6 \fb$. The events which are fit 
are required to pass relatively loose selection criteria as compared to
the selection CDF would use for a cross section measurement based on
the event yield alone. Tab.~\ref{tbl:cdf-WWExpected} shows the expected
\begin{table}
\begin{ruledtabular}
\begin{tabular}{lc}
Process           &   Events   \\\hline
$Z/\gamma^*$ (Drell-Yan)     &          79.8 $\pm$ 18.4  \\
$WZ$              &          13.8 $\pm$     1.9  \\
$W\gamma$         &          91.7 $\pm$    24.8  \\
$W+1-$jet	  &         112.7 $\pm$    31.2  \\
$ZZ$              &          20.7 $\pm$     2.8  \\
$t \bar t$        &           1.3 $\pm$     0.2  \\\hline
Total Background  &         320.0 $\pm$    46.8  \\\hline
$W^+W^-$          &         317.6 $\pm$    43.8  \\\hline
Total Expected    &         637.6 $\pm$    73.0  \\\hline
Data              &  654       \\
\end{tabular}
\caption{Expected number of signal ($WW$) and background events
  along with the total number of expected and observed events in the
  data for the CDF $WW$ cross section measurement \cite{Aaltonen:2009us}.}
\label{tbl:cdf-WWExpected}
\end{ruledtabular}
\end{table}
number of signal and background events along with the observed events
in the data used to fit for the signal. For each event passing the
signal selection criteria, four matrix-element-based (ME) event
probabilities are calculated corresponding to the production and decay
processes $WW \rightarrow \ell\nu\ell\nu$, $ZZ
\rightarrow \ell\ell\nu\nu$, $W+1-$jet $\rightarrow \ell \nu +$
1-jet, and $W\gamma \rightarrow \ell \nu + \gamma$. In the latter two
processes, the jet or $\gamma$ is assumed to have been reconstructed
as a charged lepton candidate.  The event probability for a process
$X$ is given by
\begin{equation}
\label{eqn:MEProb}
P_{X}(\vec x) = \frac{1}{\langle \sigma \rangle} \int \frac{d
  \sigma(\vec y) }{d \vec y} \; \epsilon(\vec y) \; G(\vec x, \vec y)
\; d\vec y 
\end{equation}
where $\vec x$ represents the observed lepton momenta and $\MET$ vectors, 
$G(\vec x, \vec y)$ is a transfer function representing the detector
resolution, and $\epsilon(\vec y)$ is an efficiency function
parametrized by $\eta$ which quantifies the probability for a particle
to be reconstructed as a lepton. The differential cross section
$\frac{d \sigma(\vec y) }{d \vec y}$ is calculated using leading-order
matrix elements from the {\sc mcfm} program~\cite{Campbell:1999ah} and
integrated over all possible true values of the final state particle
4-vectors $\vec y$. The normalization factor $\langle \sigma \rangle$
is determined from the leading-order cross section and detector
acceptance for each process. These event probabilities are combined
into a likelihood ratio 
\begin{equation}
\label{eqn:LR}
LR_{WW} = \frac{ P_{WW} }{ P_{WW} + \sum_{j} k_j P_j},
\end{equation}
where $j = \{ZZ,$ $W+1-$jet, $W\gamma\}$ and  $k_j$ is the relative
fraction of the expected number of events for the $j$-th process such
that $\sum_j k_j = 1$.  The templates of the $LR_{WW}$ distribution are
created for signal and each background process given in
Tab.~\ref{tbl:cdf-WWExpected}.

A binned maximum likelihood is used to extract the $WW$ production
cross section from the shape and normalization of the $LR_{WW}$
templates. The likelihood is formed from the Poisson probabilities of
observing $n_i$ events in the $i$-th bin when $\mu_i$ are expected.
Variations corresponding to the systematic uncertainties described
previously are included as normalization parameters for signal and
background, constrained by Gaussian terms. The likelihood is given by
\begin{equation}
\label{eqn:Likelihood}
\mathcal{L} = \left(\prod_i \frac{\mu_i^{n_i} e^{-\mu_i}}{n_i!}
\right) \cdot \prod_c e^{-\frac{S_c^2}{2}} 
\end{equation}
where
\begin{equation}
\label{eqn:MuI}
\mu_i = \sum_k \alpha_k \left[ \prod_c (1 + f^c_k S_c ) \right]  (N^{Exp}_k)_i ,
\end{equation}
$f^c_k$ is the fractional uncertainty for the process $k$ due to the
systematic $c$, and $S_c$ is a floating parameter associated with the
systematic uncertainty $c$. The correlations of systematic
uncertainties between processes are accounted for in the definition of
$\mu_i$. The expected number of events from process $k$ in the $i$-th
bin is given by $(N^{Exp}_k)_i$. The parameter $\alpha_k$ is an
overall normalization parameter for process $k$ and is fixed to unity
for all processes other than $WW$, for which it is freely 
floating. The likelihood is maximized with respect to the systematic
parameters $S_c$ and $\alpha_{WW}$. The $WW$ cross section is then
given by the fitted value of $\alpha_{WW}$ multiplied by $\sigma^{NLO}
(p \bar p \to WW)$. 

\begin{figure}
\includegraphics[width=\linewidth]{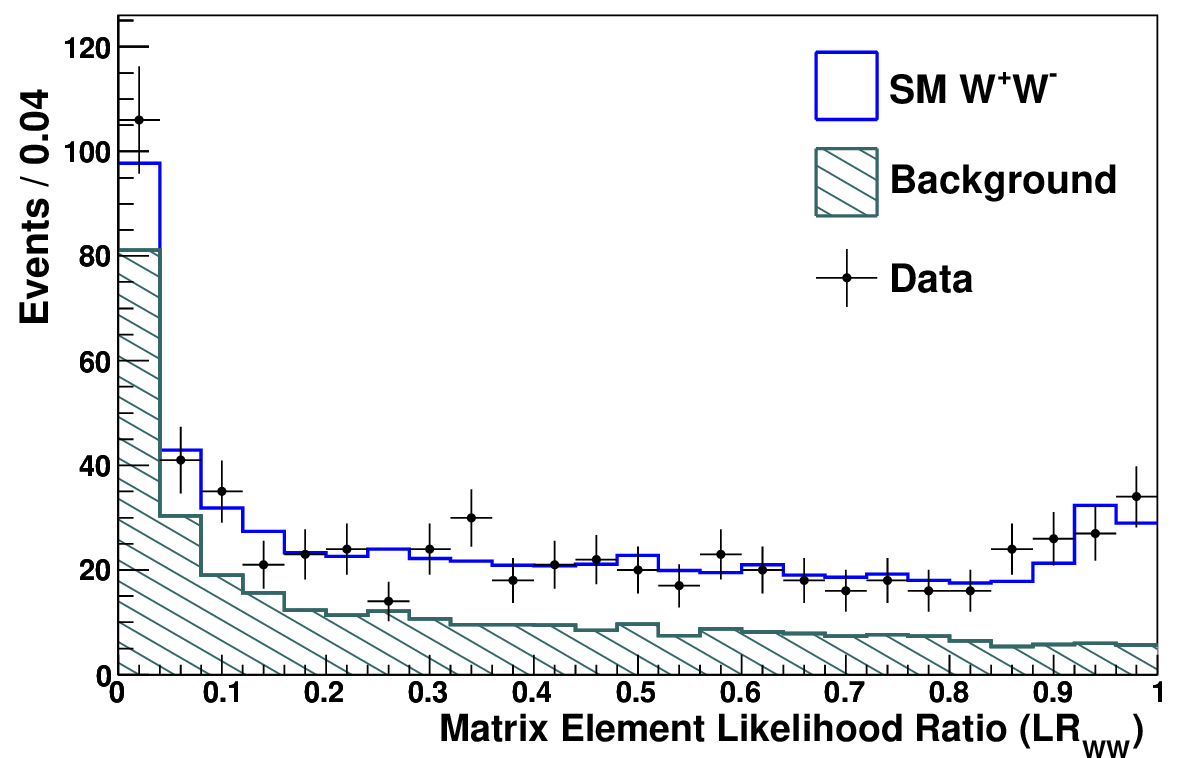}
\caption{The $LR_{WW}$ distributions for the signal ($W^+W^-$) and
  background processes after a maximum likelihood fit to the data for
  the CDF $WW$ cross section measurement \cite{Aaltonen:2009us}.}
  \label{fig:cdf-WWFit}
\end{figure}

The fit to the data of the signal and sum of the individually fitted
background templates is show in Fig.~\ref{fig:cdf-WWFit}. The measured
$WW$ production cross section is
\begin{eqnarray*}
\lefteqn{\sigma(p\bar{p}\rightarrow WW + X)} \\
& & {}  = 12.1~\pm~0.9 \rm{(stat.)}~^{1.6}_{-1.4} \rm{(syst.)}~\rm{pb} 
\end{eqnarray*}
\cite{Aaltonen:2009us} in agreement with the NLO theoretical
expectation of $11.7~\pm~0.7$ pb \cite{Campbell:1999ah}.

\begin{figure}
\includegraphics[width=\linewidth]{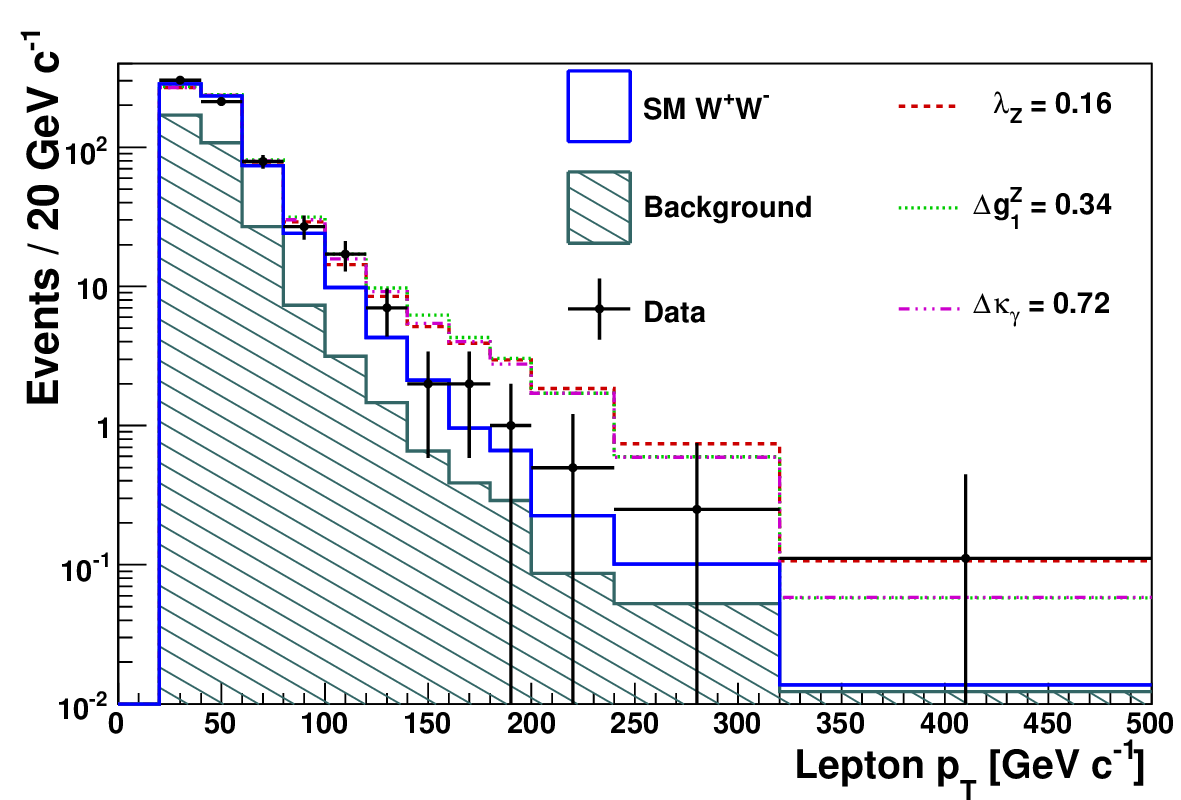}
\caption{Leading-lepton $p_T$ distribution for data compared to the SM
  expectation for the CDF $WW\rightarrow\ell\nu\ell\nu$ analysis
  \cite{Aaltonen:2009us}. Also shown is how the expectation would be
  modified by anomalous couplings near the observed limits.}
\label{fig:cdf-TGC}
\end{figure}

\begin{figure}
\mbox{
\includegraphics[width=1.62in]{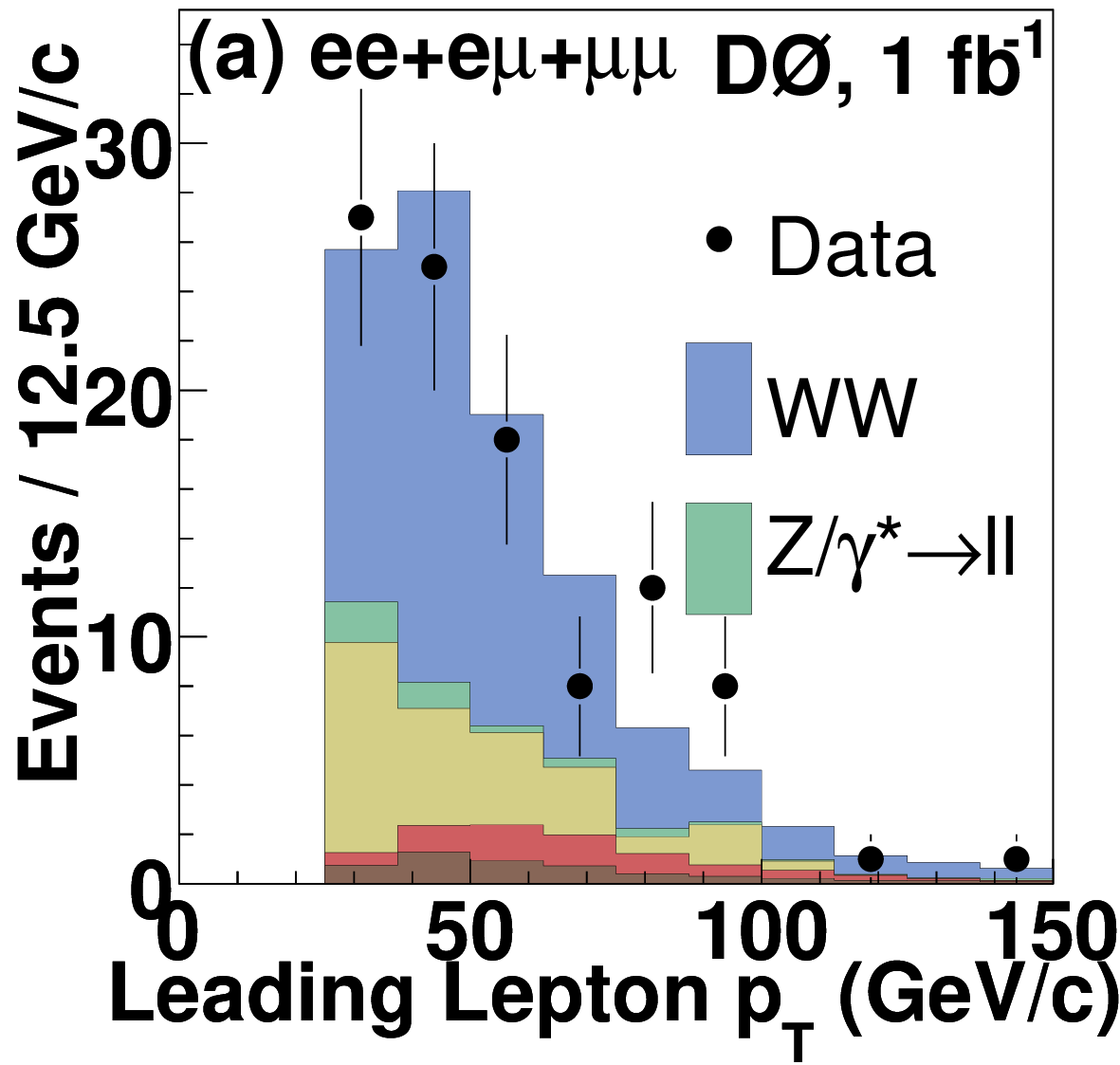} \hfill
\includegraphics[width=1.62in]{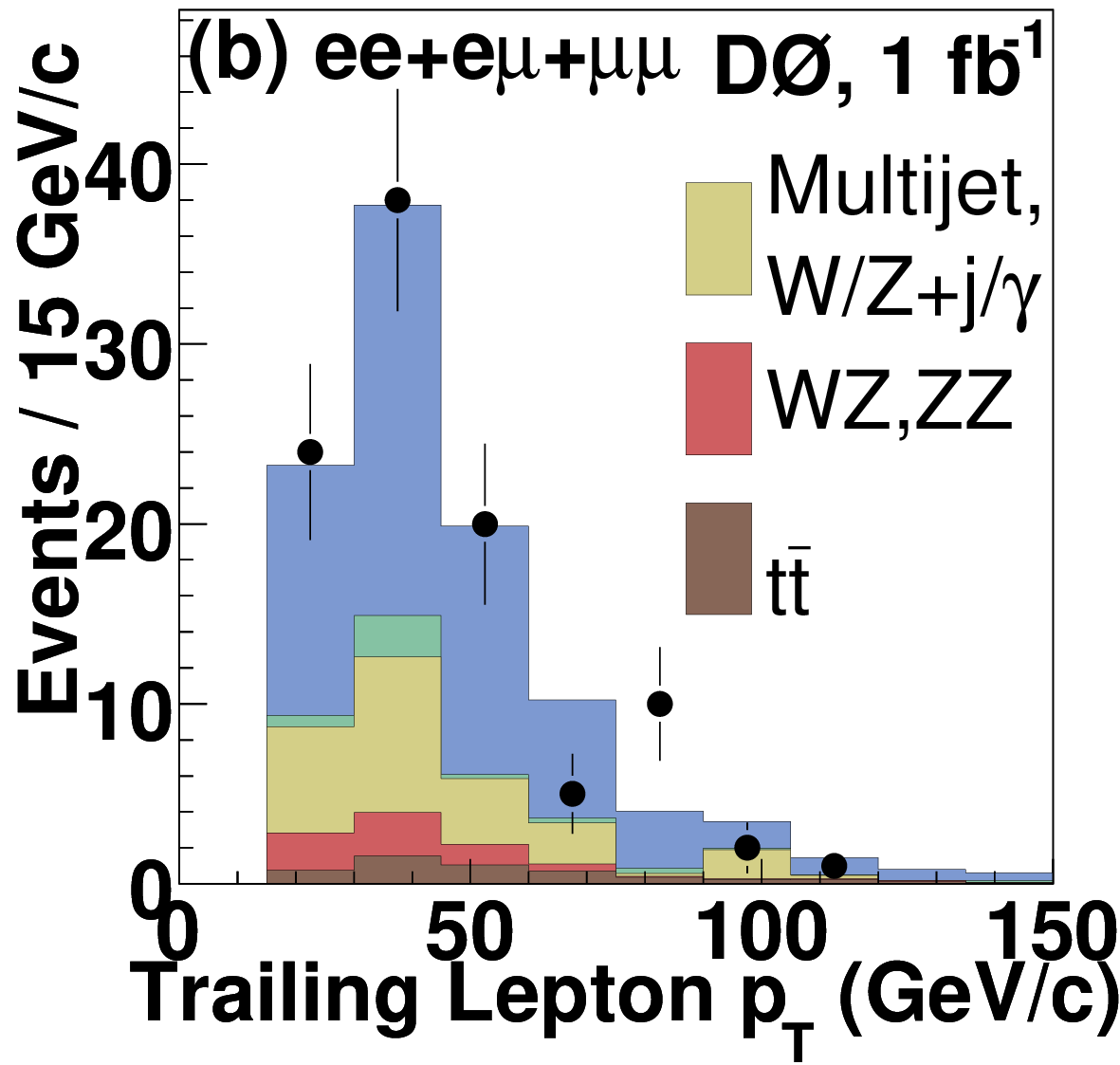}
}
\caption{\label{fig:dzero-leptonpt}
Distributions of the (a) leading and (b) trailing lepton $p_T$ for the
D0 $WW\rightarrow\ell\nu\ell\nu$ analysis \cite{Abazov:2009ys}. Data
are compared to estimated signal, $\sigma(WW)=12$~pb, and background
sum.}
\end{figure}

Both of the CDF and D0 $WW$ cross section measurements are consistent
with the SM expectations at NLO. CDF searches for anomalous $WW\gamma$
and $WWZ$ couplings using the observed leading charged lepton $\pt$
spectrum (see Fig.~\ref{fig:cdf-TGC}). In the D0 analysis, the
subleading (trailing) lepton $\pt$ is also included in a 2D histogram
with the leading lepton $\pt$ to constrain possible anomalous
couplings (see Fig.~\ref{fig:dzero-leptonpt} for the 1D
projections). 

There are several ways to relate the $WW\gamma$ and $WWZ$ couplings in
the presence of new physics. This is a convenient prescription to
reduce the number of parameters since $WW$ production involves both
$WW\gamma$ and $WWZ$ couplings. Enforcing $SU(2)_L\otimes U(1)_Y$
symmetry introduces two relationships between the remaining
parameters: $\kappa_Z = g_1^Z - (\kappa_\gamma - 1) \mathrm{tan}^2
\theta_W$ and $\lambda_Z = \lambda_\gamma$,  reducing the number of
free parameters to
three~\cite{DeRujula:1991se,Hagiwara:1993ck}. Alternatively, enforcing
equality between the $WW\gamma$ and $WWZ$ vertices ($WW\gamma$=$WWZ$)
such that $\kappa_\gamma=\kappa_Z$, $\lambda_\gamma=\lambda_Z$, and
$g^Z_1=1$ reduces the number of free parameters to two.

\begin{figure}
\mbox{
\includegraphics[width=1.62in]{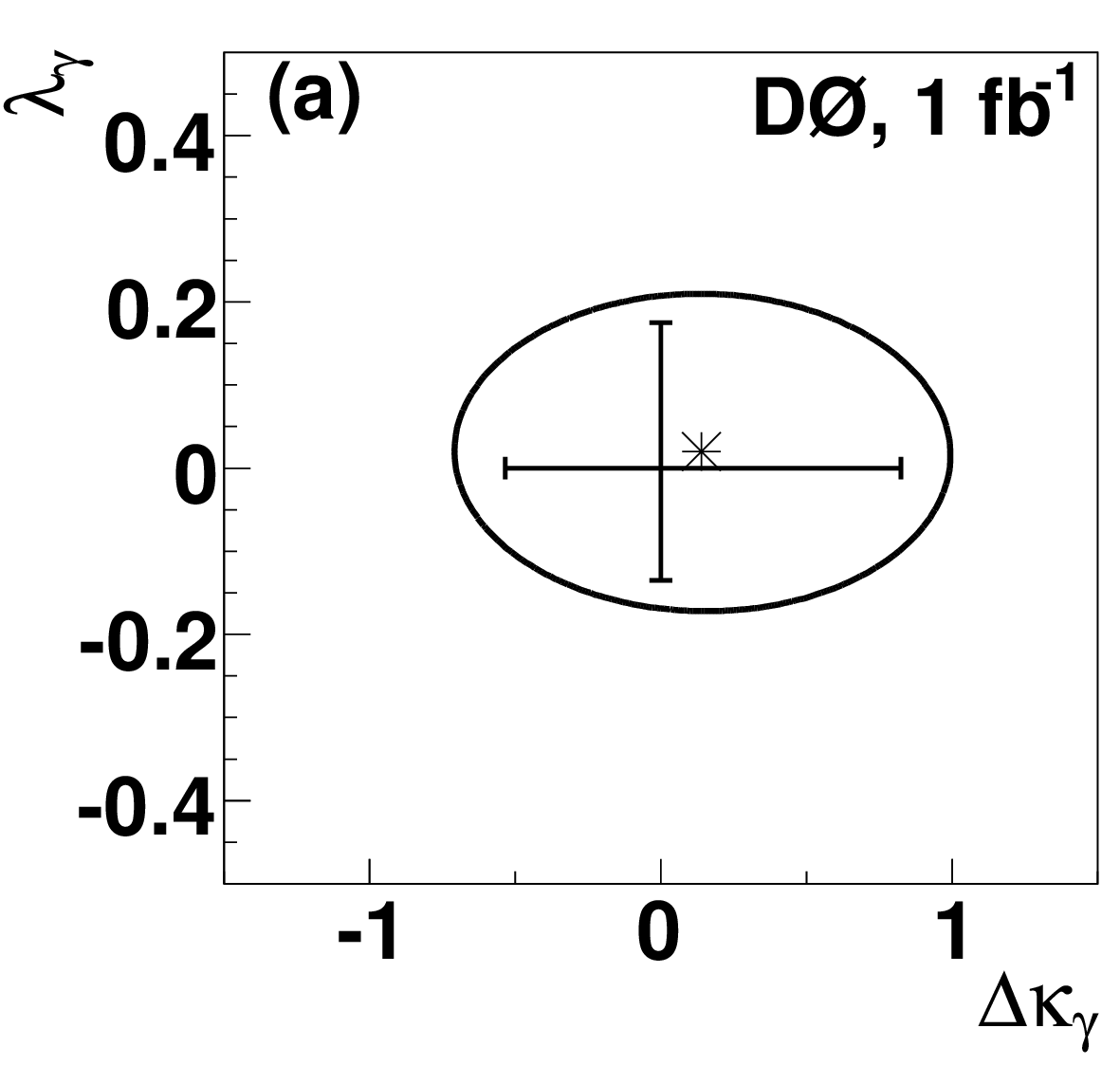} \hfill
\includegraphics[width=1.62in]{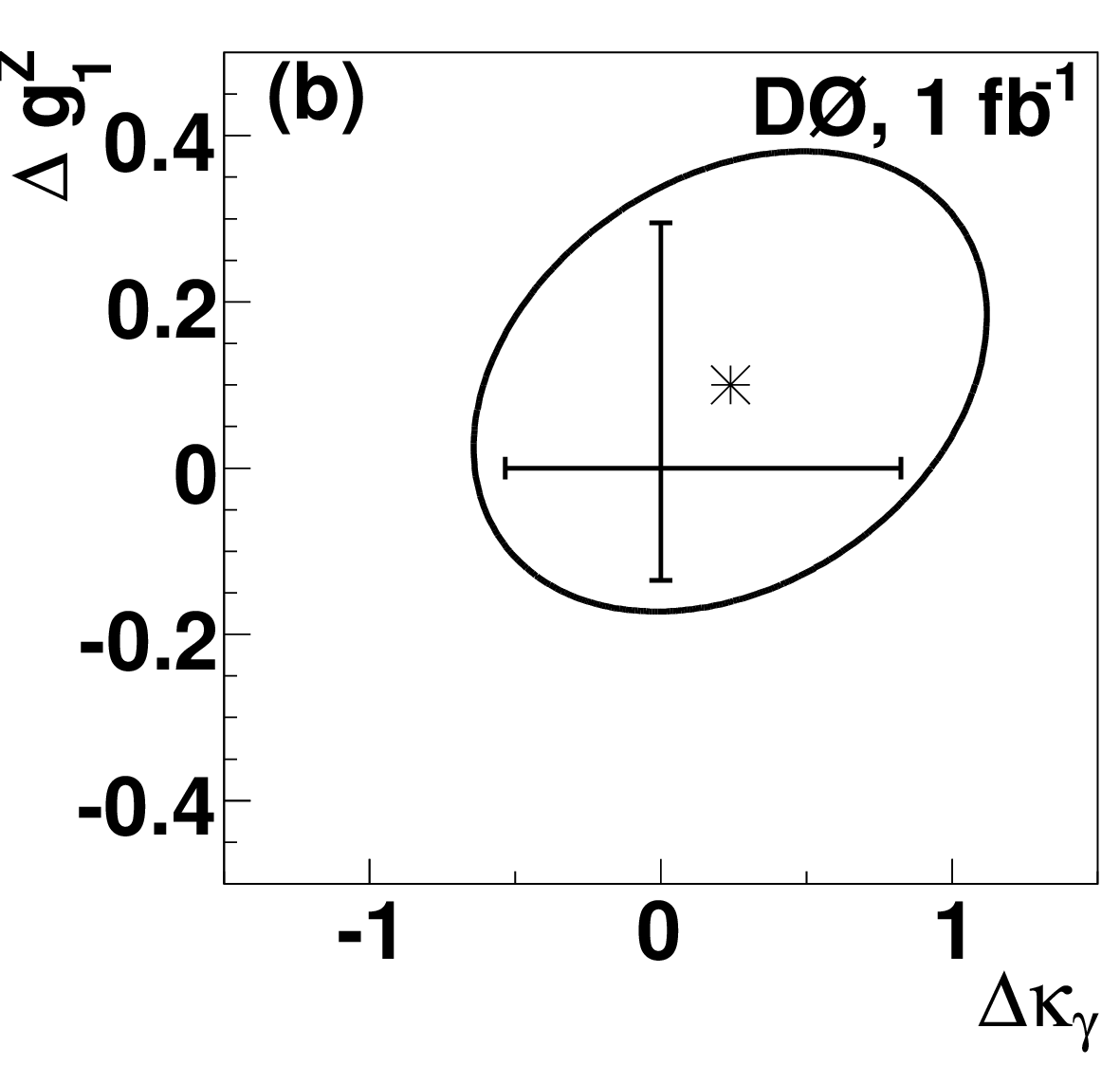}
}
\mbox{
\includegraphics[width=1.62in]{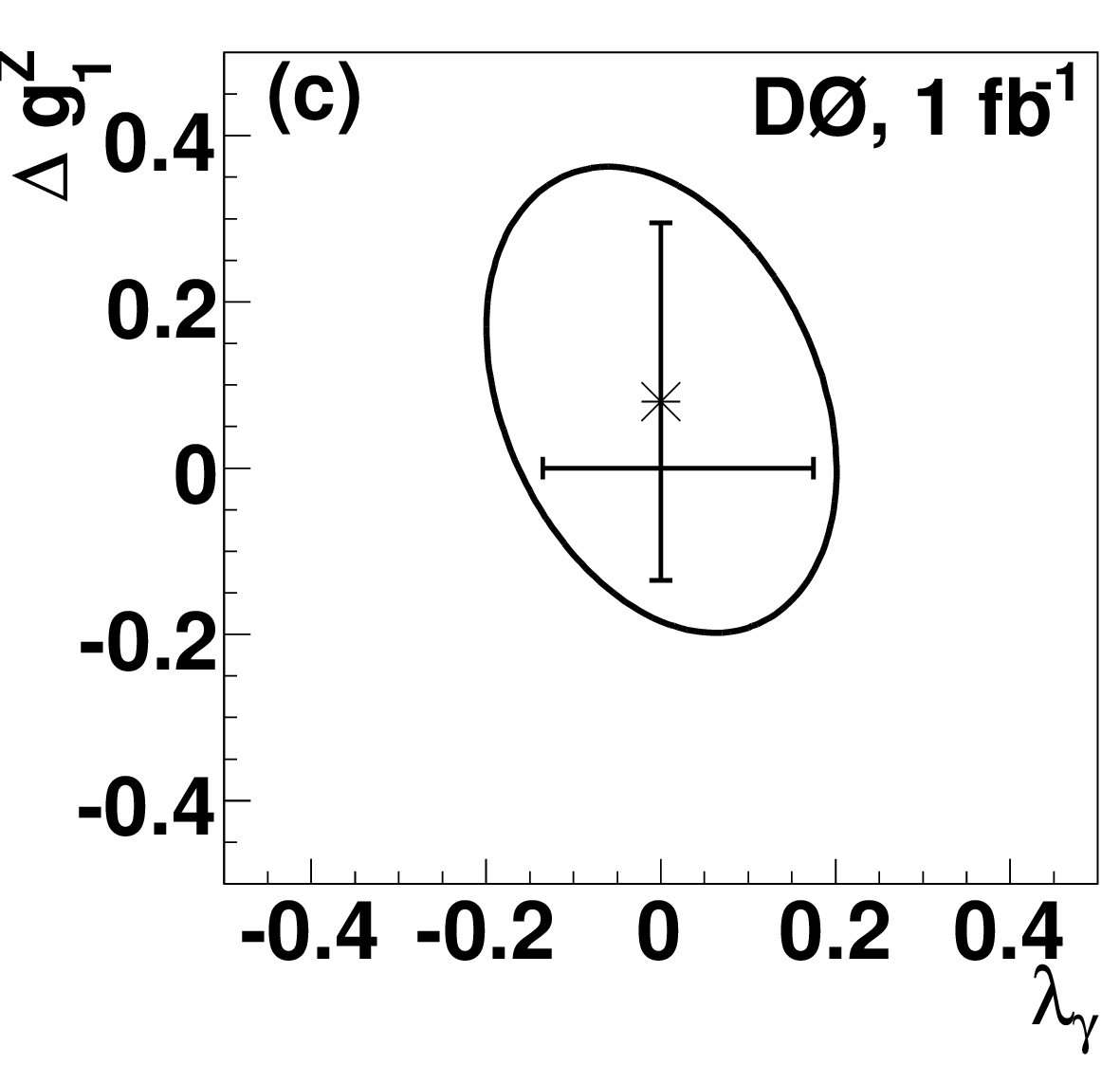} \hfill
\includegraphics[width=1.62in]{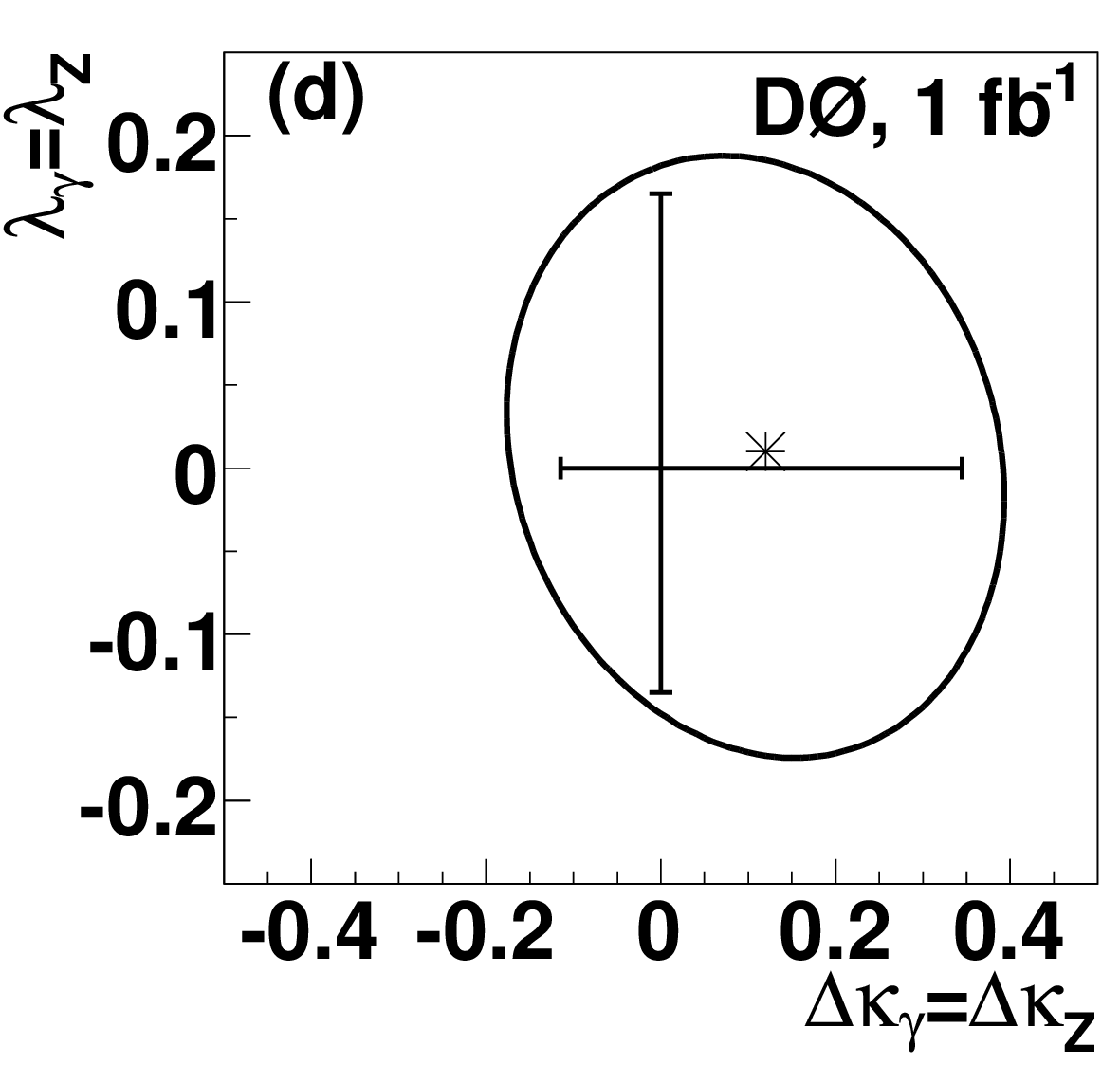}
}
\caption{\label{fig:dzero-aclimits}
One and two-dimensional 95\%~CL\ limits for the D0
$WW\rightarrow\ell\nu\ell\nu$ analysis \cite{Abazov:2009ys} when
enforcing $SU(2)_L \otimes U(1)_Y$ symmetry at $\Lambda = 2$ TeV, for
(a) $\Delta\kappa_\gamma$ vs. $\lambda_\gamma$,
(b) $\Delta\kappa_\gamma$ vs. $\Delta g_1^Z$, and
(c) $\lambda_\gamma$ vs. $\Delta g_1^Z$,
each when the third free coupling is set to its SM value; limits when
enforcing the $WW\gamma$=$WWZ$ constraints are shown in (d). The curve
represents the two-dimensional 95\%~CL\ contour and the ticks along
the axes represent the one-dimensional 95\%~CL\ limits. An asterisk
($+ \hspace{-0.8em} \times$) marks the point with the highest
likelihood in the two-dimensional plane.}
\end{figure}

In the D0 analysis, the one-dimensional 95\%~CL\ limits for $\Lambda =
2$~TeV are determined to be $-0.54 < \Delta\kappa_\gamma < 0.83$,
$-0.14 < \lambda_\gamma=\lambda_Z < 0.18$, and $-0.14 < \Delta g_1^Z <
0.30$ under the $SU(2)_L\otimes U(1)_Y$-conserving constraints, and
$-0.12 < \Delta\kappa_\gamma=\Delta\kappa_Z < 0.35$, with the same
$\lambda$ limits as above, under the $WW\gamma$=$WWZ$
constraints. One- and two-dimensional 95\%~CL\ limits are shown in
Fig.~\ref{fig:dzero-aclimits}. In the CDF analysis, only 1D limits on
the anomalous coupling parameters under the assumption of
$SU(2)_L\otimes U(1)_Y$ invariance are reported. The expected and
observed 95\% confidence limits are shown in Tab.~\ref{tbl:cdf-TGC}
where it is evident that the limits are weaker than expected. The
probability of observing these limits in the presence of only SM $WW$
production ranges from 7.1\% to 7.6\% depending on the coupling
parameters ($\lambda_Z$, $\Delta \kappa_\gamma$, $\Delta g_1^Z$) and
are consistent with a statistical fluctuation of SM physics.

\begin{table}
\begin{ruledtabular}
\begin{tabular}{llcccc}
&&$ \Lambda$ (TeV)  &   $\lambda_\gamma = \lambda_Z$   &   $\Delta
\kappa_\gamma$ &  $\Delta g_1^Z$ \\
\hline
Expected & ~~ & 1.5     & (-0.05,0.07) & (-0.23,0.31) & (-0.09,0.17) \\
Observed & ~~ & 1.5     & (-0.16,0.16) & (-0.63,0.72) & (-0.24,0.34) \\
\hline 	 
Expected & ~~ & 2.0     & (-0.05,0.06) & (-0.20,0.27) & (-0.08,0.15)  \\
Observed & ~~ & 2.0     & (-0.14,0.15) & (-0.57,0.65) & (-0.22,0.30) \\
\end{tabular}
\caption{Expected and observed limits from the CDF
  $WW\rightarrow\ell\nu\ell\nu$ analysis searching for anomalous TGCs
  assuming two different values of the form factor scale $\Lambda$
  \cite{Aaltonen:2009us}. For each coupling limit set, the two other
  couplings are fixed at their SM values. Values of the couplings
  outside of the given observed range are excluded at the 95\%
  confidence level (CL).}
\label{tbl:cdf-TGC}
\end{ruledtabular}
\end{table}

\subsubsection{$\WZ$}
\label{sec:gaugeBosons_dibosons_WZ}

\begin{figure}
\unitlength=0.33\linewidth
\includegraphics[width=0.24\textwidth]{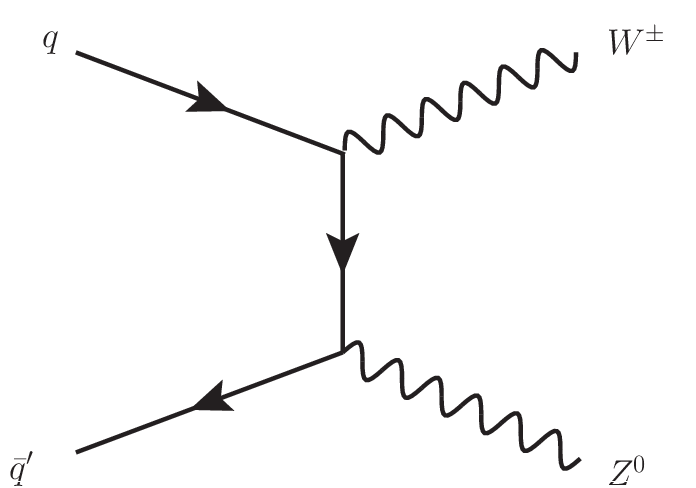}
\put(-0.80,0.0){(a)} 
\unitlength=0.33\linewidth
\includegraphics[width=0.24\textwidth]{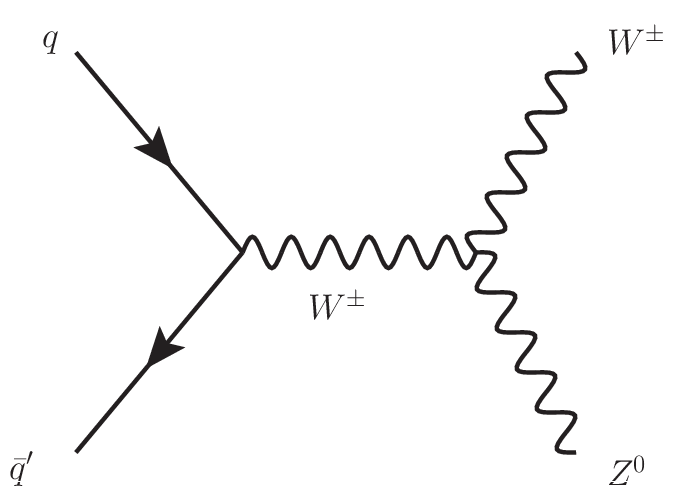}
\put(-0.80,0.0){(b)} 
\caption{Leading order (a) t-channel and (b) s-channel diagrams for
  $\WZ$ production at the Tevatron} 
\label{fig:wz}
\end{figure}

The $\WZ$ final state is not available in $\eebar$ collisions at LEP
but can be produced in $\ppbar$ collisions at the Tevatron. The study
of associated production of a $W$ and $Z$ boson is important for a
number of reasons. The production of $WZ$ involves the $WWZ$ TGC as
shown in the s-channel diagram in Fig.~\ref{fig:wz}. Unlike $WW$
production which involves a combination of the $WW\gamma$ and $WWZ$
couplings such that assumptions regarding their relation must be
invoked to interpret any anomalies observed in the data, the $\WZ$
production characteristics can be used to make a model-independent
test of the SM $WWZ$ coupling. Stated slightly differently, in $WZ$
production measurements, a direct  measure of the $WZ$ coupling
independent of the $W\gamma$ coupling can be made and compared to the
SM predictions. The fully-leptonic decay mode of $WZ$ provides a clean
SM trilepton signal which is analogous to the so-called {\it 
  golden mode} for discovering supersymmetry (SUSY) at the Tevatron via
chargino-neutralino production (${\tilde \chi}_1^\pm {\tilde
  \chi}_2^0$) and decay. Therefore, an observation of the SM $\WZ$
trilepton signal represents an important experimental milestone in
demonstrating sensitivity to the SUSY {\it golden mode} and other new
physics signatures in multileptons.

Prior to the start of Run II at the Tevatron, $\WZ$ production had
not been observed. The next-to-leading order (NLO) $\WZ$ cross section
prediction for $p\overline{p}$ collisions at $\sqrt{s} = 1.96~{\rm
  TeV}$ is 3.7 $\pm$ 0.3 pb \cite{Campbell:1999ah}. In October 2006,
$WZ$ production was first observed by the CDF Collaboration in the
three charged lepton + $\MET$ final state using 1.1$\fb$ of
integrated luminosity \cite{Abulencia:2007tu}. The most sensitive 
previous search for $\WZ$ production was reported by the $\Dzero$
Collaboration using $\intL=0.3\ {\rm fb}^{-1}$,
where three $\WZdecay$ candidates were found \cite{Abazov:2005ys}. The
observed candidates had a probability of 3.5\% to be due to background
fluctuations, corresponding to $\sigma(WZ) < 13.3$ pb at 95\% CL. The
D0 collaboration published an update to \cite{Abazov:2005ys} with
additional data to measure the $WZ$ production cross section and
search for anomalous $WWZ$ couplings \cite{Abazov:2007rab}.

As with other diboson processes, the signal for $WZ$ production is
most easily measured in fully-leptonic decay. The experimental
signature of $WZ$ production is three isolated high $\Et$ charged
leptons, at least two of which having the same flavor and opposite
charge with invariant mass consistent with decay of a $Z$ boson, and
large $\MET$ consistent with a neutrino from $W$ decay. To observe
$WZ$ in fully-leptonic decay, high acceptance for charged leptons is
required since all three must be detected to suppress backgrounds from
larger cross section processes. In the CDF analysis
\cite{Abulencia:2007tu}, a novel lepton identification strategy was
developed to maximize charged lepton acceptance while keeping the
backgrounds comparatively low by exploiting the charge and flavor
correlation of identified leptons in the events. The standard electron
and muon identification was combined with forward electron candidates
beyond the tracking acceptance and a ``track-only'' lepton category
consisting of high-quality tracks that neither project to the fiducial
regions of the calorimeters nor are identified as muons by the muon
chambers. Due to the lack of calorimeter information, electrons and
muons cannot be reliably differentiated for this category, and are
therefore treated as having either flavor in the $\WZ$ candidate
selection. For forward electrons without a matched track, both charge
hypotheses are considered when forming $WZ$ candidates, since the
charge is determined from the track curvature.

Other SM processes that can lead to three high-$\pt$ leptons include
dileptons from the Drell-Yan $Z/\gamma^{*}$ process (DY), with an
additional lepton from a photon conversion ($Z\gamma$) or a
misidentified jet ($Z$+jets) in the event; $ZZ$ production where only
three leptons are identified and the unobserved lepton results in
$\MET$; and a small contribution from $\ttbar \rightarrow \W b\W
\bar{b}$, where two charged leptons result from the $W$ boson decays
and one or more from decay of the $b$-quarks. Except for $\ttbar$,
these backgrounds are suppressed by requiring $\MET>25$ GeV in the
event, consistent with the unobserved neutrino from the leptonic decay
of a $W$ boson.

\begin{figure*}
\unitlength=0.33\linewidth
\includegraphics[width=0.333\linewidth]{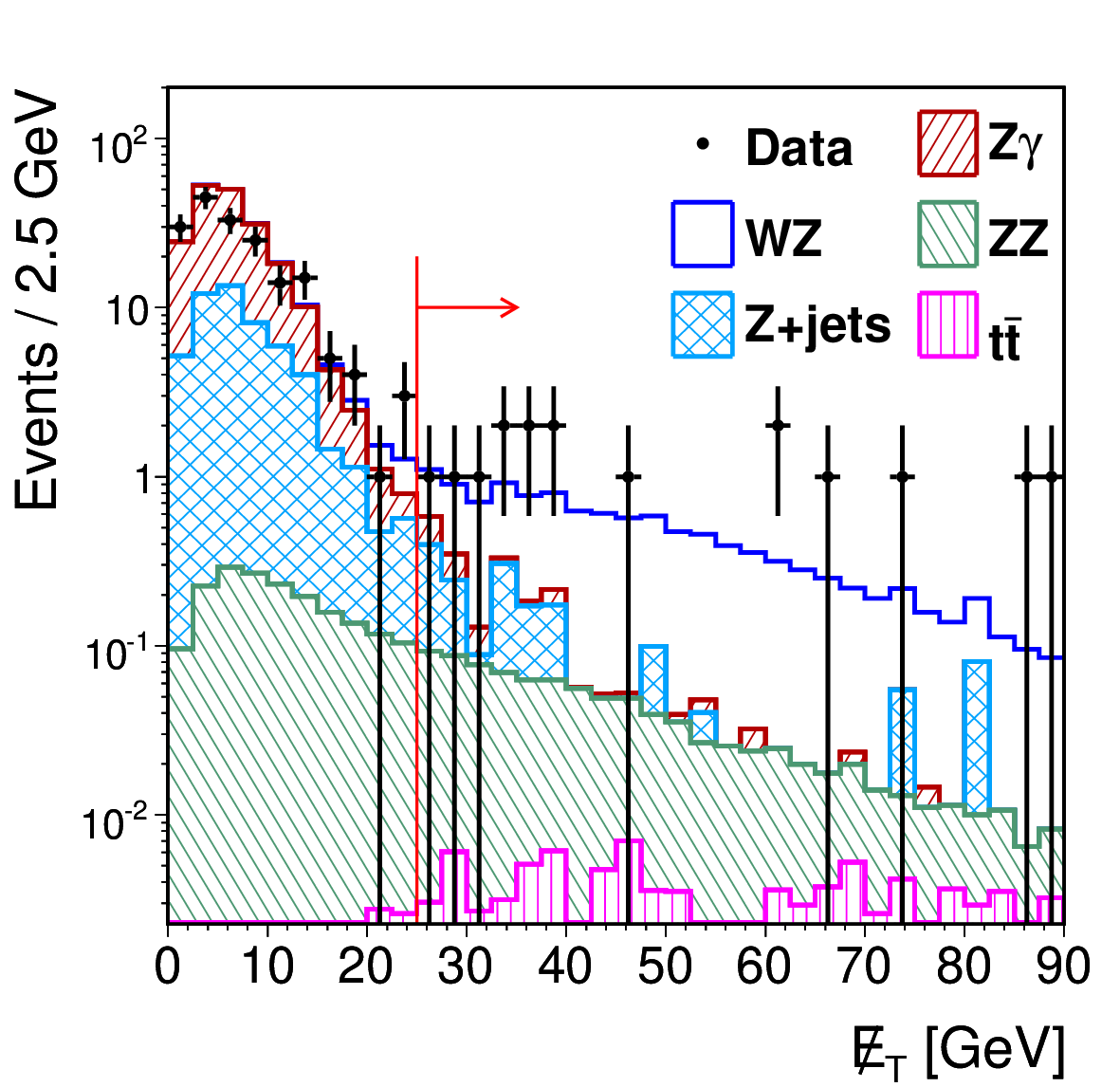}
\put(-0.70,0.80){(a)} 
\includegraphics[width=0.333\linewidth]{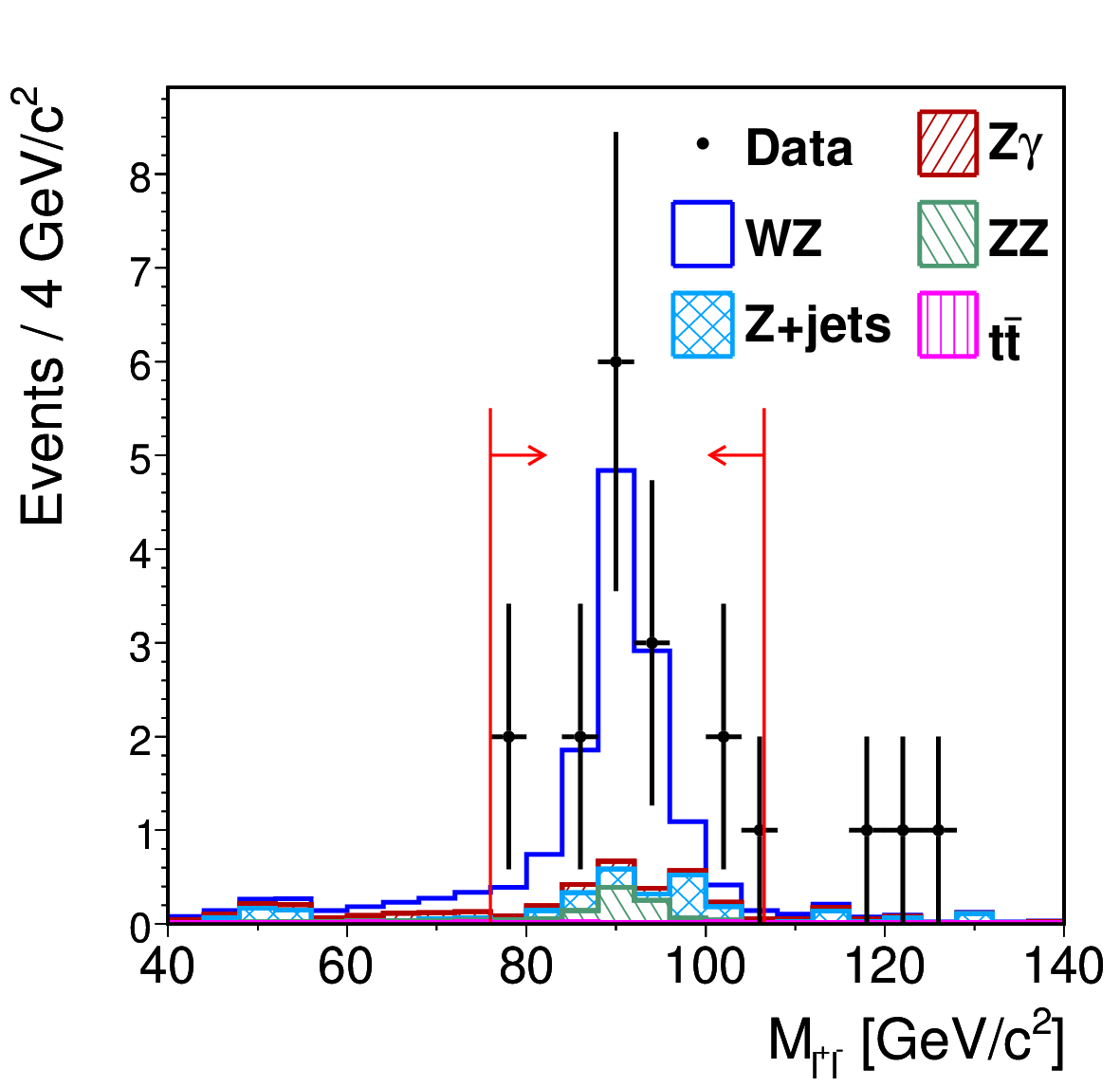}
\put(-0.70,0.80){(b)} 
\includegraphics[width=0.333\linewidth]{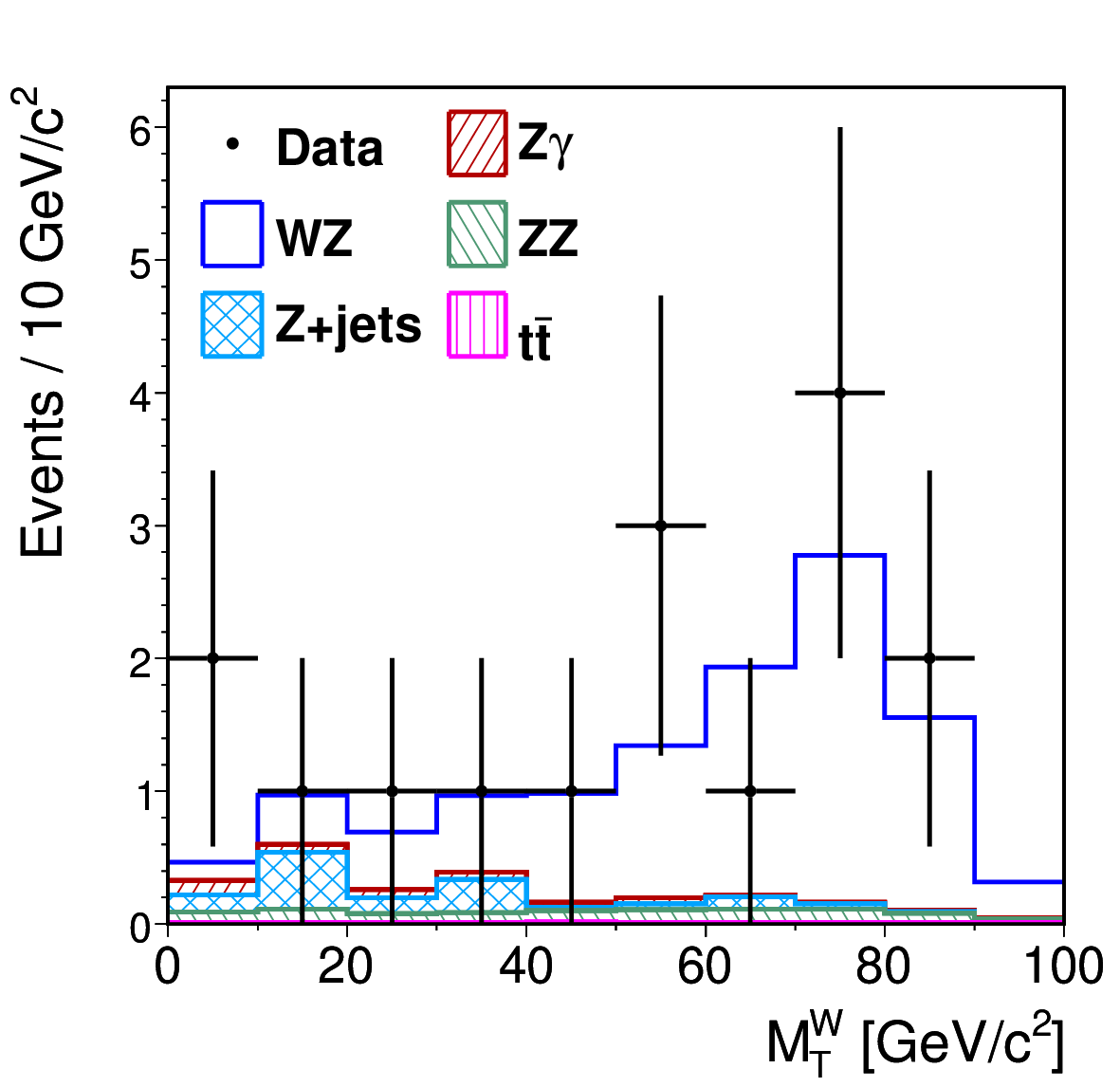}
\put(-0.19,0.80){(c)} 
\caption{Distributions for $WZ$ candidates of (a) the $\MET$ (b) the
dilepton invariant mass for the same-flavor opposite-sign dilepton
pair closest to the $Z$ mass, and (c) the $W$ transverse mass
calculated from the remaining lepton and the $\MET$, for the CDF
$WZ\rightarrow\ell\ell\ell\nu$ analysis \cite{Abulencia:2007tu}. In
(a) and (b), the arrows indicate the signal region.}
\label{fig:wz_kinematics}
\end{figure*}

Using $\intL=1.1\fb$, 16 events with an expected
background of 2.7 $\pm$ 0.4 events are observed in the CDF analysis,
as shown in Fig.~\ref{tbl:cdf-wz-results}. This corresponds to a 
6.0$\sigma$ observation of $WZ$ production when $\MET$ shape
information is included. The measured cross section is
\begin{equation*}
\sigma(p\bar{p}\rightarrow WZ + X) =
5.0^{+1.8}_{-1.6}~\rm{(stat.+syst.)}~\rm{pb},
\end{equation*}
consistent with the NLO expectation. Fig.~\ref{fig:wz_kinematics}
shows some important kinematic distributions for the 16 $WZ$
candidates, which are in good agreement with the SM.

\begin{table}
\begin{ruledtabular}
\begin{tabular}{lc}
  Source               &  Expectation $\pm$ Stat $\pm$ Syst $\pm$ Lumi \\
\hline\hline
  $Z$+jets             &  1.21 $\pm$ 0.27  $\pm$ 0.28  $\pm$ ~$-$~ \\
  $ZZ$                 &  0.88 $\pm$ 0.01  $\pm$ 0.09  $\pm$ 0.05 \\
  $Z\gamma$            &  0.44 $\pm$ 0.05  $\pm$ 0.15  $\pm$ 0.03 \\
  $t\bar{t}$           &  0.12 $\pm$ 0.01  $\pm$ 0.02  $\pm$ 0.01 \\
\hline
  Total Background     &  2.65 $\pm$ 0.28  $\pm$ 0.33  $\pm$ 0.09  \\
\hline
  $WZ$                 &  9.75 $\pm$ 0.03  $\pm$ 0.31  $\pm$ 0.59 \\
\hline
  Total Expected       &  12.41 $\pm$ 0.28  $\pm$ 0.45  $\pm$ 0.67  \\
\hline
Observed               &   16 \\
\end{tabular}
\end{ruledtabular}
\caption{Expected number of events in the signal region for $WZ$ and
the background contributions for the CDF
$WZ\rightarrow\ell\ell\ell\nu$ analysis
\cite{Abulencia:2007tu}. ``Lumi'' refers to the integrated luminosity
uncertainty, which is absent for the $Z$+jets because it is determined
from the same data set.
\label{tbl:cdf-wz-results}}
\end{table}

\begin{figure}
\includegraphics[scale=0.35]{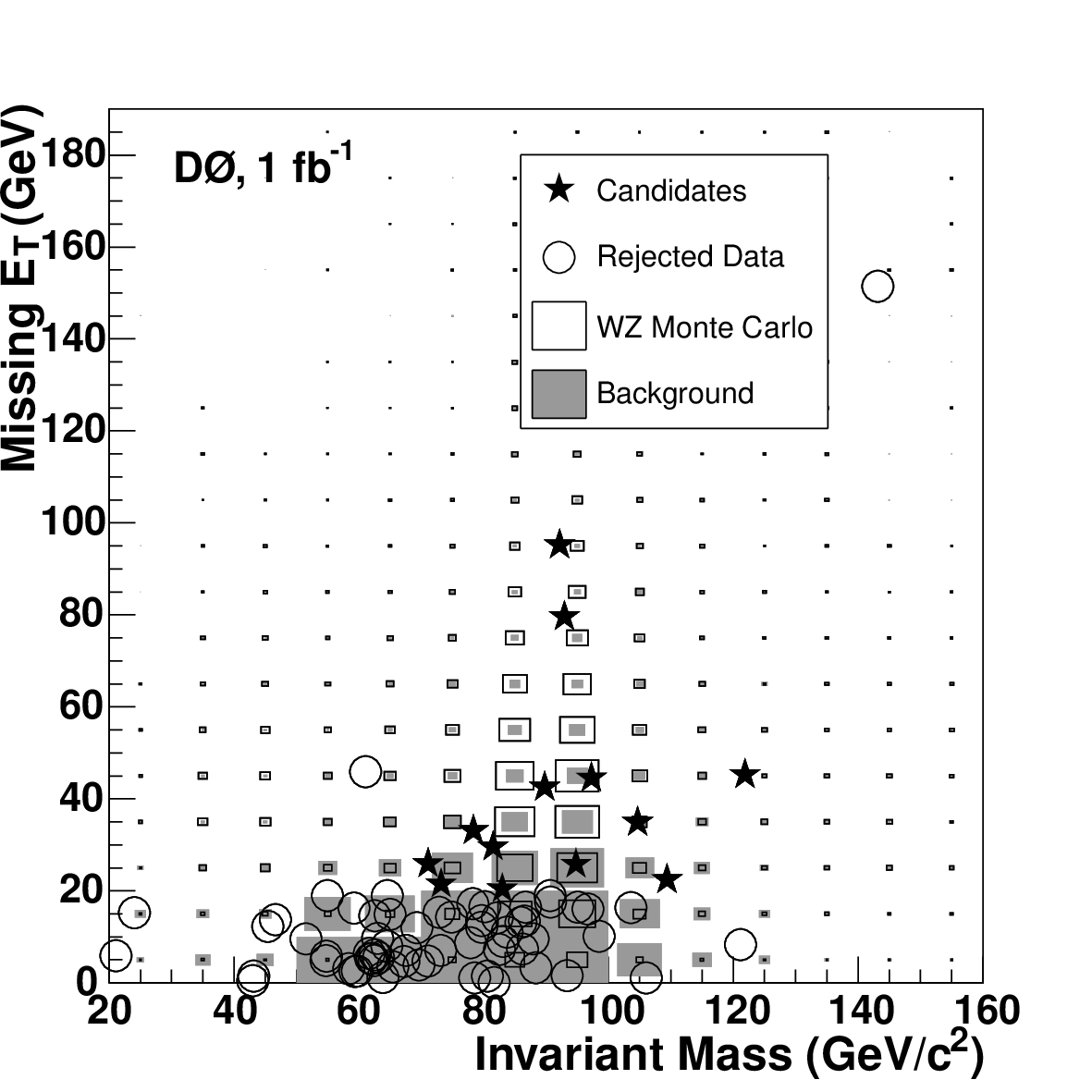}
\caption{\label{Fig:d0-massvsmet}
 \mbox{${\hbox{$E$\kern-0.6em\lower-.1ex\hbox{/}}}_T$}
versus dilepton invariant mass of $WZ\rightarrow\ell\ell\ell\nu$
candidate events in the D0 $WZ$ analysis\cite{Abazov:2007rab}. The
open boxes represent the expected $WZ$ signal. The gray boxes
represent the sum of the estimated backgrounds. The black stars are
the data that survive all selection criteria. The open circles are
data that fail either the dilepton invariant mass criterion or have 
$\MET < 20$ GeV.}
\end{figure}

In the D0 analysis \cite{Abazov:2007rab}, events with three
reconstructed, isolated charged leptons (electrons or muons) and $\MET
> 20$ GeV were used to measure the $WZ$ production cross section and
search for anomalous $WWZ$ couplings using the observed $Z$ boson
$\pt$ ($p_T^Z$) distribution. Each $WZ$ candidate event must contain a
like-flavor lepton pair with invariant mass close to the $Z$ boson
mass. For $eee$ and $\mu\mu\mu$ decay channels, the lepton pair with
invariant mass closest to that of the $Z$ boson mass is chosen to
define the $Z$ boson daughter particles. Using $\intL=1.0 \fb$, a
total of 13 $WZ$ candidates are observed with $4.5\pm 0.6$ expected
background events and $9.2\pm 1.0$ expected $WZ$ signal events,
corresponding to a 3.0$\sigma$ signal significance. The breakdown by
trilepton flavor classification for both the CDF and D0 analyses is
shown in Tab.~\ref{tbl:wz-flavors}. Fig.~\ref{Fig:d0-massvsmet} shows 
\mbox{${\hbox{$E$\kern-0.6em\lower-.1ex\hbox{/}}}_T$} versus the
dilepton invariant mass for the background, the expected $WZ$ signal,
and the data, including the candidates for the D0 analysis. The $WZ$
production cross section measured by D0 is 
\begin{equation*}
\sigma(p\bar{p}\rightarrow WZ + X) =
2.7^{+1.7}_{-1.3}~\rm{(stat.+syst.)}~\rm{pb},
\end{equation*}
where the $\pm 1 \sigma$ uncertainties are the $68\%$ CL limits from
the minimum of the negative log likelihood. The uncertainty is
dominated by the statistics of the number of observed events.

\begin{table}
\begin{ruledtabular}
\begin{tabular}{lrrrr}
Flavor & \multicolumn{2}{c}{CDF Analysis} &
\multicolumn{2}{c}{D0 Analysis}  \\
Classification          &  Expected     &  Data &  Expected  &  Data  \\
\hline\hline
$e$ $e$ $e$             & 2.7 $\pm$ 0.2 & 6  &  3.5 $\pm$ 0.2  &  2   \\
$e$ $e$ $\mu$           & 2.0 $\pm$ 0.2 & 0  &  2.7 $\pm$ 0.2  &  1   \\
$e$ $\mu$ $\mu$         & 1.5 $\pm$ 0.1 & 1  &  4.2 $\pm$ 0.5  &  8   \\
$\mu$ $\mu$ $\mu$       & 1.2 $\pm$ 0.1 & 1  &  3.4 $\pm$ 0.4  &  2   \\
$e$ $e$ $\ell_t$        & 2.0 $\pm$ 0.2 & 5  &  $-$   &  $-$ \\
$e$ $\mu$ $\ell_t$      & 1.3 $\pm$ 0.1 & 2  &  $-$   &  $-$ \\
$\mu$ $\mu$ $\ell_t$    & 1.1 $\pm$ 0.1 & 1  &  $-$   &  $-$ \\
$e$ $\ell_t$ $\ell_t$   & 0.5 $\pm$ 0.1 & 0  &  $-$   &  $-$ \\
$\mu$ $\ell_t$ $\ell_t$ & 0.2 $\pm$ 0.1 & 0  &  $-$   &  $-$ \\ 
\end{tabular}
\end{ruledtabular}
\caption{Summary of the expected and observed yields for the CDF
  \cite{Abulencia:2007tu} and D0 \cite{Abazov:2007rab}
  $WZ\rightarrow\ell\ell\ell\nu$ analyses. In the classification
  column, $\ell_t$ denotes a track-only lepton candidate having
  unknown flavor which is only relevant for the CDF analysis.
\label{tbl:wz-flavors}}
\end{table}

\begin{figure}
\includegraphics[angle=0,scale=0.4]{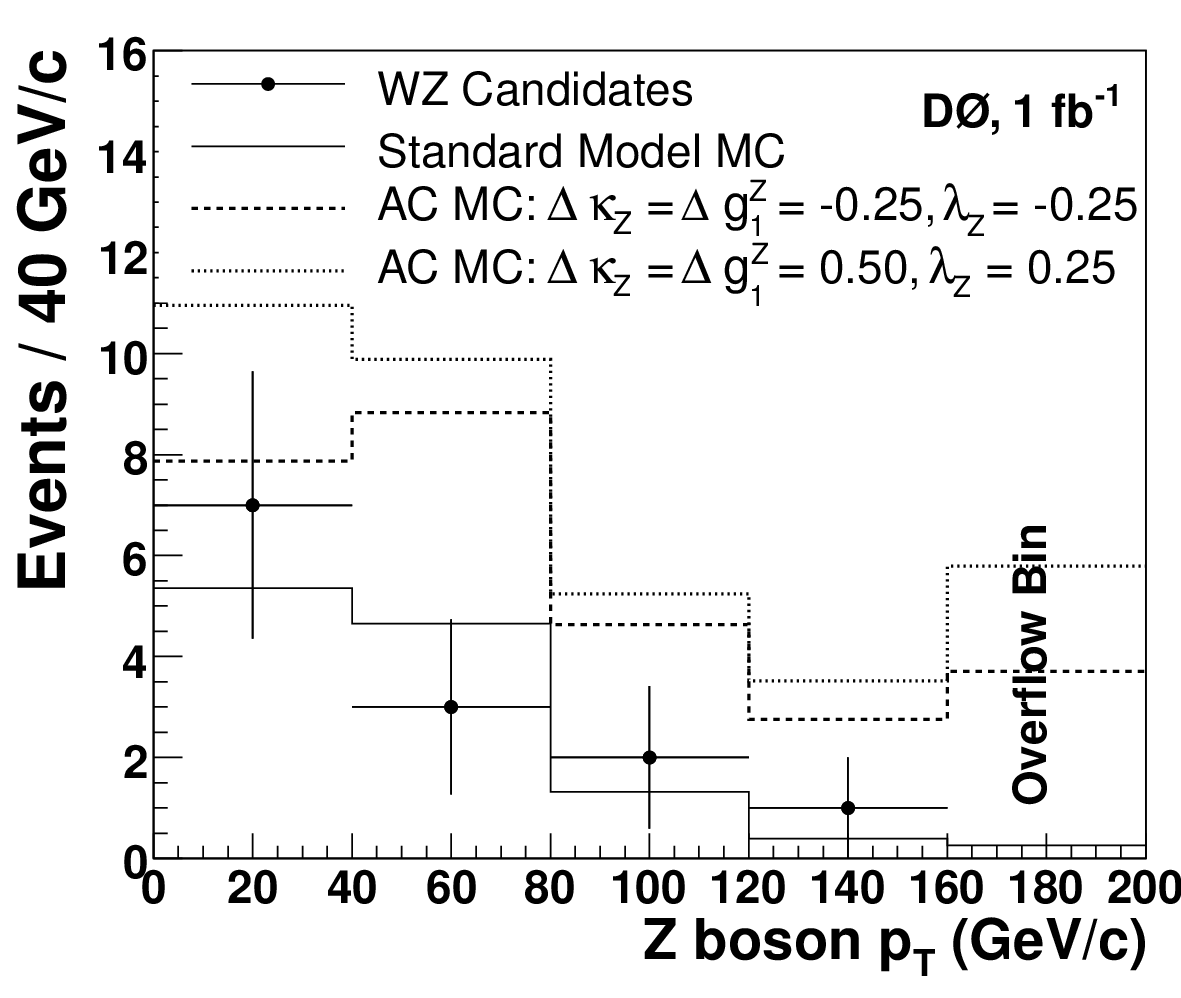}
\caption{\label{fig:d0-Zpt}The reconstructed $Z$ boson $p_T$ of the 
$WZ$ candidate events for the D0 $WZ\rightarrow\ell\ell\ell\nu$
analysis \cite{Abazov:2007rab}.  The solid histogram is the expected
sum of signal and background for the case of the $WWZ$ coupling
parameters set to their SM values. The dotted and double dotted
histograms are the expected sums of signal and background for two
different cases of anomalous $WWZ$ coupling parameter values.} 
\end{figure}

\begin{figure}
\includegraphics[scale=0.35]{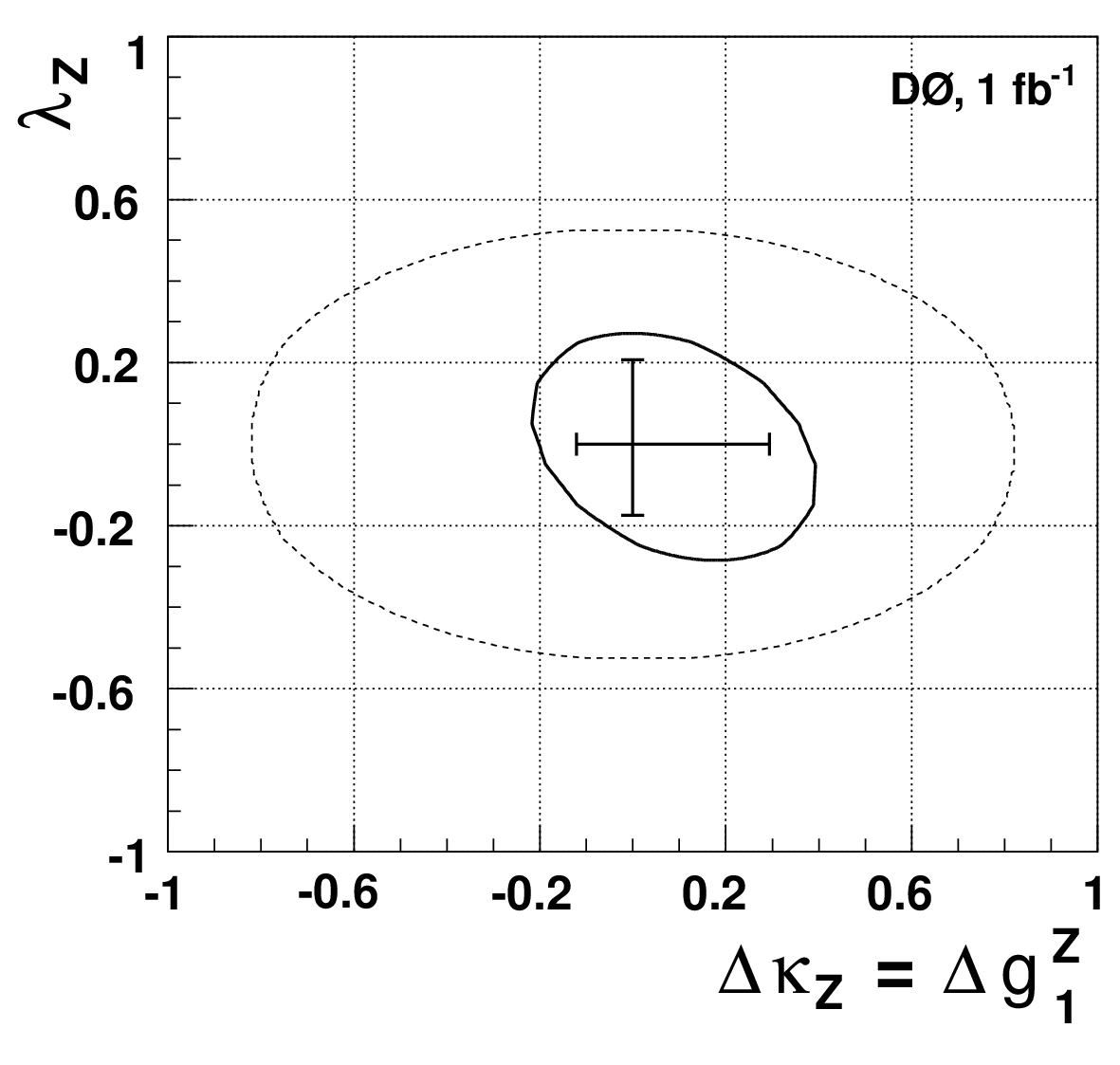}
\caption{\label{fig:twoDcontour} Two-dimensional 95\% CL contour
  limit in $\Delta g_1^Z = \Delta \kappa_Z$ versus $\Delta \lambda_Z$
  space (inner contour) for the D0 $WZ\rightarrow\ell\ell\ell\nu$
  analysis \cite{Abazov:2007rab}. The form factor scale for this
  contour is $\Lambda = 2$ TeV. The physically allowed region
  (unitarity limit) is bounded by the outer contour. The cross hairs
  are the 95\% CL one-dimensional limits.}
\end{figure}

The three $CP$-conserving $WWZ$ coupling parameters $\lambda_Z$,
$\Delta g^Z_1,$ and $\Delta \kappa_Z$ are constrained in the D0
analysis by comparing the measured cross section and $p_T^Z$
distribution to models with anomalous couplings. A comparison of the
observed $Z$ boson $p_T$ distribution with predictions from Monte
Carlo simulation is shown in Fig.~\ref{fig:d0-Zpt}. Tab.
\ref{tbl:1DLimits} presents the one-dimensional 95\% CL limits on
$\lambda_Z$, $\Delta g^Z_1$ and $\Delta
\kappa_Z$. Fig.~\ref{fig:twoDcontour} presents the two-dimensional
95\% CL limits under the assumption $\Delta g^Z_1 = \Delta \kappa_Z$
for $\Lambda=2$ TeV.

\begin{table}
\begin{ruledtabular}
\begin{tabular}{cc}
$ \Lambda = 1.5 \text{~TeV} $ & $ \Lambda = 2.0 \text{~TeV} $     \\ \hline
 $ -0.18<\lambda_Z<0.22$       & $ -0.17<\lambda_Z<0.21$          \\
 $ -0.15<\Delta g^Z_1<0.35 $   & $ -0.14< \Delta g^Z_1<0.34 $     \\
% $ -0.97<\Delta \kappa_Z <1.5$ & $ -0.88 < \Delta \kappa_Z < 1.4$\\
 $ -0.14<\Delta \kappa_Z = \Delta g^Z_1 <0.31$ 
                               & $-0.12<\Delta \kappa_Z = 
                                               \Delta g^Z_1 <0.29$ \\ 
\end{tabular}
\end{ruledtabular}
\caption{\label{tbl:1DLimits}One-dimensional 95\% CL intervals 
on $\lambda_Z$, $\Delta g^Z_1$, and $\Delta \kappa_Z$ for two sets 
of form factor scale, $\Lambda$, for the D0
$WZ\rightarrow\ell\ell\ell\nu$ analysis \cite{Abazov:2007rab}.}
\end{table}

\subsubsection{$\ZZ$}
\label{sec:gaugeBosons_dibosons_ZZ}

\begin{figure}
\includegraphics[width=0.25\textwidth]{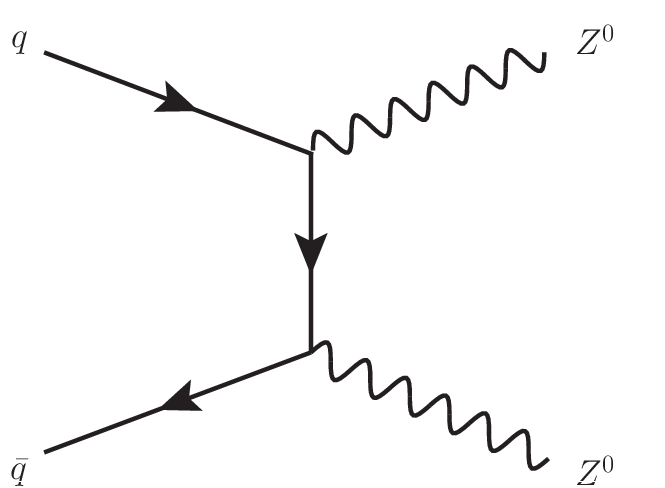}
\caption{Leading order diagram for $\ZZ$ production at the
  Tevatron. The s-channel diagram involving the neutral TGC is absent
  in the SM.} 
\label{fig:zz}
\end{figure}

The production of $Z$ pairs is predicted within the SM to have the
smallest cross section among the diboson processes. It has been been
observed in $e^+e^-$ collisions at LEP \cite{Alcaraz:2006mx}, but not
in hadron collisions as of the start of the Tevatron Run II. As a
window to new physics, $ZZ$ production is particularly interesting
because of the absence of $ZZ\gamma$ and $ZZZ$ couplings in the SM
(see Fig.~\ref{fig:zz}), and because of the very low backgrounds in
the four charged-lepton channel. Higgs boson decay can contribute to
$ZZ$ production; however, this channel is generally not competitive
with $H\rightarrow WW^{(*)}$ as a discovery channel at the Tevatron
collision energy and integrated luminosity. 

As is the case in for $WW$ and $WZ$ production, the $ZZ$ state is most
easily observed in the fully leptonic mode at a hadron collider. The
$ZZ\rightarrow \ell\ell\ell\ell$ process is rare but predicted to be nearly
background free in the SM, with $Z$+jets (jets reconstructed as
charged leptons) as the only non-negligible background. Having large 
total charged lepton acceptance in the experiments is crucial due to
the high lepton multiplicity in the final state. The $ZZ\rightarrow
\ell\ell\nu\nu$ process can also contribute sensitivity to the search
for $ZZ$ production although it suffers from large continuum $WW$
backgrounds. The full SM process is $p\overline{p}\rightarrow
Z/\gamma^{*} ~ Z/\gamma^{*}$, where the two $Z/\gamma^{*}$ interfere
with one another. For brevity, we denote $Z/\gamma^{*} ~ Z/\gamma^{*}$
as $ZZ$ throughout and indicate the dilepton invariant mass range(s),
where applicable. The next-to-leading order (NLO) $\ZZ$ cross
section for $p\overline{p}$ collisions at $\sqrt{s} = 1.96~{\rm TeV}$
is 1.4 $\pm$ 0.1~pb in the zero-width $Z$ boson approximation
\cite{Campbell:1999ah}.

\begin{figure}
\mbox{
\includegraphics[width=1.62in]{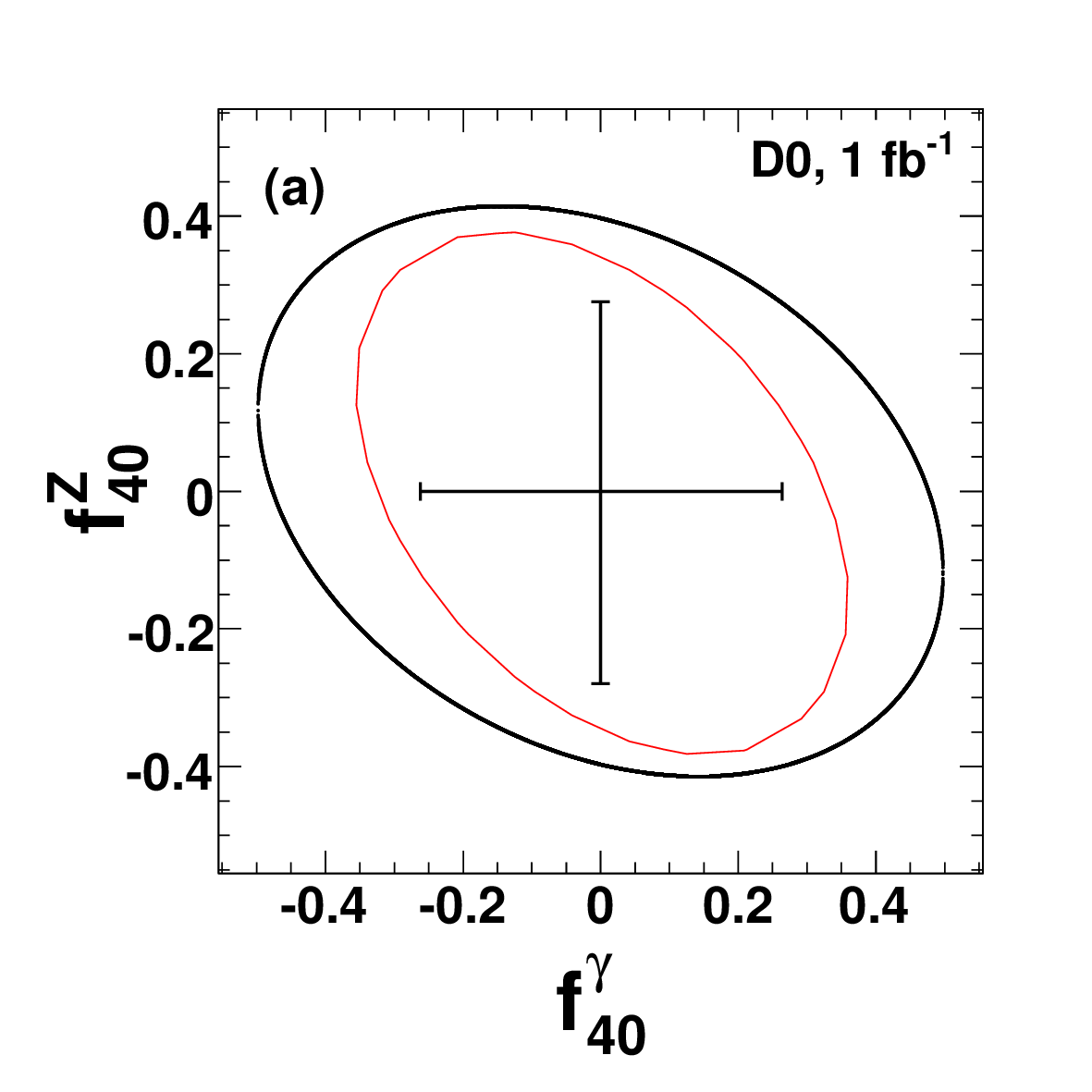} \hfill
\includegraphics[width=1.62in]{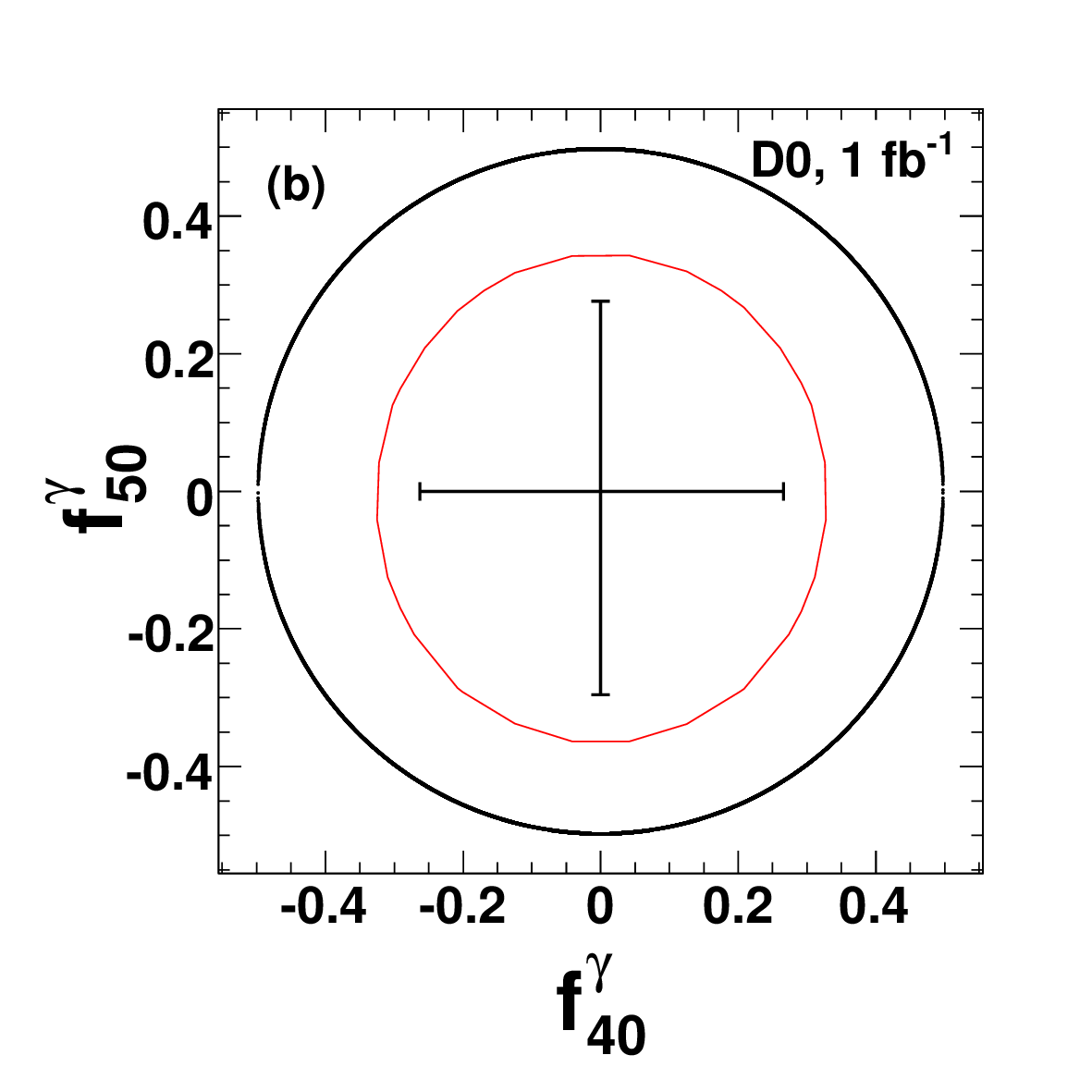}
}
\mbox{
\includegraphics[width=1.62in]{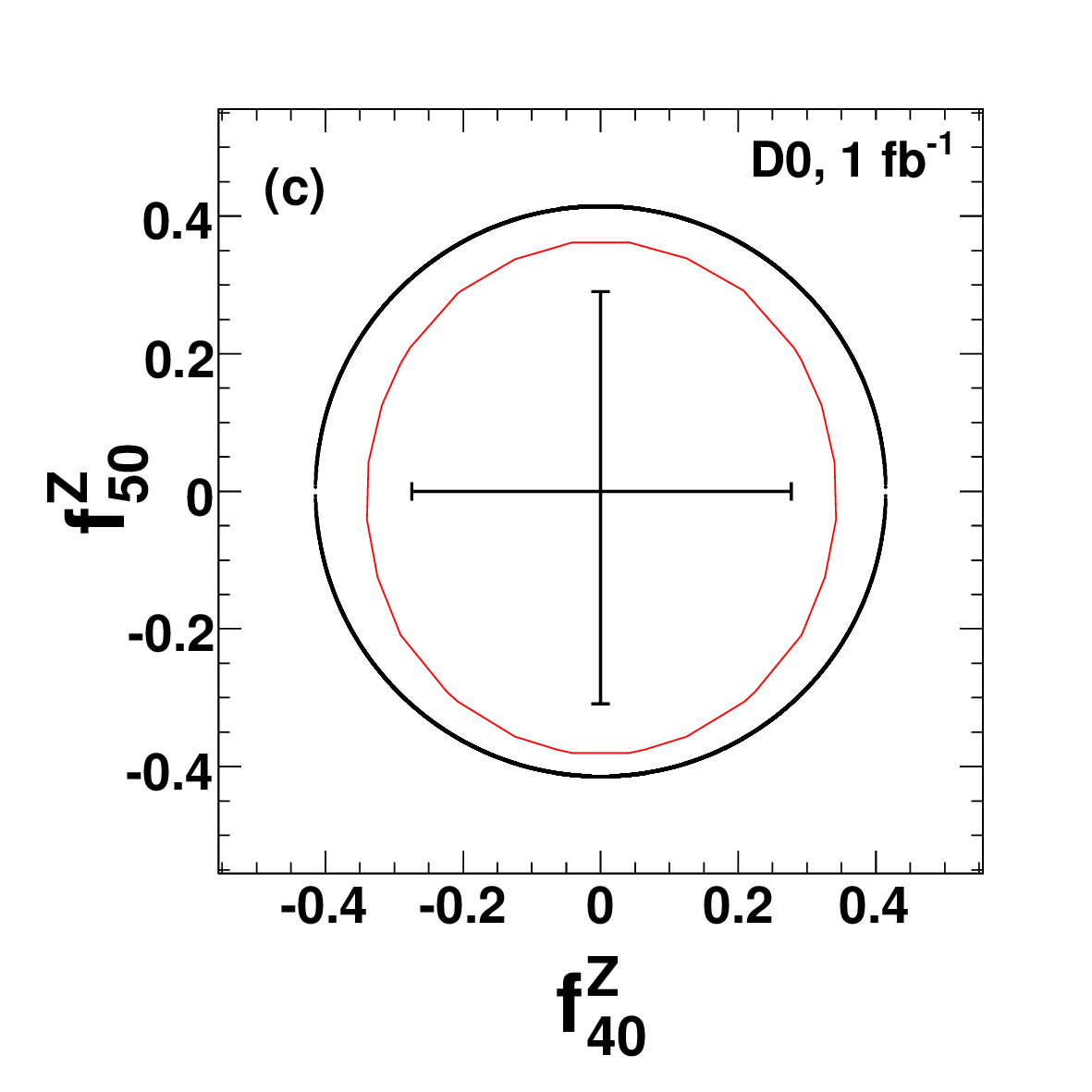} \hfill
\includegraphics[width=1.62in]{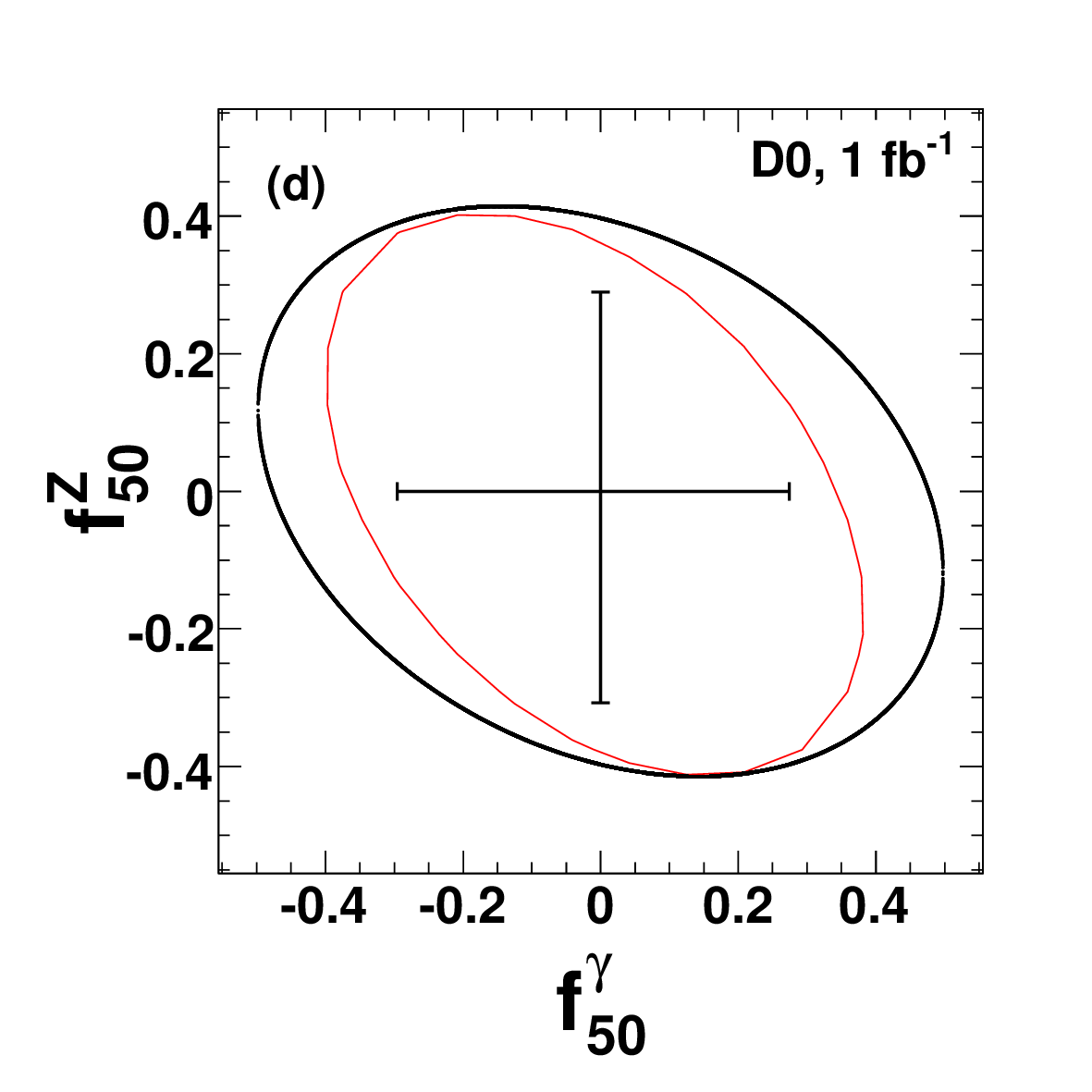}
}
\caption{Limits on anomalous couplings for $\Lambda = 1.2$ TeV for the
  D0 $ZZ\rightarrow\ell\ell\ell\ell$ analysis \cite{Abazov:2007hm}:
(a) $f_{40}^{\gamma}$ vs. $f_{40}^Z$,
(b) $f_{40}^{\gamma}$ vs. $f_{50}^{\gamma}$,
(c) $f_{40}^Z$ vs. $f_{50}^Z$, and
(d) $f_{50}^{\gamma}$ vs. $f_{50}^Z$,
 assuming in each case that the other two couplings are zero.
 The inner and outer curves are the
 $95 \%$ CL two-degree of freedom exclusion contour
 and the constraint from the unitarity condition, respectively.
 The inner cross hairs are the  $95 \%$ CL one-degree of freedom 
exclusion limits. }
\label{fig:d0-zz-combined}
\end{figure}

Using $\intL=1 \fb$, the D0 Collaboration searched for
$ZZ\rightarrow\ell\ell\ell\ell$ production and set a limit on the
cross section of $\sigma(ZZ)$ $<$ 4.4 pb at 95\% CL
\cite{Abazov:2007hm}. They constrain possible $ZZ\gamma$ and $ZZZ$
couplings based on the observed four lepton yield with the added
requirement that $M(\ell\ell) > 50\; (70)$ for electrons (muons). The
value of the cut was chosen based on the dilepton invariant mass
resolution. This lower $M(\ell\ell)$ requirement was included because
the Monte Carlo generator \cite{Baur:2000ae} used to constrain
possible anomalous couplings does not include contributions from
off-shell $Z$ bosons. Without the $M(\ell\ell)$ cut, one $\mu\mu ee$
event is observed in the data. This event is removed by the
$M(\ell\ell)$ cut used to constrain anomalous couplings. 1D and 2D
limits on anomalous $ZZ\gamma$ and $ZZZ$ couplings are determined
using $\Lambda = 1.2$~TeV. The 95\% CL 1D limits are $-0.28 < f_{40}^Z
< 0.28$, $-0.26 < f_{40}^{\gamma} < 0.26$, $-0.31 < f_{50}^Z < 0.29$,
and $-0.30 < f_{50}^{\gamma} < 0.28$. The 95\% CL 2D contours
$f_{40}^{\gamma}$ vs. $f_{40}^Z$,
$f_{40}^{\gamma}$ vs. $f_{50}^{\gamma}$,
$f_{40}^Z$ vs. $f_{50}^Z$, and 
$f_{50}^{\gamma}$ vs. $f_{50}^Z$
are shown in Fig.~\ref{fig:d0-zz-combined}. 

The CDF Collaboration uses $\intL=1.9\fb$ to search for $ZZ$
production in a combination of the $ZZ\rightarrow\ell\ell\ell\ell$ and
$ZZ\rightarrow\ell\ell\nu\nu$ channels \cite{Aaltonen:2008mv}. To
maximize the acceptance, lepton candidates are constructed out of all
reconstructed tracks and energy clusters in the EM section of the
calorimeter. This is done with the same lepton identification criteria
used in a previous CDF measurement of $\WZ$ production
\cite{Abulencia:2007tu}. The $ZZ\rightarrow\ell\ell\ell\ell$
candidates are selected from events with exactly four charged-lepton
candidates and at least two same-flavor, opposite-sign lepton pairs
are required for the event to be accepted. As in the $WZ$ analysis,
trackless electrons are considered to have either charge, and
track-only leptons either flavor. One pair must have invariant mass
$M_{\ell^{+}\ell^{-}}$ in the range [76, 106] $\GeV$, while the
requirement for the other pair is extended to [40, 140] $\GeV$ to
increase the acceptance for off-shell $Z$ decays.

The $ZZ\rightarrow\ell\ell\ell\ell$ candidate events are separated
into two exclusive categories based on whether or not they contain at
least one forward electron without a track. This is done because the
background from $Z\gamma$+jets is much larger in candidates with a
forward trackless electron. The expected signal, expected background,
and observed yields are shown in Tab.~\ref{tbl:cdf-ZZllll_results}.

\begin{table}
\begin{ruledtabular}
\begin{tabular}{lcc}
               & Candidates without a                             &     Candidates with a              \\
Category       & trackless electron                             &     trackless electron           \\  
               
\hline         
   $ZZ$             &   1.990 $\pm$ 0.013 $\pm$ 0.210           &     0.278 $\pm$ 0.005 $\pm$ 0.029      \\
   $Z$+jets       \parbox{0.0cm}{\vspace{0.55cm}}   &   ${0.014}^{+0.010}_{-0.007} \pm 0.003$   &  ${0.082}^{+0.089}_{-0.060} \pm 0.016$ \\
\hline
   Total          \parbox{0.0cm}{\vspace{0.65cm}}  &   $2.004^{+0.016}_{-0.015} \pm 0.210 $    &  $0.360^{+0.089}_{-0.060} \pm 0.033$   \\
\hline
Observed      &            2                           &     1     \\
\end{tabular}
\end{ruledtabular}
\caption{Expected and observed number of
  $ZZ\rightarrow\ell\ell\ell\ell$ candidate events for the CDF $ZZ$
  analysis \cite{Aaltonen:2008mv}. The first uncertainty is
  statistical and the second one is systematic.}
\label{tbl:cdf-ZZllll_results}
\end{table}

The $ZZ\rightarrow\ell\ell\nu\nu$ candidates are selected from events
with exactly two oppositely-charged lepton candidates excluding events
with forward electrons without a track which are contaminated by large
$W\gamma$ backgrounds. Aside from $ZZ$ production, other SM processes
that can lead to two high-$\pt$ leptons include events from DY, a
$\W$ decay with photon ($W\gamma$) or jet ($W$+jets) misidentified as
a lepton; and $\ttbar$, $WW$, and $WZ$ production.

\begin{figure}
\includegraphics[width=0.49\textwidth]{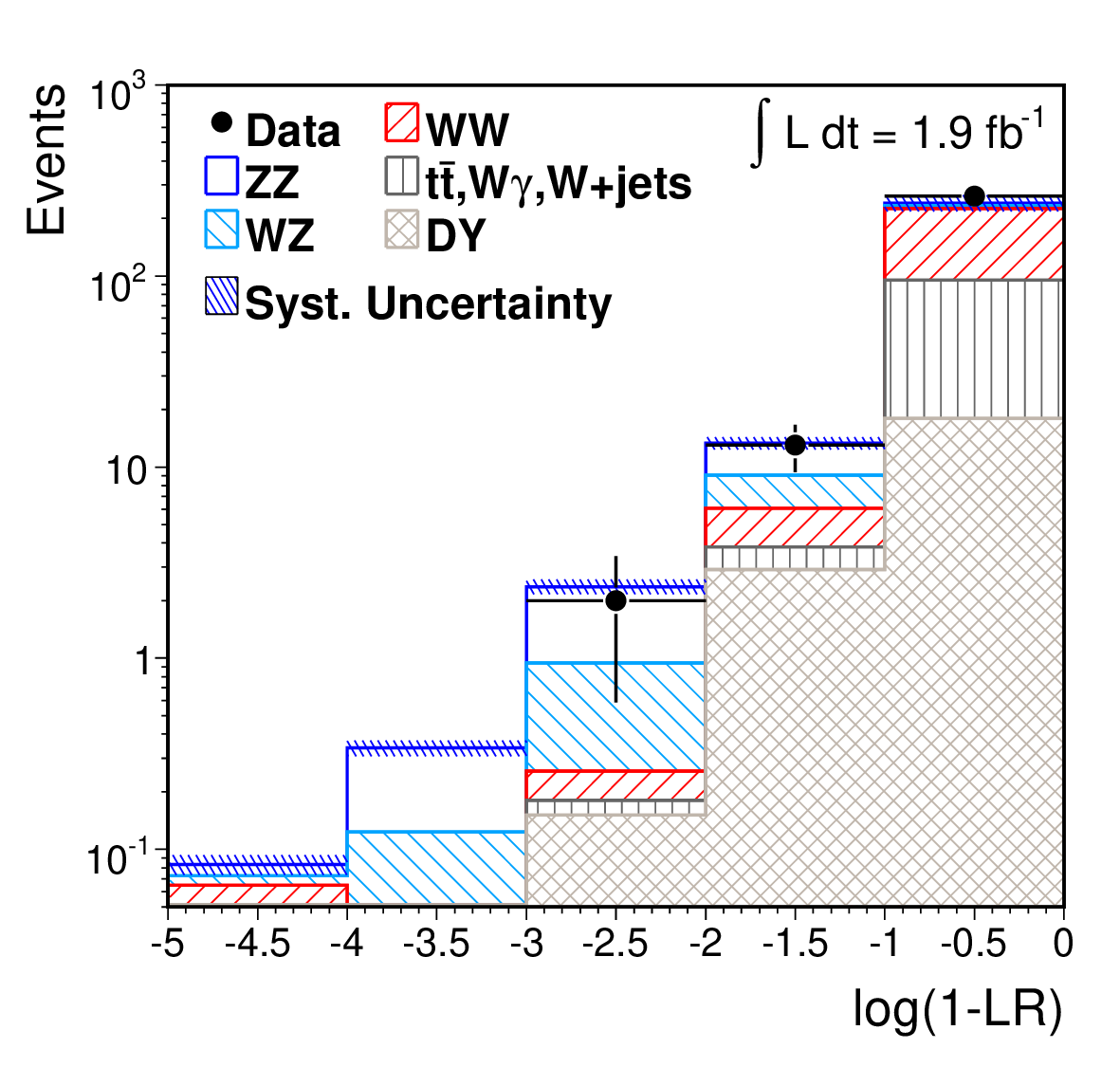}
\caption{Distribution of the discriminating variable $\log_{10}
  (1-LR)$ for the CDF $ZZ\rightarrow\ell\ell\nu\nu$ search
  \cite{Aaltonen:2008mv}.}
\label{fig:cdf-zz-logLR}
\end{figure}

There are 276 events after the event selection (which contains a
specialized high $\MET$ to suppress primarily $DY$) of which only
$14\pm2$ are expected to be from the $ZZ\rightarrow\ell\ell\nu\nu$
process in the SM. Approximately half of the yield is expected to be
due to the $WW$ process. However, $ZZ\rightarrow\ell\ell\nu\nu$ and
$WW$ have different kinematic properties which are exploited to
statistically separate the contribution of these two processes to the
data. The approach used by CDF is identical to that used in the $WW$
cross section measurement described in
Section~\ref{sec:gaugeBosons_dibosons_WW}. An event-by-event 
probability density is calculated for the observed lepton momenta and
$\MET$ using leading order calculations of the differential decay rate
for the processes \cite{Campbell:1999ah}. A likelihood ratio
discriminant $LR$ is formed which is the signal probability divided by
the sum of signal and background probabilities
$LR=P_{ZZ}/(P_{ZZ}+P_{WW})$. The distribution of $\log_{10} (1-LR)$
for the data compared to the summed signal and background expectation
is shown in Fig.~\ref{fig:cdf-zz-logLR}.

The $ZZ\rightarrow\ell\ell\ell\ell$ and $ZZ\rightarrow\ell\ell\nu\nu$
results were combined through a likelihood fit which includes the two
4-lepton bins and the $\log_{10} (1-LR)$ distribution for the
dileptons. The p-value for the $ZZ\rightarrow\ell\ell\nu\nu$ alone is 
0.12 and the combined p-value is $5.1\times10^{-6}$ corresponding to a
significance equivalent to 4.4 standard deviations. The $ZZ$
cross-section is obtained by fitting the data for the fraction of the
expected SM yield in the full acceptance and scaling the zero-width
$Z$ boson approximation cross-section by that fraction. The measured
cross section is $\sigma(p\overline{p}\rightarrow \ZZ) =
1.4^{+0.7}_{-0.6}~\mathrm{pb},$ consistent with the SM
expectation. This is the first evidence of a $ZZ$ signal with greater
than 4$\sigma$ significance in hadron collisions. 

The D0 Collaboration uses $\intL=1.7 \fb$ to search for $ZZ$
production in the $ZZ\rightarrow\ell\ell\ell\ell$ channel
\cite{Abazov:2008gya}. This analysis follows up the analysis from
\cite{Abazov:2007hm} using more data and tighter dilepton invariant
mass requirements but does not explicitly search for anomalous
couplings. Four lepton events are selected using identified muons with
$|\eta| < 2$ and electrons that are either central ($|\eta < 1.1$) or
forward $1.5 < |\eta| < 3.2$. It is required that one pair have the
same flavor with invariant mass $> 70$ $\GeV$ and another pair have
invariant mass $> 50$ $\GeV$.

\begin{figure}
  \includegraphics[scale=0.38,angle=0]{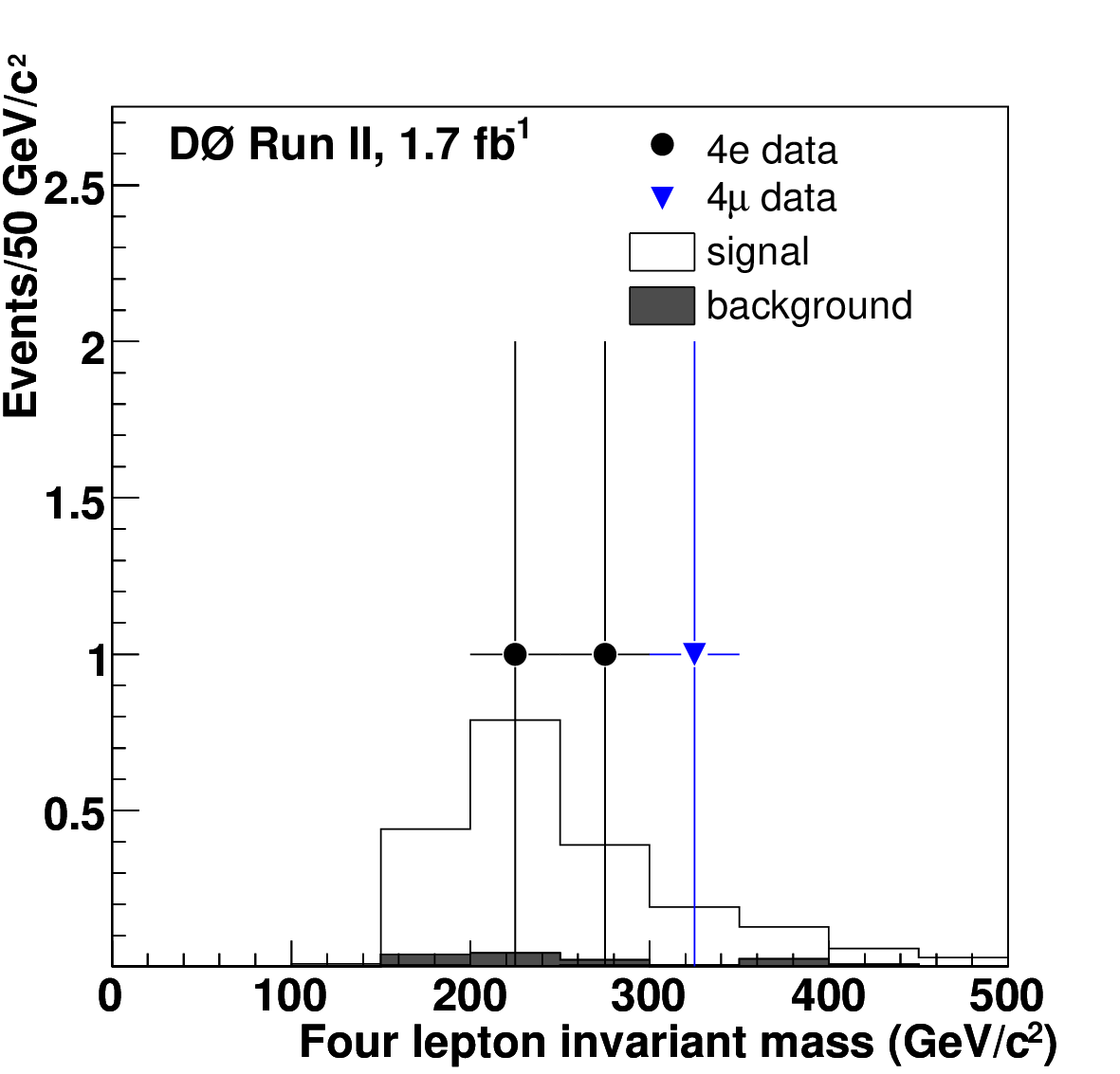}
  \caption{\label{fig:fourmass} Distribution of four lepton invariant
    mass in data, expected signal, and expected background for the D0
    $ZZ\rightarrow\ell\ell\ell\ell$ analysis \cite{Abazov:2008gya}.}
\end{figure}

Tab.~\ref{tbl:dzero-zz-channels} summarizes the expected signal and
background contributions to each subchannel, as well as the number of
candidate events in data. The total signal and background expectations
are $1.89 \pm 0.08$ events and $0.14^{+0.03}_{-0.02}$ events,
respectively. A total of three candidate events is observed, with two
in the $4e_{4C}$ subchannel (``C'' refers to the number of central
leptons) and one in the $4\mu$ subchannel. Fig.~\ref{fig:fourmass}
shows the distribution of the four lepton invariant mass for data and
for the expected signal and background. 

\begin{table*}
\vspace*{2mm}
\begin{ruledtabular}
\begin{tabular}{cccccccc} 
Subchannel 
& $4e_{2C}$ & $4e_{3C}$ & $4e_{4C}$ & $4\mu$ & $2\mu2e_{0C}$ &
$2\mu2e_{1C}$ & $2\mu2e_{2C}$ \\ 
\hline
Luminosity (fb$^{-1}$) & $1.75 \pm 0.11$ & $1.75 \pm 0.11$ & $1.75 \pm 0.11$ 
& $1.68 \pm 0.10$ & $1.68 \pm 0.10$ & $1.68 \pm 0.10$ & $1.68 \pm 0.10$ \\ 
\hline 
 & & & & & & & \\[-2mm]
Signal & $0.084 \pm 0.008$ & $0.173 \pm 0.015$ & $0.140 \pm 0.012$ 
& $0.534 \pm 0.043$ & $0.058^{+0.007}_{-0.006}$ & $0.352 \pm 0.040$ 
& $0.553^{+0.045}_{-0.044}$ \\[1mm] 
\hline
 & & & & & & & \\[-2mm]
$Z(\gamma)$+jets & $0.030^{+0.009}_{-0.008}$ & $0.018^{+0.008}_{-0.007}$ 
& $0.002^{+0.002}_{-0.001}$ & $0.0003 \pm 0.0001$ & $0.03^{+0.02}_{-0.01}$ 
& $0.05 \pm 0.01$ & $0.008^{+0.004}_{-0.003}$ \\[1mm] 
\hline
 & & & & & & & \\[-2mm]
$t\bar{t}$ 
& -- & -- & -- & -- & $0.0012^{+0.0016}_{-0.0009}$ & $0.005 \pm 0.002$ 
& $0.0007^{+0.0009}_{-0.0005}$ \\[1mm] 
\hline
Observed events       
& 0 & 0 & 2 & 1 & 0 & 0 & 0 \\ 
\end{tabular}
\caption{The integrated luminosity, expected number of signal
  ($Z/\gamma^*$ $Z/\gamma^*$) and background events ($t\bar{t}$ and
  $Z(\gamma)$+jets which includes all $W$/$Z$/$\gamma$+jets
  contributions), and the number of observed candidates in the seven
  $ZZ\to\ell^{+}\ell^{-}\ell^{'+}\ell^{'-}$ subchannels for the D0
    $ZZ\rightarrow\ell\ell\ell\ell$ analysis
    \cite{Abazov:2008gya}. Uncertainties reflect statistical and
    systematic contributions added in quadrature.} 
\label{tbl:dzero-zz-channels}
\end{ruledtabular}
\end{table*}

Using a log-likelihood ratio test statistic of the yields and
pseudo-experiments, the $p$-value is determined to be
4.3$\times$10$^{-8}$ which corresponds to a 5.3 $\sigma$ observed 
significance ($3.7\sigma$ expected). The measured cross section in the
$ZZ\rightarrow\ell\ell\ell\ell$ channel is
\begin{equation*}
\sigma(ZZ+X) = 1.75^{+1.27}_{-0.86}~\mathrm{(stat.)} \pm
0.13~\mathrm{(syst.)}~\rm{pb}.
\end{equation*}

This result is combined with the results from an independent
$ZZ\to\ell\ell\nu\nu$ search~\cite{PhysRevD.78.072002}, and the
previous $ZZ\rightarrow\ell\ell\ell\ell$ analysis \cite{Abazov:2007hm}
taking into account the systematic uncertainty correlations between 
subchannels and among analyses. The resulting $p$-value is
6.2$\times$10$^{-9}$, and  the significance for observation of $ZZ$
production increases to $5.7\sigma$ ($4.8\sigma$ expected). This is
the first observation of $ZZ$ production at a hadron collider. The
combined cross section is 
\begin{equation*}
\sigma(ZZ+X) = 1.60 \pm
0.63~\mathrm{(stat.)}^{+0.16}_{-0.17}~\mathrm{(syst.)}~\rm{pb}, 
\end{equation*}
consistent
with the SM expectation.

\subsubsection{$WV$ ($V =W,Z$)}
\label{sec:gaugeBosons_WV}

The production of vector boson pairs ($WW$, $WZ$, and $ZZ$) have been
observed at the Tevatron in decay modes where both vector bosons decay
leptonically. The semileptonic decay modes where one of the vector
boson decay hadronically has larger branching fraction as compared to
the fully leptonic modes but significantly larger backgrounds from
jets produced in association with a $W$ boson. As a result, simple
event counting above background cannot be used to observe dibosons in
semileptonic decay at the Tevatron and advanced analysis
techniques utilizing multivariate event classification are required to
statistically separate signal from background. Analysis of the $\ell\nu
jj$ final state provides an excellent testbed for such advanced
techniques to extract small signals from large backgrounds in real
hadron collision data that has great relevance to Higgs boson
(e.g. $WH\rightarrow\ell\nu b\bar{b}$) and new physics searches in
final states involving jets. In addition, semileptonic diboson decay
can provide a more sensitive search for anomalous trilinear gauge
couplings than fully leptonic decay modes since anomalous TGCs enhance
production at high gauge boson momentum where the signal-to-background
ratio improves.

The dijet mass resolution of the CDF and D0 detectors is not good
enough to distinguish hadronically decaying $W$ bosons and $Z$
bosons. As a result, the search for diboson production in the $\ell\nu
jj$ final state is a search for the sum of $WW$ and $WZ$
production. The experimental signature of the $WW+WZ$ signal in
semileptonic decay is one isolated high $\Et$ charged lepton, large
$\MET$ from the neutrino produced in $W$ decay, and at least two high
$\Et$ jets. 

In principle, one could search for SM $WZ+ZZ$ production in the
$jj\ell\ell$ channel. Large $Z$+jets backgrounds makes this a very
difficult channel to observe the SM signal, although the use of
$b$-tagging can improve the sensitivity. The use of the
$b\bar{b}\ell\ell$ channel to search for associated Higgs boson
production is described in Section~\ref{sec:higgs_direct_WZllbb}.

Both D0 and CDF have searched for $WW+WZ$ production and anomalous
$WW\gamma$, $WWZ$ couplings in the $\ell\nu jj$ final state
\cite{Abazov:2008yg,Aaltonen:2009vh,Aaltonen:2007sd,Abazov:2009tr}. In
these analyses, $W+$jets is the dominant background. Other significant
backgrounds include $Z$+jets, $t\bar{t}$, single top quark, and QCD
multijet production. Backgrounds are suppressed by requiring a
minimum $W$ transverse mass $M_T(W)$ in each event, which is the
transverse mass of the charged lepton and $\MET$ system.

\begin{figure} 
\begin{tabular}{cc}
  \multirow{1}{*}[1.0in]{(a)} & \includegraphics[height=2.0in]{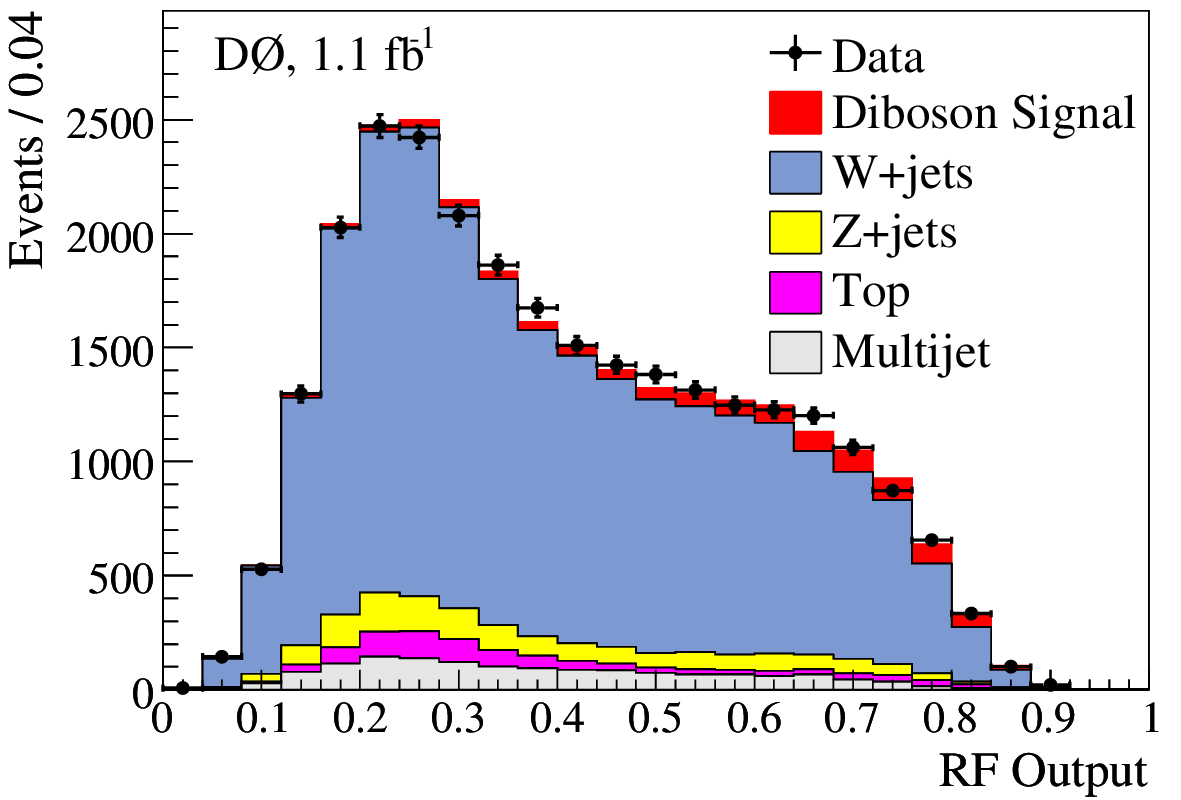}\\
  \multirow{1}{*}[1.0in]{(b)} & \includegraphics[height=2.0in]{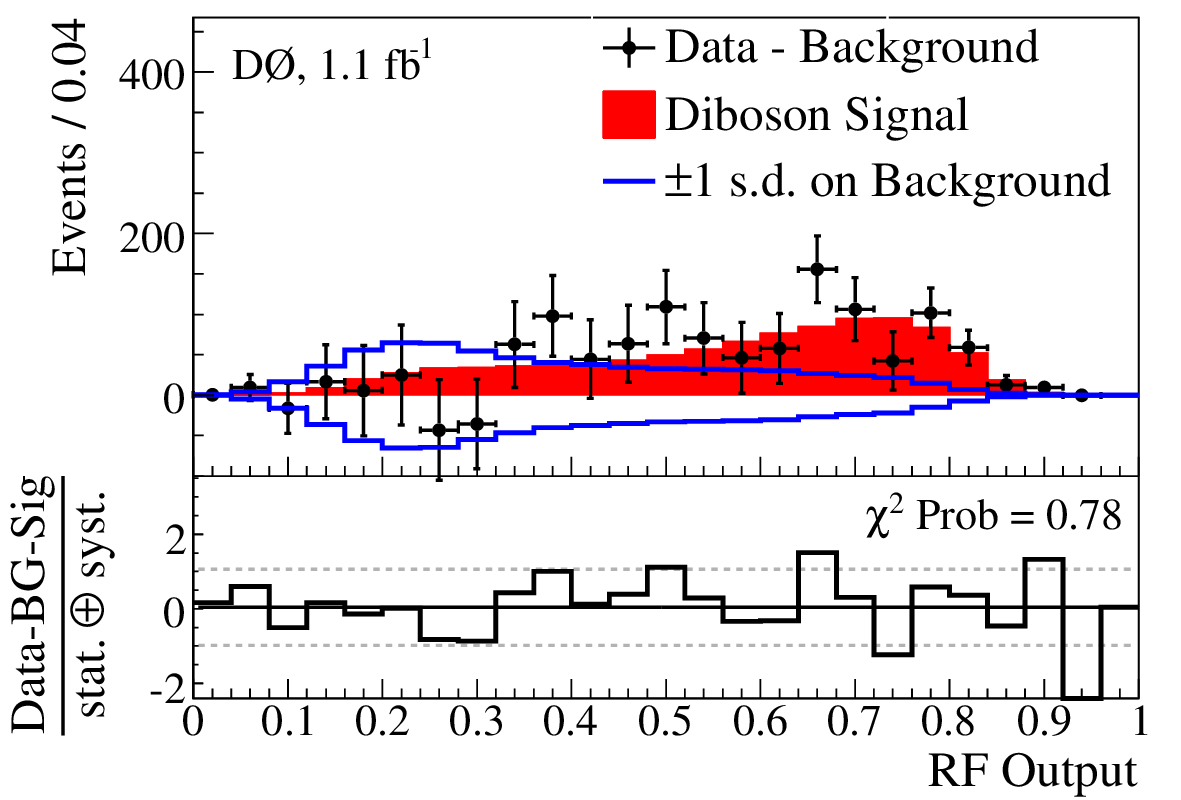}
\end{tabular}    
\caption{For the D0 analysis \cite{Abazov:2008yg}, (a) the RF output
  distribution from the combined $e\nu q\bar{q}$ and $\mu\nu q\bar{q}$
  channels for data and simulated event predictions following the fit
  of simulation to data. (b)~A comparison of the extracted signal
  (filled histogram) to background-subtracted data (points), along
  with the $\pm1$ standard deviation (s.d.) systematic uncertainty on
  the background. The residual distance between the data points and
  the extracted signal, divided by the total uncertainty, is given at
  the bottom.}
\label{fig:d0-wv-Fig1} 
\end{figure}

\begin{figure}
\begin{tabular}{cc}
  \multirow{1}{*}[1.0in]{(a)} & \includegraphics[height=2.0in]{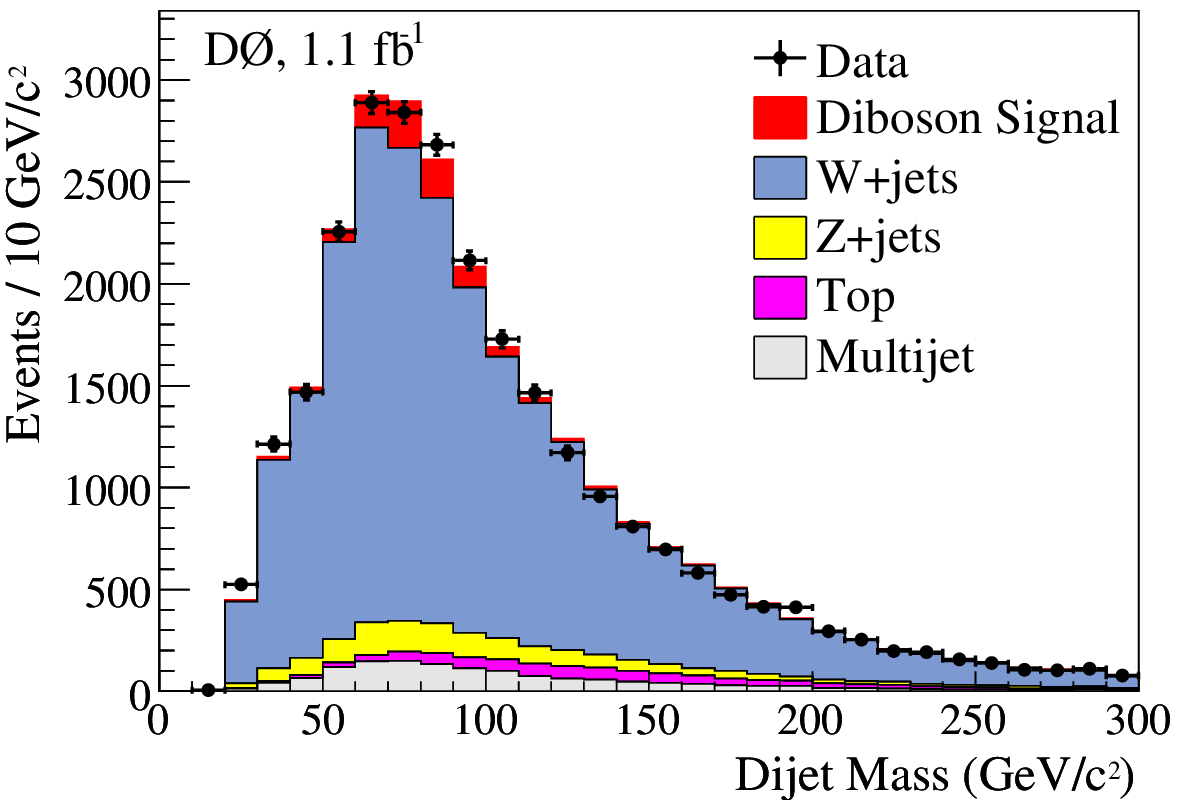}\\
  \multirow{1}{*}[1.0in]{(b)} & \includegraphics[height=2.0in]{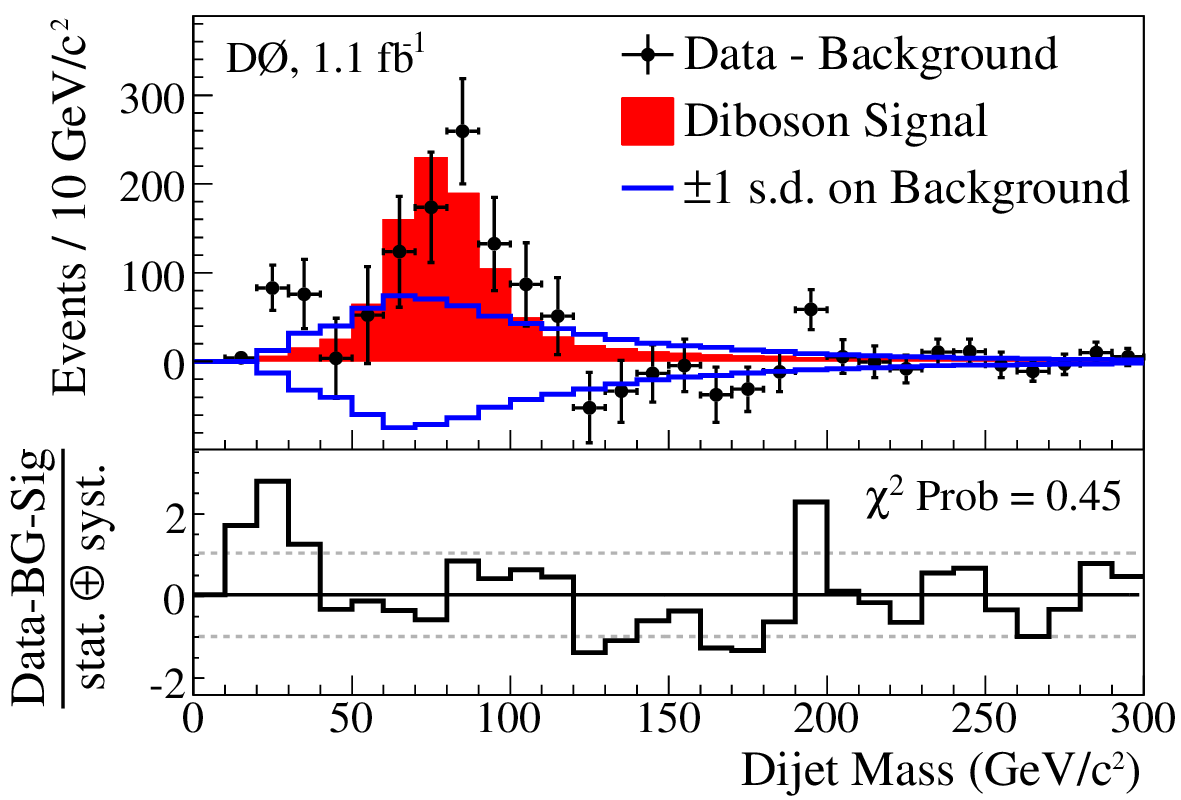}
\end{tabular}
\caption{For the D0 analysis \cite{Abazov:2008yg}, (a) the dijet mass
  distribution from the combined $e\nu q\bar{q}$ and $\mu\nu q\bar{q}$
  channels for data and simulated event predictions following the fit
  to the RF output. (b)~A comparison of the extracted signal (filled
  histogram) to background-subtracted data (points), along with the
  $\pm1$ standard deviation (s.d.) systematic uncertainty on the
  background.The residual distance between the data points and the
  extracted signal, divided by the total uncertainty, is given at the
  bottom.}
\label{fig:d0-wv-Fig2}
\end{figure}

The strategy for extracting the semileptonic $WW+WZ$ signal yield
after basic event selection differs between the D0 and CDF
analyses. In the D0 analysis \cite{Abazov:2008yg}, thirteen kinematic
variables (e.g. dijet mass) demonstrating a sensitivity to distinguish
signal and background are used as input to a Random Forest (RF)
\cite{RandomForests} multivariate event classifier using
$\intL=1.1\fb$. Fig.~\ref{fig:d0-wv-Fig1} shows the RF output
distribution after a fit to the $WV$ signal and $W+$jets background
contributions. Fig.~\ref{fig:d0-wv-Fig2} shows the dijet mass 
distribution using the results of the RF output fit. The dominant 
systematic uncertainties arise from the modeling of the $W+$jets
background and the jet energy scale. The probability for the
background to fluctuate to give an excess as large as that observed in
the data is $< 5.4\times 10^{-6}$, corresponding to a 4.4$\sigma$
signal significance. This result is the first evidence for $WV$
production is lepton $+$ jets events at a hadron collider. The
measured cross section is 
\begin{equation*}
\sigma(WV + X) = 20.2 \pm 4.5 \rm{(stat.+ syst.)} ~\rm{pb} 
\end{equation*}
consistent with the NLO SM cross section of $16.1 \pm 0.9$ pb
\cite{Campbell:1999ah}.

In the CDF analysis \cite{Aaltonen:2009vh}, two different methods are
used to extract the $WW+WZ$ signal from the data. In the first method
(``dijet method''), the dijet invariant mass $M_{jj}$ is used to
extract a signal peak from data corresponding to $\intL=3.9 \fb$.
The second method (``ME method'') takes advantage of more kinematic
information in the event by constructing a discriminant based on
calculations of the differential cross sections of the signal and
background processes data corresponding to $\intL=2.7$ fb$^{-1}$. The
ME method was discussed in Section~\ref{sec:gaugeBosons_dibosons_WW}
where it was used in the $WW\rightarrow\ell\nu\ell\nu$ analysis and
has been used in a number of different CDF and D0 analyses. 

The normalization of the $Z$+jets background is based on the measured
cross section \cite{z-xsec-cdf-short}. The $t\bar{t}$ and single top 
background normalizations are from the NLO predicted cross sections
\cite{topxsec,Harris:2002md}. The efficiencies for the $Z$+jets,
$t\bar{t}$, and single top backgrounds are estimated from
simulation. The normalization of the multijet background is estimated by 
fitting the $\MET$ spectrum in data to the sum of all contributing
processes, where the multijet and $W$+jets normalizations float in the
fit. In the final signal extractions from both methods, the 
multijet background is Gaussian constrained to the result of this $\MET$
fit and the $W$+jets background is left unconstrained.

\begin{figure}
  \includegraphics[width=1\columnwidth]{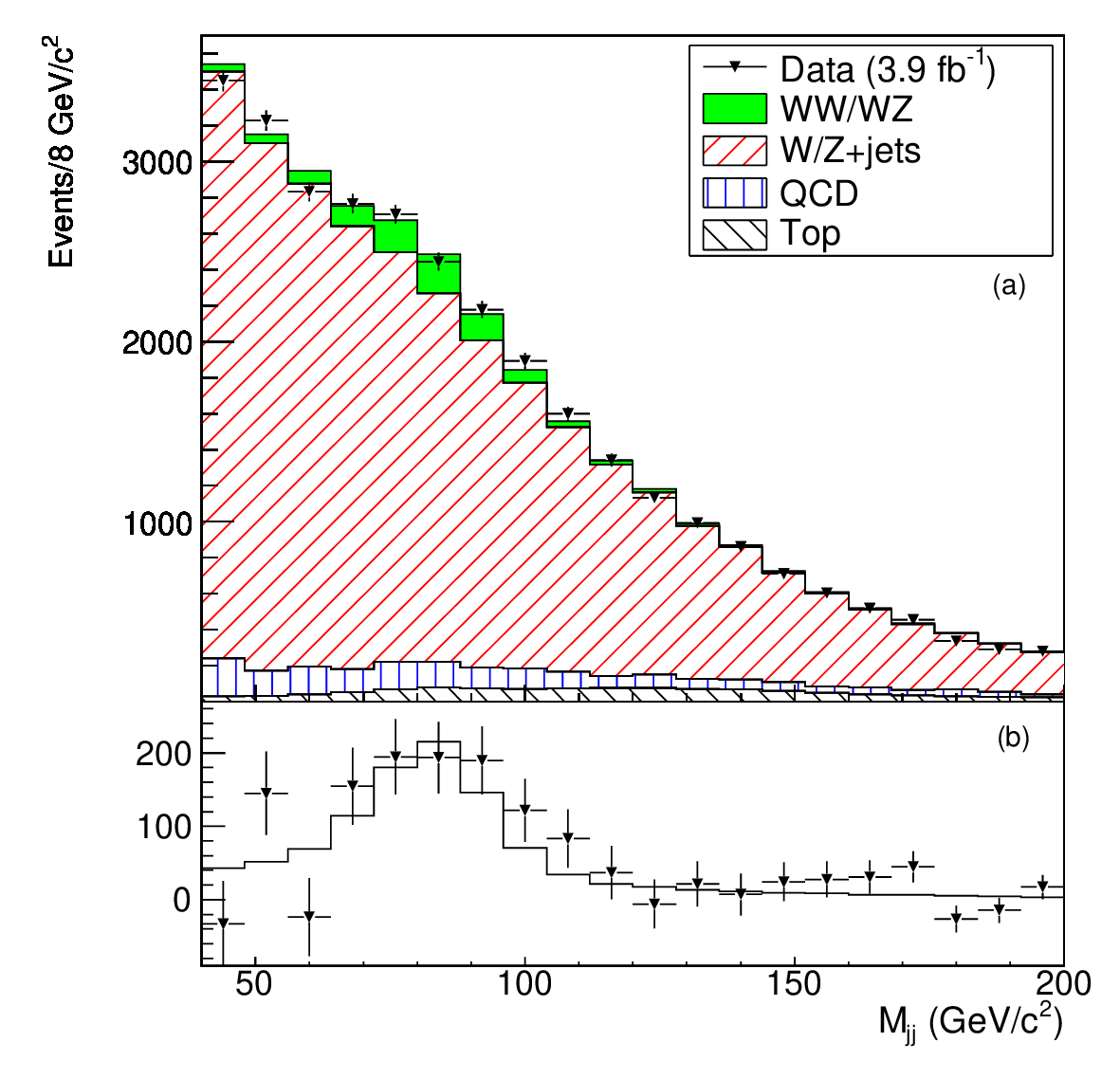}
  \caption{For the CDF analysis \cite{Aaltonen:2009vh}, the (a) dijet
    invariant mass distribution of reconstructed $W/Z \rightarrow jj$
    candidates compared to the fitted signal and background
    components, and (b) corresponding background subtracted
    distribution.}
\label{fig:cdf-wv-projections}
\end{figure}

In the dijet method, the signal fraction in the data is estimated by
performing a $\chi^2$ fit to the dijet invariant mass spectrum,
separately for electron and muon
events. Fig.~\ref{fig:cdf-wv-projections} shows the dijet invariant
mass distribution of data compared to the fitted signal and background
contributions. The observed (expected) signal significance using the
dijet method is 4.6$\sigma$ (4.9$\sigma$) for electrons and muons
combined. The measured cross section is 
\begin{equation*}
\sigma(WV + X) = 14.4 \pm 3.1 \rm{(stat.)} \pm 2.2 \rm{(syst.)}~\rm{pb} 
\end{equation*}

\begin{figure}
  \includegraphics[scale=0.5,width=1.0\columnwidth]{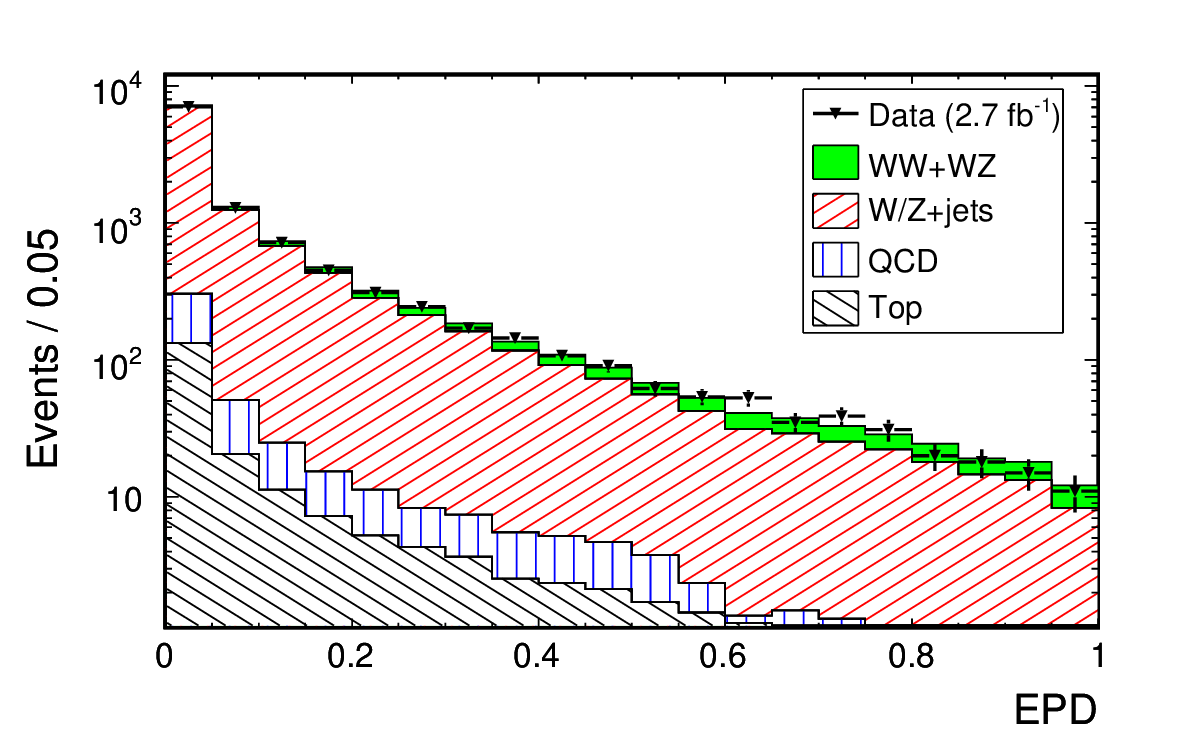}
  \caption{Observed $EPD$ distribution superimposed on distribution
    expected from simulated processes for the CDF analysis
    \cite{Aaltonen:2009vh}}.
\label{fig:cdf-wv-EPD}
\end{figure}

\begin{figure}
\includegraphics[width=0.96\columnwidth]{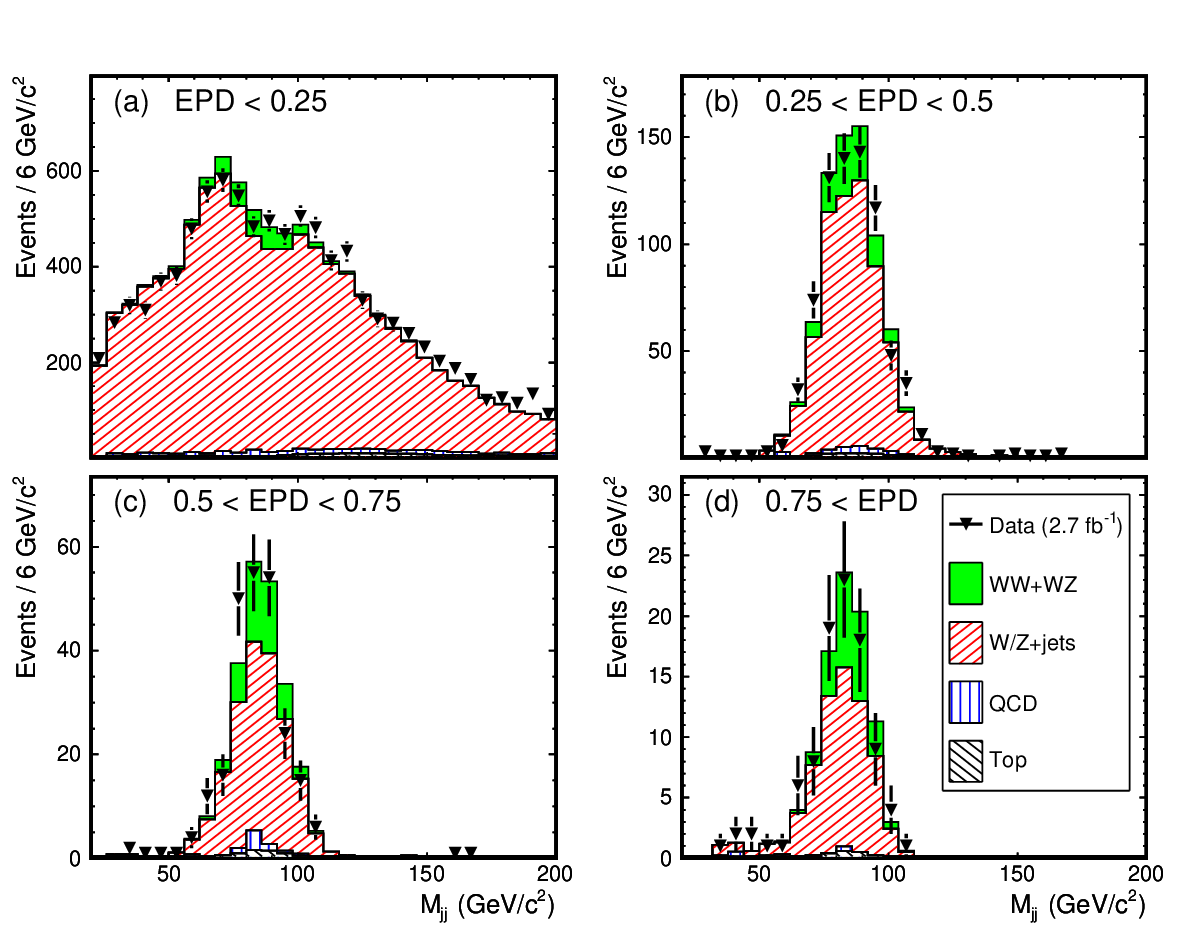}
\caption{For the CDF analysis \cite{Aaltonen:2009vh}, $M_{jj}$ for
  events with (a) $EPD < 0.25$, (b) $0.25 < EPD < 0.5$, (c) $0.5 < EPD
  < 0.75$, and (d) $EPD > 0.75$.}
\label{fig:cdf-wv-EPDbins}
\end{figure}

In the ME method, the calculated event probabilities are combined into
an event probability discriminant
$EPD=P_{signal}/(P_{signal}+P_{background})$, where $P_{signal} =
P_{WW}+P_{WZ}$ and $P_{background} =
P_{W+jets}+P_{single\;top}$. Templates of the $EPD$ generated for all
signal and background processes are used in a binned likelihood fit
for the signal yield observed in the data, as shown in 
Fig.~\ref{fig:cdf-wv-EPD}. Fig.~\ref{fig:cdf-wv-EPDbins} shows the
dijet mass in bins of $EPD$, where it can be seen that the low $EPD$
bin contains very little signal as compared to background. Events with
$EPD > 0.25$ have a dijet mass peak close to the expected $W/Z$ mass,
and the signal-to-background ratio improves with increasing $EPD$. The
observed (expected) signal significance using the ME method is
5.4$\sigma$ (5.1$\sigma$) which represents the first observation of
$WW+WZ$ production in the lepton + jets channel. The measured cross
section using the ME method is 
\begin{equation*}
\sigma(WV + X) = 17.7 \pm 3.1 ({\rm stat.}) \pm 2.4 ({\rm syst.}) ~\rm{pb}.
\end{equation*}

In the $M_{jj}$ method the largest systematic uncertainties are due to
the modeling of the EW and multijet shapes, about 8\% and 6\%
respectively. In the ME method the uncertainty in the jet energy scale
is the largest systematic uncertainty, at about 10\%, which includes
contributions both from the signal acceptance and from the shapes of
the  signal templates. In the $M_{jj}$ method this uncertainty is
about 6\%. The combined cross section for the dijet and ME methods,
with consideration of statistical and systematic uncertainties, is 
\begin{equation*}
\sigma(WV + X) = 16.0 \pm 3.3 ({\rm stat. + syst.}) ~\rm{pb}. 
\end{equation*}

\begin{figure}
\includegraphics[height=.24\textheight]{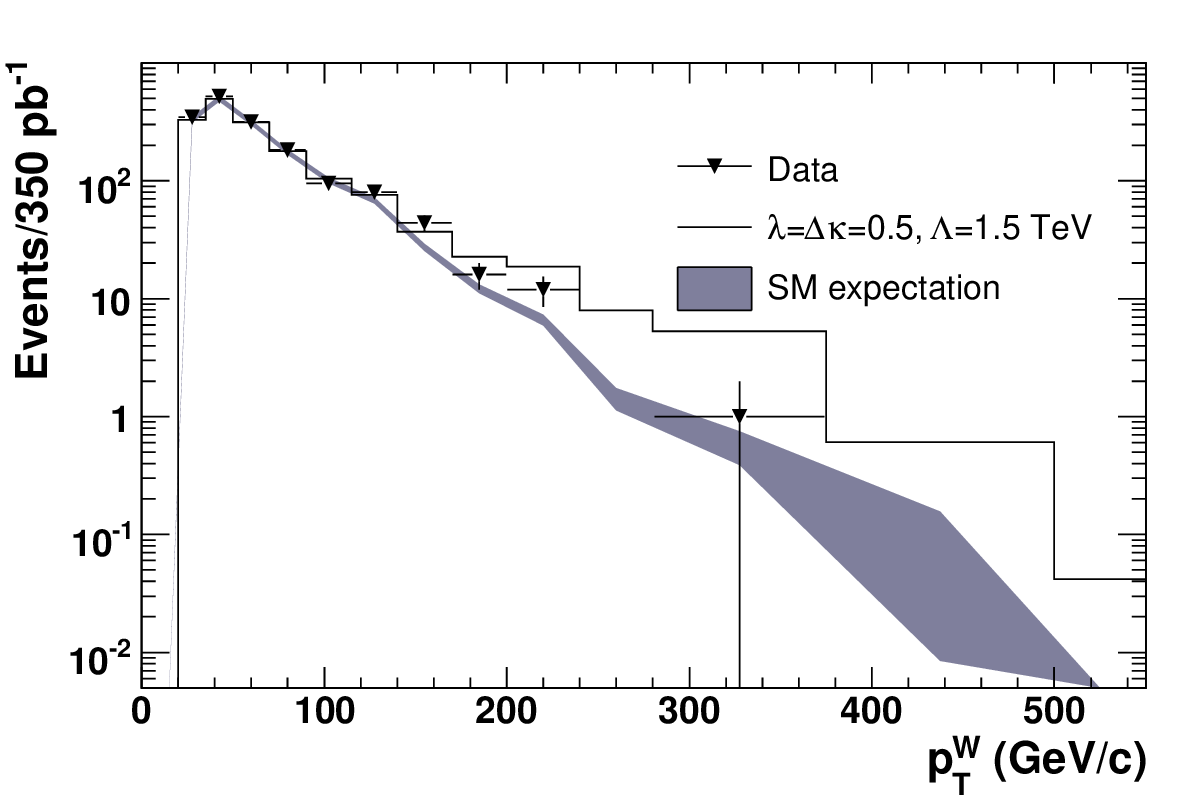} 
\caption{The $p_T^W$ distribution used for setting the anomalous
  TGC limits in the CDF analysis \cite{Aaltonen:2007sd}. Shown are
  the SM expectation, data points, and one anomalous coupling
  scenario with $\lambda=\Delta\kappa=0.5$ and $\Lambda=1.5$ TeV.
\label{fig:cdf-wv-enupt}}
\end{figure}

Both of the CDF and D0 $WW+WZ$ cross section measurements are
consistent with the SM NLO expectations. In \cite{Aaltonen:2007sd},
CDF searches for anomalous $WW\gamma$ and $WWZ$ couplings using the 
observed spectrum of the charged lepton $\pt$ from a $W$ decay and
$\intL=0.35\fb$. Fig.~\ref{fig:cdf-wv-enupt} shows good consistency
between the SM expectation for $W$ boson $\pt$ and the data which is
used to constrain possible anomalous couplings. In order to increase
the sensitivity to anomalous couplings, these data are combined with
the photon $E_T$ spectrum from the $W\gamma$ analysis
\cite{Acosta:2004it} described in
Section~\ref{sec:gaugeBosons_dibosons_Wgamma} that constrains possible
anomalous $WW\gamma$ couplings. Tab.~\ref{tbl:cdf-wv-atgc} summarizes  
the resulting anomalous coupling limits under the assumption of equal
$WW\gamma$ and $WWZ$ couplings.

\begin{table}
\begin{ruledtabular}
\begin{tabular}{ccccc}
 & $\lambda$ & $\Delta\kappa$ \\
\hline
$\ell\nu jj$ & (-0.28, 0.28) & (-0.50, 0.43)  \\
$\ell\nu\gamma$ \cite{Acosta:2004it} & (-0.21, 0.19) & (-0.74, 0.73) \\
Combined & (-0.18, 0.17) & (-0.46, 0.39) \\
\end{tabular}
\end{ruledtabular}
\caption{Allowed anomalous coupling ranges for $\Lambda = 1.5$ TeV at
  95\% CL, fixing the other coupling to the SM value and assuming equal
  $WW\gamma$ and $WWZ$ couplings for the CDF analysis
  \cite{Aaltonen:2007sd}.
\label{tbl:cdf-wv-atgc}}
\end{table}

\begin{figure} 
\includegraphics[width=8cm]{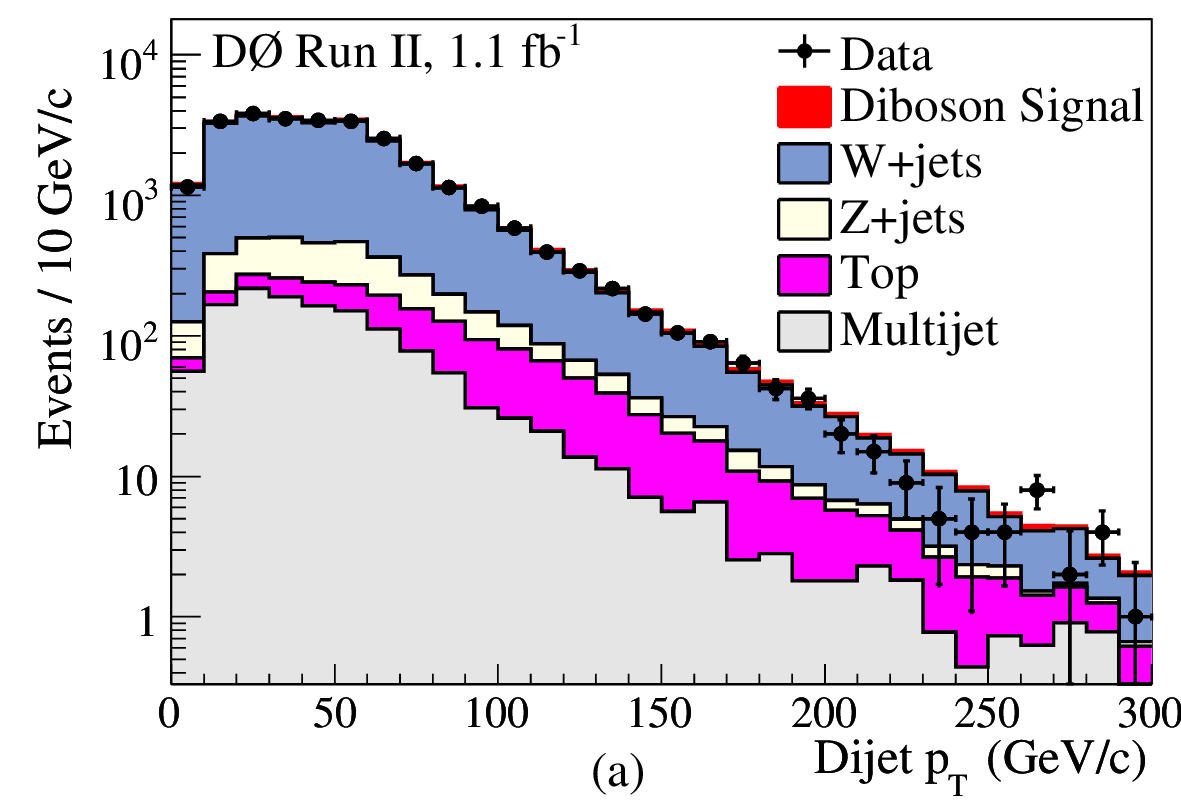}
\includegraphics[width=8cm]{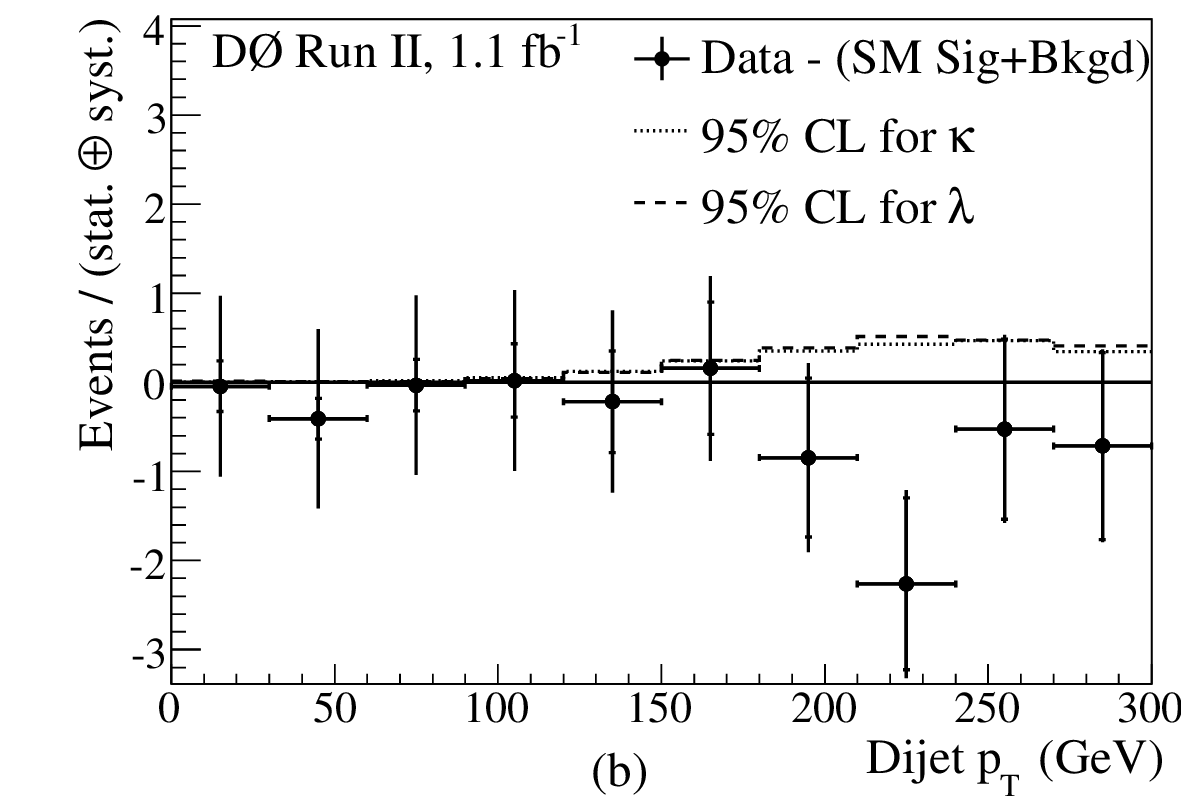}\\
    
\caption{For the D0 analysis \cite{Abazov:2009tr}, the (a) dijet
  $p_{T}$ distribution of combined (electron+muon) channels for data
  and SM predictions following the fit of simulated events to
  data and (b) difference between data and simulation divided by the
  uncertainty (statistical and systematic) for the dijet $p_{T}$
  distribution.  Also shown are the simulated signals for anomalous
  couplings corresponding to the 95\% CL limits for $\Delta\kappa$ and
  $\lambda$ in the LEP parametrization scenario. The full error bars
  on the data points reflect the total (statistical and systematic)
  uncertainty, with the ticks indicating the contribution due only to
  the statistical uncertainty.}
\label{fig:d0-wv-atgc-dijet-pt} 
\end{figure}

\begin{figure} 
\includegraphics[width=8cm]{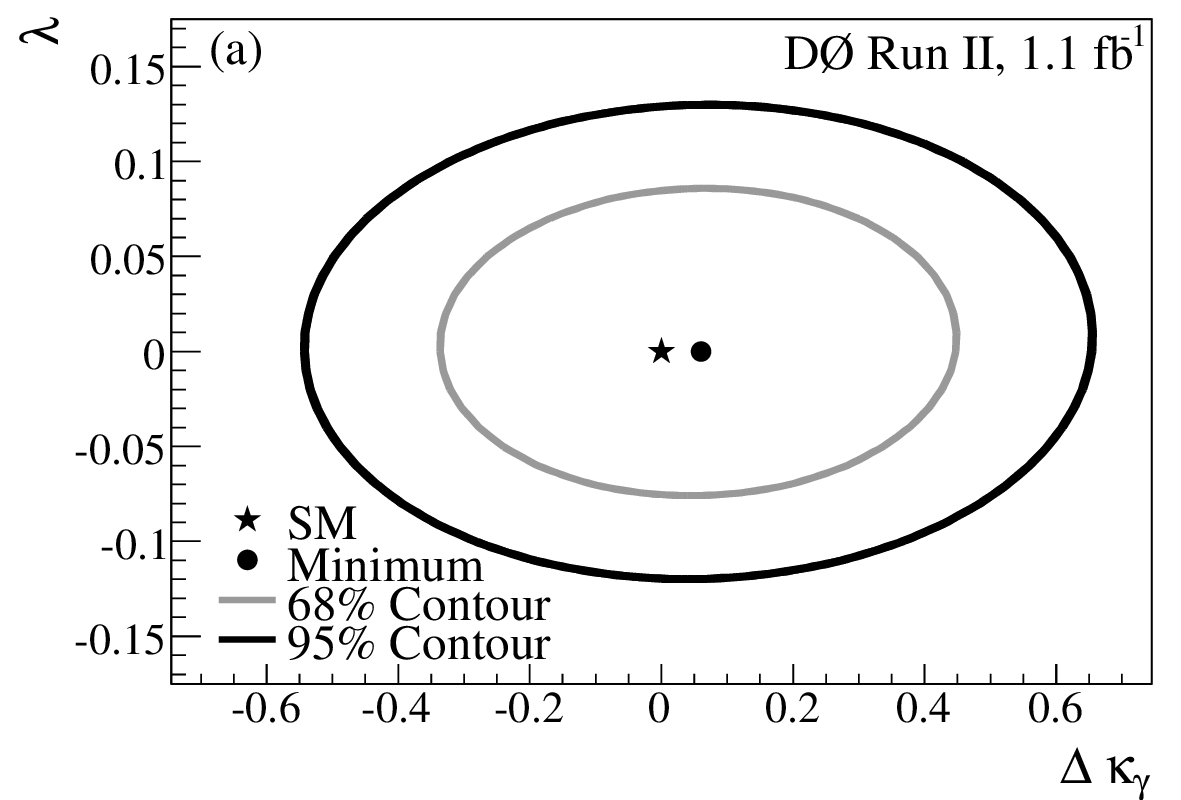}
\includegraphics[width=8cm]{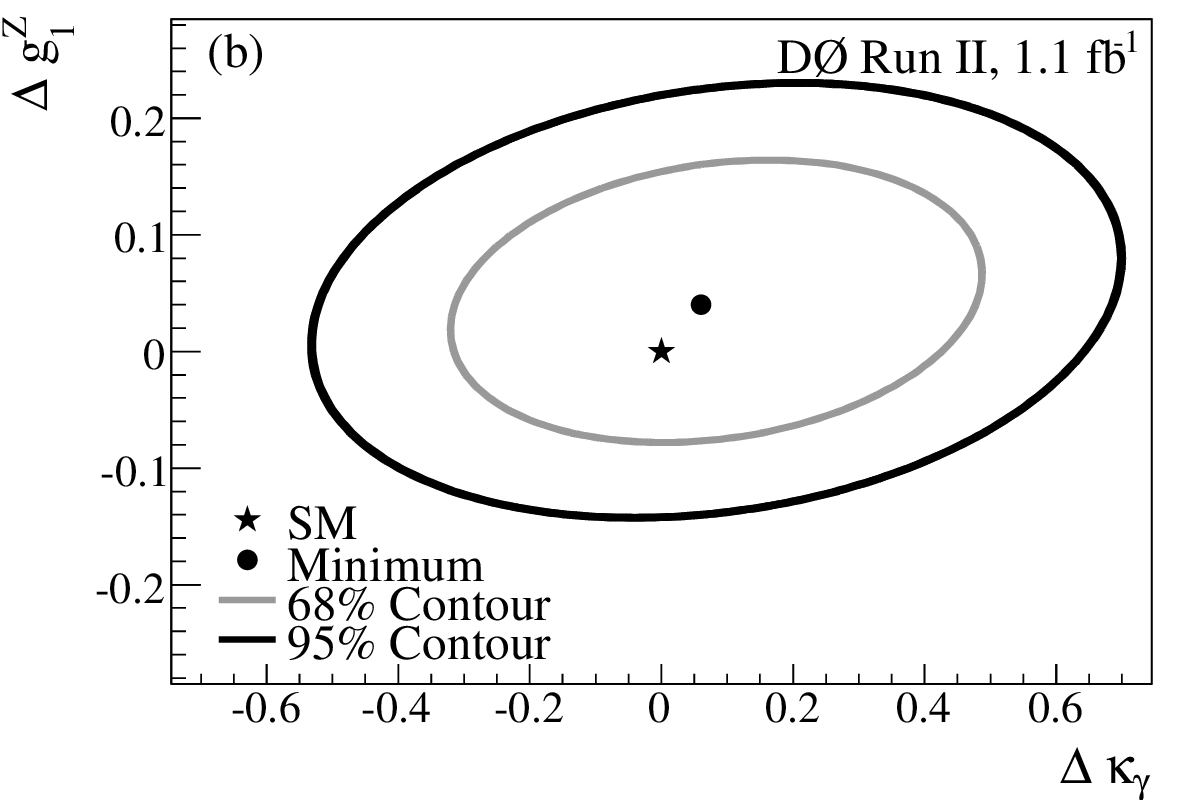} \\
\includegraphics[width=8cm]{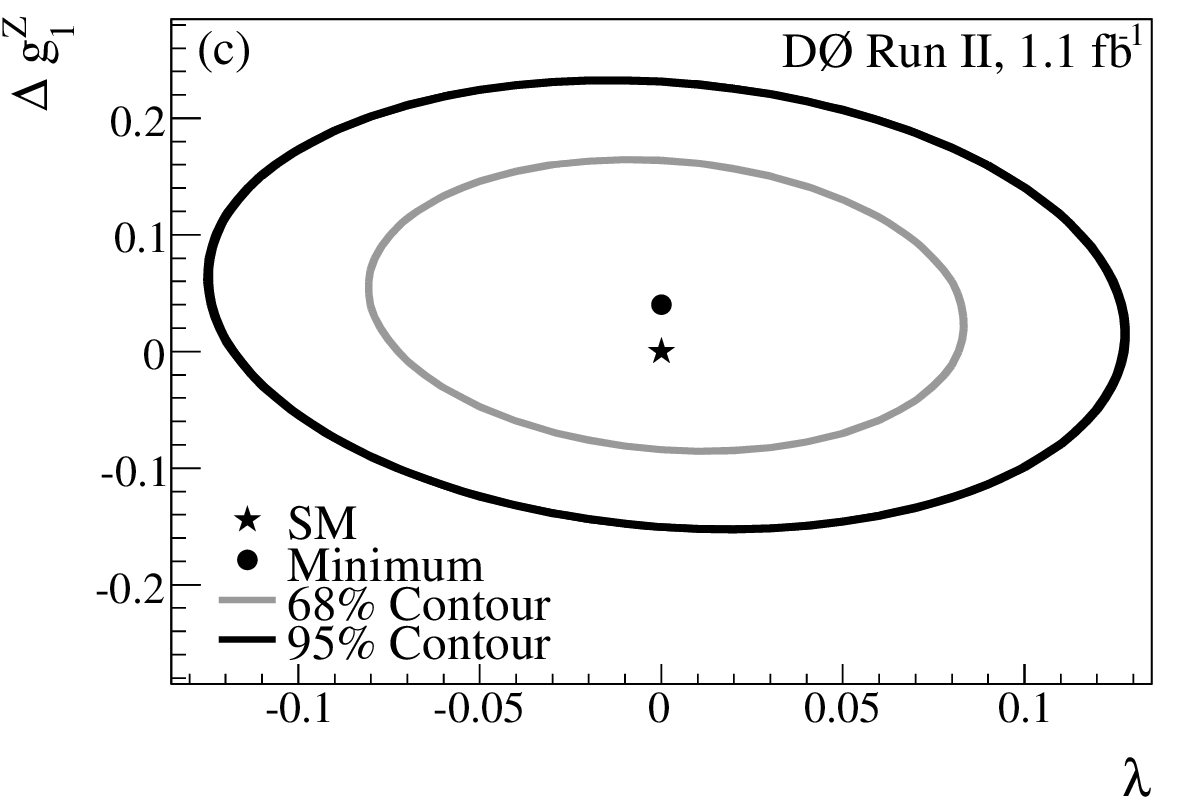}
\caption{The 68\% CL and 95\% CL two-parameter limits on the
  $\gamma WW/ZWW$ coupling parameters $\Delta\kappa_{\gamma}$,
  $\lambda$, and $\Delta g_{1}^{Z}$, in the LEP parametrization
  scenario and $\Lambda_{NP}=$ 2~TeV for the D0 analysis
  \cite{Abazov:2009tr}. The dots indicate the most probable values of
  anomalous couplings from the two-parameter combined (electron+muon)
  fit and the star markers denote the SM prediction.
\label{fig:d0-wv-atgc-limit1}} 
\end{figure}

\begin{figure} 
\includegraphics[width=8cm]{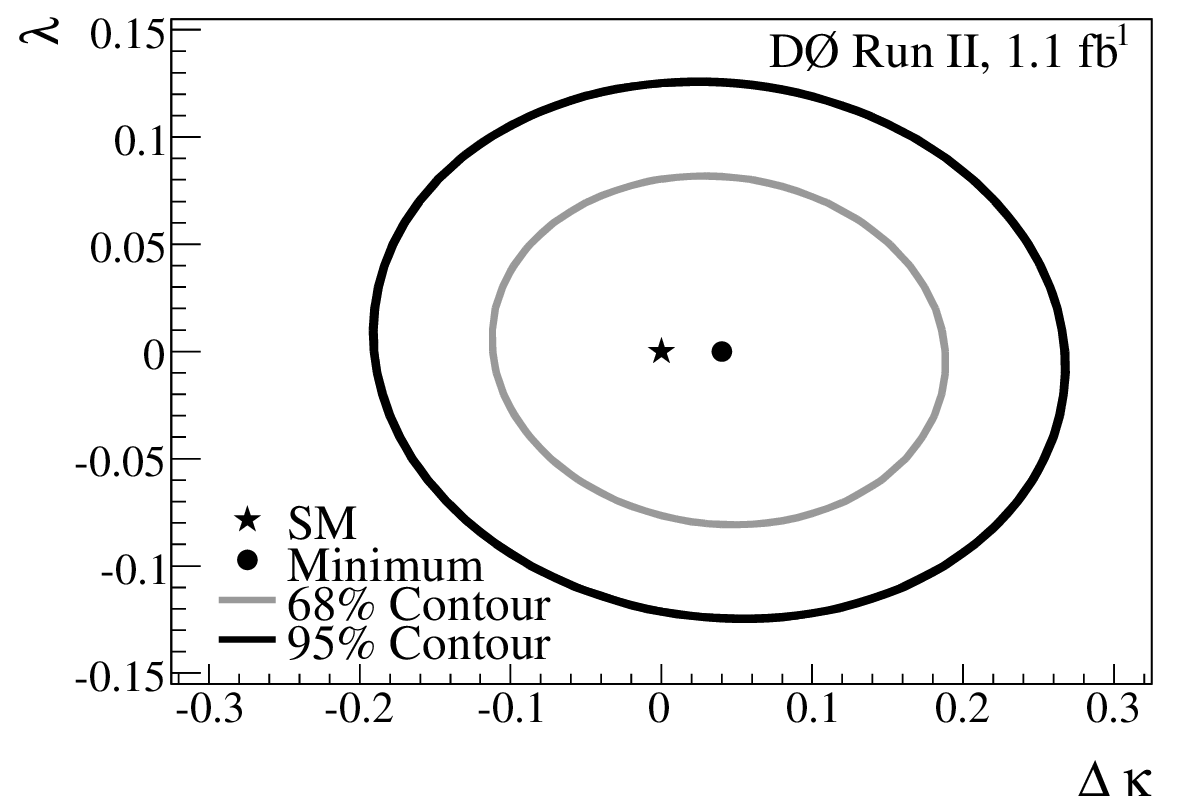}
\caption{The 68\% CL and 95\% CL two-parameter limits on the
  $\gamma WW/ZWW$ coupling parameters $\Delta\kappa$ and $\lambda$, in
  the equal couplings scenario and $\Lambda_{NP}=$ 2~TeV for the D0
  analysis \cite{Abazov:2009tr}. The dot indicates the most probable
  values of anomalous couplings from the two-parameter combined
  (electron+muon) fit and the star marker denotes the SM prediction. 
\label{fig:d0-wv-atgc-limit2}}
\end{figure}

The D0 search for anomalous couplings in the $\ell\nu jj$ channel
\cite{Abazov:2009tr} is based on the same data that was used to obtain
the first evidence for semileptonic decays of $WW/WZ$ boson pairs in
hadron collisions \cite{Abazov:2008yg}. In contrast to the CDF
analysis just described, the D0 analysis uses the dijet $\pt$
information to place limits on $WW\gamma/Z$ anomalous
couplings. Fig.~\ref{fig:d0-wv-atgc-dijet-pt} shows the dijet
$p_{T}$ distribution after a fit for the individual signal and
background contributions. No excess at high dijet $\pt$ that would
suggest that non-standard couplings are present is evident in the
data. The one-dimensional 95\% CL limits for the $WW\gamma/Z$ anomalous
coupling parameters under separate assumptions of $SU(2)_L\otimes
U(1)_Y$ symmetry and equal $WW\gamma$ and $WWZ$ couplings are shown in
Tab.~\ref{tbl:d0-wv-atgc-results}. The 2D limits under the assumption
of $SU(2)_L\otimes U(1)_Y$ symmetry \cite{Hagiwara:1993ck} is shown in
Fig.~\ref{fig:d0-wv-atgc-limit1} and for equal coupling assumption
they are shown in Fig.~\ref{fig:d0-wv-atgc-limit2}.

\begin{table*}
\begin{ruledtabular}
\begin{tabular}{lccc}
68\% CL & $\kappa_{\gamma}$ & $\lambda=\lambda_{\gamma}=\lambda_{Z}$ & $g_{1}^{Z}$ \\ \hline
$SU(2)_L\otimes U(1)_Y$ symmetry & $\kappa_{\gamma}=1.07^{+0.26}_{-0.29}$ & $\lambda = 0.00^{+0.06}_{-0.06}$ & $g_{1}^{Z}=1.04^{+0.09}_{-0.09}$\\
Equal couplings & $\kappa_{\gamma}=\kappa_{Z} = 1.04^{+0.11}_{-0.11}$
& $\lambda = 0.00^{+0.06}_{-0.06}$ &  \\ \hline \hline
 & & & \\
95\% CL & $\Delta\kappa_{\gamma}$ & $\lambda = \lambda_{\gamma}=\lambda_{Z}$ & $\Delta{g_{1}^{Z}}$ \\ \hline
$SU(2)_L\otimes U(1)_Y$ symmetry & -0.44 $< \Delta\kappa_{\gamma} <$ 0.55 & -0.10 $< \lambda <$ 0.11 & -0.12 $< \Delta g_{1}^{Z} <$ 0.20 \\
Equal couplings & -0.16 $< \Delta\kappa <$ 0.23 & -0.11 $< \lambda <$ 0.11 &  \\
\end{tabular}
\end{ruledtabular}
\caption{The most probable values with total uncertainties
  (statistical and systematic) at 68\% CL for $\kappa_{\gamma}$,
  $\lambda$, and $g_{1}^{Z}$ along with observed 95\%
  CL one-parameter limits on $\Delta\kappa_{\gamma}$, $\lambda$, and
  $\Delta g_{1}^{Z}$ measured in 1.1~fb$^{-1}$ of 
  $WW/WZ\rightarrow \ell \nu jj$ events with $\Lambda_{NP}=$ 2~TeV for
  the D0 analysis \cite{Abazov:2009tr}.
\label{tbl:d0-wv-atgc-results}}
\end{table*}

\subsubsection{$VV$ ($V =W,Z$)}
\label{sec:gaugeBosons_VV}

In
Sections~\ref{sec:gaugeBosons_dibosons_WW}-\ref{sec:gaugeBosons_dibosons_ZZ}, 
we summarized the observations of heavy vector boson pairs ($WW, WZ,
ZZ$) when each boson decays to one or more charged leptons. In
Section~\ref{sec:gaugeBosons_WV}, the measurements of a hadronically
decaying heavy vector boson produced in association with a $W$ boson
that decays to a charged lepton and a neutrino was
described. Observing diboson production in semileptonic decay mode
required sophisticated analysis techniques because of overwhelming
$W+$jets in addition to the presence of a $WW+WZ$ signal. The approach
of searching for diboson production in an more inclusive way can be
taken a step further by dropping the requirement of a reconstructed
charged lepton and instead looking at the $\MET$ + jets final
state. This approach includes an additional contribution from
$Z\rightarrow\nu\nu$ over the $WW/WZ\rightarrow\ell\nu jj$ mode so
that it measures $WW+WZ+ZZ$ production and allows for an event with a
diboson decay to be accepted even if one or more charged leptons fail
a selection cut. The experimental challenge is that this mode requires
a tighter $\MET$ cut as compared to $WW/WZ\rightarrow\ell\nu jj$ and a
better overall understanding of high reconstructed $\MET$ tails from
low true $\MET$ events with large cross section like QCD multijets.

The CDF Collaboration uses $\intL=3.5 \fb$ to search for the production
of heavy vector boson pairs ($VV$, $V$=$W,Z$) where one boson decays
to a dijet final state \cite{Aaltonen:2009fd}. The most significant
backgrounds to the diboson signal are $W$$(\ell\nu)$+jets,
$Z$$(\nu\nu)$+jets, and QCD multijet production. Other less
significant backgrounds include $Z$$(\ell\ell)$+jets, $t\bar{t}$, and
single $t$-quark production. The majority of events are collected
using an inclusive $\MET$ trigger which requires $\MET > 45$ GeV. In
the offline selection, events are required to have $\MET > 60$ GeV and
exactly two reconstructed jets with $\Et > 25$ GeV and $|\eta| < 2.0$.

\begin{figure}
  \includegraphics[width=0.49\textwidth]{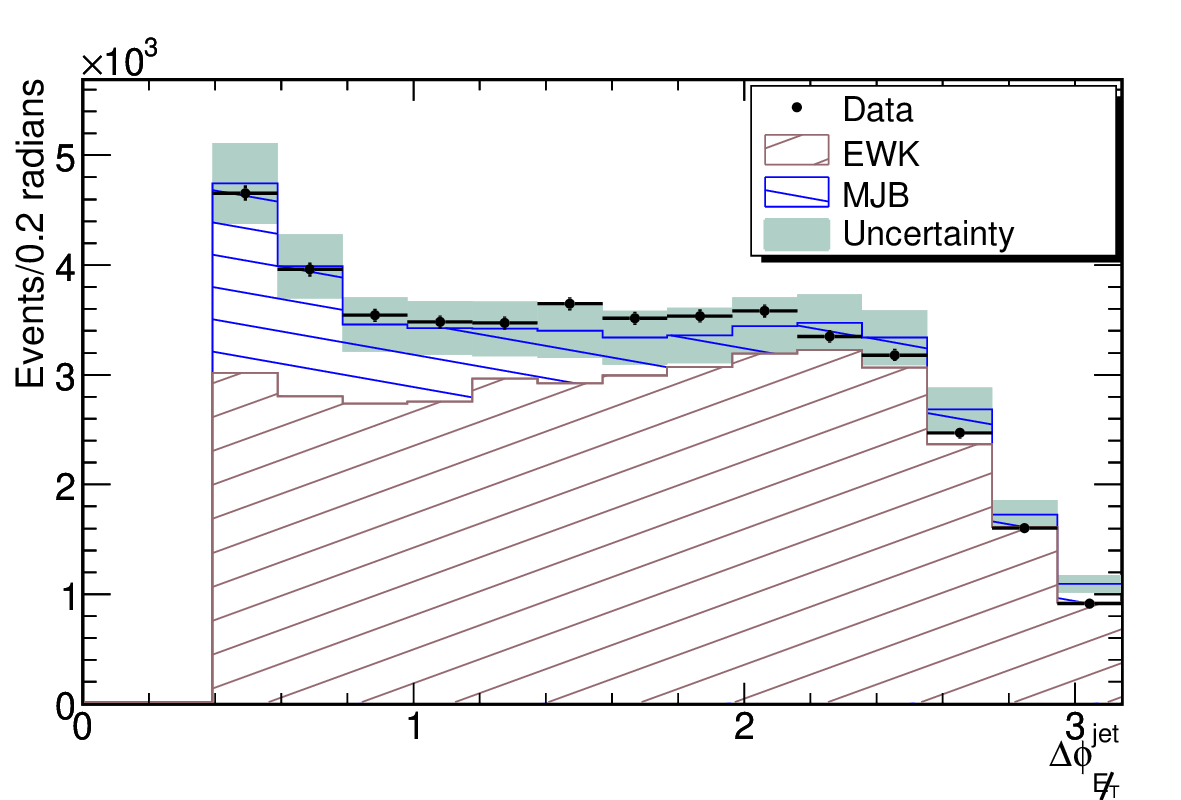} 
  \caption{Data compared with the sum of the predicted EW and multijet
    backgrounds for the $\Delta\phi^{jet}_{\MET}$ variable in the CDF
    analysis \cite{Aaltonen:2009fd}. The band represents the total
    systematic uncertainty on the background. The measured signal is
    included here in the EW contribution.}
  \label{fig:cdf-vv-dphiclosestwqcd}
\end{figure}

The shape and normalization of the multijet background are determined
from the data. A tracking-based missing momentum vector, $\mptvec$,
analogous to the calorimeter-based $\metvec$, is constructed from the
vector sum of the transverse momenta of particles measured in the
tracking system, and is largely uncorrelated to $\metvec$ for events
where jets are misreconstructed. In the absence of $\MET$ arising from
mismeasurement in the calorimeter, the $\metvec$ and $\mptvec$ will be
aligned in most events. At large values of
$\Delta\phi$(${\metvec},{\mptvec}$), multijets are expected to be
the dominant component of the data. The dijet mass shape and normalization
for the multijet background that remains after the event selection is
determined by selecting events with
$\Delta\phi$(${\metvec},{\mptvec}$)$>$1.0 and subtracting out the 
non-multijet backgrounds. The normalization is scaled up to account
for the multijet background contamination in the region
$\Delta\phi$(${\metvec},{\mptvec}$)$<$1.0. The shape of the multijet
background is fit to an exponential in $M_{jj}$ to derive a dijet mass
template for use in the $M_{jj}$ unbinned extended likelihood fit
performed to extract the diboson signal. The distribution of
$\Delta\phi_{\MET}^{jet}$ observed in the data is in good agreement
with the expectations as shown in
Fig.~\ref{fig:cdf-vv-dphiclosestwqcd}, giving confidence in the
validity of the multijet background model.

Three $M_{jj}$ template distributions are used in the fit: the first
is $V$+jets and $t$-quark production and is taken from Monte Carlo
simulation; the second is the multijet template, where the slope and
normalization are Gaussian constrained to their previously measured
values; the third template describes the signal. The signal shape is
comprised of the $WW,~WZ$, and $ZZ$ distributions obtained from a
Gaussian + polynomial fit to the signal Monte Carlo simulation where
the mean and the width of the Gaussian distribution are linearly
dependent on the jet energy scale (JES). The uncertainty associated
with the JES is the dominant source of systematic uncertainty of the
diboson cross section measurement.

\begin{figure}
\includegraphics[width=0.45\textwidth]{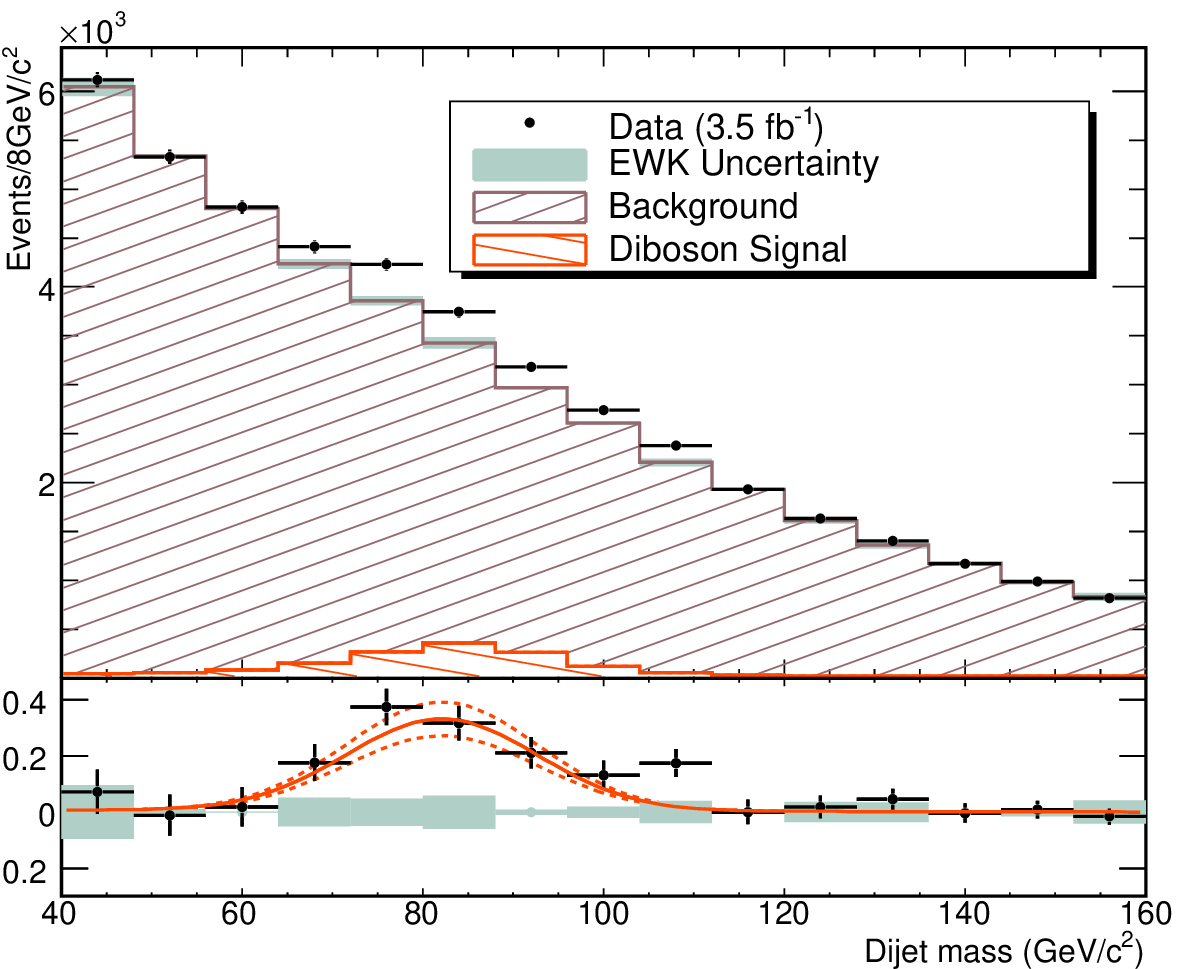}
\caption{Top: Comparison between data and fitted background only. The
  measured signal is shown unstacked. The band represents the
  systematic uncertainty due to the shape of EW background as
  described in \cite{Aaltonen:2009fd}. Bottom: Comparison of the
  diboson signal (solid line) with the background-subtracted data
  (points). The dashed lines represent the $\pm 1\sigma$ statistical
  variations on the extracted signal. The gray band represents the
  systematic uncertainty due the EW
  shape. \label{fig:cdf-vv-dijetmassfit}}
\end{figure}

Fig.~\ref{fig:cdf-vv-dijetmassfit} shows the fit result and a
comparison between the expected signal and data after background
subtraction. The signal significance is reported to be greater than
5.3$\sigma$ from the background-only hypothesis. This represents the
first observation in hadronic collisions of the production of weak
gauge boson pairs where one boson decays to a dijet final state. The
measured cross section is
\begin{eqnarray*}
\lefteqn{\sigma(WW+WZ+ZZ+X)} \\
& & {} =
18.0~\pm~2.8\mathrm{(stat.)}~\pm~2.4\mathrm{(syst.)}~\pm~1.1\mathrm{(lumi.)}
~\rm{pb}, 
\end{eqnarray*}
consistent with the SM NLO prediction of 16.8 $\pm$ 0.5 pb
\cite{Campbell:1999ah}.

\subsubsection{Cross Section and Anomalous TGC Summary}
\label{sec:gaugeBosons_summary}

\begin{figure}
\includegraphics[width=0.99\linewidth]{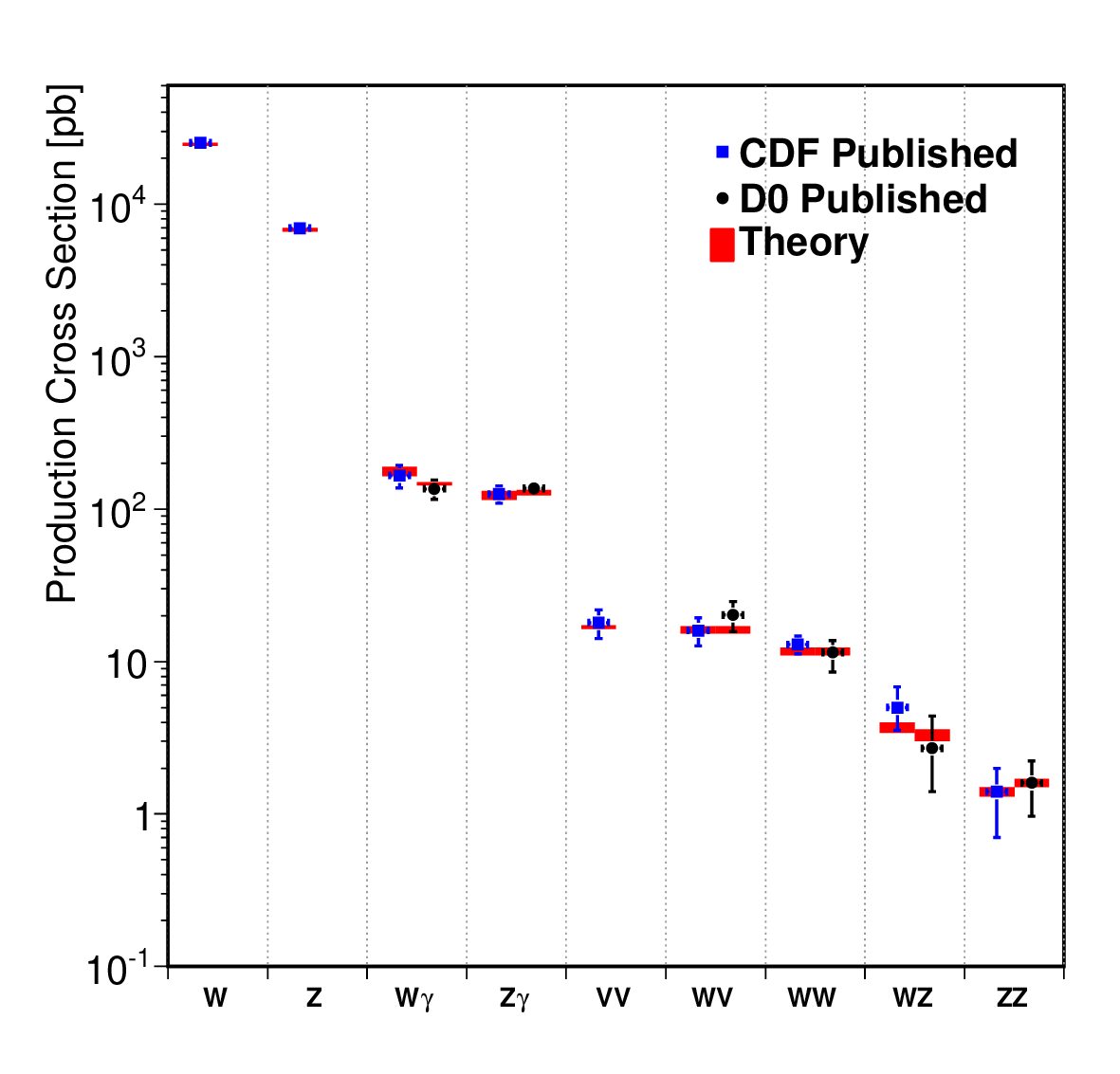}
\caption{A graphical summary of EW gauge boson cross section measurements and
  theory expectations during Run II and the Fermilab Tevatron.}
\label{fig:gaugeBosons_summary_XsecSumPlot}
\end{figure}

A summary of the diboson production cross sections at the Tevatron
with comparisons to the theory expectations can be found in
Tab.~\ref{tbl:diboson_xsect} and
Fig.~\ref{fig:gaugeBosons_summary_XsecSumPlot}. A summary of limits 
on anomalous charged and neutral TGCs at the Tevatron can be found in
Tab.~\ref{tbl:diboson_charged_aTGC} and
Tab.~\ref{tbl:diboson_neutral_aTGC}, respectively.

The experimental progress on diboson physics during Run II at the
Fermilab Tevatron has been truly remarkable. This includes the first
observation of $WZ$ production and the first observation at a hadron
collider of $ZZ$ production and of weak gauge boson pair production in
semileptonic decay. The associated diboson cross section measurements
(many of which are significantly improved from Run I) and detailed
studies of their kinematic properties provide stringent tests standard
electroweak theory and constraints on possible new physics coupling to
the EW gauge bosons.

%%%%%%%%% Cross sections summary figure

%%%%%%%%%% Cross sections table

\begin{table*}
\begin{ruledtabular}
\begin{tabular}{lccrrr}
  Process & Channel & $\int{\cal L} \ dt$ (fb$^{-1}$) 
  & \multicolumn{2}{c}{Cross section (pb)} & Expt. \\
  &         &       &  \ \ Theory \ \ & Measured \ \ \ \ \ \ \ \ \ \ \
  \ \ & Reference \\ \hline

  $p\bar{p}\rightarrow W\gamma + X \rightarrow \ell\nu\gamma + X$
  & & & & \\ 
  ~~~($E_T^\gamma > 7$ GeV, $\Delta R_{\ell,\gamma}>0.7$) 
  & $\ell\nu\gamma$ & $\sim0.2$ & $19.3\pm 1.4$ \cite{Baur:1992cd} 
  & $18.1 \pm 3.1 ~{\rm (stat. + syst.)}$
  & CDF \cite{Acosta:2004it} \\

  ~~~($E_T^\gamma > 8$ GeV, $\Delta R_{\ell,\gamma}>0.7$) 
  & $\ell\nu\gamma$ & $\sim0.2$ & $16.0\pm 0.4$ \cite{Baur:1992cd} 
  & $14.8 \pm 1.9 ~{\rm (stat. + syst.)} \pm 1.0 ~{\rm (lumi.)}$
  & D0 \cite{Abazov:2005ni} \\

  $p\bar{p}\rightarrow Z\gamma + X \rightarrow \ell\ell\gamma + X$
  & & & & \\ 
  ~~~($E_T^\gamma > 7$ GeV, $\Delta R_{\ell,\gamma}>0.7$, & & & & \\
  ~~~~~~$M_{\ell\ell} > 40$ GeV/$c^2$)
  & $\ell\ell\gamma$ & $\sim0.2$ & $4.5\pm 0.3$ \cite{PhysRevD.57.2823} 
  & $4.6 \pm 0.6 ~{\rm (stat.+syst.)}$
  & CDF \cite{Acosta:2004it} \\

  ~~~($E_T^\gamma > 7$ GeV, $\Delta R_{\ell,\gamma}>0.7$, & & & & \\
  ~~~~~~$M_{\ell\ell} > 30$ GeV/$c^2$)
  & $\ell\ell\gamma$ & $\sim1.1$ & $4.7\pm 0.2$ \cite{PhysRevD.57.2823} 
  & $5.0 \pm 0.3 ~{\rm (stat.+syst.)} \pm 0.3 ~{\rm (lumi.)}$
  & D0 \cite{Abazov:2007wy} \\

  $p\bar{p}\rightarrow Z\gamma + X \rightarrow \nu\nu\gamma + X$
  & & & & \\ 
  ~~~($E_T^\gamma > 90$ GeV)
  & $\nu\nu\gamma$ & ~~~$3.6$ & $0.039\pm 0.04$ \cite{PhysRevD.57.2823} 
  & $0.032 \pm 0.09 ~{\rm (stat.+syst.)} \pm 0.02 ~{\rm (lumi.)}$
  & D0 \cite{Abazov:2009cj} \\

  $p\bar{p}\rightarrow WW + X$ 
  & $\ell\nu\ell\nu$ & ~~~3.6 & $11.7\pm 0.7$ \cite{Campbell:1999ah} 
  & $12.9\pm 0.9 ~{\rm (stat.)} ^{+1.6}_{-1.4} ~{\rm (syst.)}$ 
  & CDF \cite{Aaltonen:2009us} \\

  & $\ell\nu\ell\nu$ & $\sim1.1$ & 
  & $11.5\pm 2.1 ~{\rm (stat.+syst.)} \pm 0.7 ~{\rm (lumi.)}$ 
  & D0 \cite{Abazov:2009ys} \\

  $p\bar{p}\rightarrow WZ + X$ 
  & $\ell\ell\ell\nu$ & ~~~1.1 & $3.7\pm 0.3$ \cite{Campbell:1999ah} 
  & $5.0 ^{+1.8}_{-1.4} ~{\rm (stat.)} \pm 0.4 ~{\rm (syst.)}$ 
  & CDF \cite{Abulencia:2007tu} \\

  & $\ell\ell\ell\nu$ & ~~~1.0 & 
  & $2.7 ^{+1.7}_{-1.3} ~{\rm (stat.+syst.)}$ 
  & D0 \cite{Abazov:2007rab} \\

  $p\bar{p}\rightarrow ZZ + X$
  & $\ell\ell\ell\ell$, $\ell\ell\nu\nu$ & ~~~1.9 & $1.4\pm 0.1$ \cite{Campbell:1999ah} 
  & $1.4 ^{+0.7}_{-0.6} ~{\rm (stat.+syst.)}$ 
  & CDF \cite{Aaltonen:2008mv} \\

  & $\ell\ell\ell\ell$, $\ell\ell\nu\nu$ & ~~~1.7 &
  & $1.6 \pm 0.6 ~{\rm (stat.)} \pm 0.2 ~{\rm (syst.)}$ 
  & D0 \cite{Abazov:2008gya} \\

  $p\bar{p}\rightarrow WV + X$ ($V\rightarrow W,Z$) 
  & $\ell\nu qq$ & ~~~2.7 & $16.1\pm 0.9$ \cite{Campbell:1999ah} 
  & $16.0\pm 3.3 {\rm (stat.+syst.)}$ 
  & CDF \cite{Aaltonen:2009vh} \\

  & $\ell\nu qq$ & ~~~1.1 & 
  & $20.2\pm 4.5 {\rm (stat.+syst.)}$ 
  & D0 \cite{Abazov:2008yg} \\

  $p\bar{p}\rightarrow VV + X $ ($V\rightarrow W,Z$) 
  & $\ell\nu qq, \nu\nu qq$ & ~~~3.5 & $16.8 \pm 0.5$ \cite{Campbell:1999ah} 
  & $18.0\pm 2.8 {\rm(stat.)} \pm 2.4{\rm(syst.)} \pm 1.1{\rm(lumi.)}$ 
  & CDF \cite{Aaltonen:2009fd} \\

\end{tabular}
\end{ruledtabular}
\caption{A tabular summary of EW gauge boson cross section measurements and
  theory expectations during Run II and the Fermilab Tevatron.}
\label{tbl:diboson_xsect}
\end{table*}

%%%%%%%%% Charged aTGC summary table

\begin{table*}
\begin{ruledtabular}
\begin{tabular}{lccccccccr}
  Mode & TGC(s) & $\int{\cal L} \ dt$ & $\Lambda$
  & \multicolumn{5}{c}{95\% CL Limits (model constraints in
    parentheses)} & Expt. \\

  & & (fb$^{-1}$) & (TeV)
  & $\lambda_\gamma$ & $\Delta\kappa_\gamma$ 
  & $\lambda_Z$ & $\Delta g_1^Z$ & $\Delta\kappa_Z$
  & Ref. \\ \hline

  $\ell\nu\gamma$ & $WW\gamma$ & $\sim0.7$ & 2.0 
  & $-0.12, 0.13$ & $-0.51, 0.51$
  & $-$ & $-$ & $-$
  & D0 \cite{Abazov:2008vja} \\

  $\ell\nu\ell\nu$ & $WW(\gamma,Z)$ & 3.6 & 1.5
  & $-0.16, 0.16$ & $-0.63, 0.72$ 
  & $\lambda_\gamma$ & $-0.24, 0.34$ & $\Delta g^Z_1 - \Delta\kappa_\gamma \tan^2\theta_W$
  & CDF \\

  & & & 2.0
  & $-0.14, 0.15$ & $-0.57, 0.65$
  & $\lambda_\gamma$ & $-0.22, 0.30$ & $\Delta g^Z_1 - \Delta\kappa_\gamma \tan^2\theta_W$ 
  & CDF \\

  & & $\sim1.1$ & 2.0
  & $-0.14, 0.18$ & $-0.54, 0.83$
  & $\lambda_\gamma$ & $-0.14, 0.30$ & $\Delta g^Z_1 - \Delta\kappa_\gamma \tan^2\theta_W$
  & D0 \cite{Abazov:2009ys} \\

  & & & 2.0
  & $-0.14, 0.18$ & $-0.12, 0.35$
  & $\lambda_\gamma$ & $0$ & $\Delta\kappa_\gamma$
  & D0 \cite{Abazov:2009ys} \\

  $\ell\ell\ell\nu$ & $WWZ$ & $\sim1.0$ & 1.5
  & $-$ & $-$
  & $-0.18, 0.22$ & $-0.15, 0.35$ & $-0.14, 0.31$ 
  & \\
  & & & & & & & & ($\Delta\kappa_Z = \Delta g_1^Z$) 
  & D0 \cite{Abazov:2007rab} \\

  & & & 2.0
  & $-$ & $-$
  & $-0.17, 0.21$ & $-0.14, 0.34$ & $-0.12, 0.29$
  & \\
  & & & & & & & & ($\Delta\kappa_Z = \Delta g_1^Z$) 
  & D0 \cite{Abazov:2007rab} \\

  $\ell\nu qq, \ell\nu\gamma$ & $WW(\gamma,Z)$ & 0.4 & 1.5
  & $-0.18, 0.17$ & $-0.46, 0.39$ 
  & $\lambda_\gamma$ & $0$ & $\Delta\kappa_\gamma$
  & CDF \cite{Aaltonen:2007sd} \\

  & & 1.1 & 2.0
  & $-0.10, 0.11$ & $-0.44, 0.55$ 
  & $\lambda_\gamma$ & $-0.12, 0.20$ & $\Delta g^Z_1 - \Delta\kappa_\gamma \tan^2\theta_W$
  & D0 \cite{Abazov:2009tr} \\

  & & &
  & $-0.11, 0.11$ & $-0.16, 0.23$ 
  & $\lambda_\gamma$ & 0 & $\Delta\kappa_\gamma$
  & D0 \cite{Abazov:2009tr} \\

  $\ell\nu\gamma$,$\ell\nu\ell\nu$,$\ell\ell\ell\nu$,$\ell\nu qq$  
  & $WW(\gamma,Z)$ & $\sim$1.0 & 2.0
  & $-0.08, 0.08$ & $-0.29, 0.38$ 
  & $\lambda_\gamma$ & $-0.07, 0.16$ & $\Delta g^Z_1 - \Delta\kappa_\gamma \tan^2\theta_W$
  & D0 \cite{Abazov:2009hk} \\

  & & &
  & $-0.08, 0.08$ & $-0.11, 0.18$ 
  & $\lambda_\gamma$ & 0 & $\Delta\kappa_\gamma$
  & D0 \cite{Abazov:2009hk} \\
\end{tabular}
\end{ruledtabular}
\caption{A summary of anomalous charged TGC limits for different diboson
  processes and decay channels measured during Run II and the Fermilab
  Tevatron.}
\label{tbl:diboson_charged_aTGC}
\end{table*}

%%%%%%%%%% Neutral aTGC summary table

\begin{table*}
\begin{ruledtabular}
\begin{tabular}{lcccccccr}
  Mode & TGC(s) & $\int{\cal L} \ dt$ & $\Lambda$
  & \multicolumn{4}{c}{95\% CL Limits} & Expt. \\

  & & (fb$^{-1}$) & (TeV) & & & & & Ref. \\ \hline

  $\ell\ell\ell\ell$ & $ZZZ, ZZ\gamma$ & 1.0 & 1.2
  & $-0.28 < f_{40}^\gamma < 0.28$
  & $-0.26 < f_{40}^\gamma < 0.26$
  & $-0.31 < f_{50}^Z < 0.29$ 
  & $-0.30 < f_{50}^Z < 0.28$ 
  & D0 \cite{Abazov:2007hm} \\

  $\ell\ell\gamma,\nu\nu\gamma$ & $Z\gamma\gamma, ZZ\gamma$ & 1.0, 3.6 & 1.5
  & $|h_{30}^\gamma | < 0.033$ 
  & $|h_{40}^\gamma | < 0.0017$ 
  & $|h_{30}^Z | < 0.033$
  & $|h_{40}^Z | < 0.0017$ 
  & D0 \cite{Abazov:2009cj} \\

\end{tabular}
\end{ruledtabular}
\caption{A summary of anomalous neutral TGC limits for different diboson
  processes and decay channels measured during Run II and the Fermilab
  Tevatron.}
\label{tbl:diboson_neutral_aTGC}
\end{table*}

%%%%%%%%%%%%%%%%%%%%%%%%%%%%%%%%%%%%%%%%%%%%%%%%%%%%%%%%%%%%%%%%%%%%%%%%%%%%
%
% Top Quark
%
%%%%%%%%%%%%%%%%%%%%%%%%%%%%%%%%%%%%%%%%%%%%%%%%%%%%%%%%%%%%%%%%%%%%%%%%%%%%

\section{TOP QUARK}
\label{sec:top}

\subsection{Top Quark Mass}
\label{sec:top_mass}
As discussed in section~\ref{sec:higgs_ewkConstraints}, constraints on
the Higgs boson mass from fits to data assuming the SM are limited by
the experimental precision of the $W$ boson and top quark masses.
Measurements of the top quark mass thus have direct implications for
the SM, and these measurements are a central component of the programs
for both CDF and D0.  As evidenced from Fig.~\ref{f-mwmt}, constraints
on the Higgs mass are limited more by the uncertainty in the $W$ boson
mass than by the uncertainty in the top-quark mass, which reflects the
tremendous progress than has been made on the top-quark mass
measurement in Run II.  In fact, the top-quark mass is now the most
accurately known quark mass (as a percentage of its value), surpassing
even its $SU(2)_{\rm L}$ partner the $b$ quark, whose mass is known to
about 2.6\% ($\bar m_b(\bar m_b)=4.24\pm 0.11\ \GeV$)
\cite{ElKhadra:2002wp}.

The top-quark mass measured at the Tevatron corresponds closely to the
pole mass.  The pole mass is the mass the quark would have in the
absence of confinement.  Although the top quark decays on a time scale
less than the time scale associated with confinement
($\Lambda_{QCD}^{-1}$), the top-quark pole mass is nevertheless
affected by confinement \cite{Smith:1996xz}.  The ambiguity in the
top-quark pole mass is of order $\Lambda_{QCD}\approx 200$ MeV, which
is considerably less than the present uncertainty.  A more precise
definition of mass will be needed if one attempts to measure the mass
to an accuracy much less than 1 $\GeV$. 

Both CDF and D0 have made measurements of the top quark mass in a
variety of final states and using different methods.  The general
procedure is to select events consistent with $\ttbar$ production in
either the dilepton or lepton plus jets final state.  In the SM, top
quarks decay via $t\to Wb$ with a branching fraction of essentially
100\%.  The dilepton final state is that in which both $W$ bosons from
the $\ttbar$ pair decay to leptons, and the lepton+jets final state is
that in which one $W$ decays to leptons and the other to quarks.
Measurements of the mass have also been made in the all jets final
state in which both $W$s decay to quarks, although this offers less
precision because of significantly higher backgrounds. 

An initial selection of candidate dilepton events is made, typically
requiring two high $p_T$ charged leptons, significant $\MET$
attributed to neutrinos, and at least two jets.  Typical thresholds
for the lepton and jet transverse momenta range between 15~$\GeVc$ and
20~$\GeVc$. The $\MET$ is attributed to the two neutrinos from
$\ttbar$ decay, and the thresholds are higher, ranging between
35 GeV and 50 GeV depending on topology.  The initial
selection of lepton+jets candidate events requires a single charged
lepton, $\MET$ and at least three (and often four) jets.  The lepton
and jet thresholds are similar to the dilepton final state, but the
$\MET$ threshold is typically relaxed giving requirements ranging from
$\MET>15$ GeV to $\MET>20$ GeV. Because the lepton+jets events have
significant background from $W$+jet events, an additional requirement
that one or more of the jets is consistent with $B$ hadron production
($b$-tagged) is often made.  The typical tagging efficiency is
50\%/jet with light quark misidentification rates of roughly 1\%. The
probability to misidentify a charm quark induced jet as
being consistent with $B$ hadron production is roughly 15\%.  Actual
values vary somewhat depending on the $b$--tagging algorithm used.
The values also depend on the signal-to-background requirements of the 
specific analysis which allow more or less restrictive tagging
requirements.  These loose initial selections have minimal top quark
mass bias and form the basis for further analysis.

Typical signal and background estimates are given in
Tab.~\ref{t:mtop-evts} for various channels.  In the selections,
trade-offs are made between signal acceptance and purity.  In
particular, the requirement that an event have one or more jets
consistent with being initiated by $b$-quarks considerably improves
the signal purity. 
\def\dzllSpace{\hphantom{$\dilep$, }}
\begin{table*}
  \begin{ruledtabular}
  \begin{tabular}{lcccccc} 
                   & $\intL$     & Expected   & Estimated  & Observed & $b$-tag    & Expt.\\
    Channel        & (fb$^{-1}$) &Signal Yield& Total Yield&  Yield   & Required?  & Ref.\\ \hline
   $\dilep$        &      2      & $43.8\pm4.4$ & $215.8\pm21.9$ & 246& $\equiv0$  & CDF\cite{b-mt-cdf-neuro}\\
                   &             & $78.0\pm6.2$ & $97.5\pm7.2$   &  98& $\ge 1$    & \\ \hline
%  $\dilep$        &     1.2     & & & & & CDF\cite{b-mt-cdf-llcomb} \\
%  $\dilep$        &      1      & & & & &  CDF\cite{b-mt-cdf-llmat} \\
   $\dilep$, $e\mu$&      1      & $36.7\pm2.4$ & $44.5\pm2.7$   & 39 &  no        &  D0\cite{b-mt-d0-llcomb} \\
   \dzllSpace $ee$       &       & $11.5\pm1.4$ & $14.8\pm1.5$   & 17 &            & \\
   \dzllSpace $\mu\mu$   &       & $8.3\pm0.5$  & $13.7\pm0.7$   & 13 &            & \\
   \dzllSpace $e$+track  &       & $9.4\pm0.1$  & $10.3\pm0.2$   &  8 & $\ge 1$    & \\
   \dzllSpace $\mu$+track &      & $4.6\pm0.1$  & $ 5.5\pm0.1$   &  6 &            & \\       \hline
   $\ljets$        &     1.9     & $183\pm25$   & $247\pm29$   & 284& $ =1$        &  CDF\cite{b-mt-cdf-ljmat} \\
                                            &             & $ 69\pm11$   & $ 75\pm11$   &  87& $\ge 2$      & \\       \hline
   $\ljets$        &      1      & $162\pm11$   & $57.2\pm4.2$(est) & 200 & $\ge 1$&  D0\cite{b-mt-d0-ljmat}  \\ 
%  $\ljets$        &     0.7     & & & & CDF\cite{b-mt-cdf-ljdl} \\
%%JDH, superceded above $\ljets$, D0\cite{}              & & & & \\
%  $\ljets$        &     0.4     & & & &  D0\cite{b-mt-d0-ideo} \\
%  $\ljets$        &     0.3     & & & &  CDF\cite{b-mt-cdf-300prl}   \\
%  All jets        &      1      & & & &  CDF\cite{b-mt-cdf-jj}       \\
%  $\mjets$        &     0.3     & & & &  CDF\cite{b-mt-cdf-mj}       \\
  \end{tabular} 
  \end{ruledtabular}
  \caption{Signal and background yields for preselected $\ttbar$ samples from
    a subset of the published results.  The purity varies considerably 
    depending on the final state and whether or not $b$-tagged
    jets are required.\label{t:mtop-evts}}
\end{table*}

After the initial selections, various methods are used to determine
the top mass. Among these are: (1) template methods in which data are
fit to the predicted signal plus background distribution for a
variable that is sensitive to the top mass, (2) ME methods similar to
those discussed in Sec.~\ref{sec:gaugeBosons_dibosons_WW} in which
$\ttbar$ event kinematics determined by matrix element calculations
are convoluted with detector resolution functions to predict measured
kinematic distributions as a function of the top mass used in the
matrix element calculations, and (3) other weighting methods in which
missing information is supplied by trying possible reconstructed jet
and true $\ttbar$ decay parton assignments and computing a probability
for the particular assignment as a function of top mass.  The ME and
various weighting methods compute event-by-event likelihoods as a
function of top mass, and then compute the total likelihood for a data
set as the products of the event likelihoods. Combinations of these
methods are also used.

For lepton+jets events, the momenta of the four jets and the lepton
are fully  measured, and the neutrino momentum in the plane transverse
to the beam is inferred from the event $\MET$.  This gives seventeen
measured values.  Two constraints are available.  The first is that
the invariant mass of one pair of reconstructed jets should be
consistent with that of the $W$ boson, and the second is that the
reconstructed mass of the two top quarks should be consistent with
each other.  This information allows a one degree-of-freedom fit to be
performed in which the top mass is a free parameter.  The resulting
event top mass values are not simply averaged to determine the mass,
but provide an input variable to the various methods described above.
The fit $\chi^2$ can also be used to determine (or weight) the most
likely parton-jet assignment in an event.  Complications to this fit
arise from the presence of intrinsic transverse momentum of the
$\ttbar$ system $p_T^{\ttbar}$ and from final state gluon radiation
giving rise to additional jets.  The $p_T^{\ttbar}$ can be
incorporated by adding an additional variable for the unknown
transverse momentum $\vec{X}$ to the fit and constraining it using
{\it a priori} estimates of the spectrum.  The impact of final state
radiation can be controlled by requiring exactly four
reconstructed (good) jets in the event.

For dilepton events, the number of measurements minus the number of
constraints does not give enough information to fully determine the
final state kinematics, and one assumption must be made.  A variety of
methods are used including ME weighting and neutrino weighting.  The
ME weighting method~\cite{b-dalitz,b-kondo1,b-kondo2} is related to
the general ME method described below (and also used for $\ell$+jets
analyses). The neutrino weighting method~\cite{b-d0mwt1,b-d0mwt2} is
unique to dilepton events.  In this method, an event weight is defined
as a function of hypothesized top quark mass $M_t$ using a comparison
of the measured $\vec{\MET}$ to $\vec{\MET}_i$ values predicted for a
set of possible neutrino pseudorapidity values
$\eta^{(1)}_i,\eta^{(2)}_i$.  Large weights correspond to situations
in which the measured and predicted $\vec{\MET}$ are similar and thus
give a probability for different $\eta$ values for the neutrinos in
each event. 

The use of the ME approach has become widespread, especially for
$\ell$+jets final states.  It is used either as the final analysis
variable used to determine the measured mass or to provide information
used in joint likelihoods for determining the top mass.  In this
approach, the probability that the the jet momenta, lepton momentum
and missing transverse energy observed in a given event, assuming it
arises from $\ttbar$ production and decay, is computed
by\cite{b-mt-d0-ljmat} 
\begin{equation*}
 P_{\ttbar} = \frac{1}{N}\int\sum d\sigma(y;M_t)\, dq1\, dq2\, f(q_1)
   \, f(q_2)\, W(y,x)
\end{equation*}
in which $d\sigma$ is the differential cross section for production of
the final state partons in (momentum) configuration $y$ for a given
$M_t$, $f(q_1)$ and $f(q_2)$ are the parton density functions for the
proton and antiproton, $W(y,x)$ is the transfer function for computing
the probability that a partonic final state $y$ gives the observed
reconstructed final state $x$ and $1/N$ is a normalization factor
equal to the expected observed cross section for a given $M_t$.  For
each event this probability, and the corresponding probability
computed assuming the event is a background event, $P_{bkg}(x)$, are
combined to give an event probability as a function of top mass 
\begin{eqnarray*}
 P(x;M_t) = A(x)(fP_{\ttbar}(x;M_t) + (1-f)P_{bkg}(x))
\end{eqnarray*}
in which $A(x)$ is a normalization factor incorporating acceptance and
efficiency effects and $f$ is the fraction of $\ttbar$ events in the total
sample.  The overall normalization of $P$ is forced to unity so that
it can be used as a probability density.  The event probability is
then used either as input to additional likelihoods or the joint
likelihood for the entire sample is formed as the product of the event
probabilities and minimized as a function of $M_t$.

\begin{figure}
  \includegraphics[width=\linewidth]{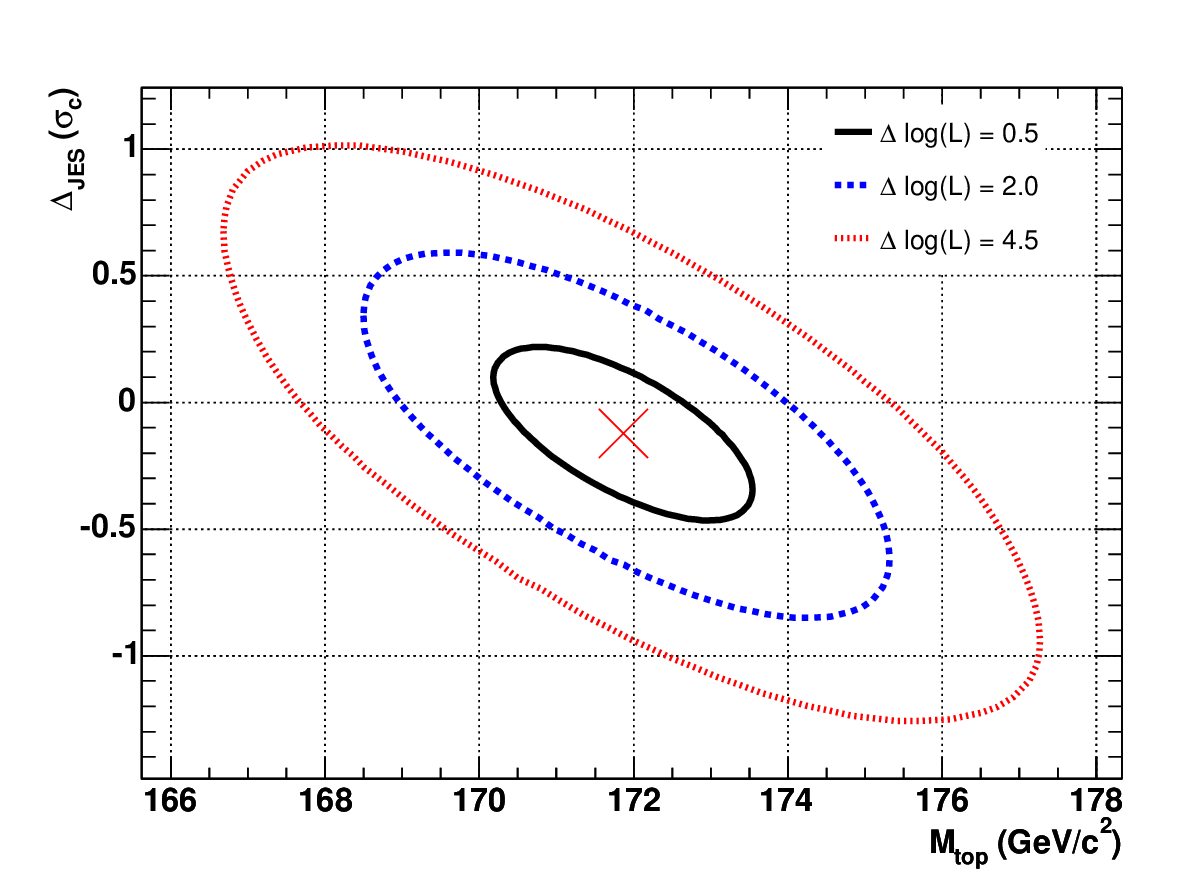}
  \caption{Contours of constant -log(L) in the $\Delta_{jes} \equiv k_{jes} -1 $ 
    versus $M_t$ plane\cite{b-mt-cdf-simult}.  A perfect {\em a priori} jet 
    energy calibration used as input to this analysis would result in 
    $k_{jes} = 1$ corresponding to $\Delta_{jes}=0$.\label{f:mt-jes}}
\end{figure}

The methods used in early Run II top mass measurements had systematic
uncertainty from the energy calibration of reconstructed jets which,
if unmodified, would have quickly surpassed the statistical
uncertainty.  A method was developed to incorporate an {\it in situ}
calibration scale factor as a second fit parameter when determining
the top mass.  As an example, the ME method event probability
definition was modified to the form now in general use.  This form is

\begin{eqnarray*}
\lefteqn{P(x;M_t,k_{jes}) =  A(x)} \\
& & {} \times \left[fP_{\ttbar}(x;M_t,k_{jes}) +
  (1-f)P_{bkg}(x;k_{jes})\right ]
\end{eqnarray*}

in which $k_{jes}$ is a scale factor applied to all jet energies.  In
this form the transfer functions also become dependent on the
calibration, $W(y,x,k_{jes})$. The negative log likelihood minimized
to determine the top mass thus becomes a function of two variables,
$m_{top}$ and $k_{jets}$, both of which are varied during the
minimization.  As an example, Fig.~\ref{f:mt-jes} shows the 2D contour
in the $\Delta_{jes} \equiv k_{jes}-1$ versus $m_{top}$ plane from the
$M_t$ measurement using the $\dilep$ and $\ljets$ final states
simultaneously.~\cite{b-mt-cdf-simult}.

Systematic uncertainties in top mass measurements arise from both
experimental sources and theory.  Although the contribution from
calibration uncertainties has been significantly reduced as described
in the previous paragraph, the dominant contributions to the
systematic uncertainty from experimental sources are the absolute
reconstructed jet energy calibration and the relative calibrations of
jets initiated by $b$-quarks and those from light quarks ($u,\ d,\
s$).  Of these, the uncertainty from the jet calibrations dominates
the experimental uncertainties.  Uncertainties arising from theory
include production model as assessed by comparing event generators and
fragmentation. The production model uncertainty becomes the dominant
theory uncertainty when various measurements are combined. 

\begin{figure}
  \includegraphics[width=\linewidth]{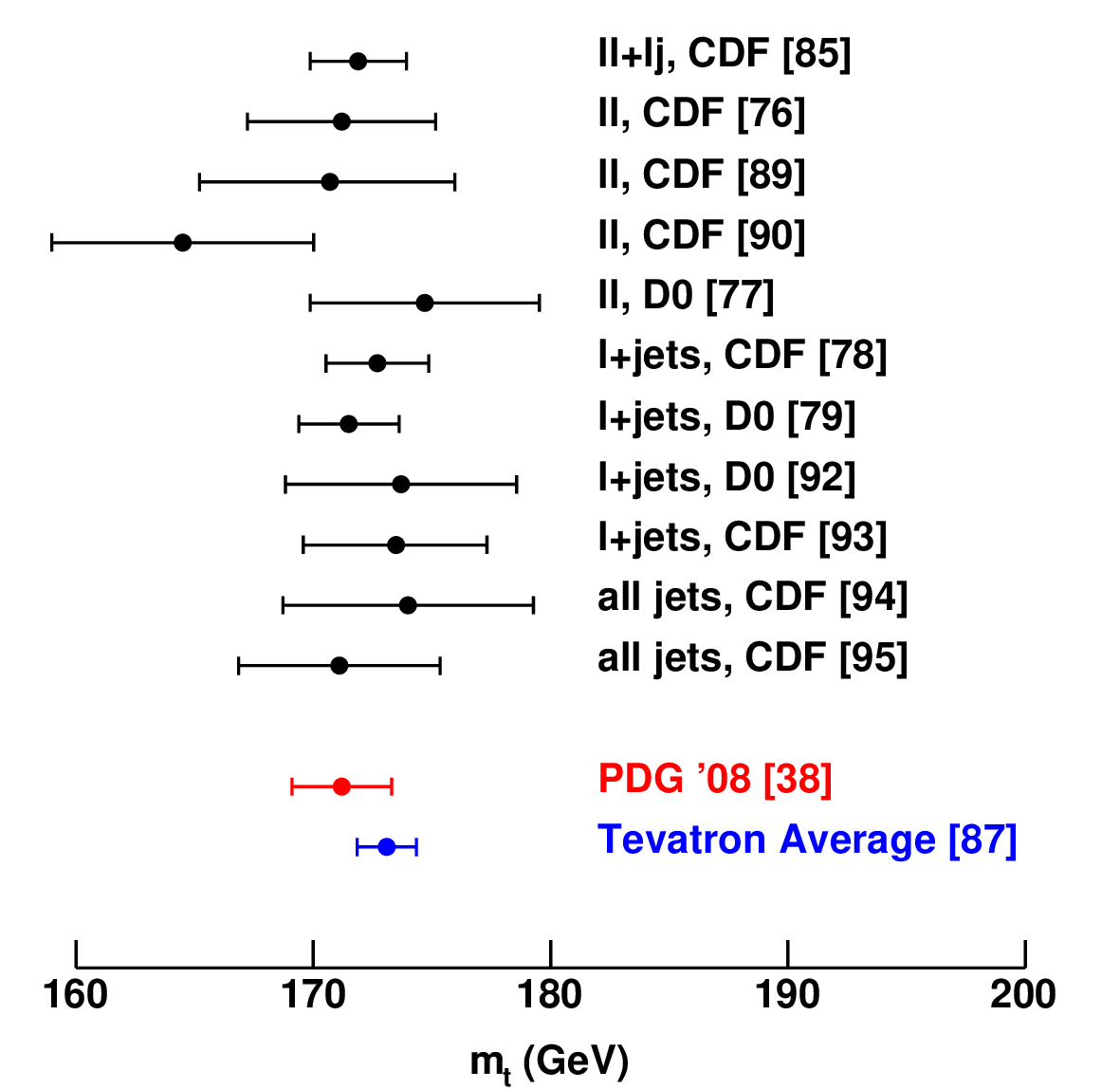}
  \caption{Published top mass measurements.  For a given experiment and
    final state, the results from different methods have a significant
    correlation.  Also shown are the most recent Tevatron 
    combination\cite{b:tevMtComb} and the PDG world average.  Both
    of these include a subset of the measurements listed in this
    table. \label{f:mtop}}
\end{figure}

Tab.~\ref{t:mtop} and Fig.~\ref{f:mtop} give a summary of the published top
mass results and a comparison with the 2008 Particle Data Group
(PDG)\cite{b-pdg08} average.  When multiple results from one
experiment using the same method and final state were available, only
the highest integrated luminosity measurement is reported.  The result
shown in Tab.~\ref{t:mtop} with highest integrated luminosity used a
$\intL \ = 2.9\,\mathrm{fb}^{-1}$ data sample. The Tevatron
experiments have now reported {\em preliminary} results using up to
4~fb$^{-1}$, and these results have been included in a world average
combination by the Tevatron Electroweak Working
Group, with the most recent such combination~\cite{b:tevMtComb} giving
\begin{equation*}
M_t = 173.1 \pm 0.6\ \mathrm{(stat.)} \pm 1.1 \mathrm{(syst.)}\ \GeV. 
\end{equation*} 
For this combination, which does include preliminary results, the
systematic uncertainty is significantly larger than the statistical
uncertainty.  The dominant components of the systematic uncertainty
arise from uncertainty on the {\it in situ} jet energy calibration,
uncertainty related to the $\ttbar$ production and decay model and
uncertainty arising for color recombination. These contribute $0.48\
\GeV$, $0.49\ \GeV$ and $0.41\ \GeV$ to the uncertainty, respectively.

An alternative method of measuring the top quark mass makes use of the
dependence of the production cross section on the mass. The top quark
$\overline{MS}$ mass may be extracted directly from such a method
\cite{PhysRevD.80.054009}.

By the end of the Tevatron running, the experiments expect to increase
the data set by at least a factor of four compared to the published
results and at least a factor of two for the data sets used in the
combination.  Thus, the statistical precision can be improved, but the
systematic uncertainties will also need to be improved if the full
statistical power of the final data set is to be realized.

\begin{table*}
  \begin{ruledtabular}
  \begin{tabular}{lllccll}
                              &  $\intL$    & Analysis&                                    & \multicolumn{2}{c}{Main Systematic Uncertainty} & Expt.\\
   Channel                    & (fb$^{-1}$) & Method         & $M_t$ ($\GeV$)   &   Source     &                   Value         & Ref. \\
                              &             &                &                             &              &                   (GeV/c$^2$)   &      \\ \hline
$\dilep+(\ljets)$ &   1.9     &                & $171.9\pm1.7\mstatJ\pm1.1\msys$    &                     &                                 &  CDF\cite{b-mt-cdf-simult} \\ \hline
   $\dilep$     &    2.9      & $\phi$ WT      & $165.5^{+3.4}_{-3.3}\mstat\pm3.1\msys$ & jet calibration & $2.2$                           &  CDF\cite{b-mt-cdf-phiwt} \\
                &      2      & Matrix         & $171.2\pm2.7\mstat\pm2.9\msys$     & jet calibration     & $2.5$                           &  CDF\cite{b-mt-cdf-neuro} \\ 
                &     1.2     & Template+$M_t$ & $170.7^{4.2}_{-3.9}\mstat\pm2.6\msys\pm2.4\theo$  & jet calibration     & $1.8?$           &  CDF\cite{b-mt-cdf-llcomb} \\ 
                &      1      & Matrix         & $164.5\pm3.9\mstat\pm3.9\msys$     & jet calibration     & $3.5$                           &  CDF\cite{b-mt-cdf-llmat} \\ 
                &      1      & Matrix+$\nu$WT & $174.7\pm4.4\mstat\pm2.0\msys$     & jet calibration     & $1.2$                           &  D0\cite{b-mt-d0-llcomb}  \\ \hline
   $\ljets$     &     1.9     & Matrix         & $172.7\pm1.8\mstat\pm1.2\msys$     & generator, jet calib& $0.6,\ 0.5$                     &  CDF\cite{b-mt-cdf-ljmat} \\
                &      1      & Matrix         & $171.5\pm1.8\mstatJ\pm1.1\msys$    & $b/u,d,s$, frag.    & $0.83$                          &  D0\cite{b-mt-d0-ljmat}   \\
                &     0.7     & Decay Length   & $180.7^{+15.5}_{-13.4}\mstat\pm8.6\msys$    & Bkg. shape       & $6.8$                           &  CDF\cite{b-mt-cdf-ljdl}  \\
%%JDH, superceded above $\ljets$, D0\cite{}&     0.4     & Multiple       & $178.1\pm4.?\mstat\pm1.?\msys$     & $b/u,d,s$ calib.    & $4.3$                           & \\
                &     0.4     & Ideogram       & $173.7\pm4.4\mstatJ\pm2.1\msys$    & $b/u,d,s$ calib.    & $1.7$                           &  D0\cite{b-mt-d0-ideo}    \\
                &     0.3     & Template+DLL   & $173.5^{+3.7}_{-3.6}\mstatJ\pm1.3\msys$    & signal model & $1.1$                       &  CDF\cite{b-mt-cdf-300prl}\\ \hline
   All jets     &      1      & Lineshape      & $174.0\pm2.2\mstat \pm4.8\msys$    & jet calibration     & $4.5$                           &  CDF\cite{b-mt-cdf-jj}    \\
                &     1.0     & Matrix         & $171.1\pm3.7\mstatJ\pm2.1\msys$    & parton shower, jet calib     & $ 0.6,\ 0.5$          &  CDF\cite{b-mt-cdf-jj2}   \\ \hline
   $\mjets$     &     0.3     & Spectra        & $172.3^{10.8}_{-9.6}\mstat\pm10.8\msys$   & jet calibration     & $9.6$                    &  CDF\cite{b-mt-cdf-mj}    \\ \hline
  \multicolumn{3}{l}{Tevatron (incl. prel.)}   & $173.1\pm0.6\mstat\pm1.1\msys$     &  &                                                    & CDF+D0\cite{b:tevMtComb}\\ \hline
  \multicolumn{3}{l}{Particle Data Group 2008} & $171.2\pm2.1$                      &  &                                                    & PDG\cite{b-pdg08}  \\
  \end{tabular}
  \end{ruledtabular}
   \caption{Summary of published Run II top quark mass measurements.  Additional
     preliminary results using up to $\intL=4$~fb$^{-1}$ have 
     been reported by both CDF and D0.  The Particle Data Group average includes
     only some of these results, and the Tevatron average is based on a subset
     of the results in this table and additional preliminary results.\label{t:mtop}}
\end{table*}

\subsection{$\W$ Helicity from Top Quark Decays}
\label{sec:top_Whelicity}

In the SM, top quarks decay with a V-A interaction through the process
$t\to W^+b$ with $B(t\to W+b)\approx 1$.  The SM predicts a fraction
$f_0 = 0.697\pm0.012$ of top quark decays give a longitudinally
polarized $W$ and a fraction $f_+ = 3.6\times 10^{-4}$ give a
right-handed $W$. This leaves $f_- = 1 - f_0 - f_+$ giving left-handed
$W$s.  Compared to the decay of lower mass quarks, the fraction of
decays with longitudinally polarized $W$ is significantly increased
because of the large Yukawa coupling between the top quark and Higgs
boson.  Changes to the Lorentz ($V-A$) structure of the decay will
result in changed values for the polarization fractions.  Indirect
results from $b\to s\gamma$ measurements~\cite{b-bsg1,b-bsg2}
constrain $f_+$ to less than a few percent. 

The angular distribution of the electron-type ($I_3 = -1/2$) $W$ decay
product in $t\to Wb$ is given by
\begin{equation*}
  \omega(c) = \frac{3}{8}[2(1-c^2)f_0 + (1-c)^2f_- + (1+c)^2f_+]
\end{equation*}
in which $c\equiv \cos\theta^*$ is the cosine of the decay angle in
the $W$ rest frame measured with respect to the top quark direction.
Both CDF~\cite{Aaltonen:2008ei} and D0~\cite{b-hel-d0} have measured the
helicity fractions in top quark decay.  The CDF analysis uses $\intL =
1.9$~fb$^{-1}$, and the D0 result is based on $\intL =
1.0$~fb$^{-1}$. 

Both experiments select events in the $\ell+$jets topology, and D0
also uses the $\ell\ell+X$ topology.  The initial selections are
similar to those described earlier in Sec.~\ref{sec:top_mass}.  CDF
uses a second selection with somewhat different requirements to
perform an independent analysis.  The CDF results are from the
combination of the two methods.   

All results rely on reconstructing the $\cos\theta^*$ distribution and
comparing that to a set of predicted distributions each of which is
generated with a different pair of $f_0$ and $f_+$ values.  Several
ambiguities arise in reconstructing $\cos\theta^*$.  These include the
assignment of the reconstructed jets to the quarks from $W$ decay (for
the hadronic side of $\ell+$jets events) and the impact from the
unmeasured $\eta$ value for the neutrino arising in leptonic $W$
decay.  These issues are handled differently in the three (2 CDF, 1
D0) analyses.  The CDF analyses use only the leptonic $W$ decay. The D0
analysis also uses the jets from the hadronic $W$, but only
$|\cos\theta^*|$ is reconstructed for these, not
$\cos\theta^*$. Although this loses the ability to distinguish $f_+$
from $f_-$ for the hadronic decay, $f_0$ is better constrained.

The results are extracted using likelihood fits to test $\cos\theta^*$
distributions.  For one of the CDF methods and the D0 analysis,
predicted reconstructed $\cos\theta^*$ distributions including signal
and background contributions are generated using simulated events.
The distributions are generated for a range of $f_0$ and $f_+$ values,
and best values are extracted using likelihood comparisons to the
data.  For the second CDF method, the result is compared to a
distribution generated by convoluting the true $\cos\theta^*$
distribution divided into six bins with a migration function which
gives the probability that an event generated in a given true bin ends
up a given reconstructed bin.  This function includes the effects of
resolution and acceptance.  Effects of helicity on the acceptance are
considered in all cases.

The experimental results are summarized in Tab.~\ref{t-hel}.  The
results are consistent with each other and with the SM
prediction. They are are generally limited by the statistical
precision of the data set and are expected to improve by a factor
between $\times 2$ and $\times 10$ by the end of Run II, depending on
the analysis and machine performance achieved.  Even if there is no
improvement in the systematic uncertainty, the final results will
still be statistics limited, but the statistical and systematic
uncertainties will be similar.  The dominant systematic uncertainty in
the D0 result comes from the $\ttbar$ production and decay model which
was tested by comparing results from events generated using the
\pythia~generator~\cite{b-pythia} and with the default events
generated using \alpgen~\cite{b-alpgen}.  The CDF result does not have
a single dominant source of systematic uncertainty.  Although the
contributions vary somewhat, the uncertainties arising from jet
calibration, the background prediction and modeling of final state
radiation are roughly equivalent to each and provide most of the
systematic uncertainty.
\begin{table}
  \begin{ruledtabular}
  \begin{tabular}{ccc}
                   & CDF                   & D0 \\ \hline
   $f_0$           & $ 0.66\pm0.16\pm0.05$ & $ 0.425\pm0.166\pm0.102$ \\
   $f_+$           & $-0.03\pm0.06\pm0.03$ & $ 0.119\pm0.090\pm0.053$ \\ \hline
   $f_0, f_+=0$    & $ 0.62\pm0.10\pm0.05$ & $ 0.619\pm0.090\pm0.052$ \\
   $f_+, f_0=0.70$ & $-0.03\pm0.04\pm0.03$ & $-0.002\pm0.047\pm0.047$
  \end{tabular}
    \caption{$W$ boson helicity fractions in $\ttbar$ decay determined from 
      fits to data.  The first two
      lines are the results from simultaneous fits for $f_0$ and $f_+$.  The
      third line is the result from the fit for $f_0$ with $f_+=0$, and the
      fourth is the from the fit for $f_+$ with $f_0$ set to the SM
      value.  In all cases, the statistical uncertainty is shown first and
      the second uncertainty is the systematic uncertainty.\label{t-hel}}
\end{ruledtabular}
\end{table}

\subsection{Single Top Production and $V_{tb}$}
\label{sec:top_singleTop}

\begin{figure}
  \includegraphics[width=0.4\linewidth]{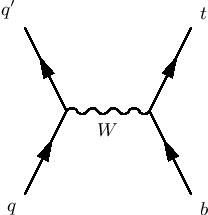}
  \includegraphics[width=0.4\linewidth]{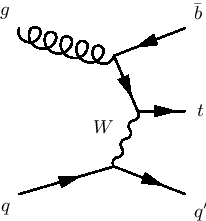}
  \caption{The $s$-channel (left) and $t$-channel (right) diagrams for
   single top production.\label{f-1top}}
\end{figure}

The top quark was first observed in reactions mediated by the strong
interaction process $\ppbar\to\ttbar$~\cite{b-top-d0,b-top-cdf}, and
nearly all measurements of the top quark properties have been made
using such events. However, top quarks can also be produced singly via
the $s$-channel~\cite{b-st-sxsec1,b-st-sxsec2,b-st-xsec} and
$t$-channel ~\cite{b-st-txsec1,b-st-txsec2,b-st-txsec3,b-st-xsec}
electroweak diagrams shown in Fig.~\ref{f-1top}.  These production
modes involve the $Wtb$ coupling and therefore cross section
measurements provide previously unmeasured information on the
electroweak sector, particularly the CKM matrix element $V_{tb}$.  In
addition to providing information about $V_{tb}$, detecting these
events demonstrates progress toward Higgs sensitivity for the $WH$
associated production mode with the subsequent decay $H\to\bbbar$. 

The NLO cross sections for these processes are
$\sigma_s=1.12\pm0.05$~pb and
$\sigma_t=2.34\pm0.13$~pb~\cite{Stelzer:1997ns,Smith:1996ij,Harris:2002md,b-sullivan,Campbell:2004ch,Cao:2004ap,Cao:2005pq,Frixione:2005vw,b-st-xsecKid,Campbell:2009ss}
for the $s$-channel and $t$-channel reactions, respectively, assuming
$|V_{tb}|=1$ and a top quark mass of $M_t = 175\ \GeV$.  While the
total single top cross section is slightly more than 40\% of the
$\ttbar$ cross section, the reduced parton multiplicity in the single
top events implies a signal-to-noise which is roughly 25 times lower
than in $\ttbar$ making these events much more difficult to identify.
In the SM with three generations, existing measurements of $|V_{ub}|$
and $|V_{cb}|$ tightly constrain $|V_{tb}|$ to almost exactly
unity~\cite{b-pdg08}. If there are more than three generations,
however, $|V_{tb}|$ is almost entirely unconstrained. 

Both D0 and CDF have long histories of searching for single top
production. The most recent results from both
D0~\cite{b-st-d0-prl,b-st-d0-prd} and CDF~\cite{b-st-cdf} give clear
evidence for single top production and include results for the cross
section and the CKM element $|V_{tb}|$.  The D0 analysis uses a data
sample corresponding to $\intL=2.3$~fb$^{-1}$. The CDF analysis uses
up to $\intL=3.2$~fb$^{-1}$. 

Candidate single top events are selected in the lepton+jets channel in
which the $W$ decays via either $W\to e\nu$ or $W\to\mu\nu$.  The
event topology is thus one high-$p_T$ charged lepton, $\MET$
corresponding to the neutrino, and two or more jets. Events are
required to have fired single electron, single muon, electron+jet or
muon+jet triggers.  Both experiments use events with 2 or 3
reconstructed jets, and D0 also makes use of 4-jet events.
Tab.~\ref{t-st-yields} shows the yields for the D0 and CDF initial
selections in which events are required to have at least one jet
having a $b$-tag.
\begin{table*}
  \begin{ruledtabular}
  \begin{tabular}{lcccccc}
                   &  \multicolumn{3}{c}{D0}         &           & \multicolumn{2}{c}{CDF} \\ 
   $\intL$         & \multicolumn{3}{c}{2.3~fb$^{-1}$}   &       & 3.2~fb$^{-1}$      & 2.1~fb$^{-1}$ \\
    Process        & $=2$ jets     & $=3$ jets     & $\ge4$ jets  & & $\ljets$           &  $\mjets$     \\     \hline
  $tb+tqb$ signal  & $139 \pm 18$  & $ 63\pm10$    & $ 21\pm5 $  & & & \\    
  $tb$ signal      &               &               &             & & $77.3 \pm 11.2$    & $29.6 \pm 3.7$    \\
  $tqb$ signal     &               &               &             & & $113.8 \pm 16.9$   & $34.5 \pm 6.1$    \\ \hline
   $W+jets$        & $1,829\pm161$ & $  637\pm61$  & $180\pm18$  & & $1551.0 \pm 472.3$ & $304.4 \pm 115.5$ \\
   Z+jets,         & $229 \pm 38$  & $ 85\pm17$    & $ 26\pm7 $  & & $52.1 \pm 8.0$     & $128.6 \pm 53.7$  \\
   Diboson         & in $Z$+jets   & in $Z$+jets   & in $Z$+jets & & $118.4 \pm 12.2$   & $42.1 \pm 6.7$    \\
   $\ttbar$        & $222 \pm 35$  & $ 436\pm66$   & $ 484\pm71$ & & $686.1 \pm 99.4$   & $184.5 \pm 30.2$  \\
   Multijet        & $196 \pm 50$  & $ 73\pm17$    & $ 30\pm6$   & & $777.9 \pm 103.7$  & $679.4 \pm 27.9$  \\ \hline
   Total Pred      & $2615\pm192$  & $1,294\pm107$ & $742\pm80$  & & $3376.5 \pm 504.9$ & $1404 \pm 172$    \\ \hline
    Yield          & 2579          &   1,216       & 724         & & 3315 & 1411 \\
  \end{tabular}
  \end{ruledtabular} 
  \caption{The integrated luminosity, predicted signal, backgrounds and event
      yields in the W+jets($b$-tag) samples used in the CDF and D0 single
      top analyses.\label{t-st-yields}}
\end{table*}

The final results for both experiments are based on combining results
from multivariate (MV) techniques which exploit correlations among
variables and event weighting methods based on calculated matrix
elements for signal and background events.  The methods are applied to
events selected by the initial offline requirements.  D0 uses three
methods: (1) a ME method, (2) a Bayesian Neural Network (BNN) method
and (3) a boosted decision tree (BDT) method, and the final result
comes from the combination of the individual results including
correlations.  CDF uses five methods: (1) a ME method, (2) a joint
likelihood (LF) method, (3) a neural network method, (4) a boosted
decision tree (BDT) and (5) projected likelihood functions (LFS), and
the final result is the combination of these including correlations
among them.

\begin{figure}
  \includegraphics[width=\linewidth]{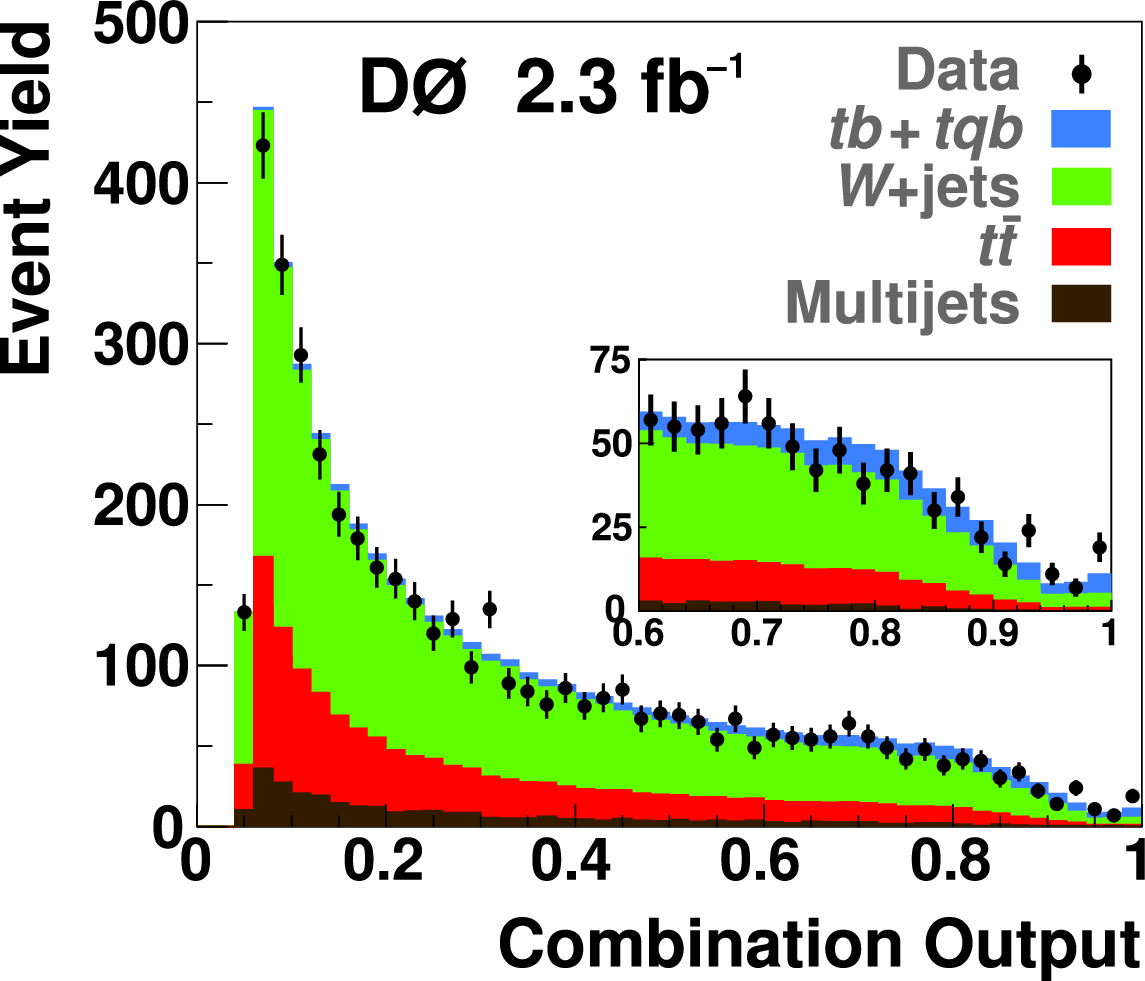}
  \caption{The combined discriminant for the D0 analysis.\label{f-st-d0}}
\end{figure}

\begin{figure}
  \includegraphics[width=\linewidth]{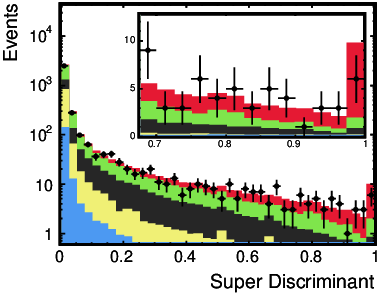}
  \caption{The combined discriminant for the CDF analysis\label{f-st-cdf}}
\end{figure}

\begin{figure}
  \includegraphics[width=0.99\linewidth]{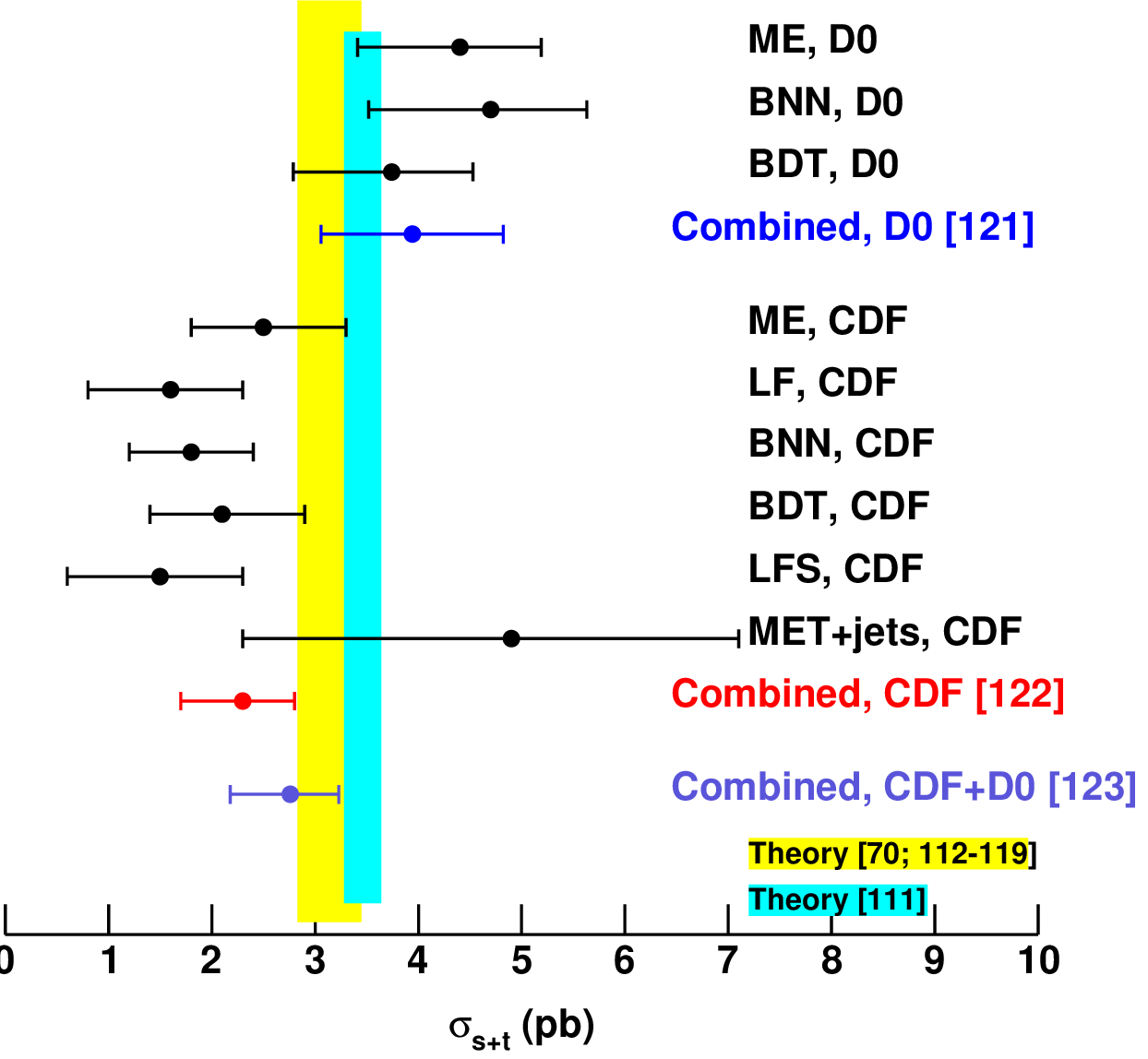}
  \caption{Ideogram of the single top cross section measurements from D0
    and CDF.\label{f-st-answ}}
\end{figure}

Final results are determined using likelihood fits.  Likelihoods are
created from the multivariate outputs for each lepton flavor, jet
multiplicity and $b$-tag multiplicity for a given multivariate
method. The joint likelihood formed by the product of the individual
likelihoods for a given MV method is then maximized as a function of
the signal cross sections with systematic correlations between lepton
species, jet multiplicity and $b$-tag multiplicity taken into account.
The BLUE method~\cite{b-BLUE} is used by D0 to combine the results
from each MV method into a single result.  The CDF results are
combined using a super discriminant~\cite{b-st-cdf}, in this case an
additional neural network trained on the outputs of the multivariate
methods. Fig.~\ref{f-st-d0} shows the output of the three multivariate
classifiers for D0, and Fig.~\ref{f-st-cdf} shows the same for three
of the CDF classifiers. The cross sections for each MV method, the
combined cross sections and the fit probabilities are given in
Tab.~\ref{t-st-answ} and illustrated in Fig.~\ref{f-st-answ}.  The
CDF and D0 results have also been combined by the Tevatron Electroweak 
Working Group.~\cite{Group:2009qk}
\begin{table}
\begin{ruledtabular}
\begin{tabular}{lcc}
        & \multicolumn{2}{c}{$\sigma_{st} = \sigma_s + \sigma_t$} \\
        &      D0                 &  CDF  \\ \hline
  ME    & $4.40^{+0.99}_{-0.79}$  & $2.5^{+0.7}_{-0.8}$  \\
  LF    & ---                     & $1.6^{+0.8}_{-0.7}$  \\
(B)NN   & $4.70^{+1.18}_{-0.93}$  & $1.8\pm0.6$          \\ 
 BDT    & $3.74^{+0.95}_{-0.79}$  & $2.1^{+0.7}_{-0.8}$  \\
  LFS   & ---                     & $1.5^{+0.9}_{-0.8}$  \\
$\mjets$& ---                     & $4.9^{+2.6}_{-2.2}$  \\ \hline 
Combined &  $3.94\pm0.88$       & $2.3^{+0.6}_{-0.5}$  \\ \hline
Predicted~\cite{b-st-xsecKid}
        & \multicolumn{2}{c}{$\sigma_{st} = 3.46 \pm 0.18$ for $M_t = 170\ \GeV$} \\
\hphantom{Predicted}~\cite{Harris:2002md}
        & \multicolumn{2}{c}{$\sigma_{st} = 3.14 \pm 0.31$} \\
  \end{tabular}
  \caption{Single top cross section measurements from the Tevatron (in
    pb). The D0 result used $M_t = 170\ \GeV$ in simulations and
    calculation. The CDF result used $M_t = 175\ \GeV$.  The predicted
    cross section changes by 0.1 pb for a 5 $\GeV$ change in $M_t$ in
    this region. \label{t-st-answ}}
\end{ruledtabular}
\end{table}

The SM cross section is proportional to $|V_{tb}|^2$, so the cross
section results are used to measure $|V_{tb}|$. The only assumption is
that $|V_{tb}|\gg |V_{ts}|,|V_{td}|$ such that there is not a
significant contribution to the signal from initial-state $s$ or $d$
quarks being transformed into top quarks via a $t$-channel $W$ boson.
These results are shown in Tab.~\ref{t:st-vtb}. The measured single
top cross section is in good agreement with the SM prediction with
$|V_{tb}|\simeq 1$. 
\begin{table}
\begin{ruledtabular}
\begin{tabular}{cc}
    D0                       &  CDF                        \\ \hline
   $|V_{tb}|>0.78$           &  $|V_{tb}|>0.71$            \\
   $|V_{tb}| = 1.07\pm 0.12$ &  $|V_{tb}| = 0.9 \pm 0.11$  \\
  \end{tabular}
  \caption{$|V_{tb}|$ values extracted from the single top cross section
    measurements. The first row is the 95\% CL result when no constraint 
    is placed on on $|V_{tb}|$, and the second row is the result with the 
    SM constraint $0 \le |V_{tb}| \le 1$.\label{t:st-vtb}}
\end{ruledtabular}
\end{table}

The precision of these results is limited by the statistics of the
data samples.  Both experiments are expected to continue these
analyses through the end of the Tevatron running.  In the most
optimistic scenario this will result in analyzed data samples of up to
$10$~fb$^{-1}$.   Assuming the results in Table~\ref{t:st-vtb} scale as 
$1/\sqrt{\intL}$, the uncertainty on $|V_{tb}|$ will be reduced approximately
two fold.

%%%%%%%%%%%%%%%%%%%%%%%%%%%%%%%%%%%%%%%%%%%%%%%%%%%%%%%%%%%%%%%%%%%%%%%%%%%%
%
% Higgs Boson
%
%%%%%%%%%%%%%%%%%%%%%%%%%%%%%%%%%%%%%%%%%%%%%%%%%%%%%%%%%%%%%%%%%%%%%%%%%%%%

\section{HIGGS BOSON}
\label{sec:higgs}

\subsection{Precision Electroweak Constraints}
\label{sec:higgs_ewkConstraints}

\begin{figure}
  \includegraphics[width=0.95\linewidth]{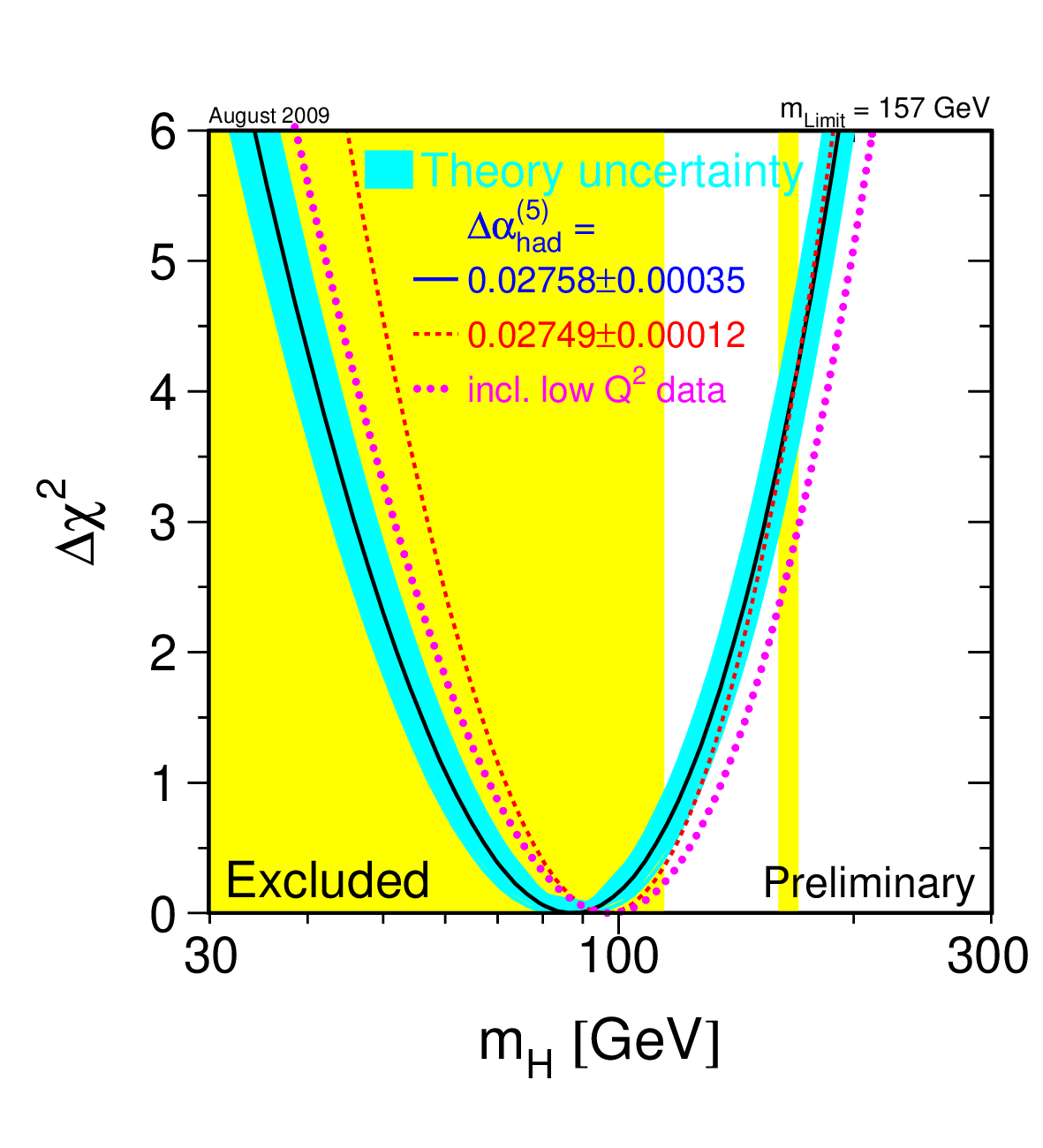}
  \caption{The $\Delta\chi^2$ as a function of assumed Higgs mass resulting 
    from fits to data assuming the SM.  The exclusion regions
    from direct searches are also shown.
    \label{higgs-blue-band}}
\end{figure}

As discussed in Sections~\ref{sec:overview_gsw} and
\ref{sec:overview_higgs}, the standard model, with the $SU(2)_{\rm L}
\otimes U(1)_{Y}$ symmetry spontaneously broken via the
vacuum-expectation value of a single Higgs-doublet field, is
consistent with precision electroweak data.  These data are so precise
that they are sensitive to the mass of the Higgs boson at one loop, 
despite the fact that the Higgs mass enters only logarithmically.  
Fig.~\ref{higgs-blue-band} shows the constraints on the Higgs-boson
mass from both precision electroweak measurements (the blue band) and
direct searches (the yellow areas).  The precision data depend on the
extrapolation of the fine-structure constant $\alpha$ from its
measured value at low energy up to high energy, which suffers from
an uncertainty associated with the contribution of low-energy QCD to
the extrapolation.  The solid black line indicates the central value, and
the blue band takes into account all uncertainties. An alternative
central value, associated with a different treatment of low-energy
QCD, is indicated by the red dashed line.  The effect on the
central value by including the NuTeV data, which has some tension with
the other precision electroweak data, is shown by the dotted
magenta line.

It is striking that the precision data strongly prefer a Higgs boson
mass in the 100 $\GeV$ region, while a purely theoretical analysis
would accommodate a Higgs boson as heavy as about 700 $\GeV$
\cite{Luscher:1988gc}.  Taken at face value, the precision data
indicate the Higgs boson is not much heavier than the current lower
bound of $m_H>114\ \GeV$. It is also remarkable that the Tevatron
experiments have succeeded in excluding the region $163\ \GeV
<m_H<166\ \GeV$.

\subsection{Direct Searches}
\label{sec:higgs_direct}

The experimental strategy and sensitivity to discover the standard
model Higgs boson at the Tevatron through direct searches depends
strongly on the value of $m_H$. For direct searches, the sensitive
range is limited to $m_H \simle 250~\GeV$ where there exists
sufficient center-of-mass collision energy at the parton-level for an
appreciable rate of Higgs boson production via the gluon fusion
($ggH$) process which proceeds via a virtual quark loop (dominated by
the top quark) \cite{Djouadi:2005gi}.

For $m_H \simle 135~\GeV$ (the ``low mass Higgs'' region), the
dominant Higgs decay is to $b$-quark pairs. The next largest
contribution to the Higgs decay in the low mass Higgs region is
$H\to\tau\tau$ which has a branching ratio of $\sim 7$\%. In the
case of $H\to\bbbar$, the $ggH$ production is not useful to the direct
Higgs search even using the $\bbbar$ invariant mass because of
overwhelming continuum $\bbbar$ production. Instead, the associated
vector boson ($VH$) process where the Higgs boson is produced in
association with a $W$ or $Z$ boson is used. The additional leptons
which can result from decay of the vector boson via $W\to\ell\nu$,
$Z\to\ell\ell$ or $Z\to\nu\nu$ with $\ell = e,\, \mu$
\cite{Stange:1993ya,Stange:1994bb}, whether detected directly or
indirectly through $\MET$, can be used suppress backgrounds at the
expense of a lower cross section for $VH$ as compared to $ggH$.

The cross section and cross section times branching fraction for
several Higgs masses in the low mass Higgs region
~\cite{b-higgs-combo} are shown in Tab.~\ref{t-hprod-low}.  In
addition to the modes listed in Tab.~\ref{t-hprod-low}, the $WH\to
WWW\to \ell^\pm\ell^{+'}\ell^{-'}$ channel provides some sensitivity
in the transition region around $m_H\approx 135\ \GeV$ and vector
boson fusion (VBF) $qq\rightarrow qqH$ and $ggH$ modes provide
sensitivity in $\tau\bar\tau$ final states.

\begin{table*}
  \begin{ruledtabular}
  \begin{tabular}{cccccc}
   $m_H$ ($\GeV$) & $\sigma(\ppbar\to WH)$ & $\sigma\times B(W\to\ell\nu)$ & $\sigma(\ppbar\to ZH)$ & $\sigma\times B(Z\to\ell\ell)$ & $\sigma\times B(Z\to\nu\nu)$ \\ \hline
   100            &  286     &  63.6  &  167  &  11.2  &  33.4 \\
   110            &  209     &  46.4  &  124  &   8.35 &  24.8 \\
   120            &  153     &  34.0  & 92.7  &   6.24 &  18.5 \\
   130            &  115     &  25.5  & 70.9  &   4.77 &  14.2 \\
  \end{tabular}
  \end{ruledtabular}
  \caption{The production cross section and cross section times branching
   fraction for the low mass ($m_H<135\ \GeV$) Higgs boson search. Cross
   sections are in femtobarns.  For the
   final states denoted $\ell\nu$ and $\ell\ell$, the branching fraction
   used is the sum of $\ell=e\, \mu$. \label{t-hprod-low}}
\end{table*}

For the low mass associated production channels $VH$ in which the $W$ or
$Z$ boson decay involves one or two charged lepton(s), an initial
sample consistent with production of a vector boson decaying to
charged and/or neutral leptons and having at least two jets is
selected.  This sample has very low signal-to-noise but is already
dominated by vector boson events and therefore provides a control
sample used to validate detector modeling and rate predictions of the
dominant backgrounds in the final analysis.

The requirement that one or more of the jets in the event is
consistent with production of a $B$ hadron is added to the selection,
dramatically reducing the contribution from the initially dominant
$Vjj$ background and increasing the signal purity. Non-linear
multivariate techniques for each final state are used to combine
kinematic properties into a (typically) single variable in which a
potential Higgs signal and backgrounds populate different
regions. The distributions of these variables are then used as input
to binned likelihood fits in which signal and background fractions are
allowed to float within constraints of the total yield, predicted
background contributions and all uncertainties. The absence (presence)
of a significant signal resulting from the fits determines the mass
limit (indicates discovery).

The low mass Higgs searches using the $ZH\to\nu\nu\bbbar$ channel and
final states involving one or two $\tau$ lepton decays follow
different strategies. The $ZH\to\nu\nu\bbbar$ channel has significant
background from multijet QCD processes without real vector boson
decay\footnote{Such backgrounds are also present in the other
  channels, but as a much lower fraction of the total background.} in
which mismeasurement of jets results in significant $\MET$.  In this
channel, an additional step (described later in this section) is added
to improve an understanding of this background. The search using final
states that involve either one or two $\tau$ leptons does not make 
use of $b$--jet identification since the $\tau$ leptons can arise
either from Higgs boson decay or from $W$ or $Z$ decay.

The latest low mass Higgs searches from CDF and D0 are described in
Section~\ref{sec:higgs_direct_HWW}.  The most recent combination of
low mass results, which includes some preliminary results not
presented here, is described in
Sections~\ref{sec:higgs_direct_comb}--~\ref{sec:higgs_direct_taus}.

For $m_H \simge 135~\GeV$ (the ``high mass Higgs'' region), the dominant Higgs
decay is to $W$ boson pairs with the next largest contribution being
$\HZZ$. Aside from a region around $m_H = 150~\GeV$ where $BR(\HZZ)$ peaks to
$\sim 10$\%, the $\HWW$ decay where one of the final state $W$ bosons is
off-shell for $M_H$ less than twice the $W$ mass completely dominates over the
sensitive region of the Tevatron.  The Higgs production cross section and
branching fractions for several Higgs masses in the high mass region are shown
in Table~\ref{t-hprod-hi}.  The standard model Higgs boson branching fraction
to $WW^{*}$ varies from 7.5\% at 115 $\GeV$ to 73.5\% at 200 $\GeV$ with a
maximum of 96.5\% at $\approx$ 170 $\GeV$ \cite{Djouadi:2005gi}. For $M_H\sim
170\; \GeV$, the standard model Higgs boson decays almost exclusively to two
on-shell $W$ bosons, making this a region where the Tevatron has the best
chance at a discovery or an exclusion.
\begin{table*}
  \begin{ruledtabular}
  \begin{tabular}{cccccc}
   $m_H$ ($\GeV$) & $\sigma(gg\to H)$ & $\sigma($VBF$)$ & $B(H\to WW)$ \\
   150            &  548  &  45.7  & 0.682 \\
   160            &  439  &  38.6  & 0.901 \\
   170            &  349  &  33.6  & 0.965 \\
   180            &  283  &  28.6  & 0.934 \\
  \end{tabular}
  \end{ruledtabular}
  \caption{The production cross section and cross section times branching
   fraction for the high mass ($m_H>135\ \GeV$) Higgs boson search.  Cross 
   sections are in femtobarns.\label{t-hprod-hi}}
\end{table*}

At the Tevatron, the most sensitive Higgs search channel over the
range $135~\GeV \simle m_H \simle 200~\GeV$ is $gg\rightarrow
H\rightarrow WW^{(*)}\rightarrow \ell\nu\ell\nu$, where the two
charged leptons in the final state are of opposite charge
\cite{PhysRevLett.82.25,PhysRevD.59.093001}. The $\ell\nu\ell\nu$
final state represents 6.0\% of all $WW^{*}$ decays, where $\ell$ is
either an electron or a muon, including those from $\tau$ leptons
produced in the $W$ decays. Over the last few years, the experimental
search for a high mass Higgs in decay to $WW$ at the Tevatron has
evolved substantially to include powerful multivariate techniques and
additional $\HWW$ processes such as VBF $\HWW$ and $VH(\rightarrow
WW^{(*)})$. The latest high mass Higgs searches in the $\HWW$ channel
from CDF and D0 are described in Section~\ref{sec:higgs_direct_HWW}. A
combination of these these results are described in
Section~\ref{sec:higgs_direct_comb}.

The following sections describe the individual channels used in the
Higgs search starting with the low mass channels, $WH\to\ell\nu bb$,
$ZH\to\ell\ell bb$, $ZH\to \nu\nu bb$ and $\tau$ final states, and
finishing with the high mass $H\to WW^{(*)}\to\ell\nu\ell\nu$ search.
The Higgs boson section concludes with a presentation of the most
recent set of CDF and D0 combined limits. 

\subsubsection{$WH\to\ell\nu\bbbar$ Final State}
\label{sec:higgs_direct_WHlvbb}
 
This final state has the largest cross-section times branching ratio
of the entries shown in Tab.~\ref{t-hprod-low}, and it also gives the
most sensitivity to Higgs boson production for $M_H\simle 135\; \GeV$.
Both CDF~\cite{b-cdf-wh} and D0~\cite{b-d0-wh} have carried out
searches in this final state. The currently published results for D0
use a sample of $\intL=1.0$~fb$^{-1}$. The CDF published results use
a sample of $\intL=2.1$~fb$^{-1}$, and preliminary results have been
reported by both collaborations which use up to approximately
$\intL=4$~fb$^{-1}$. The search strategies used by the two
collaborations in the $WH$ channel are generally similar.

The initial event selection requires one high-$p_T$ electron or muon,
large $\MET$ and two or more jets. This selection mirrors the $\ell$,
$\nu$ and $\bbbar$ pair present in the final state. CDF requires the
events to be selected by a trigger based on the presence of a
high-energy electron or muon.\footnote{The event selection in recent
  preliminary results from CDF include additional events in this
  channel which are selected by $\MET$ triggers.}  The D0 analysis
requires $\ell= e$ events to be selected by single--electron or
electron plus jet triggers.  For the muon final state D0 uses a two
phase trigger selection. In the first phase, the events are required
to be selected based on at least one of a set of single lepton or
single-lepton plus jet triggers, and all yields are predicted and
compared with data. After establishing agreement in this pass, the
analysis is repeated allowing events to be selected by any trigger.
The yield increases by an amount predicted by the inefficiency of the
triggers for the first pass and all kinematic distributions remain in
agreement after a simple yield scaling to 100\% trigger efficiency.

At this stage, the dominant event source is $W$+jets with the jets
arising from light quark ($u$, $d$, $s$, $c$) production.  The
signal-to-background is then improved by requiring one or two of the
jets in the event to be identified as consistent with $b$ quark
production. The D0 $b$ identification algorithm~\cite{b-D0-btag} uses
a Neural Network (NN). The NN is trained and verified on a combination
of data and simulation, with the critical efficiencies and
misidentification rates determined from data control samples. The CDF
analysis used three different $b$ identification algorithms. Jets are
$b$-tagged by one or more algorithms based on the presence of a
secondary vertex ({\sc secvtx})~\cite{b:cdf-secvtx}, a neural
network~\cite{b:cdf-nnttag}, or signed impact parameters
(JP)\cite{b-cdf-jetprob}.

The signal purity is greatest when two $b$ jets are required, but this
introduces significant efficiency loss compared to the case in which
only one $b$ jet is required.  To gain the most sensitivity, both
categories of events, single $b$-tagged and double $b$-tagged, are
retained, but they are analyzed separately. Optimization studies
indicate that for the single-tag channel, rather restrictive tagging
requirement is needed to control the background contribution from
events with light-flavor (lf) jets misidentified as $b$ jets while for the
double-tag channel, a less stringent requirement suffices thereby
giving increased per--jet tagging efficiency. The D0 single-tag
analysis uses a $b$--tag operating point which gives a typical
efficiency of 48\% with misidentification rate of 0.5\% and an
operating point for the double-tag analysis which gives an efficiency
of 59\% and a misidentification rate of 1.7\%. The CDF event
selection divides events into three exclusive categories based on
which algorithms identify a jet as $b$-tagged. The first category
contains events which have two jets tagged by the {\sc secvtx}
algorithm. The second category contains events not selected into the
first category which have one jet identified by the {\sc secvtx}
algorithm and one by the jet probability algorithm. The third
category contains events which are not selected into either of the
other two categories and have one jet identified by both the {\sc
  secvtx}\ and NN algorithms.

The sensitivity of the analyses are further improved by using
kinematic properties of the events to distinguish signal and
background events. The most important variable is the mass of the
dijet system corresponding to the Higgs decay. For single-tagged
events, the mass is computed using the tagged jet and the highest
$p_T$ jet remaining, and for the double-tagged events the mass is
computed using the two tagged jets.  In addition to the mass, other
variables can also distinguish signal and background on a statistical
basis. To make best use of these, both experiments use a neural
network to enhance the separation of signal and background in the
final step of the analysis. The CDF neural network has six input
variables and the D0 network have seven input
variables.\footnote{Recent preliminary results has also used the
  matrix element method adapted from the top mass measurement and
  decision trees for the multivariate technique instead of neural
  networks.} The variables used by each experiment are listed in
Tab.~\ref{t-nn-wh-in} for comparison. Because of limited statistics in
the $W$+three-jet sample, D0 uses the NN only for $W$+2 jet case and
uses the dijet mass for the $W$+3 jet case.
\begin{table*}
  \begin{tabular} {cc} 
     {\bf CDF} & {\bf\hphantom{00000000000000} D0 \hphantom{00000000000000}} \\ \hline
%CDF
      \hphantom{0000}
      $M_{JJ+}$ - invariant mass of $J_1$ and $J_2$ 
      \hphantom{0000} &
%D0
      $M_{JJ}$ - invariant mass of $J_1$ and $J_2$ \\
%CDF, cont
        and the closest loose jet if $\Delta R(J,J_{loose})<0.9$  &  
 \\ \hline
%CDF 
      $\Sigma E_T$(loose jets) &
%D0
      $E_T(J_1)$ 
 \\ \hline      
%CDF 
     $p_T(J_1)+p_T(J_2)+pT(\ell)-\MET$ &
%D0
     $E_T(J_2)$ \\ \hline
%CDF 
     $M_{\ell\nu j}^{min}$ &
%D0 
     $\Delta R(J_1,J_2)$ \\ \hline 
%CDF 
     $\Delta R(\ell,\nu_{max})$ &
%D0
     $\Delta\phi(J_1,J2)$ \\ \hline
%CDF 
     $|\vec{p}_T(\ell)+\vec{p}_T(\MET)+\vec{p}_T(J_1)+\vec{p}_T(J_2)|$ &
%D0
     $|\vec{p}_T(\ell)+\vec{p}_T(\MET)|$ \\ \hline      
%D0 
 & $|\vec{p}_T(J_1)+\vec{p}_T(J_2)|$ \\
  \end{tabular}
  \caption{The neural network inputs for the CDF and D0 $WH$ analyses.  Here
    $J_i$ denotes the $i$-th jet in a list ordered by jet $E_T$ in which $J_1$
    is the highest $E_T$ jet in the event.  
    $M_{\ell\nu j}^{min}$ is the mass of the lepton, $\MET$ and the
    jet ($J_1$ or $J_2$) which gives the lower mass value.  Finally,
    $\nu_{max}$ is the three momentum of the neutrino in which $|p_Z|$ is
    the larger of the two values calculated when forcing the lepton and 
    neutrino system to have a mass equal to the $W$ boson mass.
    \label{t-nn-wh-in}}
\end{table*}
Backgrounds to the $WH$ search include $W$+jets production (including
jets arising from heavy flavor (hf) production), $\ttbar$, single top,
diboson production ($WW$, $WZ$ and $ZZ$) and a small contribution from
multijet events in which either the lepton is actually a jet
misidentified as a lepton or the lepton arises from heavy flavor decay
and the jet energy is not reconstructed so the lepton appears to be
isolated.  Both experiments estimate the $\ttbar$, single top and
diboson yields using the product of the theoretical (N)NLO
cross--section for each process, the luminosity and acceptance times
efficiency for each process.  Corrections are applied based on
comparison of data and simulated control samples.  The $W$+jets
background cross sections are poorly known, and data--driven
approaches are used to estimate these yields. The multijet background
is hard to model from simulation, so data--driven methods are also
used for this background. The two experiments use different
data--driven methods as outlined below. 

The total $W$+jets background after $b$--tagging in the CDF result is
estimated separately for $W$+lf events and $W$+hf events.  The
$W$+lf contribution is estimated by applying data--derived mistag
probabilities to untagged $W$+jets samples. Three different tagging
algorithms are used in the analysis, and the mistag probability
determination differs for each of these.  For the {\sc secvtx} and JP
tagging algorithms, the mistag
probability~\cite{b:cdf-secvtx,b-cdf-jetprob} is derived using events
with a negative decay length with a correction applied to account for
the heavy-flavor content of the control sample used to determine the
mistag probability. For the JP algorithm, the mistag probability is
parametrized as a function of $\eta$, primary vertex $z$ position,
jet $E_T$, scalar transverse energy, and vertex and track
multiplicity. For the $NN$--based tagging algorithm, a light--flavor 
rejection factor derived from control data samples is used. The
$W$+hf contribution is determined by measuring the heavy flavor 
faction in $W$+jets events and applying a $b$--tagging efficiency to
these events. The initial heavy flavor fraction is derived from
\alpgen\ and \pythia\ simulation and is corrected by a factor of
$1.4\pm0.4$ derived from a jet control data sample. The $b$--tagging
efficiencies are determined from simulation and checked using control
data samples. 

The $W$+jets background yields after $b$--tagging in the D0 result are
fixed by normalizing the simulated events to the untagged $W$+jets
data after subtracting the other backgrounds from the data.  The
relative contributions of the $W$+lf, $W+\ccbar$, and $W+\bbbar$ in
the untagged sample are fixed to the cross-section ratios predicted by
{\sc mcfm}.  The yields in the $b$--tagged sample are then computed by
applying flavor--based (mis)identification probabilities to the jets
in simulated events.  The probabilities (one for $u$-, $d$- and $s$-
quark initiated-jets, one for $c$-quark initiated jets and for for
$b$-quark initiated jets) are derived using data control
samples~\cite{b-D0-btag} and are parametrized as functions of jet
$p_T$ and $\eta$.

The small background from multijet events is difficult to simulate
accurately, so this component is also determined from control data
samples.  For CDF, a control sample is selected using events which
have non-isolated leptons and low $\MET$, and the yield in the signal
sample is determined by extrapolating the yield from this sample into
the signal region having isolated leptons and high $\MET$. For D0,
this background is determined by selecting a multijet dominated
control sample with kinematics similar to the $WH$ events, and then
applying a probability that these events would be misidentified and
appear in the signal sample.  The multijet control sample is selected
by requiring lepton candidates which pass very loose isolation
requirements, and the background yield is computed by applying the
event--by-event probability that these loose--isolation events would
pass the standard isolation requirement and thus appear in the signal
sample. 

Tab.~\ref{t-wh-data-cdf} shows the data yields and background and
signal predictions for the CDF analysis, and Tab.~\ref{t-wh-data-d0}
shows the yields for the D0 analysis. The predicted Higgs yield
includes not only $WH$ events, but also a small contribution from
$ZH\to\ell\ell\bbbar$ events in which one of the leptons ($\ell$) is
not identified and thus generates $\MET$ and a final state consistent
with the $WH$ final state. 

\begin{table*}
\begin{ruledtabular}
\begin{tabular}{ccccccc}
                    &  \multicolumn{3}{c}{Central region}   &       \multicolumn{3}{c}{ Plug region} \\ \hline     
       Pretag Events&    \multicolumn{3}{c}{32242} &    \multicolumn{3}{c}{5879} \\ \hline		  
       $b$-tagging  &  ST+ST & ST+JP & ST+NN & ST+ST & ST+JP & ST+NN \\ \hline 
              Mistag&   3.88$\pm$0.35&   11.73$\pm$0.92&  107.1$\pm$9.38&   1.00$\pm$0.18 &  3.18$\pm$0.49& 28.47$\pm$3.30 \\ 
         $Wbb$&   37.93$\pm$16.92& 31.15$\pm$14.03& 215.6$\pm$92.34&  7.40$\pm$3.96 &  6.23$\pm$3.37& 43.09$\pm$12.33\\		   
         $Wcc$&   2.88$\pm$1.25&   7.87$\pm$3.43&   167.0$\pm$62.14&  0.96$\pm$0.49 & 1.53$\pm$0.81&  33.37$\pm$9.55\\		   
    $t\bar{t}$(6.7pb)&   19.05$\pm$2.92&  15.56$\pm$2.39&  60.68$\pm$9.30&   2.14$\pm$0.34 &  1.79$\pm$0.31&  7.17$\pm$1.00\\		   
    Single top(s-ch)&   6.90$\pm$1.00&   5.14$\pm$0.75&   14.38$\pm$2.09&   0.69$\pm$0.10 &  0.51$\pm$0.08& 1.53$\pm$0.20\\		   
    Single top(t-ch)&   1.60$\pm$0.23&   1.87$\pm$0.27&   29.57$\pm$4.33&   0.22$\pm$0.04 &  0.24$\pm$0.04& 3.54$\pm$0.47\\		   
                $WW$&   0.17$\pm$0.02&   0.93$\pm$0.11&   15.45$\pm$1.91&   0.01$\pm$0.01 &   0.12$\pm$0.04& 3.00$\pm$0.20\\		   
                $WZ$&   2.41$\pm$0.26&   1.84$\pm$0.20&   7.59$\pm$0.81&    0.58$\pm$0.06 &  0.42$\pm$0.05& 1.62$\pm$0.09\\		   
                $ZZ$&   0.06$\pm$0.01&   0.08$\pm$0.01&   0.31$\pm$0.03&    0.00$\pm$0.01 &  0.01$\pm$0.01&0.02$\pm$0.00\\		   
    $Z\rightarrow\tau\tau$& 0.25$\pm$0.04& 1.29$\pm$0.20&   7.27$\pm$1.12&    0.00$\pm$0.01 &   0.01$\pm$0.01&  0.24$\pm$0.03\\		   
         non-$W$ QCD&   5.50$\pm$1.00&   9.55$\pm$1.73&  184.7$\pm$33.04&   1.16$\pm$0.44 &  1.51$\pm$0.55& 18.34$\pm$5.54\\ \hline		   
    Total Background&   80.6$\pm$18.8& 87.0$\pm$18.0& 809.6$\pm$159.4& 14.2$\pm$4.0 & 15.5$\pm$3.6& 140.4$\pm$16.9\\ \hline	   
    $WH$ signal (120 $\GeV$)&  0.85$\pm$0.10&   0.60$\pm$0.07&   1.70$\pm$0.14  &   0.09$\pm$0.01 &   0.06$\pm$0.01&  0.20$\pm$0.01\\  \hline	   
     Observed Events&   83&               90&               805&              11&   13& 138 \\
  \end{tabular}
  \caption{ The predicted and observed yields for the CDF 2~fb$^{-1}$ $WH$ search
     separated by detector region.  This table is from Ref.~\cite{b-cdf-wh}\label{t-wh-data-cdf}}
\end{ruledtabular}
\end{table*}

\begin{table}
\begin{ruledtabular}
  \begin{tabular}{lrclrclrclrcl}
                    &  $W $&+&$$ 2 jet   &  $W $&+&$ 2$ jet   &  $W $&+&$ 3$ jet   &  $W $&+&$ 3$ jet   \\
         & \multicolumn{3}{c} { 1$b$-tag }& \multicolumn{3}{c} {  2 $b$-tag  }
                                                & \multicolumn{3}{c} { 1 $b$-tag } & \multicolumn{3}{c} { 2 $b$-tag }\\
    \hline
     {\it WH}      &   2.8 &$\pm$& 0.3 &    1.5 &$\pm$& 0.2  &   0.7&$\pm$&0.1   &  0.4 &$\pm$& 0.1 \\
    \hline
    $WZ$       &   34.5 &$\pm$&3.7 &    5.3 &$\pm$& 0.6  &    9.1&$\pm$& 1.0 &   1.7 &$\pm$& 0.2 \\
    $Wb\bar{b}$&  268   &$\pm$& 67 &   54   &$\pm$& 14   &   87&$\pm$&22     &   22.7  &$\pm$&  5.7 \\
    $W$+jets   &  347   &$\pm$& 87 &   14.0 &$\pm$&  4.4 &   96&$\pm$&24     &   8.5 &$\pm$&  2.7 \\
    $t\bar{t}$ &   95   &$\pm$& 17 &   37.4 &$\pm$&  7.0 &  156&$\pm$&29     &   81  &$\pm$& 15 \\
    single $t$ &   49.4 &$\pm$&9.0 &   12.4 &$\pm$&  2.3 &  15.7&$\pm$&2.9   &   6.7 &$\pm$&  1.2 \\
    m-jet      &  104   &$\pm$& 29 &    8.9 &$\pm$&  2.1 &   54&$\pm$&15     &   8.7 &$\pm$&  2.1 \\
    \hline
    Total      &  896   &$\pm$&177 &  132   &$\pm$& 27    &  418&$\pm$&76    &  129  &$\pm$& 24   \\
    Data       &    885          &&&    136            &&&    385            &&&   122   &&         \\
  \end{tabular}
  \caption{The predicted and observed yields for the D0 1~fb$^{-1}$ $WH$
    search.  This table is from Ref.~\cite{b-d0-wh}\label{t-wh-data-d0}}
\end{ruledtabular}
\end{table}

Both CDF and D0 derive their final result by comparing the predicted
and observed spectra using binned likelihoods.  For both experiments,
the single-tag and double-tag samples are separated and their
likelihoods combined for the final results. D0 further separates the
samples into events with exactly two jets and events with three jets.
When determining the final result, CDF uses the dijet mass as the
input to their likelihood. D0 uses the NN output for the two jet
events and the dijet mass for three jet events. The likelihood methods
are described in section~\ref{sec:higgs_direct_comb}.

Systematic uncertainties are evaluated by both experiments for trigger
efficiency, lepton identification efficiency, $b$--jet
(mis)identification efficiency, the jet identification efficiency and
jet energy calibration, multijet background calculation method,
luminosity, and theory cross sections used for background and signal
event yield calculations. CDF additionally reports uncertainties from
parton density functions and initial and final state radiation
modeling. D0 reports an additional systematic from the $W$+jets
simulation derived by comparing shapes of distributions of data and
simulated events before $b$--tagging.

\begin{figure}
  \includegraphics[width=0.45\textwidth]{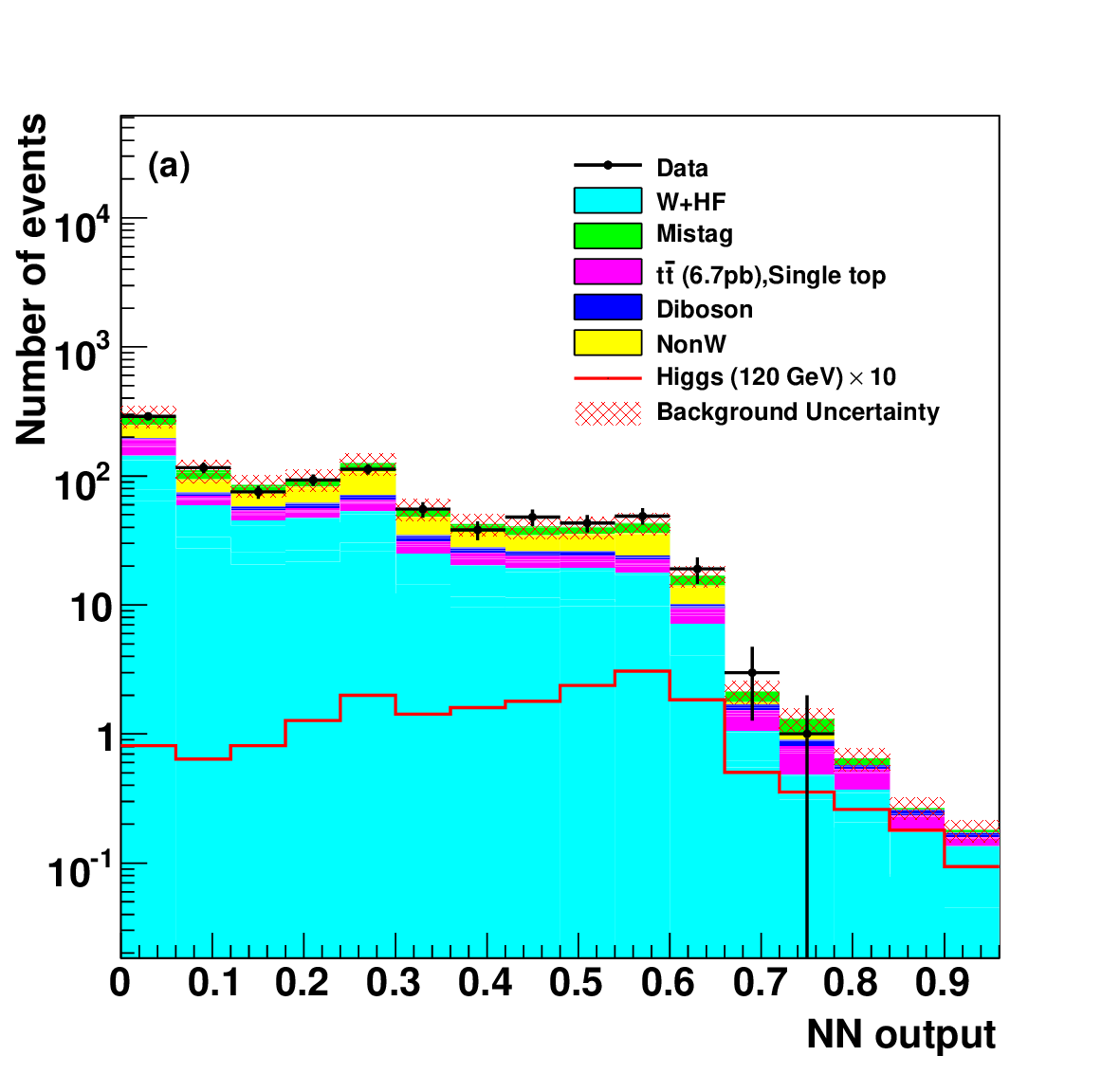}
  \includegraphics[width=0.45\textwidth]{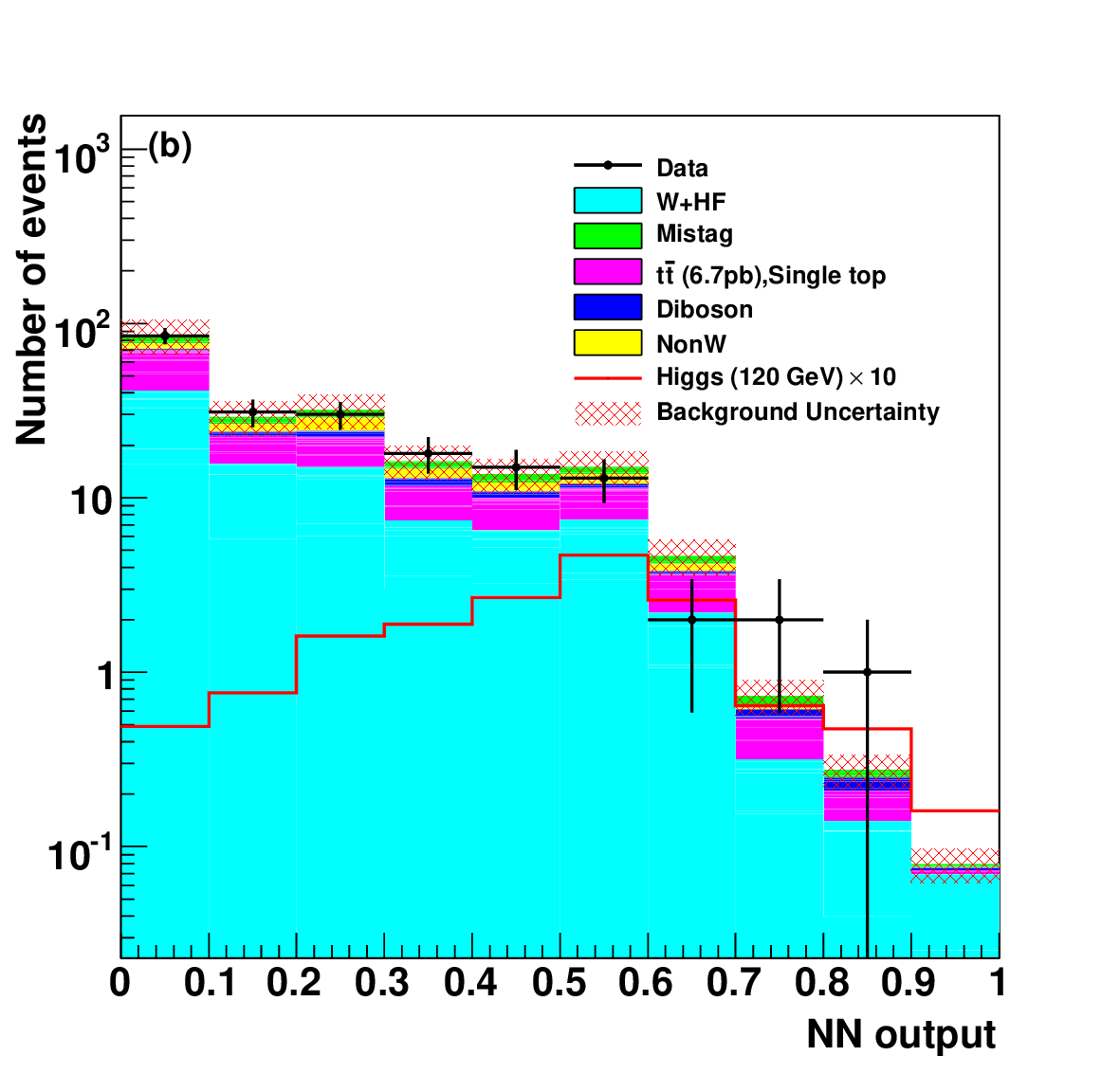}
  \caption{The NN output distributions for (a) single-- (ST+NN) and (b) double--tagged
  (ST+ST, ST+JP) events from the CDF 2~fb$^{-1}$ $WH$ search.\label{f-wh-cdf-final}}
\end{figure}

\begin{figure}
  \includegraphics[width=0.4\textwidth]{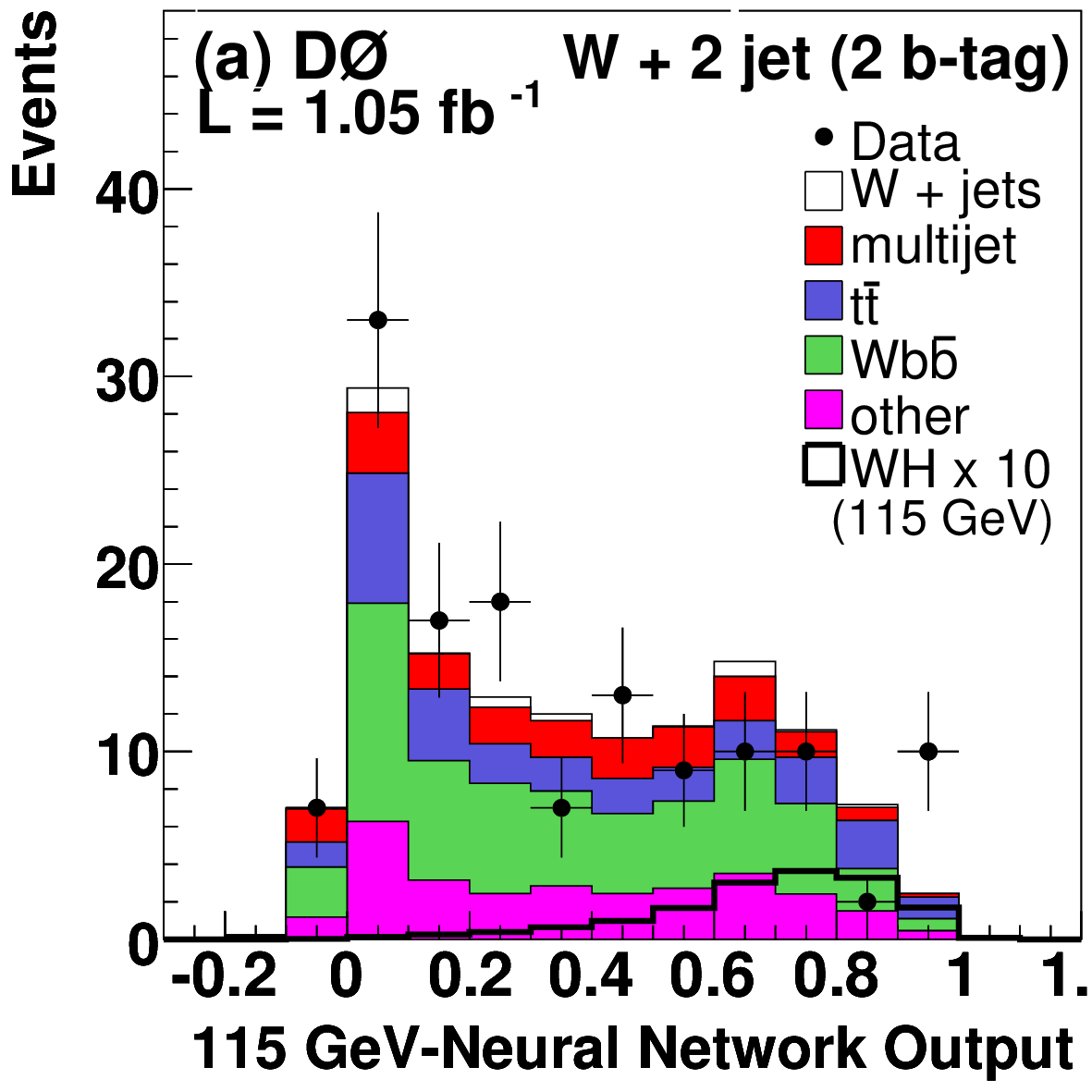}
  \includegraphics[width=0.4\textwidth]{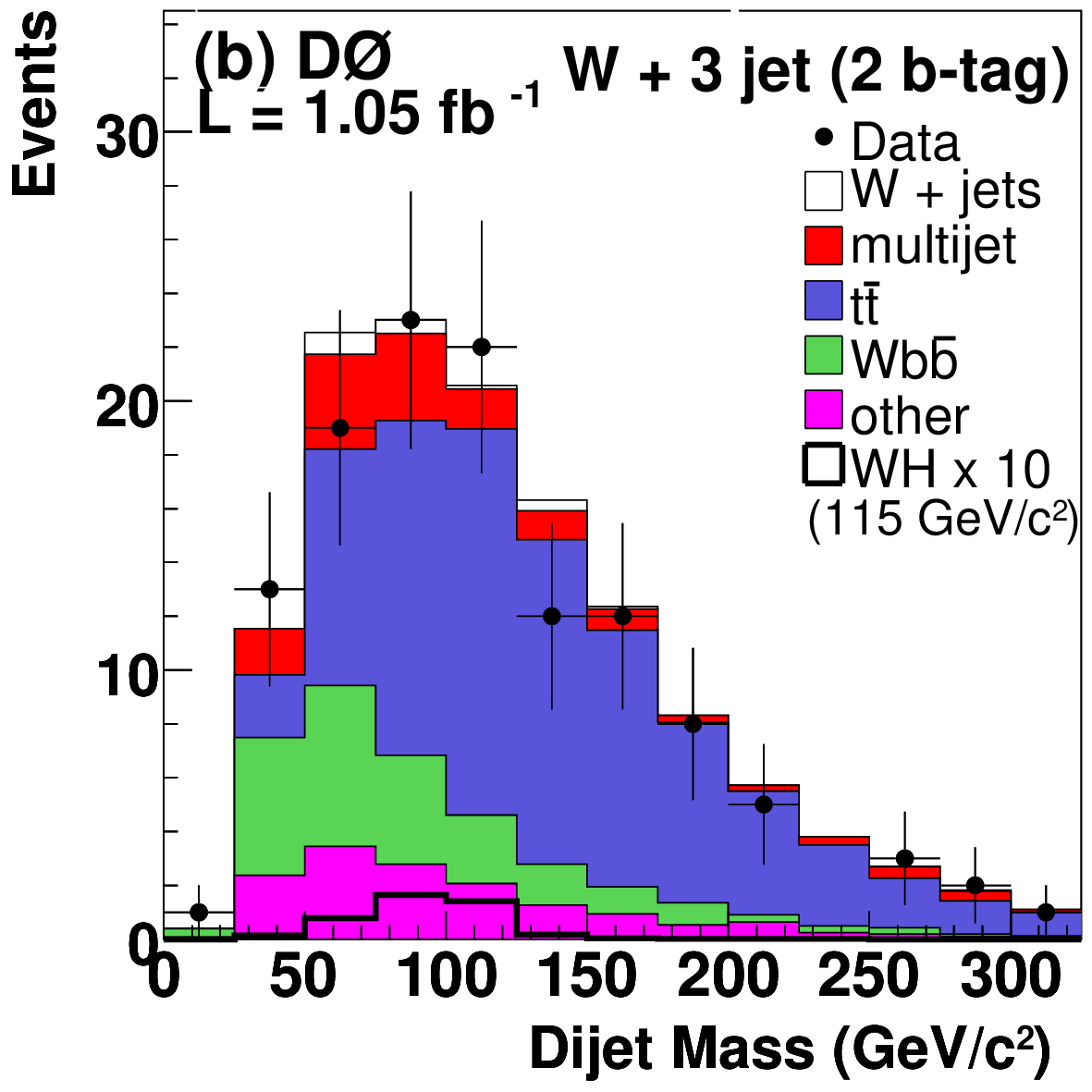}
  \caption{(a) The NN output for double-tagged dijet events and (b) the dijet mass
    for double-tagged three jet events from the D0 $WH$ search. 
    \label{f-wh-d0-final}}
\end{figure}

\begin{figure}
  \includegraphics[width=0.45\textwidth]{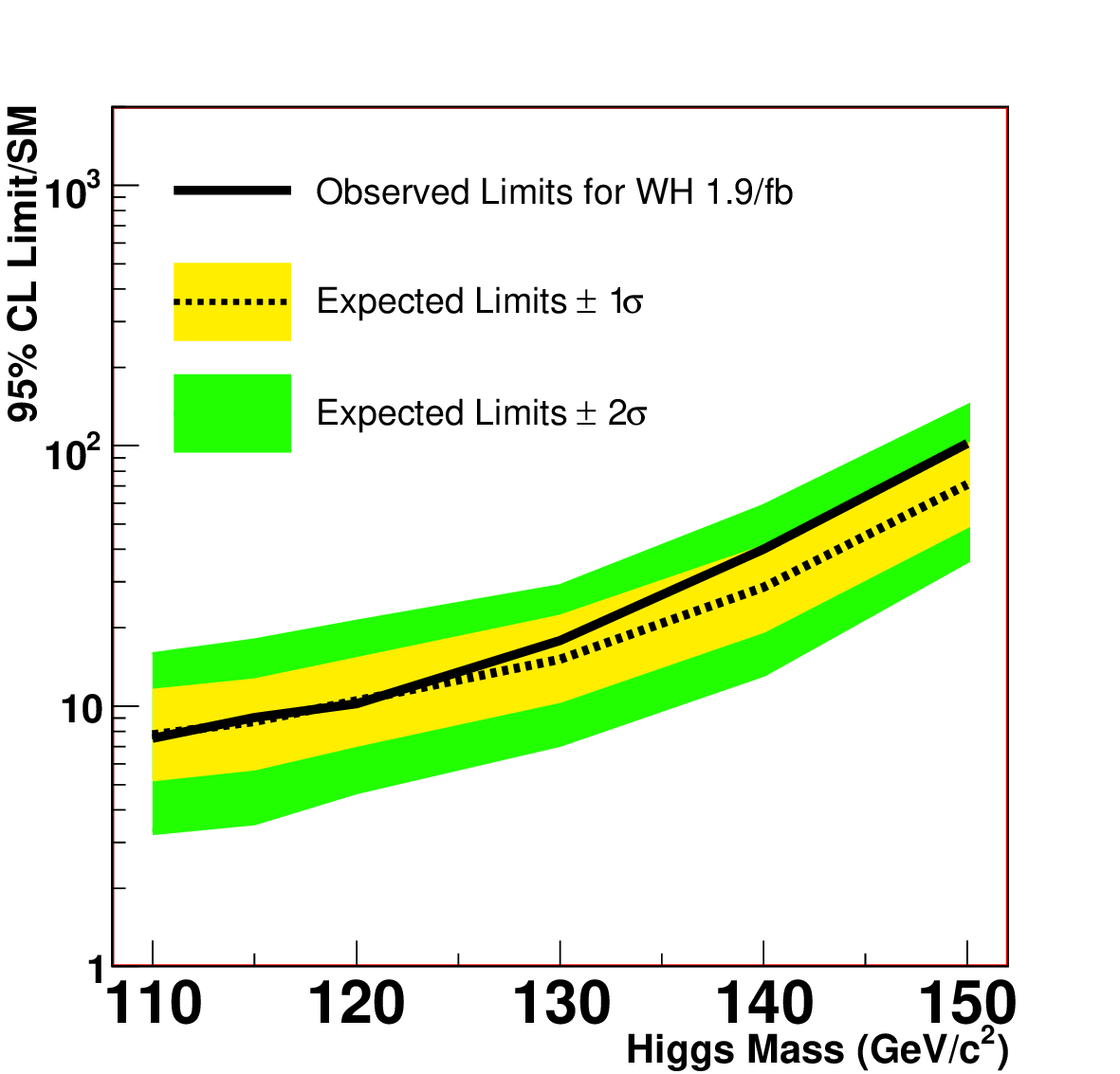}
  \caption{The cross section limits from the CDF $WH$ search.\label{f-wh-cdf}}
\end{figure}

\begin{figure}
  \includegraphics[width=0.46\textwidth]{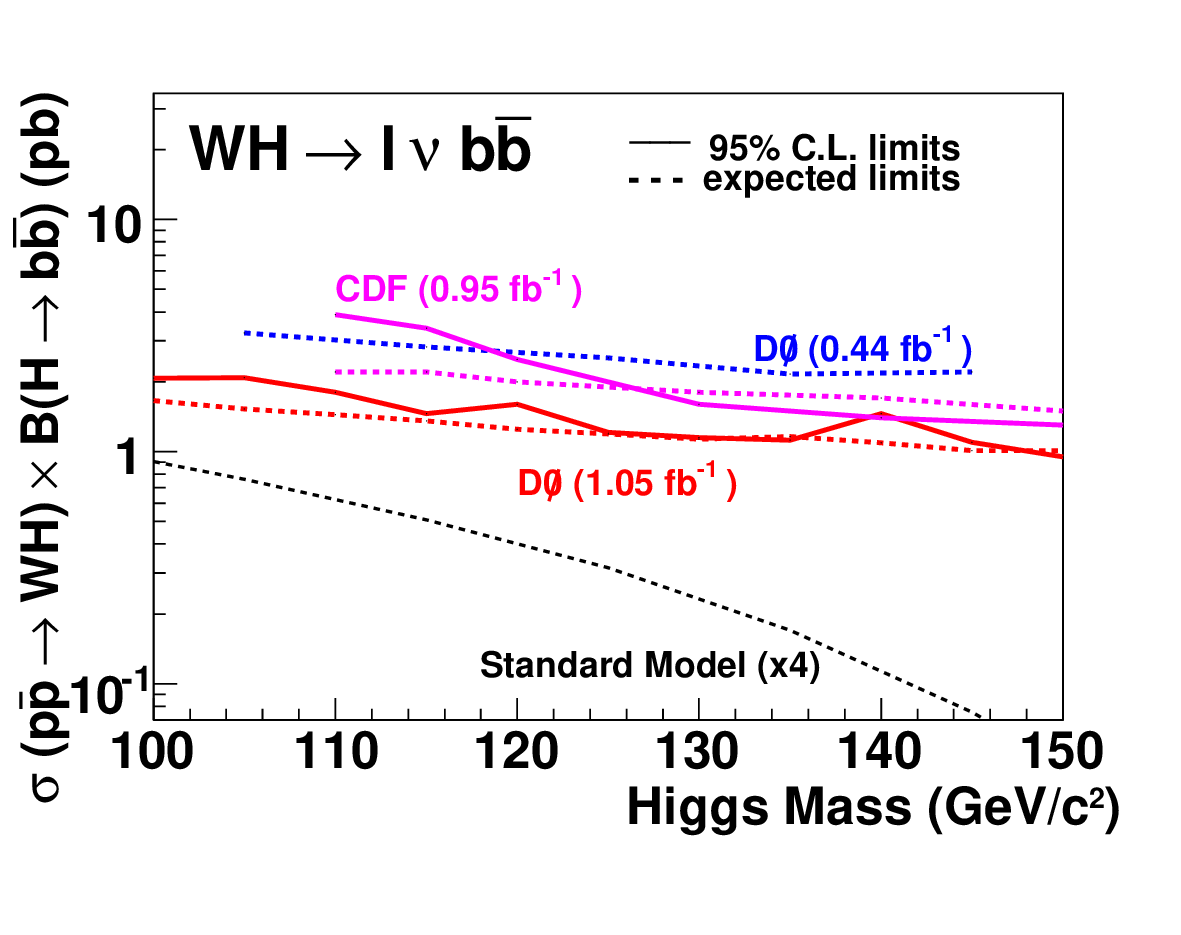}
  \caption{The cross section limits from the D0 $WH$ search.\label{f-wh-d0}}
\end{figure}

The CDF NN output distributions for the tagged event selections are
shown in Fig..~\ref{f-wh-cdf-final}. The D0 NN output distribution
(two jet events) and dijet mass distribution (three jet events) are
shown in Fig.~\ref{f-wh-d0-final}.  The CDF and D0 cross section
limits for the $WH$ final state only are shown if Figs.~\ref{f-wh-cdf}
and~\ref{f-wh-d0} respectively. The result from combining the results
for all low mass final states is shown in Section~\ref{sec:higgs_direct_comb}. 

\subsubsection{$ZH\to\ell\ell\bbbar$ Final State}
\label{sec:higgs_direct_WZllbb}

The final state from $ZH\to\ell\ell\bbbar$ events has the lowest
production rate among those shown in Tab.~\ref{t-hprod-low} but has
the highest purity. A similar strategy is taken for this final state
as for the $WH$ final state. An initial sample of $Z$+dijet events is
selected, the purity is improved by requiring $b$ jets, and the final
result is determined by using either the invariant mass distribution
or the output of a neural network as inputs to a shape--based limit
setting program.  Both D0~\cite{b-zh-llbb-d0} and
CDF~\cite{b-zh-llbb-cdf} have published results in this final state.
The D0 publication uses a data sample corresponding to
$\intL=0.45$~fb$^{-1}$. The CDF published results corresponds to
$\intL=1.0$~fb$^{-1}$. 

\begin{figure}
  \includegraphics[width=0.46\textwidth]{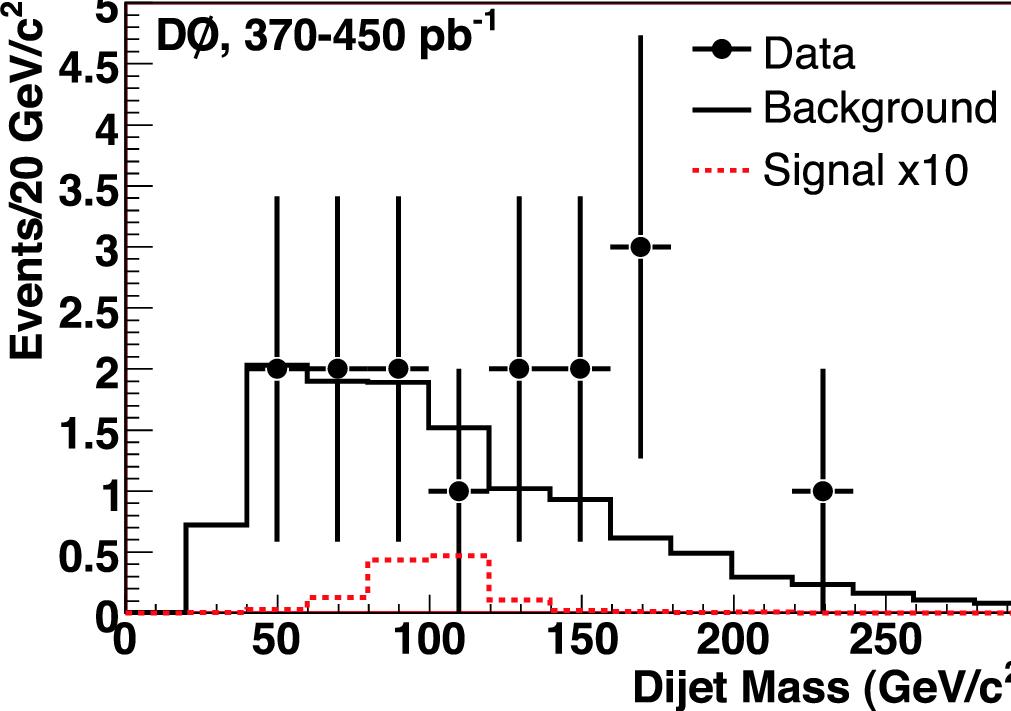}
  \caption{Dijet mass distribution for the D0 1.0~fb$^{-1}$
    $ZH\to\ell\ell\bbbar$ analysis.\label{f-zh-llbb-d0-dijet}}
\end{figure}

\begin{figure}
  \includegraphics[width=0.45\textwidth]{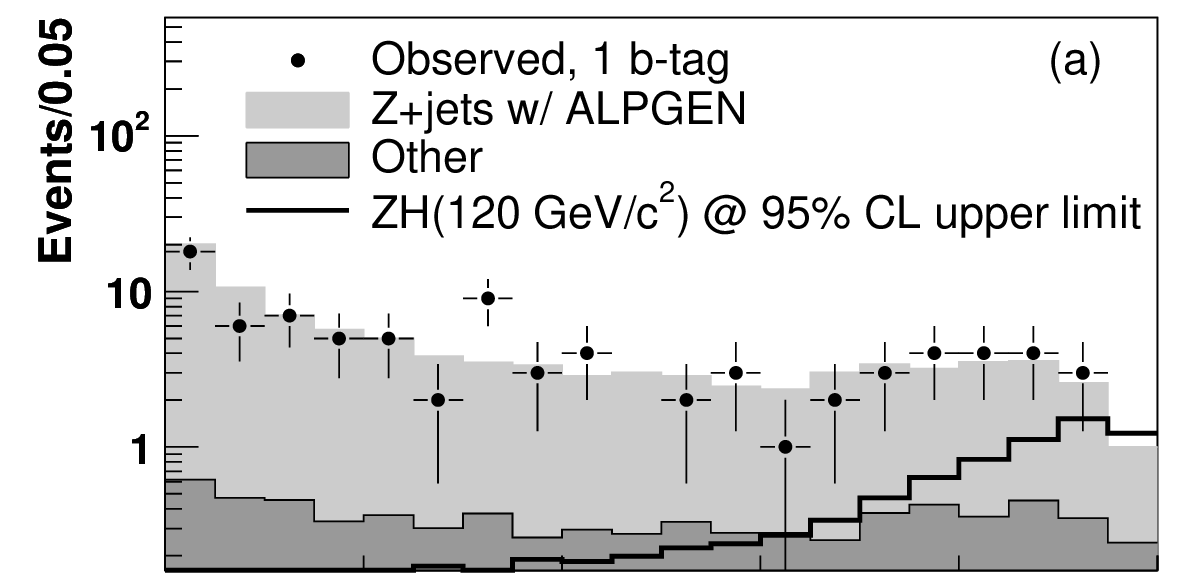}
  \includegraphics[width=0.45\textwidth]{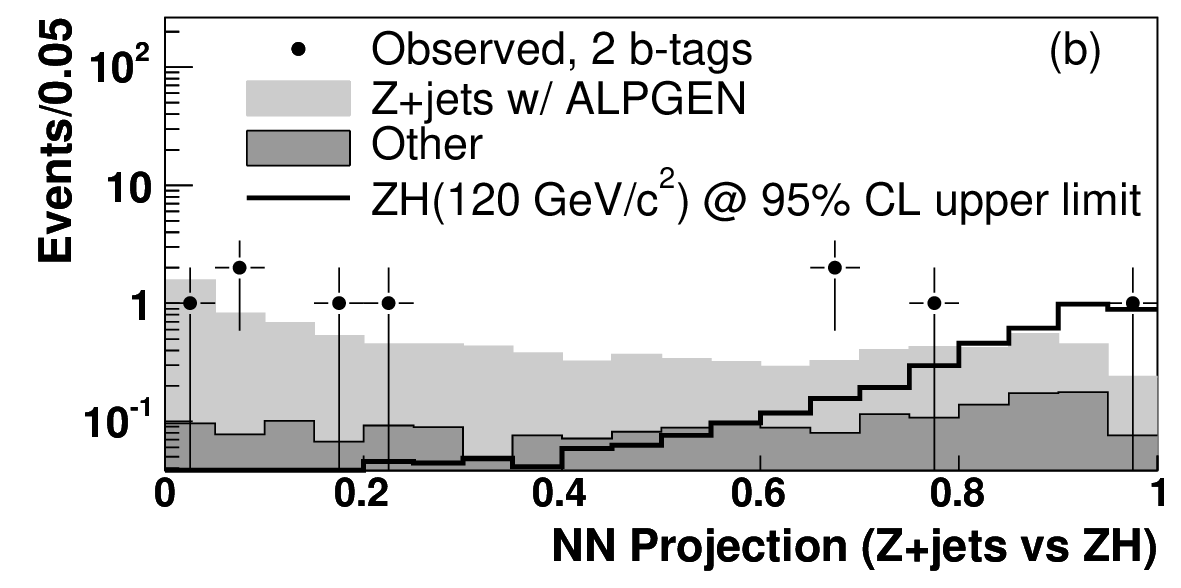}
  \caption{NN distributions for the (a) single-tagged and (b) double-tagged
    samples from the CDF 1.0~fb$^{-1}$ $ZH\to\ell\ell\bbbar$ 
    analysis.  The 95\% CL upper bound cross section is 19$\times$ the
    SM prediction.\label{f-zh-llbb-cdf-nn}}
\end{figure}

The initial selections for both experiments require events with two
leptons whose invariant mass is consistent with a $Z$ boson and two
additional jets. The backgrounds in this channel arise from $Z$+jets,
$\ttbar$ events, diboson production, and from jets misidentified as
leptons or a lepton+jet system being misidentified as an isolated
lepton. They are determined using a combination of simulated events
and data control samples in essentially the same manner as for the
$WH$ channel.  One difference with respect to the $WH$ calculation is
that sidebands of the $\ell\ell$ invariant mass distribution are used
by D0 to determine the misidentification and false isolation
background. Both experiments identify $b$ jets using secondary vertex
algorithms\footnote{For D0 this differs from the $WH$ case because the
  published $ZH$ results predate the availability of the NN
  tagger. More recent D0 preliminary results use the same NN tagger
  described in the $WH$ analysis.}. The CDF result splits the final
tagged sample into single- and double-tagged samples as in the $WH$
analysis, but the D0 result does not. The D0 result is derived using
the dijet mass distribution only, while the CDF result is based on the
binned likelihood of a NN output. The sources of systematic
uncertainties for these channels are the same as for the $WH$
result. Tab.~\ref{t-zh-llbb-cdf-yields} shows the yields for the
single- and double-tag analyses from CDF, and
Tab.~\ref{t-zh-llbb-d0-yields} shows the yields for the dijet and
double-tagged samples for D0
respectively. Figure~\ref{f-zh-llbb-d0-dijet} (\ref{f-zh-llbb-cdf-nn})
shows the dijet invariant mass (neural network output) distribution
from the D0 (CDF) analysis.  This distribution is used as the input to
the limit setting program.

\begin{table}
\begin{ruledtabular}
  \begin{tabular}{lcc} 
    Source        & Single-Tagged & Double-Tagged \\ \hline
   $Z + \bbbar$   & $35.1\pm14.6$ &  $6.3\pm2.5$ \\
   $Z + \ccbar$   & $21.8\pm8.5$  &  $1.0\pm0.4$ \\
   $Z + \qqbar$   & $32.3\pm5.5$  &  $1.0\pm0.2$ \\
   $\ttbar$       & $5.2\pm1.0$   &  $2.8\pm0.6$ \\
   $ZZ$           & $4.0\pm0.8$   &  $1.3\pm0.3$ \\
   $WZ$           & $1.2\pm0.2$   &  $0.04\pm0.01$ \\
   Non-$Z$        & $1.9\pm1.4$   &  $0.2\pm0.2$ \\ \hline
   Total Expected & $101.5\pm32$  &  $12.7\pm4.1$ \\
   Observed       &   100         &  11 \\ \hline
   Signal Yield   &   $0.44$      &  $0.23$ \\
  \end{tabular}
  \caption{Single- and double-tagged yields for the CDF 1.0~fb$^{-1}$
    $ZH\to\ell\ell\bbbar$ search.
\label{t-zh-llbb-cdf-yields}}
\end{ruledtabular}
\end{table}

\begin{table}
\begin{ruledtabular}
  \begin{tabular}{ccc} 
   Source      &   Dijet  &  Double-tagged \\ \hline
   $Z+\bbbar$  &   17.4   &   3.3          \\
   $Z+jj$      &  851     &   3.8          \\
   $\ttbar$    &   12.3   &   3.9          \\
   $WZ+ZZ$     &   30.6   &  0.74          \\
   Non-$Z$     &   44.1   &  0.59          \\ \hline
   Total Exp.  &  956     &  12.3          \\
   Observed    & 1008     &  15            \\
  \end{tabular}
  \caption{Dijet and double-tagged sample yields for the D0 0.45~fb$^{-1}$
    $ZH\to\ell\ell\bbbar$ search.
\label{t-zh-llbb-d0-yields}}
\end{ruledtabular}
\end{table}

Systematic uncertainties are included for trigger and lepton
identification efficiencies, parton density functions, background
cross sections, $b$--tagging efficiencies, jet energy reconstruction
and the methods used to estimate instrumental backgrounds. The CDF
result also includes a systematic from the top mass uncertainty. The
systematic uncertainties for the D0 measurement range between 2\% and 20\%
expressed as a fraction of the total background. The systematic
uncertainties for the CDF measurement range between 1\% and roughly
25\% of the total background. The largest contribution for both
experiments is from the background cross sections.

\begin{figure}
  \includegraphics[width=0.46\textwidth]{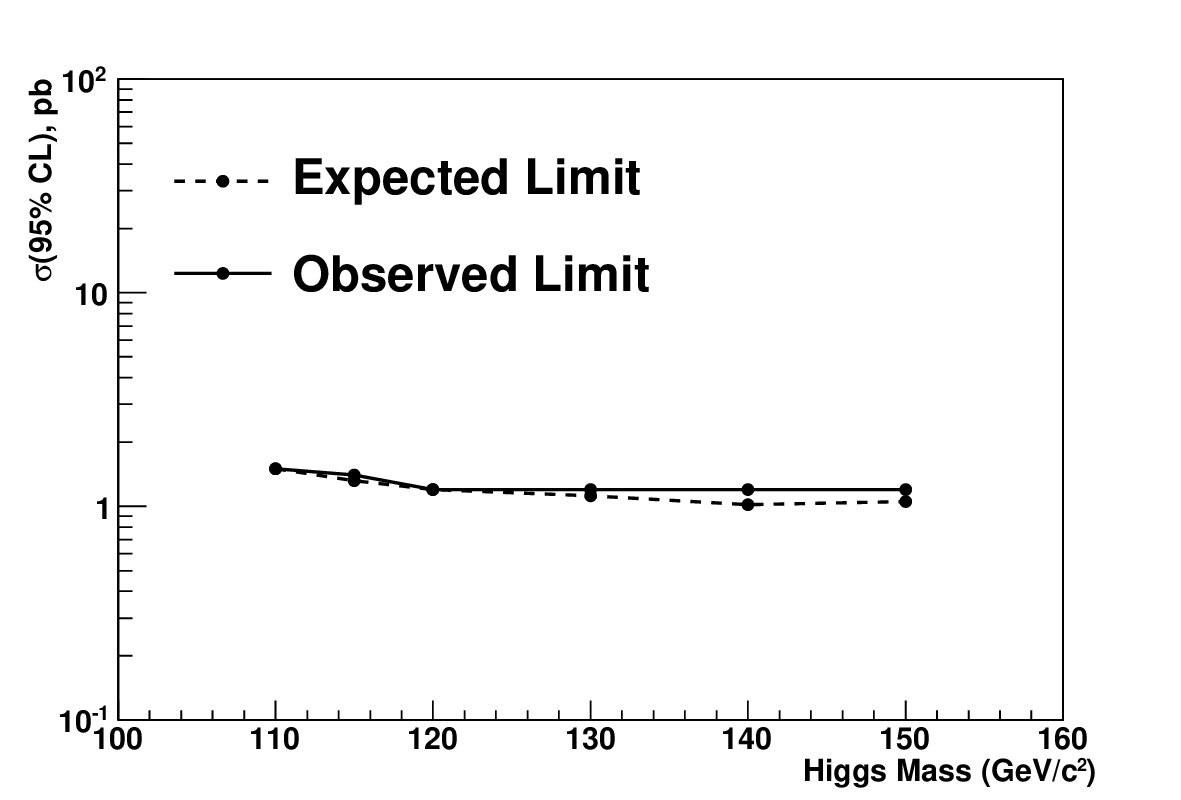}
  \caption{Expected and Observed limits for the CDF 1.0~fb$^{-1}$
    $ZH\to\ell\ell\bbbar$ analysis.\label{f-zh-llbb-cdf}}
\end{figure}

\begin{figure}
  \includegraphics[width=0.46\textwidth]{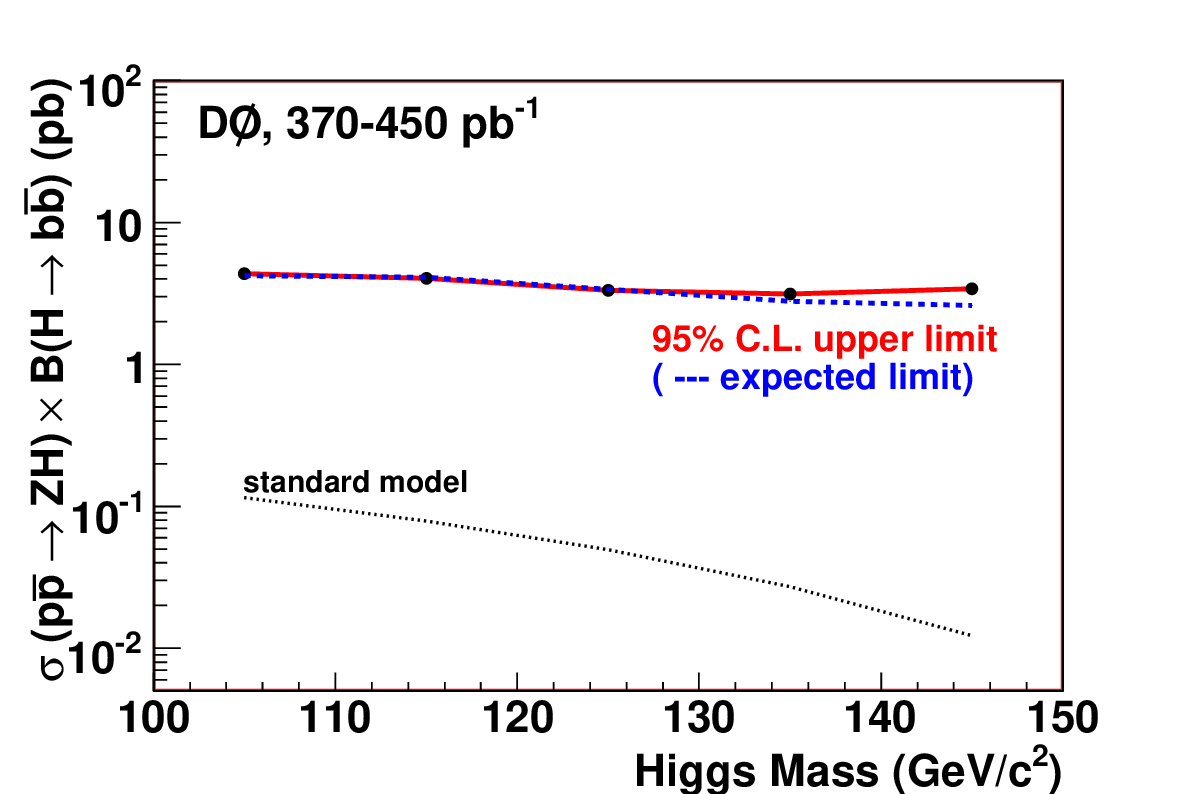}
  \caption{Expected and Observed limits for the D0 0.45~fb$^{-1}$
    $ZH\to\ell\ell\bbbar$ analysis.\label{f-zh-llbb-d0}}
\end{figure}

The final limits, derived using the same methods as the $WH$ results,
are shown for CDF in Fig.~\ref{f-zh-llbb-cdf} and for D0 in
Fig.~\ref{f-zh-llbb-d0}.  As expected these channels have considerably
less sensitivity than the $WH$ channels because of the significantly
lower signal cross-section times branching fraction.

\subsubsection{$ZH\to\nu\nu\bbbar$ and Related Final States}
\label{sec:higgs_direct_ZHvvbb}

The $ZH\to\nu\nu\bbbar$ final state has a production rate intermediate
between the $WH$ and $ZH\to\ell\ell\bbbar$ final states. This final
state also has a significant contribution from the process
$WH\to\ell\nu\bbbar$ in which the charged lepton $\ell$ escapes
detection. This is particularly true for the case $\ell=\mu$ because
the muon leaves very little energy in the calorimeter and thus results
in event $\MET$ similar to that from $Z\to\nu\nu$ decay.

Unlike either of the previously discussed final states, the
$ZH\to\nu\nu\bbbar$ final state has no charged leptons from vector
boson decay. This implies a significantly increased background from SM
multijet events in which $\MET$ arises from mismeasurement. This
background is difficult to model from simulation, and analyses of this
final state must develop techniques to measure it using data control
samples. Both CDF~\cite{b-zh-vvbb-cdf} and D0~\cite{b-zh-vvbb-d0} have
published results in this final state. The two experiments have
developed different methods for controlling and estimating the
multijet background.

\paragraph{CDF Search}
The CDF analysis uses a data sample corresponding to 1~fb$^{-1}$ and
begins with selection of events passing a $\MET$ trigger with level
one $\MET>25$ GeV, a level two requirement of two jet clusters
having $E_T>10$~GeV and a level three requirement of $\MET>35$ GeV.
At least one of the level two jets must also satisfy $\eta<1.1$.  The
initial offline selection (``pre-tag'') requires events to have
$MET>50$ GeV and exactly two jets with $E_T>20$~GeV. One of the
jets must have $E_T>35$~GeV, and the other must have
$E_T>25$~GeV. Additionally, one of the jets must satisfy $|\eta|<0.9$, 
and the other jet must satisfy $|\eta|<2.4$. The two jets must also
have $\Delta\phi>1.0$~rad, and events with high $p_T$, isolated
leptons are vetoed.  Finally, at least one of the jets is required to
have a secondary vertex $b$--tag

All non-multijet backgrounds; $\ttbar$, $W$+jets, $Z$+jets and diboson
production are modeled using simulated events.  The multijet background is
studied by dividing the sample into two control regions and a signal
region. The regions are defined in Tab.~\ref{t-zh-vvbb-cdf-regs}. The
multijet background in all regions is divided into two components: (1)
events with only light-flavor ($u$, $d$ and $s$) quarks in which one
or more of the jets is misidentified with a secondary vertex and (2)
events with $c$ and $b$ quarks. The contribution from light flavor
jets is determined using a control sample with no tags to which a
misidentification factor is applied. The heavy flavor contribution is
determined using simulated events with normalization factors for
single-tag and double-tag topologies determined by forcing the data
yields and the sum of all backgrounds to agree (before dividing the
sample into the three regions). The scale factors are $1.30\pm0.4$
($1.47\pm0.07$) for the single(double) $b$--tagged events. 

\begin{table}
\begin{ruledtabular}
  \begin{tabular}{ccc}
           & Dominant     & \\
    Region & Source       & Selection         \\ \hline
     CR1   & Multijet     &  Leptons Vetoed   \\
           &              &  $\Delta\Phi(\vec{E}_{T2},\vec\MET)<0.4$ \\ \hline
     CR2   & EW, $\ttbar$ &  Lepton Required  \\
           &              &  $\Delta\Phi(\vec{E}_{T2},\vec\MET)>0.4$ \\ \hline
    Signal &              &  Leptons Vetoed   \\
           &              &  $\Delta\Phi(\vec{E}_{T1},\vec\MET)>0.4$ \\
           &              &  $\Delta\Phi(\vec{E}_{T2},\vec\MET)>0.4$ \\
  \end{tabular}
  \caption{Definitions of the three regions (CR1, CR2, and Signal) in
    the CDF $VH\to\MET\bbbar$ search.  $\Delta\Phi(\vec{E}_{T1(2)},\vec\MET)$
    is the angle between the $\MET$ and the jet with the highest(second 
    highest) $E_T$.\label{t-zh-vvbb-cdf-regs}}
\end{ruledtabular}
\end{table}

The final analysis selection is determined by optimizing $S/\sqrt(B)$
where $S$ is the total signal yield, including the contribution from
$WH\to\ell\nu\bbbar$ in which the lepton is not identified, and $B$ is
the total background. The optimization is carried out only for the
signal region.  The final selection then requires
$\Delta\phi(j_1,j_2)>0.8$, $\HET/H_T>0.454$, $E_T^{j_1}>60$~GeV and
$\MET>70$ GeV. Here $H_T$ is the scalar sum of the jet $E_T$
values, $\HET$ is the magnitude of the vector sum and $j_1$($j_2$)
denotes the jet with the highest(second highest) $E_T$.

The systematic uncertainties arise from a number of sources. For the
CDF analysis the dominant uncertainty in the jet energy calibration
which varies between 10\% - 26\% for multijet and $V$+jets (including
heavy flavor) backgrounds, but is only 8\% for signal events.  The
other dominant systematic arises from the calculated cross sections
used to normalize backgrounds. This ranges between 11\% - 40\% for a
given source depending on the samples. The total systematic is 17\%
for the single--tagged analysis and 19\% for the double-tagged
analysis.

\begin{figure}
   \includegraphics[width=0.49\textwidth]{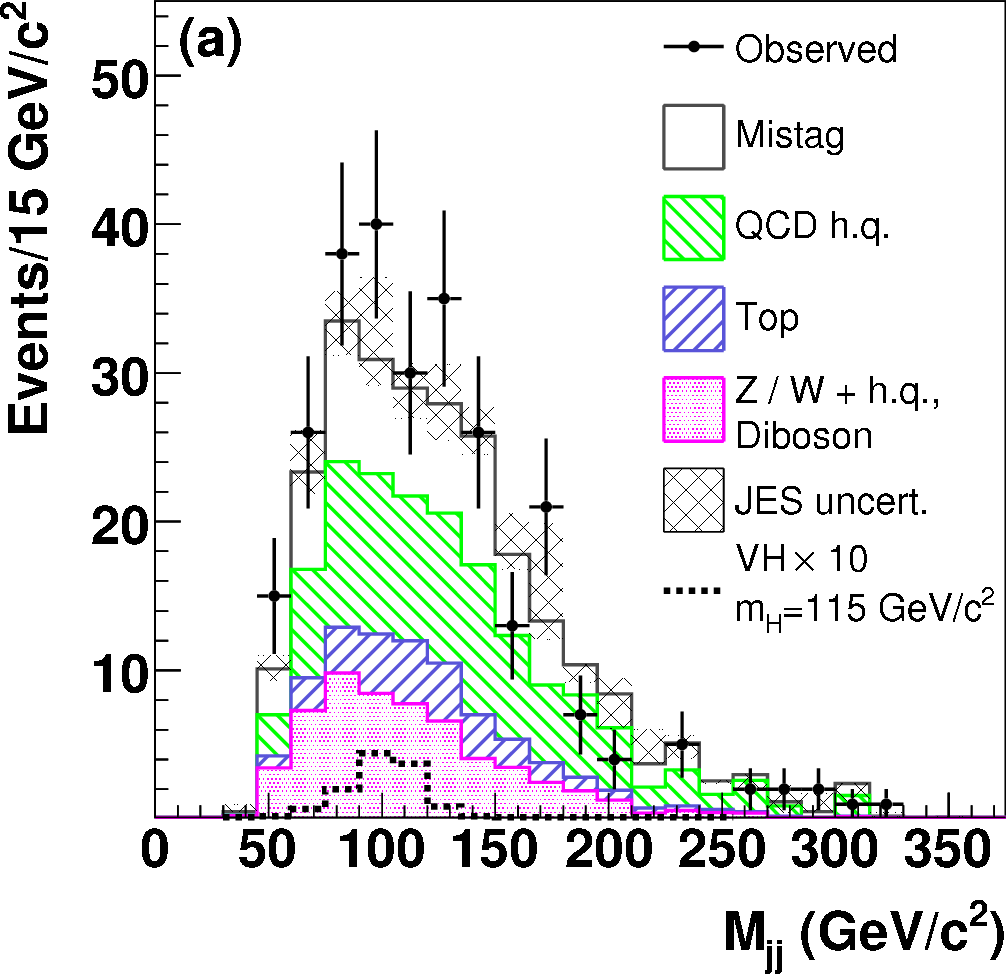}
   \includegraphics[width=0.49\textwidth]{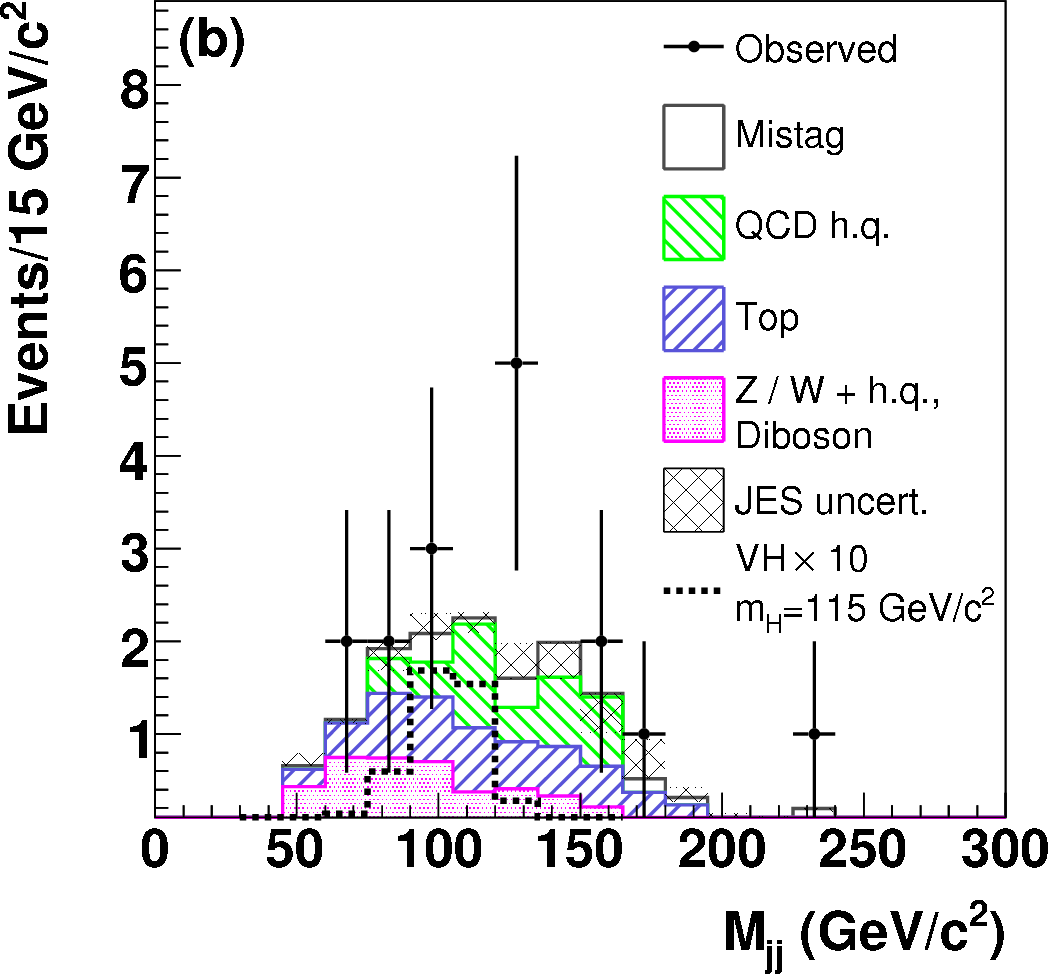}
   \caption{The dijet invariant mass distributions for the (a) single-tagged
   and (b) double--tagged CDF searches in the $VH\to\MET\bbbar$ final state.
   \label{f-zh-vvbb-cdf-m}}
\end{figure}

\begin{figure}
   \includegraphics[width=0.49\textwidth]{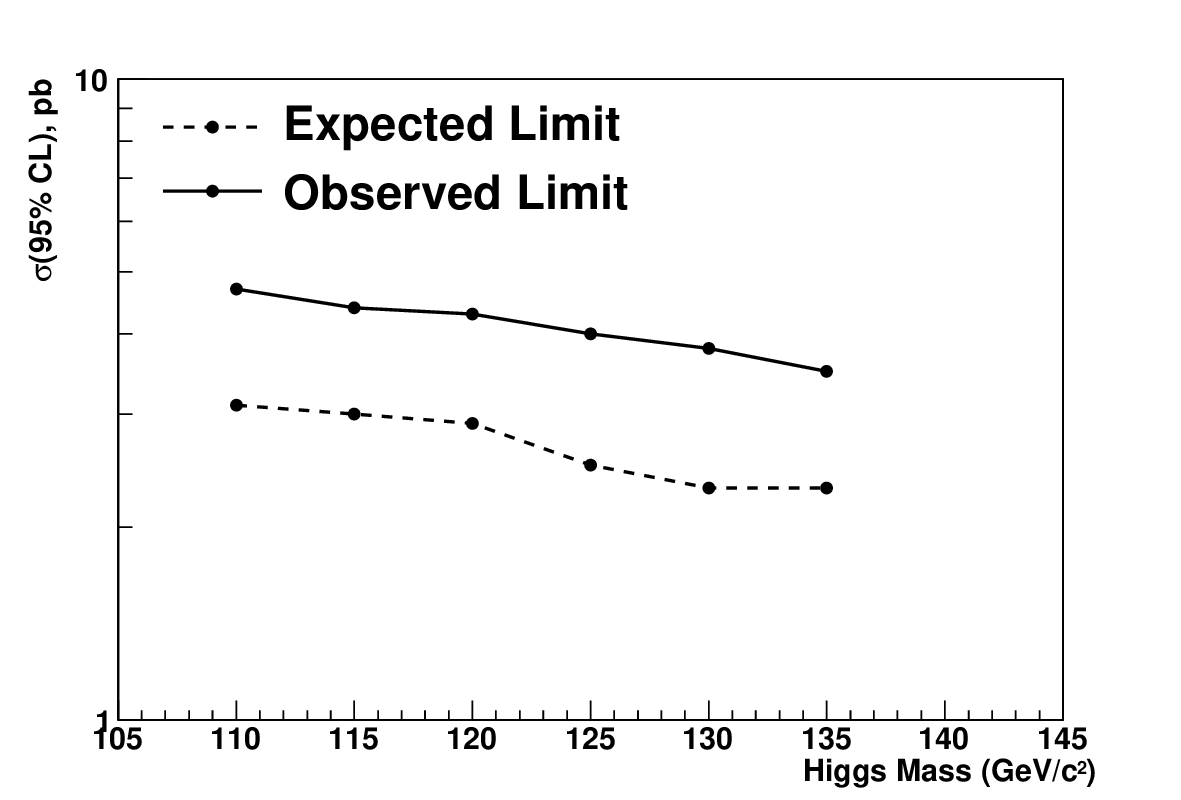}
   \caption{Expected and observed 95\% CL upper bounds on the $VH$ production
   cross section for the CDF $\MET+\bbbar$ search.\label{f-zh-vvbb-cdf}}
\end{figure}

Fig.~\ref{f-zh-vvbb-cdf-m} shows dijet invariant mass for the
single- and double-tagged events in the signal region. The yields are 
given in Tab.~\ref{t-zh-vvbb-cdf-yield}.  The limits on Higgs
production are calculated using the same procedure as for the previous 
two channels, and they are shown in Fig.~\ref{f-zh-vvbb-cdf}.

\begin{table}
\begin{ruledtabular}
  \begin{tabular}{ccc} 
    Source   & Signal--tagged & Double--tagged \\ \hline
    Multijet &   $93\pm23$    & $3.74\pm1.27$   \\
    $\ttbar$ & $27.3\pm3.8$   & $4.88\pm0.8$    \\
    Diboson  & $7.0\pm1.4$    & $0.79\pm0.19$   \\
    $W+$hf & $33.4\pm16.2$  & $1.65\pm0.86$   \\
    $Z$+hf & $18.3\pm8.1$   & $1.67\pm0.77$   \\
    Mistags  & $69\pm9$       & $1.64\pm0.48$   \\ \hline
    Total    & $248\pm43$     & $14.4\pm2.7$    \\
    Observed &  268           &    16           \\
  \end{tabular}
  \caption{Yields in the CDF $VH\to\MET\bbbar$ analysis.\label{t-zh-vvbb-cdf-yield}}
\end{ruledtabular}
\end{table}

\paragraph{D0 Search}
The published D0 analysis uses a data sample corresponding to
$\intL=5.2$~fb$^{-1}$. Events selected for this analysis must satisfy
a trigger requiring two acoplanar jets and $\MET$ in which the
kinematic thresholds varied as a function of the instantaneous
luminosity. The primary requirements of the initial offline event
preselection are at least two jets with $E_T>20$~GeV and $|\eta|<2.5$,
$\Delta\phi(j_1,j_2)<165^o$, and $\MET>20$ GeV. This is divided
into four samples: (1) a signal sample defined by the additional
requirements of $\MET>40$ GeV, $S>5$, $D<\pi/2$ and a veto of
events with high-$p_T$, isolated leptons, (2) an electroweak test
sample enhanced in $W\to\mu\nu$ events and defined similarly as for
the signal sample, but requiring the presence of a high $p_T$,
isolated muon, (3) a multijet modeling sample defined similarly to the
signal sample except $D>\pi/2$ and (4) a multijet enriched sample used to
confirm the background model predicted using the multijet modeling sample
and defined by $\MET>30$ GeV and no requirement on $S$. Here $S$ is
the significance of the $\MET$ and $D \equiv
\Delta\phi(\vec{\MET},\vec{\TET)} = \phi_{\MET} - \phi_{\TET}$ in
which $\TET$ is the magnitude of the vector sum of the transverse
momentum of the tracks in the event and $\phi_i$ is the azimuthal
angle of either $\MET$ or $\TET$.

The presence of $b$ hadrons from the $H\to\bbbar$ signal decay is used
to improve the signal to noise. Events were additionally required to
have a least one of the two highest $E_T$ jets identified as being
consistent with a $b$ hadron jet as determined by a high purity
(tight) $b$-jet identification requirement. Events are then
categorized as double--tagged if the remaining one of the two highest
$E_T$ jets satisfies a lower purity (loose) identification
requirement. Events without an additional tag are denoted
single--tagged.  For the tight requirement, the per jet identification
efficiency is 50\% with a misidentification probability of 0.5\%. The
loose requirement has an efficiency of 70\% with a misidentification
probability of 6.5\%. Tab.~\ref{t-zh-vvbb-d0-yield} shows the data
yield and prediction for the initial preselection and for the single--
and double--tagged samples and the expected Higgs boson signal for a
Higgs boson of mass $115~\GeV$.

\begin{table*}
  \begin{ruledtabular}
  \begin{tabular}{lccccccccc}
   Sample & $ZH$  & $WH$ & $W$+jets & $Z$+jets & Top & $VV$ & Multijet & Total background & Observed \\\hline
   Preselection   & 13.73 $\pm$ 1.37 & 11.64 $\pm$ 1.17 & 19\,069 & 9432 & 1216 & 1112 & 1196 & 32\,025 $\pm$ 4037 & 31\,718 \\
   Single--tagged & ~\,4.16 $\pm$ 0.42 & ~\,3.60 $\pm$  0.37 & 802 & 439 & 404 & 60 & 125 & ~1830 $\pm$ 255 & 1712 \\
   Double--tagged & ~\,4.66 $\pm$ 0.58 & ~\,4.00 $\pm$  0.50 & 191 & 124 & 199 & 24 & $<8$ & ~538 $\pm$ 87 & 514 \\
  \end{tabular}
  \end{ruledtabular}
  \caption{The predicted and observed yields for the D0 $VH\to\MET\bbbar$
  analysis for the untagged control sample and the single-- and double--tagged 
  analysis samples after applying the multijet BDT selection.\label{t-zh-vvbb-d0-yield}}
\end{table*}

The backgrounds from $\ttbar$, $W$+jets, $Z$+jets and diboson
processes are estimated using simulated events. These results are
validated using the EW enriched control sample. The shape of the multijet
background is taken from the multijet modeling sample, and the normalization
is determined by forcing the number of multijet events plus the number of SM
predicted background events to equal the data yield in the
preselection sample. This procedure is validated by comparing the
prediction with the multijet enriched sample.

\begin{figure}
  \includegraphics[width=0.49\textwidth]{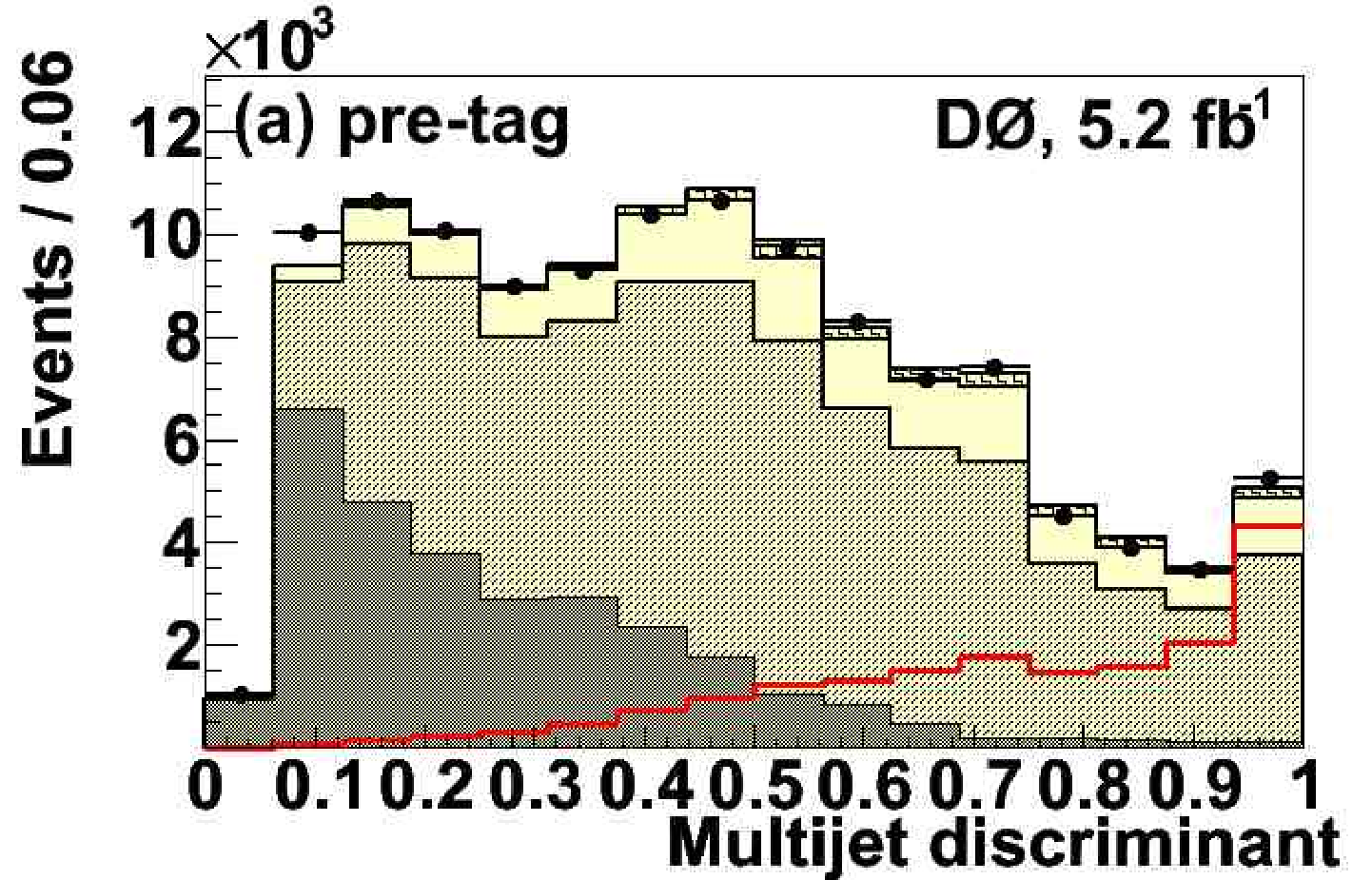}
  \includegraphics[width=0.49\textwidth]{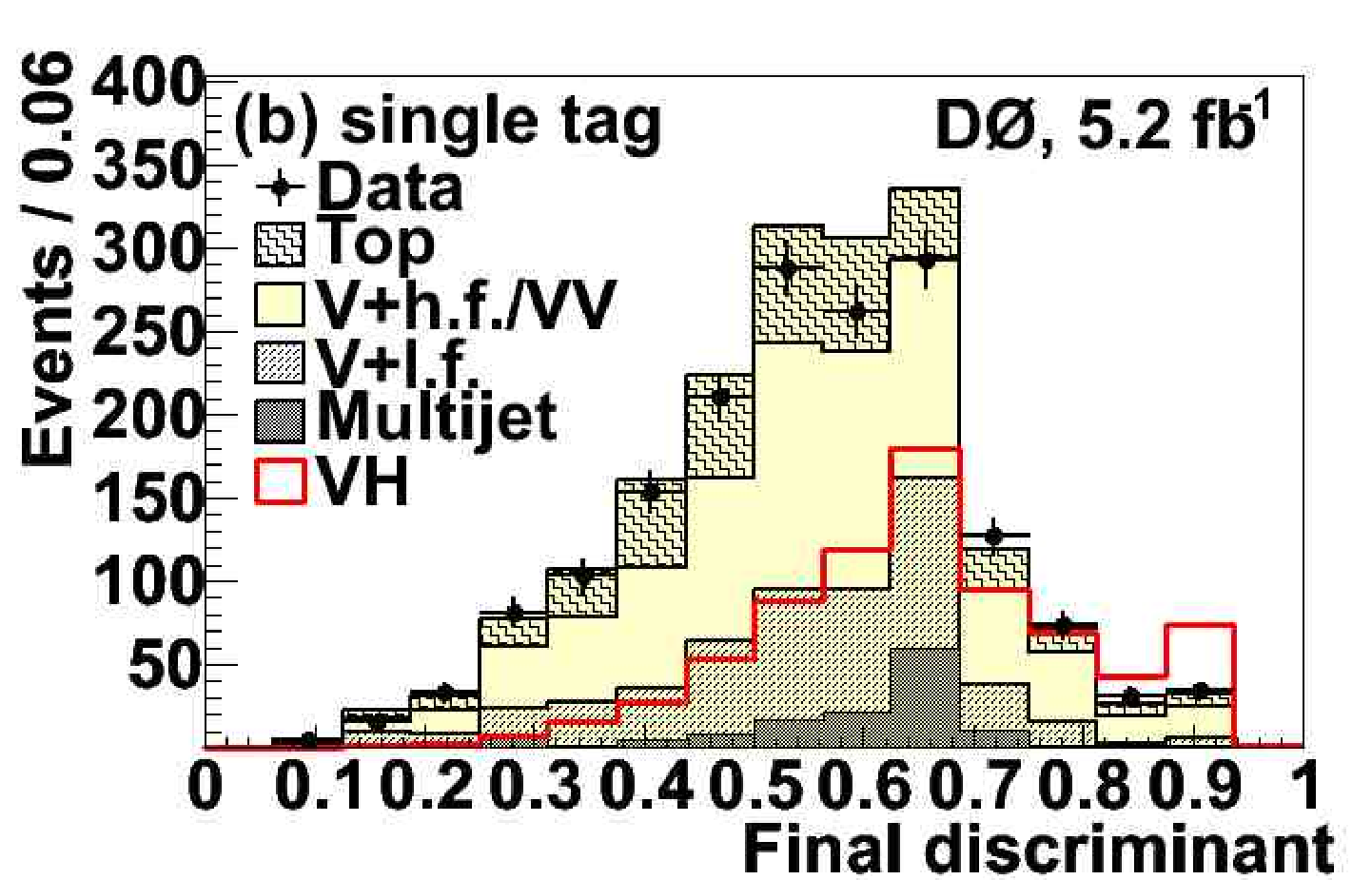}
  \includegraphics[width=0.49\textwidth]{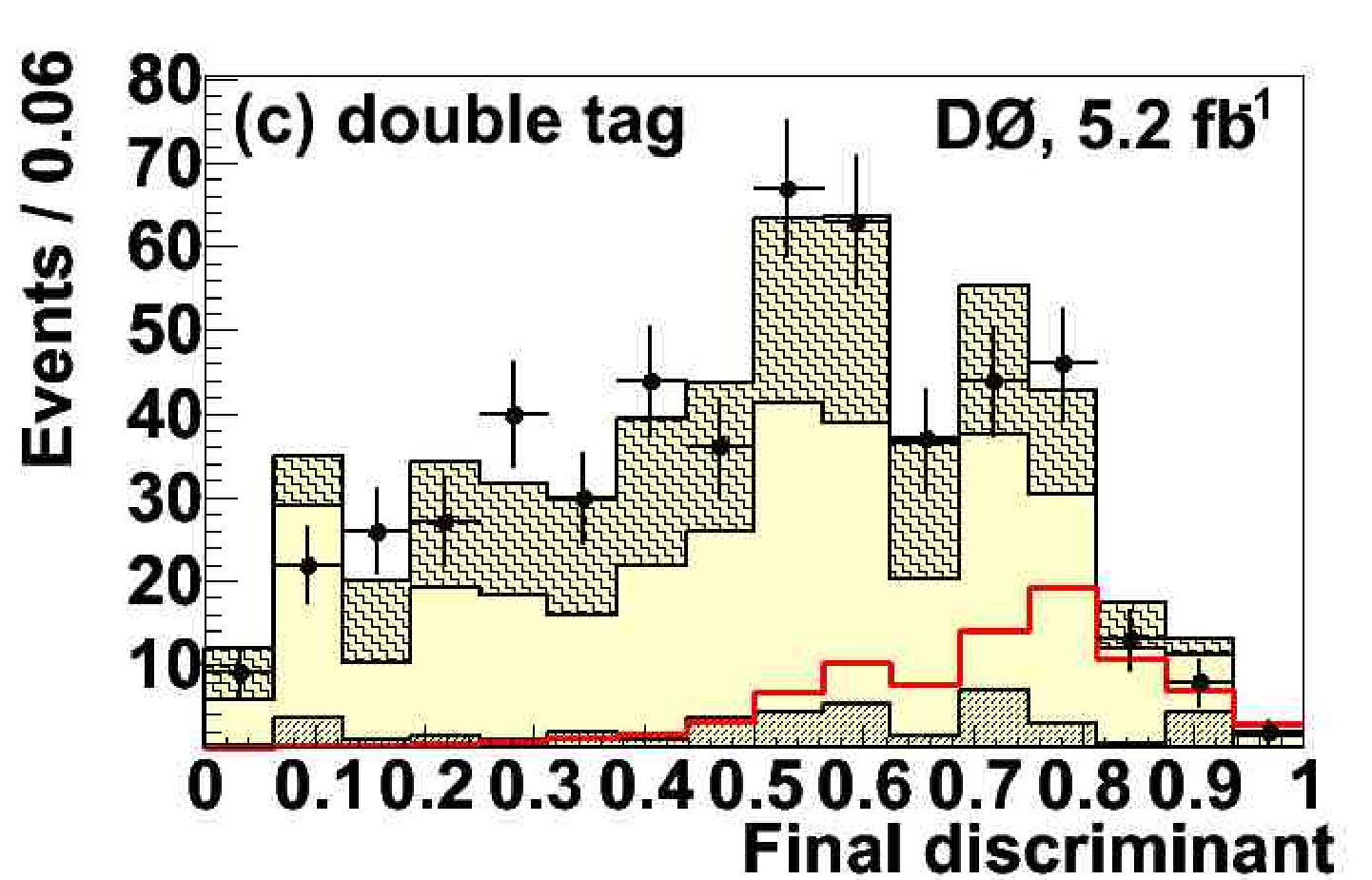}
  \caption{The outputs for the (a) MJ BDT, (b) the SM BDT for the
  single--tagged channel and (c) the SM BDT for the double--tagged channel
  for the D0 $VH\to\MET\bbbar$ analysis.\label{f-zh-vvbb-d0-nn}}
\end{figure}

\begin{figure}
  \includegraphics[width=0.49\textwidth]{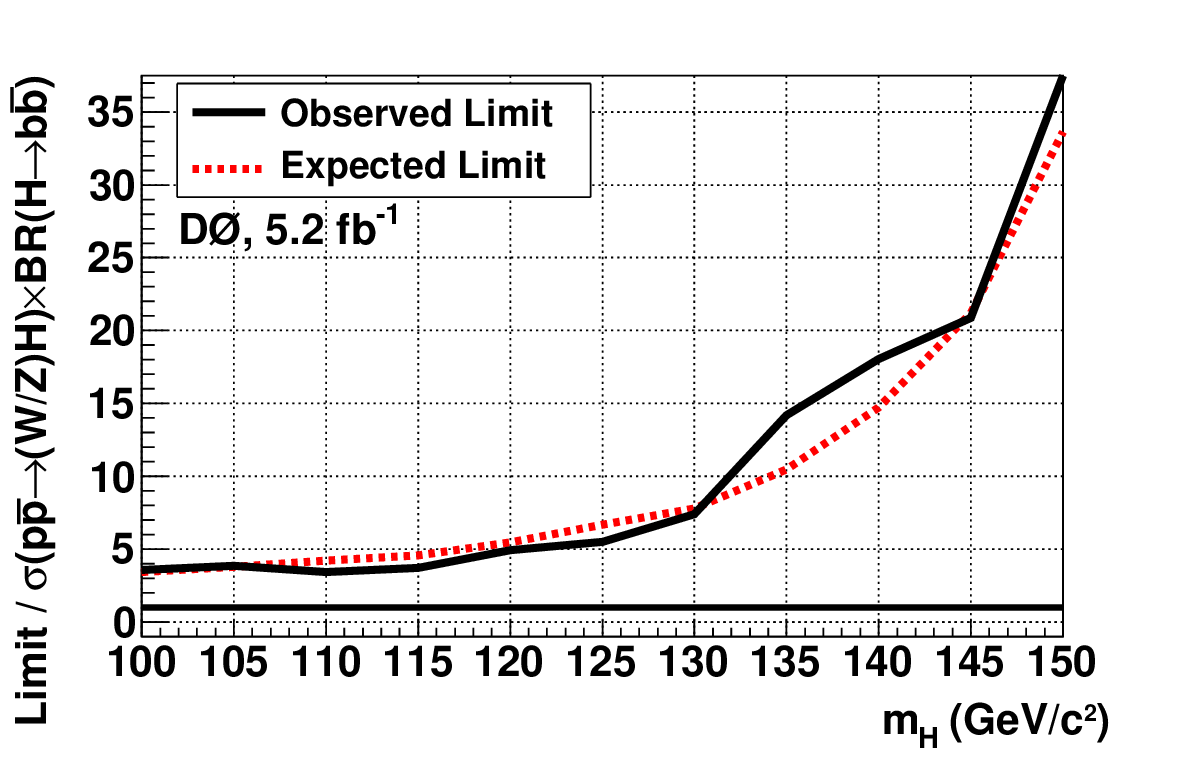}
  \caption{The expected and observed 90\% CL upper bounds on the Higgs boson
    production cross section in the D0 $VH\to\MET\bbbar$ search.\label{f-zh-vvbb-d0}}
\end{figure}

The signal-to-background separation is then further improved using
boosted decision trees (BDTs). For each $m_H$ considered, a multijet
boosted decision tree (MJ BDT) with 23 input variables is trained on
Higgs signal and multijet backgrounds. Events which have a MJ BDT output
greater than 0.6 are retained. These events are then input to a second
BDT (SM BDT) trained on the remaining backgrounds and Higgs signal
events using the same 23 variables input to the MJ BDT.
Fig.~\ref{f-zh-vvbb-d0-nn} shows the BDT outputs for the data,
predicted background and signal. The agreement between data and
prediction is good. Limits are extracted by fitting signal and
background SM BDT outputs to the data distribution using the same
modified frequentist algorithm as was used for the previously described
D0 results.  Fig.~\ref{f-zh-vvbb-d0} shows the Higgs cross section
limits from the D0 analysis.


The systematic uncertainties for the D0 result are dominated by
similar sources as for the CDF analysis. The D0 jet energy calibration
systematic is $<10$\%, and the background normalization systematic
varies between 6\% - 20\%. D0 also reports additional systematic
uncertainties from luminosity, trigger and identification efficiencies
and $b$-tagging efficiency. These sources have an uncertainty of
roughly 5\% for each.

\subsubsection{$\tau+$jets and $\tau\tau$+jets Final States}
\label{sec:higgs_direct_taus}

The searches described earlier for the processes $WH\to\ell\nu\bbbar$
and $ZH\to\ell\ell\bbbar$ required $\ell = e,\,\mu$, but corresponding
channels exist with $\ell = \tau$ leptons in the final state. In
addition, B$(H\to\tau\tau) \approx 7\%$ in the SM, the most important
mode after $H\to\bbbar$.  D0\cite{b:d0-taus} has published searches
for Higgs production in the modes $VH\to\tau\bbbar \to \tau_h\nu$ and
$VH\to\tau\tau jj \to \mu\nu\nu_\tau \tau_h jj$, $\ppbar\to VVjj \to
Hjj \to\tau\tau jj \to \mu\nu\nu_\tau \tau_h jj$, and $gg\to H
jj\to\tau\tau jj \to \mu\nu\nu_\tau \tau_h jj$ using a data sample
corresponding to 1.0~$fb^{-1}$. Here $\tau_h$ represents a $\tau$
decay to hadrons.

The primary signal contribution to the $\tau\nu\nu\bbbar$ final state
comes from $WH\to\tau\nu\bbbar$, but roughly 7\% of the final signal
acceptance comes from $ZH\to\tau\tau\bbbar$ in which one $\tau$ is not
identified. The identified $\tau$ is required to decay to
hadrons~\cite{Abazov:2009cj}. Candidate events are initially selected
using a trigger requiring a high $p_T$ jet and large $\MET$. The
offline selection requires a hadronic $\tau$ decay, with a minimum
transverse energy from 12 $\GeVc$ to 20 $\GeVc$, with the range
depending in which of two decay topologies the $\tau$ lepton is
reconstructed. The topologies correspond roughly to single
charged-hadron $\tau$ lepton decays (type 1) and single charged-hadron with 
multiple neutral-hadron decays (type 2). In addition to the $\tau$ lepton
candidate, the basic offline selection requires $\MET>30$ GeV, the
presence of at least two candidate $b$-jets with $p_T>20\ \GeVc$ and
$|\eta|<2.5$, $\le 3$ jets with $p_T\ \GeVc$, and a veto of events
with a high-$p_T$ electron or muon to keep orthogonality with the
other searches. Events are also required to pass
$\Delta\phi(\MET,\TET)<\pi/2$, $H_T<200$~GeV, $50<M_{JJ}<200\ \GeV$,
and $\Delta\phi(\tau,\MET) < 0.02(\pi-2)(\MET-30)+2$. In addition, for
events with type 2 tau decay, the transverse mass of the $\vec\tau$
and $\vec{\MET}$ must satisfy $M_T<80\ \GeV$.

The $\tau\tau jj$ final state search has signal contributions from a
variety of processes: $ZH\to\tau\tau\bbbar$, $ZH\to \qqbar\bbbar$ (in
which $H\to\tau\tau$, VBF $\ppbar\to VVjj \to Hjj \to\tau\tau jj \to
\mu\nu\nu_\tau \tau_h jj$, and gluon fusion $gg\to H jj\to\tau\tau jj
\to \mu\nu\nu_\tau\tau_h jj$. Unlike other low mass channels, some of
these processes have no $b$--quarks in the decay chain. Consequently,
no $b$-jet identification is required for this search. One of the
$\tau$ leptons is required to decay to hadrons, and the other $\tau$ lepton is
required to decay to a muon. Here three topologies are used. Two of
these are the same as in the $\tau\nu\bbbar$ search, and the third
corresponds roughly to $\tau$ lepton decays to three charged hadrons. Candidate
events are initially selected by requiring at least one
trigger from a set of single muon and muon plus jet triggers. The
initial offline selection requires a muon with $p_T>12\ \GeVc$ and
$\eta<2.0$, a hadronic $\tau$ lepton decay candidate with a minimum transverse
energy from 15~GeV to~20 GeV (again depending on topology) and at
least two jets with $p_T>15\ \GeVc$ and $\eta|<2.5$.  No $b$-jet
identification is used.  Signal and background separation is then
improved using a set of neutral networks. A separate neutral network
is trained for each of the four signal sources ($WH$, $ZH\
(H\to\bbbar)$, $ZH\ (H\to\tau\tau)$ and $VBF$), paired with each of
the main backgrounds ($W+$jets, $Z+$jets, $\ttbar$ and multijet). In
addition separate networks are trained for low mass (105, 115 and 125
$\GeV$) and high mass (135, 145 $\GeV$) giving a total of 32 networks.
Each network uses six or seven input variables chosen such that each
one individually improves the $S/\sqrt{B}$ when it is added as the
last variable.  The same choice of input variables is made at all
masses for each signal and background pairing.

All backgrounds except that from multijet events in which jets are
misidentified as $\tau$ leptons are estimated using simulated events processed
through a detailed detector simulation with corrections based on
comparison of data and simulated control samples. The absence of 
electrons and muons from direct $W$ or $Z$ decay results in a
significantly higher multijet background in this search than in most. The multijet
background is estimated from using a control data sample enriched in
$\tau$-like hadronic jet events.  Systematic uncertainty sources
include trigger efficiency, lepton and jet identification efficiencies,
jet energy calibration and production cross section uncertainties for
backgrounds determined using simulated events.

As for the other analyses, systematic uncertainties are divided into
those which affect only normalization and those which affect the
shape. Sources affecting the normalization include: the integrated
luminosity, the trigger efficiency, muon identification efficiency,
$\tau$ lepton identification efficiency, jet identification
efficiency, the $\tau$ lepton
energy calibration and background cross sections. The systematic
uncertainties which affect the shape for the $\tau\nu\bbbar$ analysis
are the jet energy resolutions, the jet energy calibration and the
$b$--tagging efficiencies. For the $\tau\tau jj$ analysis only the multijet
background had a shape dependence. The systematic uncertainties,
expressed as a fraction of the related source, range between 3\% and
30\%. For example, the $W$+hf cross section systematic is 30\% of the
$W$+hf background, but because the background is roughly 50\% of the
total background in the $\tau\nu\bbbar$ channel, the effective
systematic is roughly 15\%.

\begin{figure}
  \includegraphics[width=0.49\textwidth]{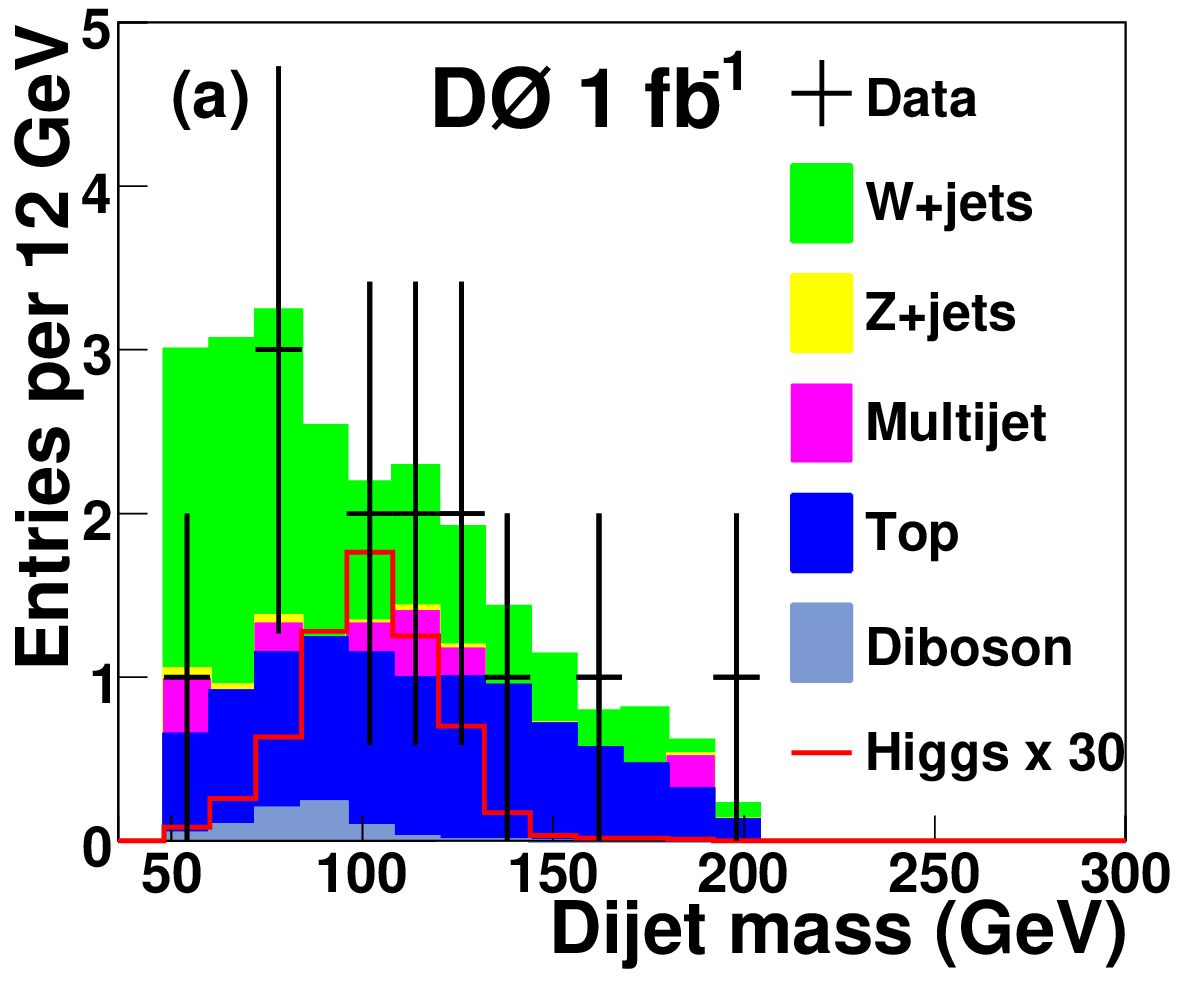}
  \includegraphics[width=0.49\textwidth]{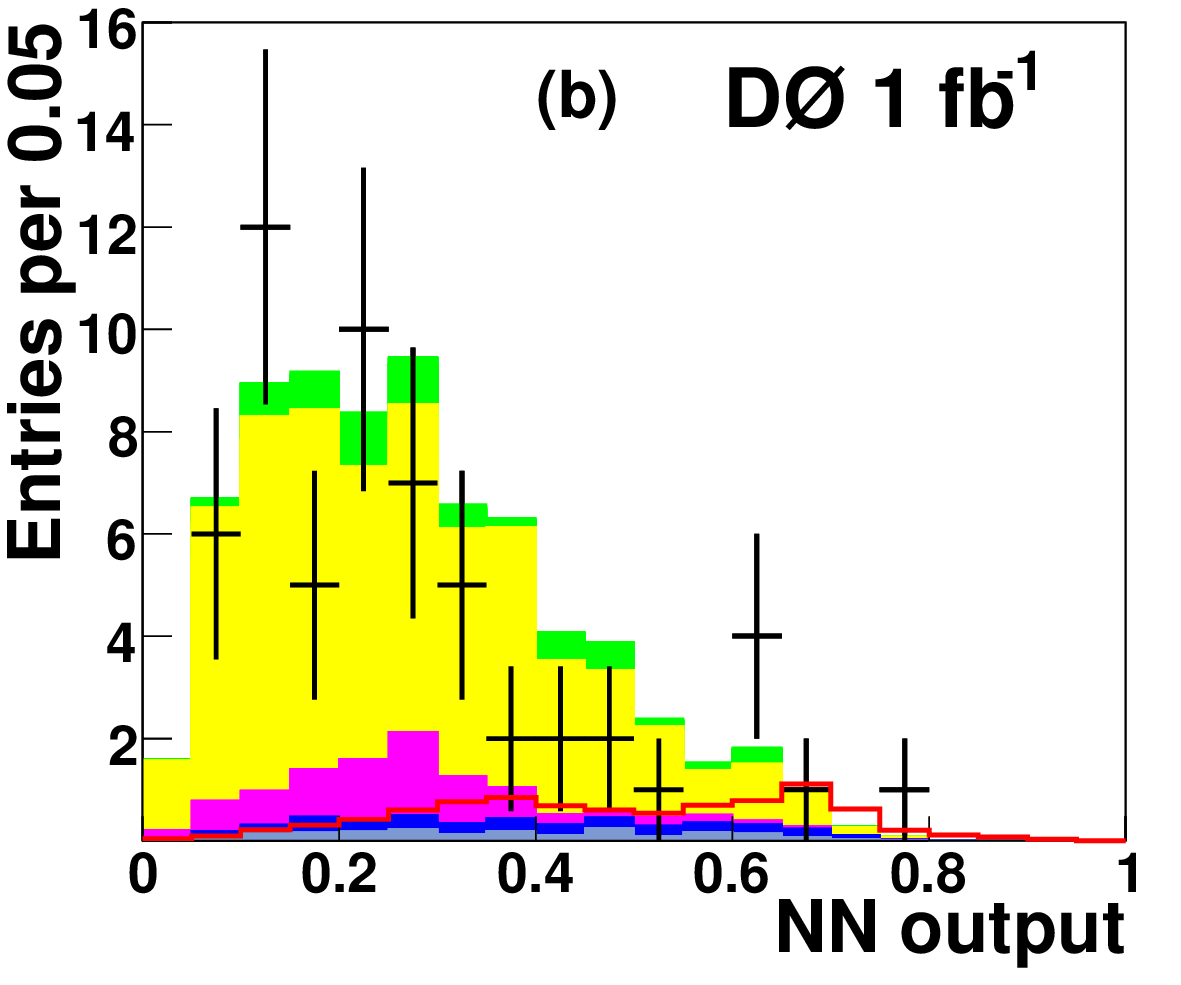}
  \caption{The dijet invariant mass distribution for the D0 $VH\to\tau\nu\bbbar$
    search, and the NN output for the $VH\to\tau\tau jj$ search.\label{f-tau-dists}}
\end{figure}

Predicted background and signal levels and observed yields for the two
searches are shown in Tab.~\ref{t-tau-evts} assuming a signal mass of
$M_H = 115\ \GeV$. The distributions or $M_{JJ}$ and the output from
the $Z+$jets NN $NN_Z$ are used as inputs to the limit setting program
for the $\tau\nu\bbbar$ and $\tau\tau jj$ searches respectively. The
distributions are shown in Fig.~\ref{f-tau-dists}, and the
resulting limits are shown in Tab.~\ref{t-tau-limits}. 

\begin{table}
\begin{ruledtabular}
  \begin{tabular}{ccc} 
           & \multicolumn{2}{c}{Search Channel} \\ 
    Source & $\tau\nu\bbbar$       & $\tau\tau jj$ \\ \hline
 $W+$lf    &  $0.5\pm0.0$          &   $5.1\pm0.3$  \\
 $W+$hf    &  $10.9\pm0.3$         &   $0.9\pm0.1$  \\
 $Z+$lf    &  $<0.2$               &   $43.8\pm0.6$  \\
 $Z+$hf    &  $0.4\pm0.0$          &   $10.1\pm0.7$  \\
 $\ttbar$  &  $9.5\pm0.1$          &   $2.8\pm0.0$  \\
 Diboson   &  $0.7\pm0.0$          &   $2.1\pm0.2$  \\
 Multijet  &  $1.3\pm0.1$          &   $6.5\pm2.8$  \\  \hline
 Total     &  $23.3\pm0.4$         &   $71.2\pm3.0$  \\
 Data      &  $13$                 &   $58$  \\  \hline
 Signal    &  $0.216$              &   $0.293$  \\ \hline
  \end{tabular}
  \caption{Expected and observed yields for the D0 $VH\to\tau\nu\bbbar$ and
    $VH\to\tau\tau jj$ searches.  Only statistical uncertainties are shown.
    \label{t-tau-evts}}
\end{ruledtabular}
\end{table}

\begin{table}
\begin{ruledtabular}
  \begin{tabular}{ccccccc}
                &  \multicolumn{2}{c}{$\tau\nu\bbbar$} & 
                    \multicolumn{2}{c}{$\tau\tau jj$} & 
                     \multicolumn{2}{c}{Combined} \\
    $M_H$ ($\GeV$) & exp. & obs. & exp. & obs. & exp. & obs. \\ \hline
    105            & 33   &  27  &  39  &  36  &  24  &  20  \\
    115            & 42   &  35  &  43  &  47  &  28  &  29  \\
    125            & 62   &  60  &  60  &  65  &  40  &  44  \\
    135            & 105  &  106 &  87  &  61  &  63  &  50  \\
%   145            & 226  &  211 & 158  &  95  & 120  &  82  \\
  \end{tabular}
  \caption{Expected and observed 95\% CL signal cross--section upper limits for the
     D0 $VH\to\tau\nu\bbbar$ and $VH\to\tau\tau jj$ searches. The limits are
     given as a ratio of the cross--section limit to the SM
     cross--section prediction.\label{t-tau-limits}}
\end{ruledtabular}
\end{table}

\subsubsection{$H\rightarrow WW^{*}$}
\label{sec:higgs_direct_HWW}

For $m_H \simge 135~\GeV$, the dominant Higgs decay is to $W$ boson
pairs. Below the $W$ pair production threshold, one of the $W$ bosons
will be off-shell which leads to lower $p_T$ leptons from the $W^{*}$
decay. The most sensitive high mass Higgs channel is $gg\rightarrow
H\rightarrow WW^{(*)}\rightarrow \ell\nu\ell\nu$, where the two
charged leptons in the final state of are of opposite charge. In
addition to the dominant $ggH$ process, other Higgs
production processes that contribute at the Tevatron are VBF and $VH$
where $V \equiv (W,Z$).

Early Tevatron Run II searches for a high mass Higgs boson
\cite{Abulencia:2006aj},\cite{Abazov:2005un} focused mainly on the $ggH$
process and exploiting the angular correlation between final state
charged leptons due to the scalar (spin-0) nature of the SM Higgs. In
a more recent analysis \cite{Aaltonen:2008ec}, CDF used a NN technique
to search for a Higgs boson signal in dilepton + $\MET$ events having
either zero or one reconstructed jet. Several kinematic variables,
including the results of matrix element calculations that combined
charged lepton and $\MET$ information, were used as inputs to the NN.

The CDF and D0 searches for $\HWW$ in \cite{PhysRevLett.104.061803}
and \cite{PhysRevLett.104.061804}, respectively, significantly increase
their sensitivities compared to previous results through the use of
additional data, topologies arising from VBF and $VH$ channels, and
analysis improvements such as increased charged lepton acceptance (D0). 

\begin{figure*}
  \begin{tabular}{lcr}
  \includegraphics[width=0.675\columnwidth]{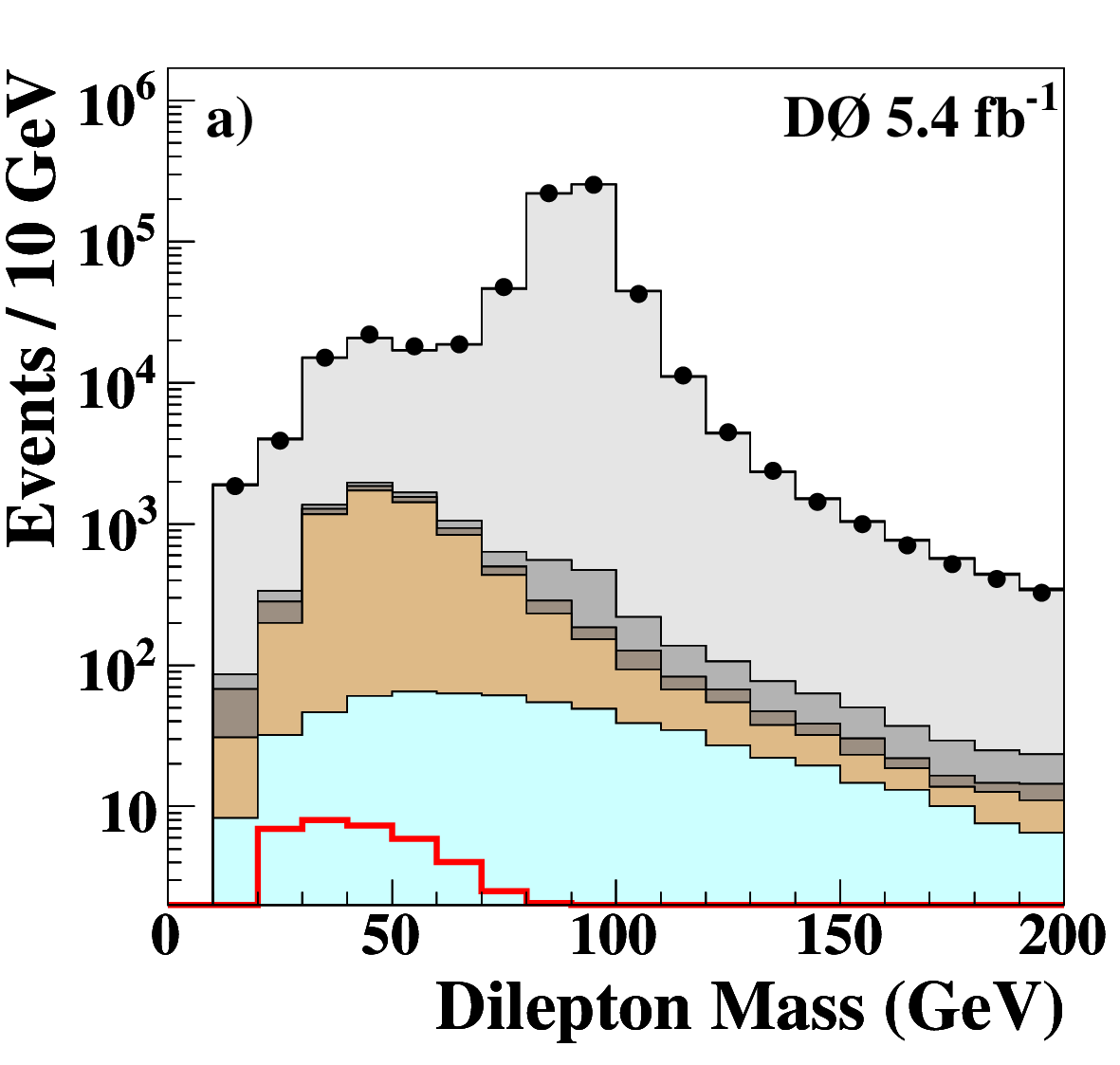} &
  \includegraphics[width=0.675\columnwidth]{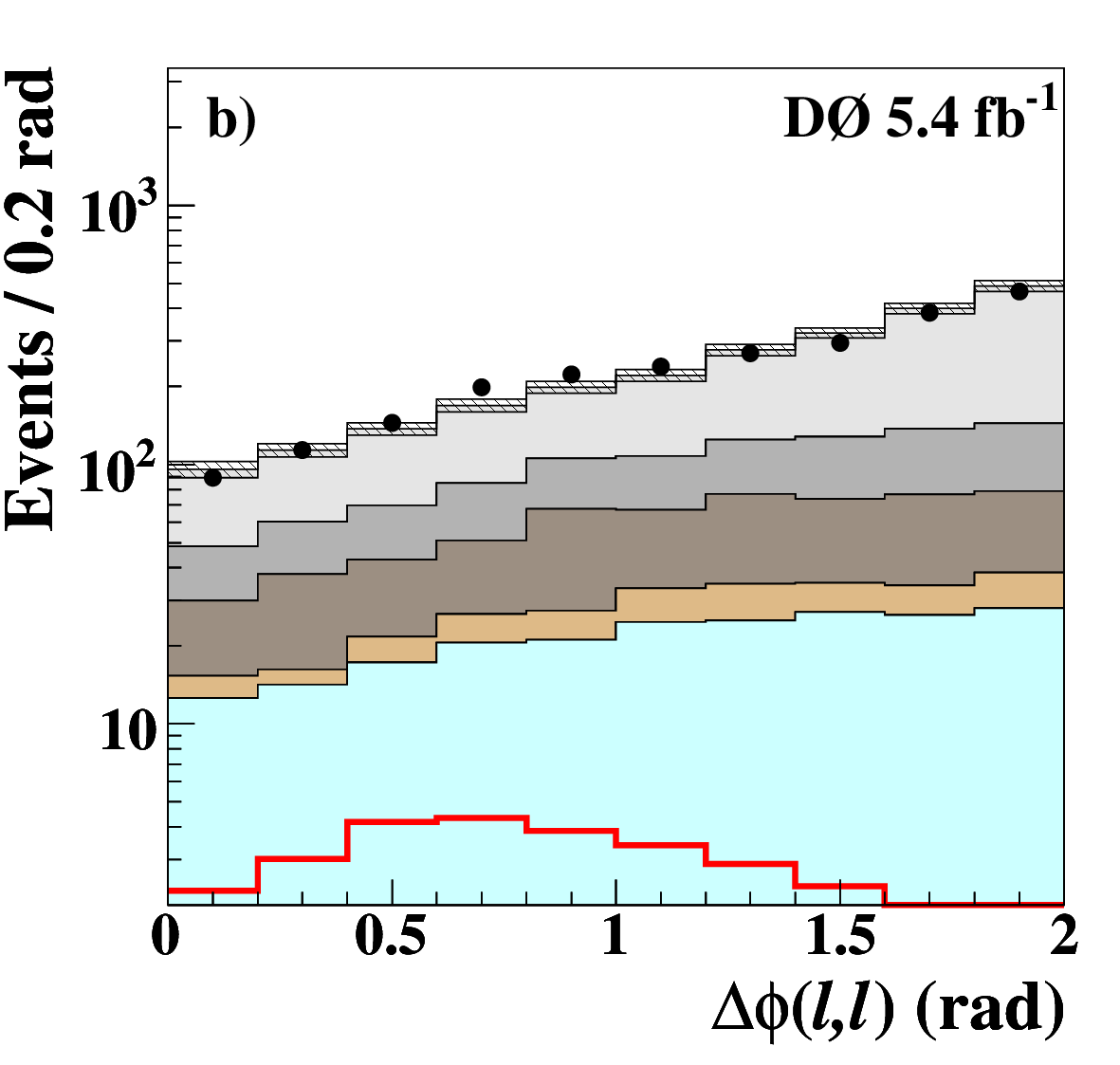} &
  \includegraphics[width=0.675\columnwidth]{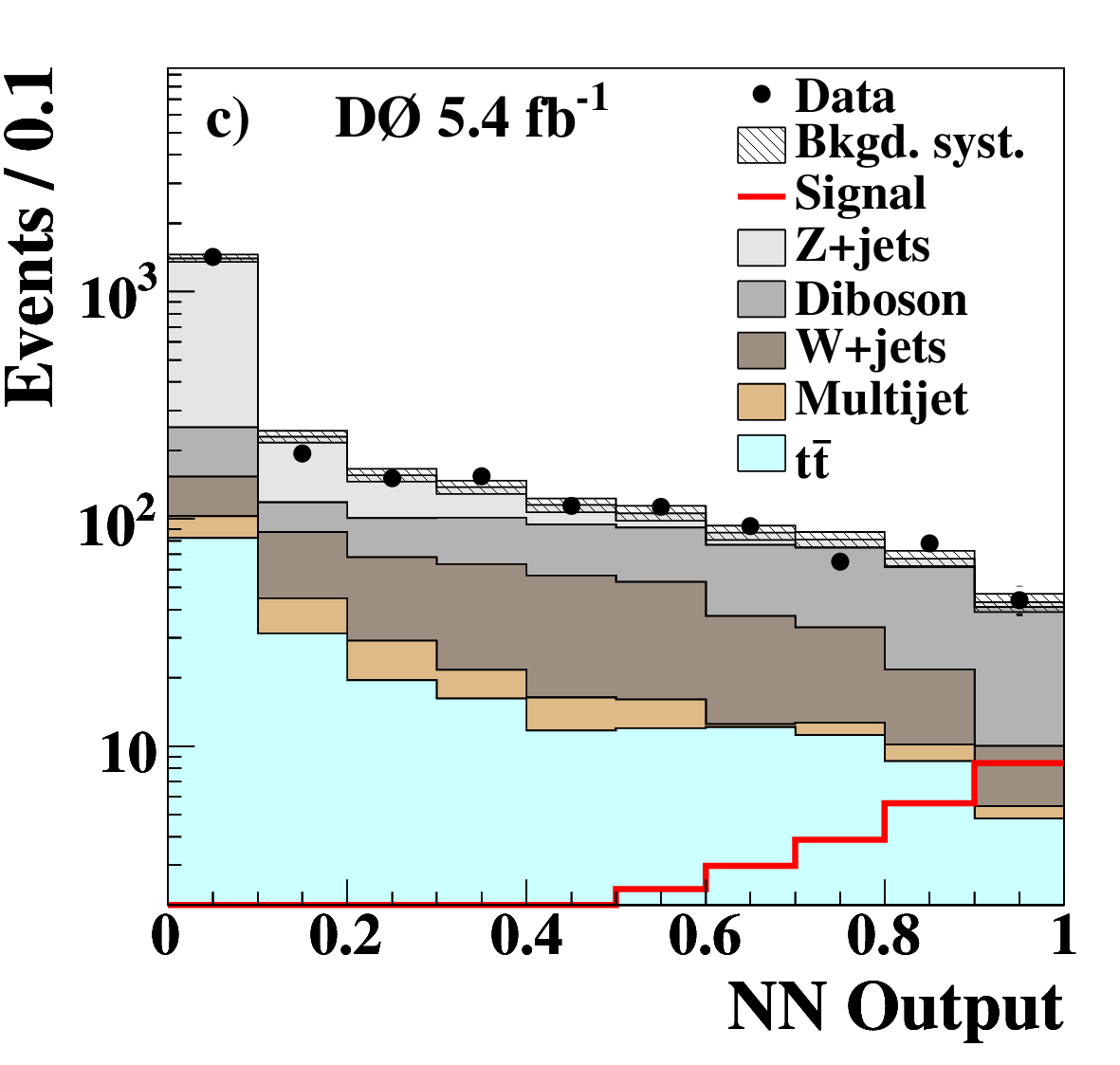} \\
  \end{tabular}
  \caption{For the D0 $\HWW$ analysis \cite{PhysRevLett.104.061804}, the (a)
    dilepton invariant mass after preselection; (b)
    $\Delta\phi(\ell,\ell)$ angle after final selection; and (c)
    neural network output after final selection. The signal is shown 
    for $m_H$=165~GeV. The systematic uncertainty is shown after the fit.
  \label{fig:d0-hww-datamc}}
\end{figure*}

In the D0 analysis \cite{PhysRevLett.104.061804}, an integrated
luminosity of 5.4 $\fb$ is used to search for $\HWW$ in events with
two oppositely-charged leptons $e^+e^-$, $e^{\pm} \mu^{\mp}$, or
$\mu^+\mu^-$. Electrons are required to have $|\eta|<2.5$ ($<2.0$ in
the $e^+e^-$ channel) and $E_T^{e} >15$ GeV. Muons are required to have
$|\eta|<2.0$ and $p_T^{\mu}>10~\GeVc$ (in the $\mu^+\mu^-$ channel, one
of the two muons is required to have $p_T^{\mu} >20~\GeVc$). In
addition, the dilepton invariant mass is required to exceed
$15~\GeV$. Reconstructed jets are required to have $E_T^{\rm jet}
>15$~GeV and $|\eta| < 2.4$, however no jet-based event selection is
applied since the number of jets in the event is used as input to an
NN to help discriminate signal from background. The dilepton
invariant mass after this ``preselection'' is shown in
Fig.~\ref{fig:d0-hww-datamc}(a).

Additional cuts are made to suppress $Z/\gamma^*$ production. These
include requiring $\MET>20$ GeV ($>25$~GeV in the $\mu^+\mu^-$
channel), high quality $\MET$ measurement, a minimum $W$ transverse
mass to be $>20$~GeV ($>30$~GeV in the $e^+e^-$ channel), and
azimuthal opening angle between the two leptons $\Delta
\phi(\ell,\ell)<2.0$~rad.

To improve the separation between signal and background, a NN
optimized for several $m_H$ values between $115~\GeV$ and $200~\GeV$ 
is used in each of the three channels. Several discriminant variables
are used as inputs to the NN: the transverse momenta of the leptons,
a variable indicating the quality of the leptons' identification, the
transverse momentum and invariant mass of the dilepton system, minimum
$W$ transverse mass, $\MET$, $\MET$ quality, $\Delta \phi(\ell,\ell)$,
$\Delta \phi(\ell_1,\MET)$, $\Delta \phi(\ell_2,\MET)$, the number of
identified jets, and the scalar sum of the transverse momenta of the
jets.

Fig.~\ref{fig:d0-hww-datamc} shows the agreement between data
and Monte Carlo simulation after final selection for (b) the
$\Delta\phi(\ell,\ell)$ angle and (c) the neural network output. The
expected and observed event yields are shown in
Tab.~\ref{tbl:d0-hww-yields}.

\begin{table*}
\begin{ruledtabular}
\begin{tabular}{c|r@{$\, \,$}lr@{$\,\pm \,$}l|r@{$\, \,$}lr@{$\,\pm \,$}l|r@{$\, \,$}lr@{$\,\pm \,$}l}
                & \multicolumn{4}{c|}{$e^{\pm}\mu^{\mp}$}&
                \multicolumn{4}{c|}{$e^+e^-$}          &
                \multicolumn{4}{c}{$\mu^+\mu^-$} \\                & \multicolumn{2}{c}{Preselection} & \multicolumn{2}{c|}{ Final selection}& \multicolumn{2}{c}{Preselection}    & \multicolumn{2}{c|}{Final selection}          & \multicolumn{2}{c}{Preselection} & \multicolumn{2}{c}{Final selection} \\
\hline
$Z/\gamma^*\to e^+e^-$       & \multicolumn{2}{c}{120} & \multicolumn{2}{c|}{$<0.1$} & \multicolumn{2}{c}{274886}& 158 & 13 &     \multicolumn{2}{c}{$-$}         &   \multicolumn{2}{c}{$-$}  \\
$Z/\gamma^*\to \mu^+\mu^-$   & \multicolumn{2}{c}{89}     & 4.3&0.3    &     \multicolumn{2}{c}{$-$}            &    \multicolumn{2}{c|}{$-$
}            & \multicolumn{2}{c}{373582}& 1247& 37\\
$Z/\gamma^*\to \tau^+\tau^-$ & \multicolumn{2}{c}{3871}       & 7.1&0.5    & \multicolumn{2}{c}{1441}       & 0.7 &0.1 & \multicolumn{2}{c}{
2659}   & 12.0 & 0.7\\
$\ttbar$        &  \multicolumn{2}{c}{312}    & 93.8 & 8.3   & \multicolumn{2}{c}{159}     & 47.0&4.4     & \multicolumn{2}{c}{184}    & 74.
6 & 6.8 \\
$W+{\rm jets}/\gamma$        &  \multicolumn{2}{c}{267}   & 112 &9  & \multicolumn{2}{c}{308}    & 122&11   & \multicolumn{2}{c}{236} & 91.5
 & 6.5 \\
$WW$            & \multicolumn{2}{c}{455}    & 165&6  & \multicolumn{2}{c}{202}    & 73.9 &6.4    & \multicolumn{2}{c}{272} & 107 & 9 \\
$WZ$            & \multicolumn{2}{c}{23.6} & 7.6&0.2 & \multicolumn{2}{c}{137} & 11.5 &1.0 & \multicolumn{2}{c}{171}    & 21.5&2.0 \\
$ZZ$            & \multicolumn{2}{c}{5.4}  & 0.6&0.1 & \multicolumn{2}{c}{117} & 9.3 &0.9  & \multicolumn{2}{c}{147} & 18.0&1.8 \\
Multijet       & \multicolumn{2}{c}{430}  & 6.4&2.5    & \multicolumn{2}{c}{1370}       & 1.0&0.1    & \multicolumn{2}{c}{408}      & 53.8&10.3 \\
\hline Signal
($m_H=165$~GeV) & \multicolumn{2}{c}{18.8} & 13.5 & 1.5 & \multicolumn{2}{c}{11.2}    & 7.2 &0.8    & \multicolumn{2}{c}{12.7}   & 9.0 & 1.0
 \\
\hline
Total background& \multicolumn{2}{c}{5573}   & 397 & 14 & \multicolumn{2}{c}{278620}    & 423 &19  & \multicolumn{2}{c}{377659}   & 1625 & 41 \\
\hline\hline
Data            &  \multicolumn{2}{c}{5566}        &  \multicolumn{2}{c|}{390}        &  \multicolumn{2}{c}{278277}          & \multicolumn{
2}{c|}{421}           &  \multicolumn{2}{c}{384083}        & \multicolumn{2}{c}{1613} \\
\end{tabular}
\end{ruledtabular}
\caption{Expected and observed event yields in each channel after
  preselection and at the final selection for the D0 $\HWW$ analysis
  \cite{PhysRevLett.104.061804}. The systematic uncertainty after
  fitting is shown for all samples at final selection.} 
\label{tbl:d0-hww-yields}
\end{table*}

\begin{figure}
  \includegraphics[width=0.45\textwidth]{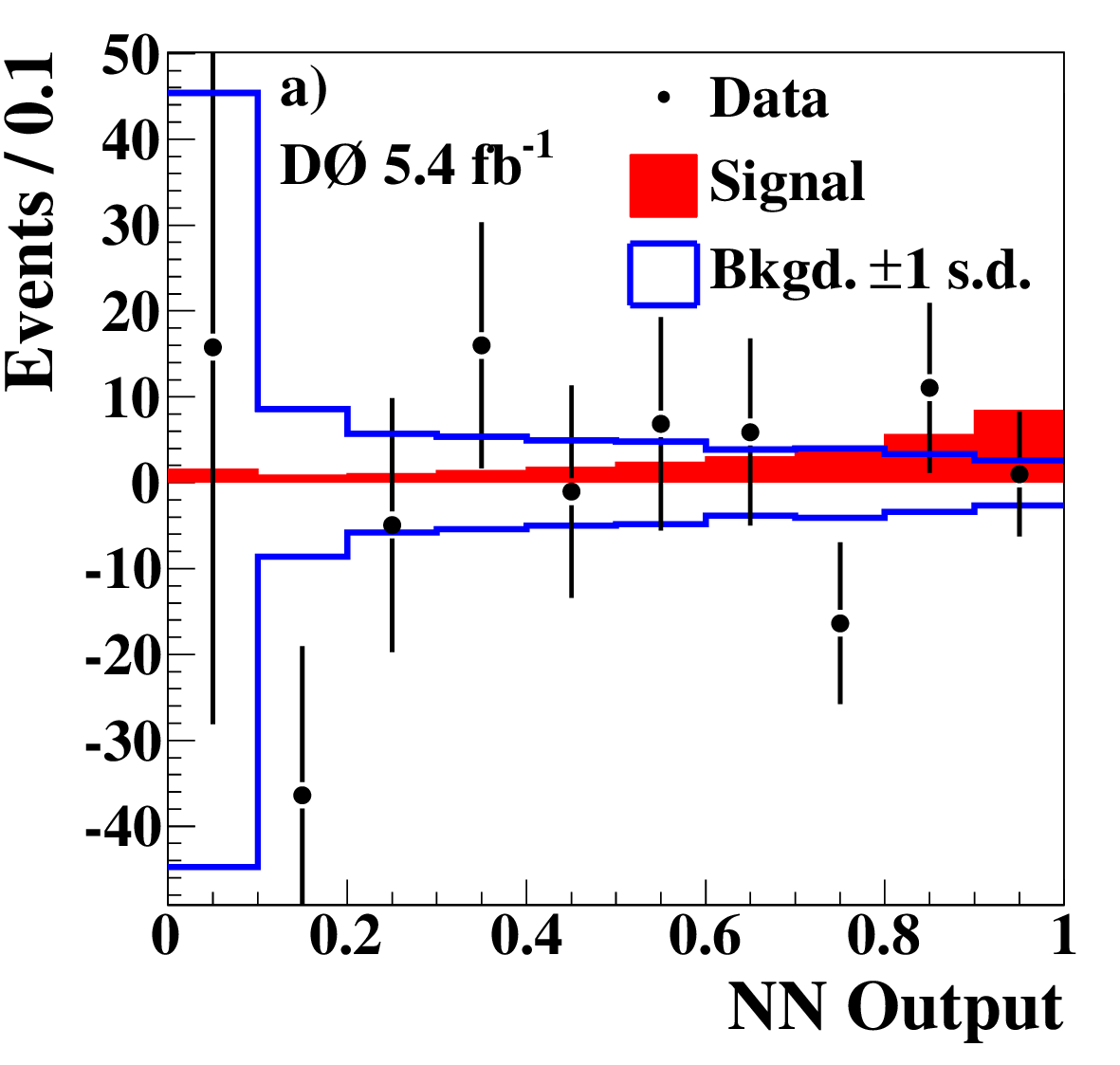}
  \includegraphics[width=0.45\textwidth]{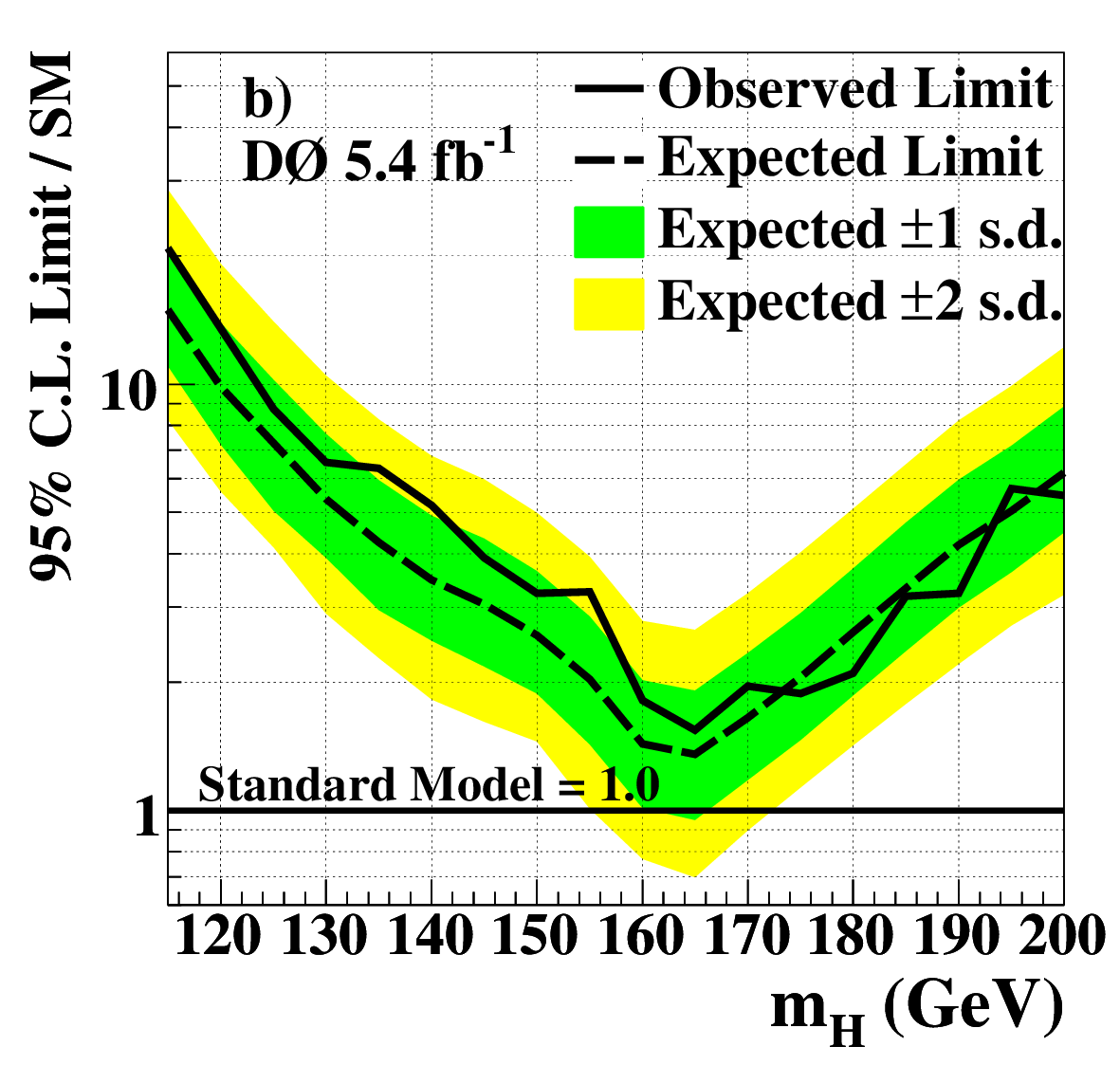}
  \caption{For the D0 $\HWW$ analysis \cite{PhysRevLett.104.061804},
    the (a) data after subtracting the fitted background (points) and
    SM signal expectation (filled histogram) as a function of the NN
    output for $m_H=165~\GeV$. Also shown is the $\pm1$ standard
    deviation (s.d.) band on the total background after fitting. (b)
    Upper limit on Higgs boson production cross section at 95\% CL
    expressed as a ratio to the SM cross section. The one and two
    s.d.~bands around the curve corresponding to the background-only
    hypothesis are also shown.
    \label{fig:d0-hww-results}}
\end{figure}

No significant excess of signal-like events is observed for any test
value of $m_H$ after the final selection. The NN output distributions
are used to set upper limits on the SM Higgs boson production cross
section. Fig.~\ref{fig:d0-hww-results}(a) shows a comparison of the NN
distribution between background-subtracted data and the expected
signal for $m_H=165~\GeV$ hypothesis. Fig.~\ref{fig:d0-hww-results}(b)
shows the expected and observed upper limits as a ratio to the
expected SM cross section. Assuming $m_H = 165~\GeV$, the observed
(expected) upper limit at 95\% CL on Higgs boson production is a
factor of 1.55 (1.36) times the SM cross section.

In the CDF analysis \cite{PhysRevLett.104.061803}, an integrated
luminosity of 4.8 $\fb$ is used to search for $\HWW$ in events with
either an opposite sign (OS) or same-sign (SS) charged lepton
pair. This is an inclusive search that expands the signal acceptance
by 50\% for $m_H = 160 ~\GeV$ compared to searching for only the
$ggH$ production process as published previously by
CDF~\cite{Aaltonen:2008ec}.

The CDF analysis uses physics objects identified as jets, electrons, and
muons as well as the $\MET$ in events. The search is based on the
requirement that events contain two charged leptons resulting from the
decays of the final-state vector bosons which are OS except in the
case of the $VH$ channel where they can be SS. At least one charged
lepton is required to match the lepton found in the trigger and have
$\Et(\pt) > 20$ GeV ($\GeVc$) for electrons (muons). The second
charged lepton is required to have $\Et(\pt) > 10$ GeV ($\GeVc$)
except in events with same charge leptons, where both leptons are
required to have $\Et(\pt) > 20$ GeV ($\GeVc$). Requirements on the
event $\MET$ indicative of the presence of neutrinos from $W$ boson
decay are made in opposite-charge dilepton events. Backgrounds due to
Drell-Yan and heavy flavor are suppressed by requiring that the
invariant mass of the lepton pair be greater than $16~\GeV$. The Higgs
boson signature can also involve jets of hadrons produced from the
decay of one of the vector bosons in the $VH$ process, forward quarks
in the VBF process, or from the radiation of gluons.

In order to increase the sensitivity to the various Higgs production
processes in the SM, candidate events are subdivided into six analysis
channels based on jet multiplicity, lepton categories, and lepton
charge combinations. Five of the channels have signatures with OS
leptons and the other is for SS leptons.

Discrimination of signal from background is based on NNs trained in
each analysis channel and at each of 14 hypothesized $m_H$ values in
the range $110 \le m_H \le 200~\GeV$. The NN inputs are based on
kinematic quantities selected to exploit features such as the spin
correlation between the $W$ bosons in Higgs boson decay, the presence
of large $\MET$ from the neutrinos; the transverse mass of the Higgs
boson (calculated using the leptons' four-momenta and $\MET$ vector);
and the modest total energy of the Higgs boson decay products compared
to $\ttbar$ decay. In the zero jet categories, we additionally
classify events by evaluating the observed kinematic configuration in
a likelihood ratio of the signal probability density divided by the
sum of the signal and background probability densities. These
probability densities are determined from LO matrix element
calculations of the cross sections of each process
\cite{Aaltonen:2008ec} (see also
Section~\ref{sec:gaugeBosons_dibosons_WW} for the case of SM $WW$).

\begin{figure}
\includegraphics[width=0.49\textwidth]{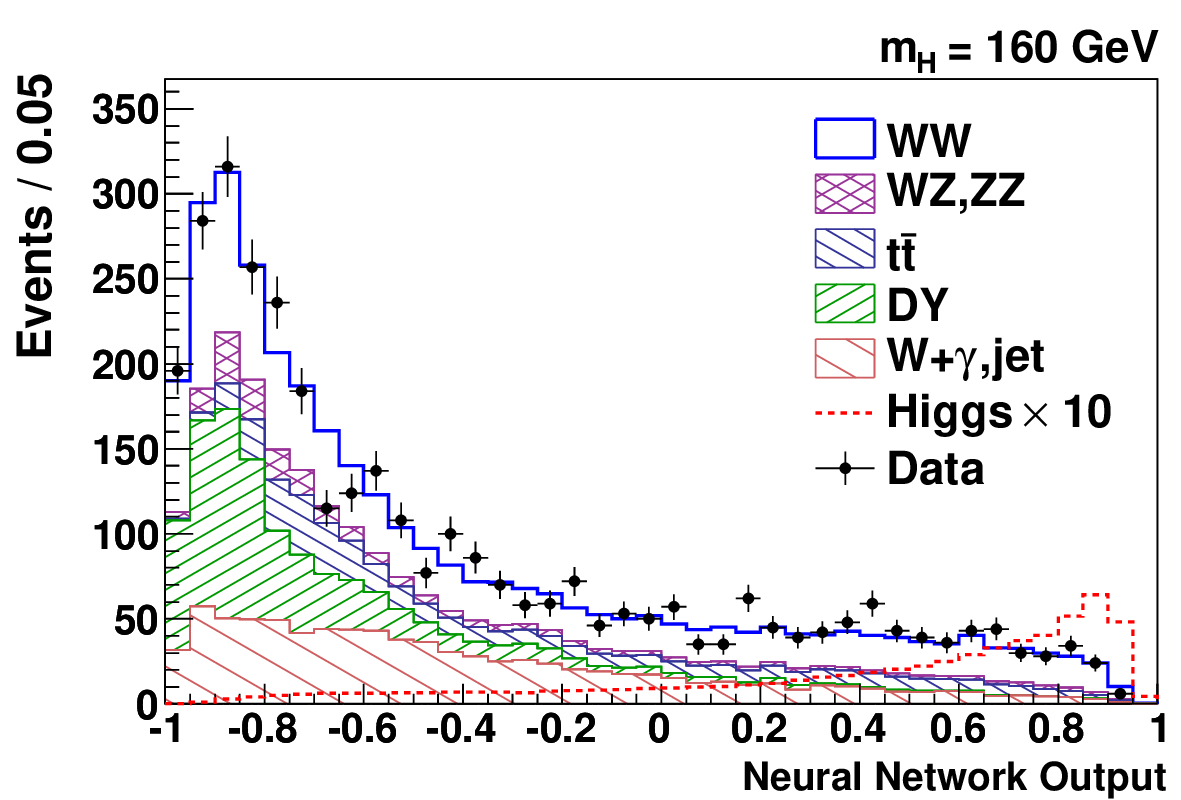}
\caption{The combined distribution of NN scores for backgrounds and a
  $M_H=160~\GeV$ Higgs boson compared to the observed data shown with
  statistical uncertainties for the CDF $\HWW$ analysis
  \cite{PhysRevLett.104.061803}. The Higgs boson distribution is
  normalized to ten times the SM expectation.}
\label{fig:cdf-hww-NN}
\end{figure}

An example NN discriminant distribution for the combination of all
categories is shown in Fig.~\ref{fig:cdf-hww-NN}, where signal and
background expectations for a $160~\GeV$ Higgs boson are compared to
the observed data.  

\begin{figure}
\includegraphics[width=0.49\textwidth]{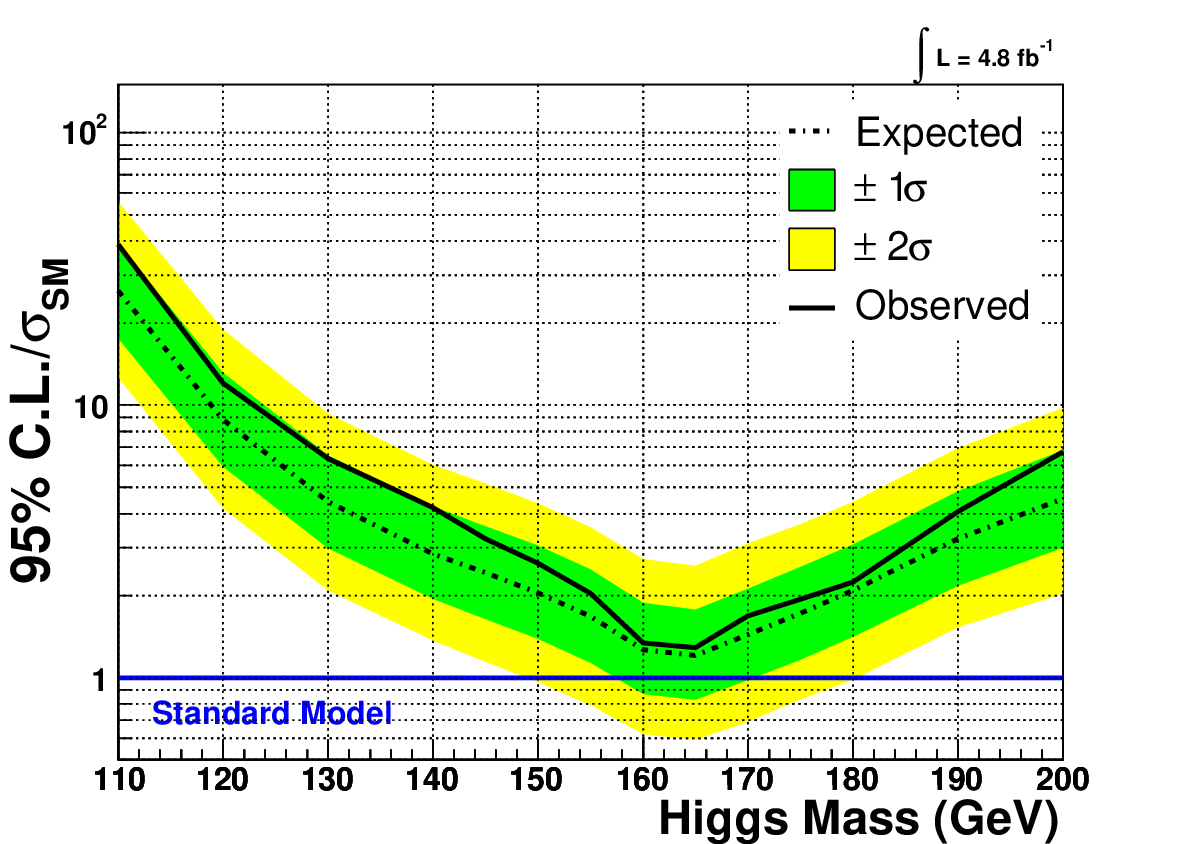}
\caption{Expected and observed upper limits at the 95\% CL on $\sigma_{H}$ 
presented as a ratio to the predicted SM values as a function of $M_H$
for the CDF $\HWW$ analysis \cite{PhysRevLett.104.061803}. The dashed
line represents the median expected limits, the green and yellow bands
the estimated one and two sigma probability bands for the distribution
of expectations, and the solid line the observed limit.}
\label{fig:cdf-hww-limits}
\end{figure}

No significant excess of events beyond SM background expectations in
the NN discriminant are observed. The 95\% confidence limits on
$\sigma_{H}$, expressed as a ratio to the expected SM rate as a
function of $m_H$, are determined from the data. The median expected
and observed upper limits on $\sigma_{H}$ as a function of $m_H$ is 
shown in Fig.~\ref{fig:cdf-hww-limits}.

The expected and observed upper limits at 95\% CL for Higgs boson
production cross section expressed as a ratio to the cross section
predicted by the SM as a function of $m_H$ for the CDF and D0 searches
for $\HWW$ in \cite{PhysRevLett.104.061803} and
\cite{PhysRevLett.104.061804}, respectively, is shown in
Tab.~\ref{tbl:hww-alllimitbay}.

\begin{table*}
\caption{\label{tbl:hww-alllimitbay} Expected and observed upper limits
at 95\% CL for Higgs boson production cross section expressed as a
ratio to the cross section predicted by the SM for a range of test
Higgs boson masses.}
\begin{ruledtabular}
\begin{tabular}{rcccccccccccccc}
& $m_H$ (GeV)  & 120     & 130    & 140    & 145    & 150    & 155    & 160    & 165    & 170    & 175    & 180    & 190    & 200 \\
\hline
D0~\cite{PhysRevLett.104.061804} & Limit (exp.) & 9.74    & 5.40   & 3.48   & 3.07   & 2.58   & 2.02   & 1.43   & 1.36   & 1.65   & 2.06   & 2.59   & 4.20   & 6.23  \\
   & Limit (obs.) & 13.6    & 6.63   & 5.21   & 3.94   & 3.29   & 3.25   & 1.82   & 1.55   & 1.96   & 1.89   & 2.11   & 3.27   & 5.53  \\
\hline
CDF \cite{PhysRevLett.104.061803} & Limit (exp.) & 8.85    & 4.41   & 2.85   & 2.43   & 2.05   & 1.67   & 1.26   & 1.20   & 1.44   & 1.72   & 2.09   & 3.24   & 4.53  \\
  & Limit (obs.) & 12.04   & 6.38   & 4.21   & 3.23   & 2.62   & 2.04   & 1.34   & 1.29   & 1.69   & 1.94   & 2.24   & 4.06   & 6.74  \\
\end{tabular}
\end{ruledtabular}
\end{table*}

\subsection{Combined Limits}
\label{sec:higgs_direct_comb}
Results for individual search final states are combined by the
Tevatron New--Phenomena and Higgs Working Group\cite{b:tevNPHWG}.  The
combinations are updated roughly every six months to one year using
the most up to date preliminary and published results. The combination
is performed using a joint probability density incorporating each
channel as an individual result with systematic uncertainties and
their correlations included.  Two algorithms are used for the
combination.  One is a Bayesian approach\cite{PhysRevLett.104.061803},
and the other is a modified frequentist
approach\cite{Junk:1999kv,Read2002,Fisher:2006zz}.  In both
algorithms, the impact of systematic uncertainties is reduced by
treating these as nuisance parameters. Results for the 95\% CL upper 
bound on the Higgs production cross section from the two algorithms
agree to within 10\%.  

\begin{figure}
  \includegraphics[width=0.52\textwidth]{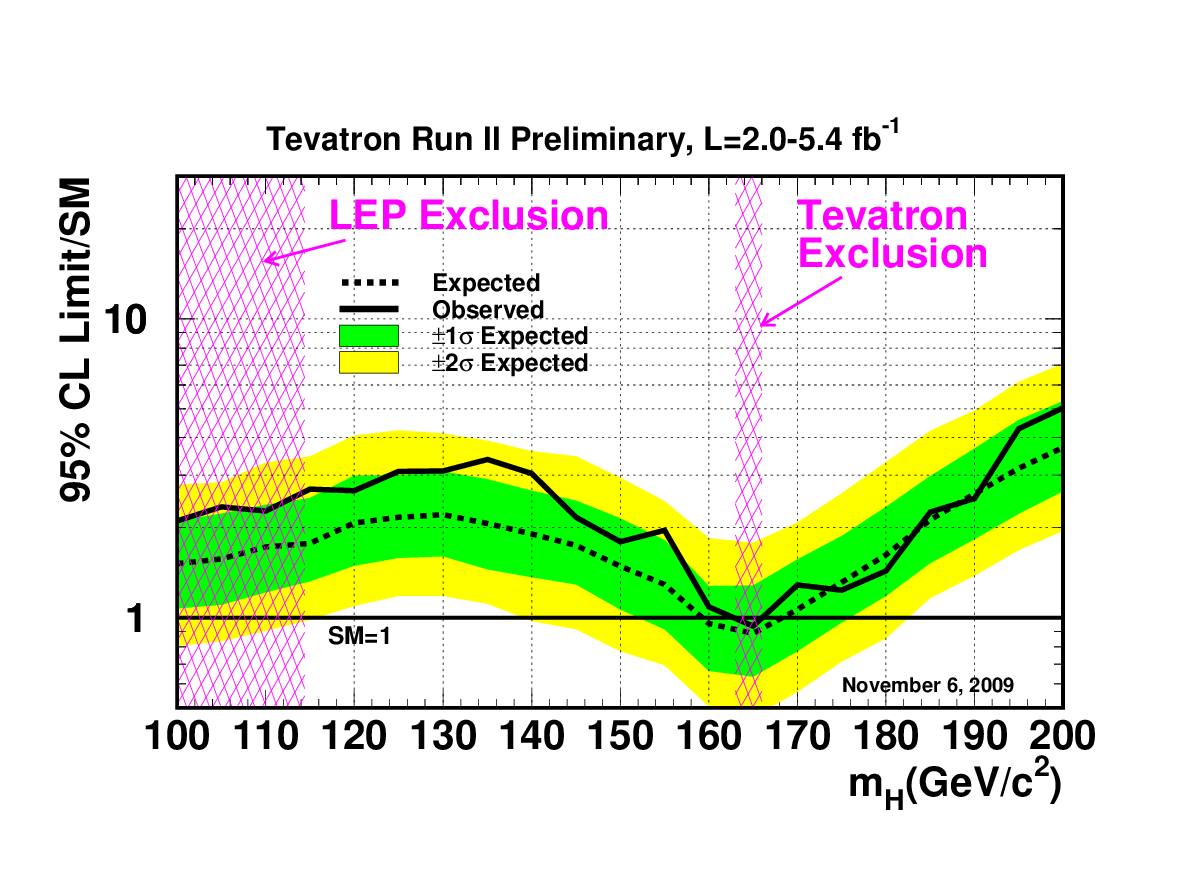}
  \caption{The 95\% CL upper bounds on the Higgs production cross section 
    divided by the predicted cross section.  This result
    includes both published and preliminary results from CDF and D0.
    \label{f:tevhiggs}}
\end{figure}

The most recent combination\cite{b-higgs-combo} incorporates published
and preliminary results based on samples ranging from $\intL =
2.1$~fb$^{-1}$ to $\intL = 5.4$~fb$^{-1}$, and combined $H\rightarrow
WW$ limits are at  \cite{PhysRevLett.104.061802}. The cross section
upper bounds are a factor of 2.7 higher than the SM prediction for
$m_H = 115$~$\GeV$ and production of a SM Higgs is excluded at the
95\% CL in the region $163 < m_H < 166\ \GeV$.  The results are shown
in Fig.~\ref{f:tevhiggs} expressed as a ratio of the 95\% CL upper
bound cross section divided by the SM prediction.

%%%%%%%%%%%%%%%%%%%%%%%%%%%%%%%%%%%%%%%%%%%%%%%%%%%%%%%%%%%%%%%%%%%%%%%%%%%%
%
% Summary
%
%%%%%%%%%%%%%%%%%%%%%%%%%%%%%%%%%%%%%%%%%%%%%%%%%%%%%%%%%%%%%%%%%%%%%%%%%%%%

\section{SUMMARY AND OUTLOOK}
\label{sec:summary}

The SM has been tested extensively over the last forty years. The only
significant deviation from its predictions is the existence of
neutrino mass. Ongoing studies of the SM are being carried out by the
Tevatron experiments, CDF and D0.  This paper presents results of
these tests and discusses prospects for the remainder of the Tevatron
running and the transition from the Tevatron to the LHC.

CDF and D0 have presented a wide variety of results relating to the EW
sector. Among the results are observation of all SM diboson processes
involving $W$, $Z$ and $\gamma$, including for the first time $WZ$
production, observation of EW production of single top quarks and the
corresponding measurement of $|V_{tb}|$, world's best measurements of
the top quark and $W$ boson masses allowing improved indirect
constraints on the Higgs boson mass, and ongoing searches for the
Higgs boson which have begun to exclude mass values outside of the
existing limits.  In addition to testing the SM, some of these are
benchmark measurements eventually to be compared with corresponding
results from the LHC.  Others, like the top quark and $W$ boson
masses, are likely to be legacy measurements from the Tevatron.

With additional luminosity still to come for CDF and D0, many of these
measurements will significantly improve before the end of Run II at
the Tevatron. This is an exciting time in particle physics as the field
transitions from the era of the Tevatron to an era of the LHC as the
machine to probe physics at the energy frontier.

%%%%%%%%%%%%%%%%%%%%%%%%%%%%%%%%%%%%%%%%%%%%%%%%%%%%%%%%%%%%%%%%%%%%%%%%%%%%
%
% Acknowledgments
%
%%%%%%%%%%%%%%%%%%%%%%%%%%%%%%%%%%%%%%%%%%%%%%%%%%%%%%%%%%%%%%%%%%%%%%%%%%%%

\section{ACKNOWLEDGMENTS}
\label{sec:acknowledgements}

The many beautiful experimental measurements summarized in this paper are the
result of the work of thousands of people from the Fermilab accelerator
division, CDF and D0.  The authors wish to thank those involved in making these
possible.  In addition, the authors thank Andrew Askew, Tom Diehl, Reinhard
Schwienhorst and Junjie Zhu for useful discussion and clarification.

%%%%%%%%%%%%%%%%%%%%%%%%%%%%%%%%%%%%%%%%%%%%%%%%%%%%%%%%%%%%%%%%%%%%%%%%%%%%
%
% Bibliography
%
%%%%%%%%%%%%%%%%%%%%%%%%%%%%%%%%%%%%%%%%%%%%%%%%%%%%%%%%%%%%%%%%%%%%%%%%%%%%

%\cleardoublepage
%\bibliographystyle{apsrmp}
\bibliographystyle{unsrt}
\bibliography{rmp}

\begin{thebibliography}{100}

\bibitem{Aaltonen:2007ps}
T.~Aaltonen et~al.
\newblock {\em Phys. Rev. D}, 99:151801, 2007.

\bibitem{Dzero}
V.~M. Abazov et~al.
\newblock {\em Nucl. Instrum. Methods Phys. Res., Sect. A}, 565:463, 2006.

\bibitem{Weinberg:1979sa}
S.~Weinberg.
\newblock {\em Phys. Rev. Lett.}, 43:1566, 1979.

\bibitem{Buchmuller:1985jz}
W.~Buchmuller and D.~Wyler.
\newblock {\em Nucl. Phys.}, B268:621, 1986.

\bibitem{DeRujula:1991se}
A.~De~Rujula, M.~B. Gavela, P.~Hernandez, and E.~Masso.
\newblock {\em Nucl. Phys.}, B384:3, 1992.

\bibitem{Hagiwara:1993ck}
K.~Hagiwara, S.~Ishihara, R.~Szalapski, and D.~Zeppenfeld.
\newblock {\em Phys. Rev.}, D48:2182, 1993.

\bibitem{Arnison:1983rp}
G.~Arnison et~al.
\newblock {\em Phys. Lett.}, B122:103, 1983.

\bibitem{Banner:1983jy}
M.~Banner et~al.
\newblock {\em Phys. Lett.}, B122:476, 1983.

\bibitem{z-disco-ua1}
G.~Arnison et~al.
\newblock {\em Phys. Lett.}, B126:398, 1983.

\bibitem{z-disco-ua2}
P.~Bagnaia et~al.
\newblock {\em Phys. Lett.}, B129:130, 1983.

\bibitem{Z-Pole}
G.~P. Abbiendi et~al.
\newblock {\em Phys. Rept.}, 427:257, 2006.

\bibitem{Kober:2007bc}
M.~Kober, B.~Koch, and M.~Bleicher.
\newblock {\em Phys. Rev.}, D76:125001, 2007.

\bibitem{b-Z-xsec-theo}
R.~Hamberg, W.~L. van Neerven, and T.~Matsuura.
\newblock {\em Nucl. Phys.}, B359:343, 1991.

\bibitem{b-Z-mrst}
A.~D. Martin, R.~G. Roberts, W.~J. Sterling, and R.~S. Thorne.
\newblock {\em Phys. Lett.}, B604:61, 2004.

\bibitem{b-Z-cteq}
J.~Pumplin et~al.
\newblock {\em J. High Energy Phys.}, 0207:12, 2002.

\bibitem{Abulencia:2005ix}
A.~Abulencia et~al.
\newblock {\em J. Phys.}, G34:2457, 2007.

\bibitem{Sirlin:1980nh}
A.~Sirlin.
\newblock {\em Phys. Rev.}, D22:971, 1980.

\bibitem{Schael:2006mz}
S.~Schael et~al.
\newblock {\em Eur. Phys. J.}, C47:309, 2006.

\bibitem{:2008xh}
J.~Abdallah et~al.
\newblock {\em Eur. Phys. J.}, C55:1, 2008.

\bibitem{Achard:2005qy}
P.~Achard et~al.
\newblock {\em Eur. Phys. J.}, C45:569, 2006.

\bibitem{Abbiendi:2005eq}
G.~Abbiendi et~al.
\newblock {\em Eur. Phys. J.}, C45:307, 2006.

\bibitem{b:CDF-mWI}
A.~Affolder et~al.
\newblock {\em Phys. Rev.}, D64:052001, 2001.

\bibitem{b:D0-mWI}
V.~M. Abazov et~al.
\newblock {\em Phys. Rev.}, D66:012001, 2002.

\bibitem{Alcaraz:2009jr}
J.~Alcaraz et~al.
\newblock {\em arXiv:0911.2604 [hep-ex]}, 2009.

\bibitem{b-BLUE}
L.~Lyons, D.~Gibaut, and P.~Clifford.
\newblock {\em Nucl. Instrum. Meth.}, A270:110, 1988.

\bibitem{b-BLUE2}
A.~Valassi.
\newblock {\em Nucl. Instrum. Meth.}, A500:391, 2003.

\bibitem{Abazov:2009cp}
V.~M. Abazov et~al.
\newblock {\em Phys. Rev. Lett.}, 103:141801, 2009.

\bibitem{:2009nu}
Tevatron Electroweak~Working Group.
\newblock {\em arXiv:0908.1374 [hep-ex]}, 2009.

\bibitem{Rosner:1993rj}
J.~Rosner, M.. Worah, and T.~Takeuchi.
\newblock {\em Phys. Rev.}, D49:1363, 1994.

\bibitem{Affolder:2000mt}
T.~Affolder et~al.
\newblock {\em Phys. Rev. Lett.}, 85:3347, 2000.

\bibitem{Abazov:2002xj}
V.~M. Abazov et~al.
\newblock {\em Phys. Rev.}, D66:032008, 2002.

\bibitem{Aaltonen:2007mg}
T.~Aaltonen et~al.
\newblock {\em Phys. Rev. Lett.}, 100:071801, 2008.

\bibitem{Collaboration:2009vsa}
V.~M. Abazov et~al.
\newblock {\em Phys. Rev. Lett.}, 103:231802, 2009.

\bibitem{b:d0HadNIM}
V.~M. Abazov et~al.
\newblock {\em Nucl. Instrum. Meth.}, A609:250, 2009.

\bibitem{b-zasymm-cdf}
D.~Acosta et~al.
\newblock {\em Phys. Rev. D.}, 971:052002, 2005.

\bibitem{b-zasymm-d0}
V.~M. Abazov et~al.
\newblock {\em Phys. Rev. Lett.}, 101:191801, 2008.

\bibitem{b-pdg08}
C.~Amsler et~al.
\newblock {\em Phys. Lett. B}, 667:1, 2008.

\bibitem{b-nutev}
G.~P. Zeller et~al.
\newblock {\em Phys. Rev. Lett.}, 88:091802, 2002.

\bibitem{Hagiwara:1986vm}
K~Hagiwara, R.~D. Peccei, D.~Zeppenfeld, and K.~Hikasa.
\newblock {\em Nucl. Phys.}, B282:253, 1987.

\bibitem{PhysRevD.41.2113}
K.~Hagiwara, J.~Woodside, and D.~Zeppenfeld.
\newblock {\em Phys. Rev. D}, 41(7):2113, 1990.

\bibitem{PhysRevD.37.1775}
D.~Zeppenfeld and S.~Willenbrock.
\newblock {\em Phys. Rev. D}, 37(7):1775, 1988.

\bibitem{Baur1988383}
U.~Baur and D.~Zeppenfeld.
\newblock {\em Physics Letters B}, 201(3):383, 1988.

\bibitem{Baur:1992cd}
U.~Baur and E.~L. Berger.
\newblock {\em Phys. Rev.}, D47:4889, 1993.

\bibitem{Baur:2000ae}
U.~Baur and D.~L. Rainwater.
\newblock {\em Phys.Rev.}, D62:113011, 2000.

\bibitem{QuiggReview}
C.~Quigg.
\newblock {\em Annual Review of Nuclear and Particle Science}, 59:505, 2009.

\bibitem{Acosta:2004it}
D.~Acosta et~al.
\newblock {\em Phys. Rev. Lett.}, 94:041803, 2005.

\bibitem{Abazov:2005ni}
V.~M. Abazov et~al.
\newblock {\em Phys. Rev.}, D71:091108, 2005.

\bibitem{Abazov:2008vja}
V.~M. Abazov et~al.
\newblock {\em Phys. Rev. Lett.}, 100:241805, 2008.

\bibitem{Baur:1989gk}
U.~Baur and E.~L. Berger.
\newblock {\em Phys. Rev.}, D41:1476, 1990.

\bibitem{Baur:1993ir}
U.~Baur, T.~Han, and J.~Ohnemus.
\newblock {\em Phys. Rev.}, D48:5140, 1993.

\bibitem{PhysRevD.20.1164}
R.~W. Brown, D.~Sahdev, and K.~O. Mikaelian.
\newblock {\em Phys. Rev. D}, 20(5):1164, 1979.

\bibitem{PhysRevLett.43.746}
K.~O. Mikaelian, M.~A. Samuel, and D.~Sahdev.
\newblock {\em Phys. Rev. Lett.}, 43(11):746, 1979.

\bibitem{PhysRevD.23.2682}
C.~J. Goebel, F.~Halzen, and J.~P. Leveille.
\newblock {\em Phys. Rev. D}, 23(11):2682, 1981.

\bibitem{PhysRevLett.49.966}
S.~Brodsky and R.~Brown.
\newblock {\em Phys. Rev. Lett.}, 49(14):966, 1982.

\bibitem{Baur:1994sa}
U.~Baur, S.~Errede, and G.~L. Landsberg.
\newblock {\em Phys.Rev.}, D50:1917, 1994.

\bibitem{Abazov:2007wy}
V.~M. Abazov et~al.
\newblock {\em Phys. Lett.}, B653:378, 2007.

\bibitem{PhysRevD.57.2823}
U.~Baur, T.~Han, and J.~Ohnemus.
\newblock {\em Phys. Rev. D}, 57(5):2823, 1998.

\bibitem{Abazov:2009cj}
V.~M. Abazov et~al.
\newblock {\em Phys. Rev. Lett.}, 102:201802, 2009.

\bibitem{PhysRevLett.78.4536}
F.~Abe et~al.
\newblock {\em Phys. Rev. Lett.}, 78(24):4536, 1997.

\bibitem{PhysRevLett.94.151801}
V.~M. Abazov et~al.
\newblock {\em Phys. Rev. Lett.}, 94(15):151801, 2005.

\bibitem{PhysRevLett.94.211801}
D.~Acosta et~al.
\newblock {\em Phys. Rev. Lett.}, 94(21):211801, 2005.

\bibitem{Aaltonen:2009us}
T.~Aaltonen et~al.
\newblock {\em Phys. Rev. Lett.}, 104(20):201801, 2010.

\bibitem{Abazov:2009ys}
V.~M. Abazov et~al.
\newblock {\em Phys. Rev. Lett.}, 103:191801, 2009.

\bibitem{Campbell:1999ah}
J.~M. Campbell and R.~K. Ellis.
\newblock {\em Phys. Rev. D}, 60(11):113006, 1999.

\bibitem{Abulencia:2007tu}
A.~Abulencia et~al.
\newblock {\em Phys. Rev. Lett.}, 98:161801, 2007.

\bibitem{Abazov:2005ys}
V.~M. Abazov et~al.
\newblock {\em Phys. Rev. Lett.}, 95:141802, 2005.

\bibitem{Abazov:2007rab}
V.~M. Abazov et~al.
\newblock {\em Phys. Rev.}, D76:111104, 2007.

\bibitem{Alcaraz:2006mx}
J.~Alcaraz et~al.
\newblock {\em hep-ex/0612034}, 2006.

\bibitem{Abazov:2007hm}
V.~M. Abazov et~al.
\newblock {\em Phys. Rev. Lett.}, 100:131801, 2008.

\bibitem{Aaltonen:2008mv}
T.~Aaltonen et~al.
\newblock {\em Phys. Rev. Lett.}, 100:201801, 2008.

\bibitem{Abazov:2008gya}
V.~M. Abazov et~al.
\newblock {\em Phys. Rev. Lett.}, 101:171803, 2008.

\bibitem{PhysRevD.78.072002}
V.~M. Abazov et~al.
\newblock {\em Phys. Rev. D}, 78(7):072002, 2008.

\bibitem{Abazov:2008yg}
V.~M. Abazov et~al.
\newblock {\em Phys. Rev. Lett.}, 102:161801, 2009.

\bibitem{Aaltonen:2009vh}
T.~Aaltonen et~al.
\newblock {\em Phys. Rev. Lett.}, 104(10):101801, 2010.

\bibitem{Aaltonen:2007sd}
T.~Aaltonen et~al.
\newblock {\em Phys. Rev.}, D76:111103, 2007.

\bibitem{Abazov:2009tr}
V.~M. Abazov et~al.
\newblock {\em Phys. Rev.}, D80:053012, 2009.

\bibitem{RandomForests}
L.~Breiman.
\newblock {\em Mach. Learn.}, 45:5, 2001.

\bibitem{z-xsec-cdf-short}
D.~Acosta et~al.
\newblock {\em Phys. Rev. Lett.}, 94:091803, 2005.

\bibitem{topxsec}
M.~Cacciari et~al.
\newblock {\em J. High Energy Phys.}, 09:127, 2008.

\bibitem{Harris:2002md}
B.~Harris, E.~Laenen, L.~Phaf, Z.~Sullivan, and S.~Weinzierl.
\newblock {\em Phys. Rev.}, D66:054024, 2002.

\bibitem{Aaltonen:2009fd}
T.~Aaltonen et~al.
\newblock {\em Phys. Rev. Lett.}, 103:091803, 2009.

\bibitem{Abazov:2009hk}
V.~M. Abazov et~al.
\newblock {\em arXiv:0907.4952 [hep-ex]}, 2009.

\bibitem{ElKhadra:2002wp}
A.~X. El-Khadra and M.~Luke.
\newblock {\em Ann. Rev. Nucl. Part. Sci.}, 52:201, 2002.

\bibitem{Smith:1996xz}
M.~Smith and S.~Willenbrock.
\newblock {\em Phys. Rev. Lett.}, 79:3825, 1997.

\bibitem{b-mt-cdf-neuro}
T.~Aaltonen et~al.
\newblock {\em Phys. Rev. Lett.}, 102(15):152001, 2009.

\bibitem{b-mt-d0-llcomb}
V.~M. Abazov et~al.
\newblock {\em Phys. Rev.}, D80:092006, 2009.

\bibitem{b-mt-cdf-ljmat}
T.~Aaltonen et~al.
\newblock {\em Phys. Rev.}, D79:072001, 2009.

\bibitem{b-mt-d0-ljmat}
V.~M. Abazov et~al.
\newblock {\em Phys. Rev. Lett.}, 101:182001, 2008.

\bibitem{b-dalitz}
R.~Dalitz and G.~Goldstein.
\newblock {\em Phys. Rev.}, D45:1531, 1992.

\bibitem{b-kondo1}
K.~Kondo.
\newblock {\em J. Phys. Soc. Jpn.}, 57:4126, 1988.

\bibitem{b-kondo2}
K.~Kondo.
\newblock {\em J. Phys. Soc. Jpn.}, 60:836, 1991.

\bibitem{b-d0mwt1}
B.~Abbott et~al.
\newblock {\em Phys. Rev. Lett.}, 80:2063, 1998.

\bibitem{b-d0mwt2}
B.~Abbott et~al.
\newblock {\em Phys. Rev.}, D60:052001, 1999.

\bibitem{b-mt-cdf-simult}
T.~Aaltonen et~al.
\newblock {\em Phys. Rev. D}, 79(9):092005, 2009.

\bibitem{b:tevMtComb}
Tevatron Electroweak~Working Group.
\newblock {\em arXiv:0903.2503 [hep-ex]}, 2009.

\bibitem{PhysRevD.80.054009}
U.~Langenfeld, S.~Moch, and P.~Uwer.
\newblock {\em Phys. Rev. D}, 80(5):054009, 2009.

\bibitem{b-mt-cdf-phiwt}
T.~Aaltonen et~al.
\newblock {\em Phys. Rev.}, D79:072005, 2009.

\bibitem{b-mt-cdf-llcomb}
T.~Aaltonen et~al.
\newblock {\em Phys. Rev. Lett.}, 100:062005, 2008.

\bibitem{b-mt-cdf-llmat}
A.~Abulencia et~al.
\newblock {\em Phys. Rev. D}, 75:031105, 2007.

\bibitem{b-mt-cdf-ljdl}
A.~Abulencia et~al.
\newblock {\em Phys. Rev. D}, 75:071102, 2007.

\bibitem{b-mt-d0-ideo}
V.~M. Abazov et~al.
\newblock {\em Phys. Rev. D}, 75:092001, 2007.

\bibitem{b-mt-cdf-300prl}
A.~Abulencia et~al.
\newblock {\em Phys. Rev. Lett.}, 96:022004, 2006.

\bibitem{b-mt-cdf-jj}
T.~Aaltonen et~al.
\newblock {\em Phys. Rev. D}, 79(7):072010, 2009.

\bibitem{b-mt-cdf-jj2}
T.~Aaltonen et~al.
\newblock {\em Phys. Rev.}, D79:072010, 2009.

\bibitem{b-mt-cdf-mj}
T.~Aaltonen et~al.
\newblock {\em Phys. Rev. D}, 75:111103, 2007.

\bibitem{b-bsg1}
K.~Fujikawa and A.~Yamada.
\newblock {\em Phys. Rev. D}, 49:5890, 1994.

\bibitem{b-bsg2}
P~Cho and M.~Misiak.
\newblock {\em Phys. Rev. D}, 49:5894, 1994.

\bibitem{Aaltonen:2008ei}
T.~Aaltonen et~al.
\newblock {\em Phys. Lett.}, B674:160, 2009.

\bibitem{b-hel-d0}
V.~M. Abazov et~al.
\newblock {\em Phys. Rev. Lett.}, 100:062004, 2008.

\bibitem{b-pythia}
T~Sjostrand et~al.
\newblock {\em Comput. Phys. Commun.}, 135:238, 2001.

\bibitem{b-alpgen}
M.~L. Mangano, M.~Moretti, F.~Piccinini, R.~Pittau, and P.~Polosa.
\newblock {\em J. High Energy Phys.}, 07:1, 2003.

\bibitem{b-top-d0}
S.~Abachi et~al.
\newblock {\em Phys. Rev. Lett.}, 74:2632, 1995.

\bibitem{b-top-cdf}
F.~Abe et~al.
\newblock {\em Phys. Rev. Lett.}, 74:2626, 1995.

\bibitem{b-st-sxsec1}
S.~Cortese and R~Petronzio.
\newblock {\em Phys. Lett. B}, 253:494, 1991.

\bibitem{b-st-sxsec2}
T.~Stelzer and S.~Willenbrock.
\newblock {\em Phys. Lett. B}, 357:125, 1995.

\bibitem{b-st-xsec}
{A.P.~Heinson {\it et al.}}
\newblock {\em Phys. Rev. D}, 56:3114, 1997.

\bibitem{b-st-txsec1}
S.~Willenbrock and D.~Dicus.
\newblock {\em Phys. Rev. D}, 34:155, 1986.

\bibitem{b-st-txsec2}
C.~P. Yuan.
\newblock {\em Phys. Rev. D}, 41:42, 1990.

\bibitem{b-st-txsec3}
K.~Ellis and S.~Parke.
\newblock {\em Phys. Rev. D}, 46:3785, 1992.

\bibitem{Stelzer:1997ns}
T.~Stelzer, Z.~Sullivan, and S.~Willenbrock.
\newblock {\em Phys. Rev.}, D56:5919, 1997.

\bibitem{Smith:1996ij}
M.~Smith and S.~Willenbrock.
\newblock {\em Phys. Rev.}, D54:6696, 1996.

\bibitem{b-sullivan}
Z.~Sullivan.
\newblock {\em Phys. Rev. D}, 70:114012, 2004.

\bibitem{Campbell:2004ch}
J.~Campbell, R.~K. Ellis, and F.~Tramontano.
\newblock {\em Phys. Rev.}, D70:094012, 2004.

\bibitem{Cao:2004ap}
Q.-H. Cao, R.~Schwienhorst, and C.-P. Yuan.
\newblock {\em Phys. Rev.}, D71:054023, 2005.

\bibitem{Cao:2005pq}
Q.-H. Cao, R.~Schwienhorst, J.~A. Benitez, R.~Brock, and C.-P. Yuan.
\newblock {\em Phys. Rev.}, D72:094027, 2005.

\bibitem{Frixione:2005vw}
S.~Frixione, E.~Laenen, P.~Motylinski, and B.~R. Webber.
\newblock {\em JHEP}, 03:092, 2006.

\bibitem{b-st-xsecKid}
N.~Kidonakis.
\newblock {\em Phys. Rev. D}, 74:114012, 2006.

\bibitem{Campbell:2009ss}
J.. Campbell, R.~Frederix, F.~Maltoni, and F.~Tramontano.
\newblock {\em Phys. Rev. Lett.}, 102:182003, 2009.

\bibitem{b-st-d0-prl}
V.~M. Abazov et~al.
\newblock {\em Phys. Rev. Lett.}, 98:181802, 2007.

\bibitem{b-st-d0-prd}
V.~M. Abazov et~al.
\newblock {\em Phys. Rev. D}, 78:012005, 2008.

\bibitem{b-st-cdf}
T.~Aaltonen et~al.
\newblock {\em Phys. Rev. Lett.}, 101:252001, 2008.

\bibitem{Group:2009qk}
Tevatron Electroweak~Working Group.
\newblock {\em arXiv:0908.2171 [hep-ex]}, 2009.

\bibitem{Luscher:1988gc}
M.~Luscher and P.~Weisz.
\newblock {\em Phys. Lett.}, B212:472, 1988.

\bibitem{Djouadi:2005gi}
A.~Djouadi.
\newblock {\em Phys. Rept.}, 457:1, 2008.

\bibitem{Stange:1993ya}
A.~Stange, W.~J. Marciano, and S.~Willenbrock.
\newblock {\em Phys. Rev.}, D49:1354, 1994.

\bibitem{Stange:1994bb}
A.~Stange, W.~J. Marciano, and S.~Willenbrock.
\newblock {\em Phys. Rev.}, D50:4491, 1994.

\bibitem{b-higgs-combo}
{The Tevatron New--Phenomena and Higgs Working Group}.
\newblock {\em arXiv:0911.3930 [hep-ex]}, 2009.

\bibitem{PhysRevLett.82.25}
T.~Han and R.-J. Zhang.
\newblock {\em Phys. Rev. Lett.}, 82(1):25, 1999.

\bibitem{PhysRevD.59.093001}
T.~Han, A.~S. Turcot, and R.-J. Zhang.
\newblock {\em Phys. Rev. D}, 59(9):093001, 1999.

\bibitem{b-cdf-wh}
T.~Aaltonen et~al.
\newblock {\em Phys. Rev.}, D80:012002, 2009.

\bibitem{b-d0-wh}
V.~M. Abazov et~al.
\newblock {\em Phys. Rev. Lett.}, 102:051803, 2009.

\bibitem{b-D0-btag}
T.~Scanlon.
\newblock {\em FERMILAB-THESIS-2006-43}, 2006.

\bibitem{b:cdf-secvtx}
D.~Acosta et~al.
\newblock {\em Phys. Rev.}, D71:052003, 2005.

\bibitem{b:cdf-nnttag}
T.~Aaltonen et~al.
\newblock {\em Phys. Rev.}, D78:032008, 2008.

\bibitem{b-cdf-jetprob}
A.~Abulencia et~al.
\newblock {\em Phys. Rev.}, D74:072006, 2006.

\bibitem{b-zh-llbb-d0}
V.~M. Abazov et~al.
\newblock {\em Phys. Lett.}, B655:209, 2007.

\bibitem{b-zh-llbb-cdf}
T.~Aaltonen et~al.
\newblock {\em Phys. Rev. Lett.}, 101:251803, 2008.

\bibitem{b-zh-vvbb-cdf}
T.~Aaltonen et~al.
\newblock {\em Phys. Rev. Lett.}, 100:211801, 2008.

\bibitem{b-zh-vvbb-d0}
V.~M. Abazov et~al.
\newblock {\em Phys. Rev. Lett.}, 104(7):071801, 2010.

\bibitem{b:d0-taus}
V.~M. Abazov et~al.
\newblock {\em Phys. Rev. Lett.}, 102:251801, 2009.

\bibitem{Abulencia:2006aj}
A.~Abulencia et~al.
\newblock {\em Phys. Rev. Lett.}, 97:081802, 2006.

\bibitem{Abazov:2005un}
V.~M. Abazov et~al.
\newblock {\em Phys. Rev. Lett.}, 96:011801, 2006.

\bibitem{Aaltonen:2008ec}
T.~Aaltonen et~al.
\newblock {\em Phys. Rev. Lett.}, 102:021802, 2009.

\bibitem{PhysRevLett.104.061803}
T.~Aaltonen et~al.
\newblock {\em Phys. Rev. Lett.}, 104(6):061803, 2010.

\bibitem{PhysRevLett.104.061804}
V.~M. Abazov et~al.
\newblock {\em Phys. Rev. Lett.}, 104(6):061804, 2010.

\bibitem{b:tevNPHWG}
Tevatron New-Phenomena and Higgs~Working Group.
\newblock {\em {{\tt http://tevnphwg.fnal.gov}}}.

\bibitem{Junk:1999kv}
T.~Junk.
\newblock {\em Nucl. Instrum. Meth.}, A434:435, 1999.

\bibitem{Read2002}
A.~L. Read.
\newblock {\em J. Phys. G: Nucl. Part. Phys.}, 28:2693, 2002.

\bibitem{Fisher:2006zz}
W.~Fisher.
\newblock 2006.
\newblock FERMILAB-TM-2386-E.

\bibitem{PhysRevLett.104.061802}
T.~Aaltonen et~al.
\newblock {\em Phys. Rev. Lett.}, 104(6):061802, 2010.

\end{thebibliography}

%%%%%%%%%%%%%%%%%%%%%%%%%%%%%%%%%%%%%%%%%%%%%%%%%%%%%%%%%%%%%%%%%%%%%%%%%%%%
%
% End document
%
%%%%%%%%%%%%%%%%%%%%%%%%%%%%%%%%%%%%%%%%%%%%%%%%%%%%%%%%%%%%%%%%%%%%%%%%%%%%

\end{document}